\documentclass{aastex61}

\newcommand\aastex{AAS\TeX}

\received{\today}
\revised{}
\accepted{}
\submitjournal{ApJ}

\shorttitle{\aastex\ Stellar Yields of Rotating First Stars. II.}
\shortauthors{Takahashi et al.}

\begin{document}

\title{Stellar Yields of Rotating First Stars. II. Pair Instability Supernovae and Comparison with Observations}

\correspondingauthor{Koh Takahashi}
\email{ktakahashi@astro.uni-bonn.de}

\author{Koh Takahashi}
\affil{Argelander-Institut f\"{u}r Astronomie, Universit\"{a}t Bonn,
D-53121 Bonn, Germany}
\author{Takashi Yoshida}
\affil{Department of Astronomy, Graduate School of Science, University of Tokyo,
113-0033 Tokyo, Japan}
\author{Hideyuki Umeda}
\affil{Department of Astronomy, Graduate School of Science, University of Tokyo,
113-0033 Tokyo, Japan}

\begin{abstract}
Recent theory predicts that a first star is born with a massive initial mass of $\gtrsim$ 100 $M_\odot$.
Pair instability supernova (PISN) is a common fate for such a massive star.
Our final goal is to prove the existence of PISN
and thus the high mass nature of the initial mass function
in the early universe by conducting {\it abundance profiling},
in which properties of a hypothetical first star is constrained by
metal-poor star abundances.
In order to determine reliable and useful abundances,
we investigate the PISN nucleosynthesis
taking both rotating and non-rotating progenitors for the first time.
We show that the initial and CO core mass ranges for PISNe depend on the envelope structures:
non-magnetic rotating models developing inflated envelopes
have a lower-shifted CO mass range of $\sim$ 70--125 $M_\odot$,
while non-rotating and magnetic rotating models with deflated envelopes
have a range of $\sim$ 80--135 $M_\odot$.
However, we find no significant difference in explosive yields from rotating and non-rotating progenitors,
except for large nitrogen production in non-magnetic rotating models.
Furthermore, we conduct the first systematic comparison between theoretical yields
and a large sample of metal-poor star abundances.
We find that the predicted low [Na/Mg] $\sim$ $-1.5$ and high [Ca/Mg] $\sim$ $0.5$--$1.3$ abundance ratios
are the most important to discriminate PISN signatures from normal metal-poor star abundances,
and confirm that no currently observed metal-poor star matches with the PISN abundance.
Extensive discussion on the non-detection is finally made.
\end{abstract}
\keywords{nuclear reactions, nucleosynthesis, abundances
--- stars: abundances
--- stars: Population III
--- stars: rotation
--- supernovae: general}

\section{Introduction}

In the early universe, matters are only composed of light elements
that are synthesized in the Big Bang nucleosynthesis.
First stars, also known as Population III (Pop. III) stars, are born from the primordial gas clouds.
They synthesize heavier elements than $^7$Li for the first time,
introducing chemical diversity to the universe.
Not only the first nucleosynthesis, but also their photon emission during the lifetime
and energy and matter ejection at the final explosion affect ambient environments.
As a consequence, formation and evolution of first stars dramatically change the phase of matter evolution,
which finally leads to the formation of a complex structure observed in the present universe.
Hence, to understand the nature of first stars is one of the most important mission
for the modern astronomy and astrophysics.

As they exist in the most far-away universe,
it is extremely challenging to directly observe first stars by present-day telescopes. 
Then in order to get some information about the nature,
indirect but reasonable and powerful method called {\it abundance profiling}
has been done \citep[see][and references therein]{nomoto+13, tominaga+14}.
The key idea is that metal-poor (MP) stars detected in the nearby universe
would be born from chemically primitive, but already slightly metal-polluted,
gas clouds in the early universe.
If the amount of metals in a MP star is small enough, one can assume that
the metal pollution is a consequence of supernova explosion of a massive first star.
On this line of reasoning, nucleosynthesis outcome of first supernovae can be restored
by observing surface abundances of MP stars.
This enables to conduct an observational test 
for a theoretical prediction of first star nucleosynthesis.

This is a second paper of a series, in which
we aim to find nucleosynthetic signatures
that are useful to constrain properties of massive first stars.
In our previous paper, evolution of massive first stars of 12--140 M$_\odot$,
which are considered to lead to iron core collapse at the end, are investigated \citep{takahashi+14}.
By focusing the peculiar nucleosynthesis taking place in a helium layer,
the possibility to derive masses and rotational properties of massive first stars
by observing the abundances of intermediate-mass elements
from a surface of carbon-enhanced MP (CEMP) stars has been shown.
As a demonstration, existence of a massive first star of $\sim$ 50--80 $M_\odot$ has been indicated
from the characteristic abundance pattern of the most iron-deficient star known so far,
SMSS 0313-6708 \citep{keller+14}.
As a next step, we move our scope to the heavier side in the initial mass in this paper.

\citet{barkat+67} and \citet{rakavy+67} firstly pointed out that
a star having a massive enough CO core of $>$ 65 M$_\odot$
finally explodes after the hydrodynamical collapse induced by electron-positron pair production.
The pair instability supernova (PISN, sometimes also referred to as pair creation supernova)
becomes a highly energetic thermonuclear explosion, which injects
at least ten times more explosion energy than the canonical core collapse supernovae
to surroundings \citep{heger&woosley02, umeda&nomoto02, takahashi+16}.
The explosion has also been confirmed by multidimensional simulations \citep{chatzopoulos+13b, chen+14b}.
Under current theoretical understanding, a star having a massive CO core of
$\sim$ 65--130 M$_\odot$ inevitably explodes as a PISN,
although an observational confirmation of the existence has not been done
except for the only candidate of the luminous and energetic PISN explosion at low-redshift of $z$ = 0.1279
(SN 2007bi, \citealt{gal-yam+09}).
As well as the high explosion energy, the large $^{56}$Ni yield makes the explosion luminous.
Therefore PISN is one of the most promising candidates as an observable object at high redshift
\citep{scannapieco+05, kasen+11, kozyreva+14a, kozyreva+14b, chatzopoulos+15, whalen+13, smidt+15}.

Formation of a massive star that can form the required high mass CO core is
considered to be difficult for metal-rich environments due to the efficient wind mass loss
(\citealt{Yoshida&Umeda11, Yusof+13, Yoshida+14}, also see \citealt{langer+07} for the highest possible metallicity for PISNe).
On the other hand, expectation to have a PISN is considered to be much higher for
metal-free environment in the early universe.
Recent cosmological simulation indicates that $\sim$ 25\% of first stars in number
would explode as PISNe \citep{hirano+15}.
The first reason of this high percentage is that
a typical initial mass of first stars can be as large as $\sim$ 100 M$_\odot$
due to the absent of line cooling during its formation \citep[e.g.][]{bromm&larson04}.
The second is the estimated rate of the line-driven wind mass loss is too small
to reduce the total mass during the evolution \citep[e.g.][]{krticka&kubat09}.
As well as the high number fraction, the ejected metal mass by one explosion of the order of $\sim$ 10 M$_\odot$
is much larger than that of a canonical core-collapse supernova of the order of $\sim$ 1 M$_\odot$.
Therefore it can be reasonable to consider that
the chemical pattern of the predominant PISN ejecta in the early universe is
conserved on a surface of numbers of MP stars.

So far, no candidate MP stars have been discovered for PISN children
except for the work by \citet{aoki+14} \citep{nomoto+13, frebel&norris15}.
The reason of the non-detection may be explained as a result of the observational bias.
Because of the large metal yield, PISN children may be born having a relatively large [Ca/H] $\sim$ $-2.5$
\citep[e.g.,][]{greif+10}.
Then this can be missed from metal-poor star surveys which utilize Ca${\rm I\hspace{-.1em}I}$ K line
as the indicator of the metal poorness \citep{karlsson+08}.
However, as a growing number of metal-poor stars have been discovered by recent observations
\citep[e.g.,][]{hollek+11, bonifacio+12, cohen+13, yong+13b, roederer+14a},
and moreover a systematic observation to find the PISN signatures in metal-poor stars
has been undertaken \citep{ren+12},
it is possible to expect that a candidate of a PISN child will be discovered near future.
Indeed, by applying the PISN fraction for the first stars of $\sim$ 25\% to
the theoretical model constructed by \citet{karlsson+08},
the number fraction of PISN children from all the MP stars can be estimated as
$\sim$ 1/400, which is about the current observational limit for the detection.

In this work, we newly perform sequence of numerical simulations of the evolution,
the explosion, and the nucleosynthesis, for first stars with initial masses of $\sim$ 140--300 M$_\odot$.
Explosive nucleosynthesis of PISNe has been calculated by
\citet{heger&woosley02} and by \citet{umeda&nomoto02}.
However, their stellar simulations neglect the effects of stellar rotation.
It has been shown that stellar rotation can affect all important outputs of stellar evolution
\citep{meynet&maeder00, heger+00}, and moreover
first stars are suggested to posses a large angular velocity at its birth \citep{stacy+11b, stacy+13}.
Therefore we firstly consider the effect of stellar rotation for the yields of PISNe,
applying a moderate rotation speed of 30\% of the kinetic critical velocity at ZAMS stages.
Through the systematic calculations, we aim to find characteristic abundance patterns
that can be used to constrain the progenitor's properties, namely, the initial mass and the rotation.

Furthermore, we conduct the first systematic comparison between PISN theoretical yields
and observations using the big stellar abundance data compiled in {\it SAGA} database
({\bf http://sagadatbase.jp/}, \citealt{suda+08, suda+11, yamada+13, suda+17}).
The purpose of the comparison is,
firstly, to confirm the (non-)existence of PISN signatures on the current MP stellar sample,
and secondly, to validate what are the fundamentally reliable and practically useful abundance ratios
to discriminate PISN signatures.

This paper is organized as follows.
In the next section, code description is given for the evolution, explosion, and post-processing codes used in this work.
Results of stellar evolution calculations are analyzed in \S 3,
in which effects of stellar rotation are mainly discussed.
Section 4 is attributed for the discussion on the initial and CO core mass ranges for PISNe.
Results of PISN nucleosynthesis are analyzed and
discussions on characteristics of PISN abundance patterns are given in \S 5.
Comparison between the theoretical models and the observed abundances of MP stars is conducted in \S 6.
Finally, discussion and conclusion are presented in \S 7.

\section{Computational Method}

\subsection{Stellar Evolution Code}

\begin{table}
\centering
	\begin{tabular}{ccc|ccc}
	\hline
	Element & \multicolumn{2}{c}{mass number} & Element & \multicolumn{2}{c}{mass number} \\
	\hline
	n	&	1		&	1		&	Ar	&	34-40	&	33-42	\\
	H	&	1-3		&	1-3		&	K	&	37-41	&	36-43	\\
	He	&	3-4		&	3-4		&	Ca	&	38-43	&	37-48	\\
	Li	&	6-7		&	6-7		&	Sc	&	41-45	&	40-49	\\
	Be	&	7-9		&	7-9		&	Ti	&	43-48	&	41-51	\\
	B	&	8-11		&	8-11		&	V	&	45-51	&	44-52	\\
	C	&	12-13	&	11-14	&	Cr	&	47-54	&	46-55	\\
	N	&	13-15	&	12-15	&	Mn	&	49-55	&	48-56	\\
	O	&	14-18	&	13-20	&	Fe	&	51-58	&	50-61	\\
	F	&	17-19	&	17-21	&	Co	&	53-59	&	54-62	\\
	Ne	&	18-22	&	18-24	&	Ni	&	55-62	&	56-66	\\
	Na	&	21-23	&	20-25	&	Cu	&	57-63	&	59-67	\\
	Mg	&	22-26	&	21-27	&	Zn	&	60-64	&	62-70	\\
	Al	&	25-27	&	23-29	&	Ga	&	-		&	65-73	\\
	Si	&	26-32	&	24-32	&	Ge	&	-		&	69-76	\\
	P	&	29-33	&	27-34	&	As	&	-		&	71-77	\\
	S	&	30-36	&	29-36	&	Se	&	-		&	73-79	\\
	Cl	&	33-37	&	31-38	&	Br	&	-		&	76-80	\\
	\hline
	\end{tabular}
	\caption{Isotopes included in the stellar evolution and hydrodynamic code (153 isotopes, left)
	and in the post-processing code (300 isotopes, right).}
	\label{tab-isotope}
\end{table}
Stellar evolution of zero-metallicity stars are calculated
using the stellar evolution code described in \citet{takahashi+16}.
Result of Big bang nucleosynthesis by \citet{steigman07} is used for the initial chemical composition,
and 153 isotopes are considered in the reaction network (Tab.\ref{tab-isotope}, left column).
The reaction rates are taken from the current version of {\it JINA REACLIB} \citep{Cyburt+10}
except the rate of $^{12}$C($\alpha$, $\gamma$)$^{16}$O is
taken from \citet{caughlan&fowler88} multiplied by a factor of 1.2.

The Ledoux criterion is used for convective instability.
Inside the convective region, a diffusion coefficient is estimated by
$D_{\rm cv} = v_{\rm MLT} l_{\rm MLT} /3$,
where $v_{\rm MLT}$ and $l_{\rm MLT}$ are the velocity of the convective blob and the mixing length
calculated by the mixing-length theory.
To describe the chemical mixing by convective overshoot, an exponentially decaying coefficient,
\begin{eqnarray}
	D_{\rm cv}^{\rm ov} = D_{\rm cv,0} \exp \Bigl( -2\frac{\Delta r}{f_{\rm ov} H_{p,0}} \Bigl),
\end{eqnarray}
where $f_{\rm ov}$ is an adjustable parameter,
$D_{\rm cv,0}$ and $H_{p,0}$ are the convective diffusion coefficient and the pressure scale hight
at the edge of the convective region, and $\Delta r$ is a distance from the edge,
is added to the diffusion coefficient.
A very small constant mass loss rate of $10^{-14}$ M$_\odot$/yr,
the effect of which is practically negligible,
is considered in non-rotating models \citep{yoon+12}.

Effects of stellar rotation are taken into account \citep{heger+00, meynet&maeder00}.
Deformation factors are included in the equations of pressure and temperature balances \citep{endal&sofia78}.
Diffusion approximation is applied to transportation of angular momentum using the diffusion coefficient $\nu_{\rm eff}$.
For the diffusion coefficient, $\nu_{\rm eff, non-mag} = \nu_{\rm ES} + \nu_{\rm GSF}
+ \nu_{\rm SH} + \nu_{\rm DS} + \nu_{\rm SS}$, in which
coefficients owing to the Eddington-Sweet circulation (the meridional circulation),
the Goldreich-Schubert-Fricke instability, the Solberg-H\o iland instability, and 
the dynamical and secular shear instabilities are summed up \citep{heger+00, pinsonneault+89},
is used in a non-magnetic model.
While in a magnetic model, $\nu_{\rm eff, mag} = \nu_{\rm eff, non-mag} + \nu_{\rm TS}$,
where $\nu_{\rm TS}$ is the viscosity owing to the Tayler-Spruit dynamo (TS dynamo, \citealt{spruit02}),
is applied.
To account for the rotation induced mixing, additional diffusion coefficients of
$D_{\rm rot, non-mag} = f_c \times \nu_{\rm eff, non-mag}$ for the non-magnetic model and 
$D_{\rm rot, mag} = D_{\rm rot, non-mag} + D_{\rm TS}$,
where $D_{\rm TS}$ is the diffusion coefficient of the Tayler-Spruit dynamo, for the magnetic model
are included in the diffusion equation of chemical species.
Mass loss is enhanced due to the nearly-critical rotation at the surface
\citep[the $\Omega-\Gamma$ limit,][]{langer98, maeder&meynet00}.
According to \citet{yoon+10, yoon+12}, the enhanced mass loss rate is calculated as
\begin{eqnarray}
	\dot{M} = -\min \Bigl[ |\dot{M}(v_{\rm rot}=0)| \times
	\Bigl( 1-\frac{v_{\rm rot}}{v_{\rm crit}} \Bigl)^{-0.43},
	0.1 \frac{M}{\tau_{\rm KH}} \Bigl],
\end{eqnarray}
where $v_{\rm rot}$, $v_{\rm crit}\equiv \sqrt{GM(1-L/L_{\rm Edd})/R}$,
$\tau_{\rm KH}$, $L_{\rm edd}$ are the surface rotation velocity, the critical rotation velocity,
the Kelvin-Helmholtz timescale, and the Eddington luminosity, respectively.

Important note here is that there is still large uncertainty
in treatment of rotation induced mixing in spite of vigorous efforts over the years.
For example, the diffusive treatment for the meridional circulation
that is essentially an advective process is debatable \citep{Maeder&Zahn98, Chieffi&Limongi13},
although a detail comparison between the different treatments has not been done.
The treatment of interplay between stellar rotation and magnetic field is even more disputable.
Different treatments for TS dynamo are known \citep{Maeder&Meynet04, Denissenkov&Pinsonneault07},
and moreover, numerical simulation by \citet{Zahn+07} has found
no dynamo process in their differentially rotating stellar model.
Our rotating results should thus be regarded as representative results with rotation induced mixing:
A case with efficient diffusion of angular momentum will be represented by the magnetic models.
Therefore the different models will cover a reasonable range of theoretical uncertainty
involved in modeling for PISN progenitors.

The same calibration to the recent grid calculations with {\it GENEC}
\citep{ekstroem+12, georgy+12, georgy+13} has been done to fix adjustable parameters in the code.
Parameters are 
the mixing length parameter $\alpha_{\rm mix}$,
the overshoot parameter $f_{\rm ov}$,
the ratio between the mixing coefficient and the effective viscosity $f_c$,
and a parameter showing the reduction efficiency of $\mu$-barrier $f_\mu$.
Values of ($\alpha_{\rm mix}$, $f_{\rm ov}$, $f_c$, $f_\mu$) = 
(1.8, 0.01, 1/32, 0.1) are used for non-magnetic models and 
(1.8, 0.01, 1/8, 0.1) are used for magnetic models.

\subsection{Explosion Code}

A 1D-spherical general-relativistic Lagrangian hydrodynamic code
developed by \citet{yamada97} is used for explosion simulations.
The code integrates the time in an implicit manner,
iteratively solving equations of the metric and the hydrodynamics.
In order to utilize the code for general purpose simulations,
\citet{takahashi+16} introduced a reaction network and a non-NSE EOS,
which are also used in the stellar evolution code, into the hydrodynamic code.
153 isotopes are included in the reaction network.
Although the code is capable of directly solving the Boltzmann equation for the neutrino transport
\citep{yamada+99, sumiyoshi+00}, the complicated transport equation is not treated in this work.
Instead, the thermal neutrino energy loss rate \citep{itoh+89, itoh+96},
which is also imported from the stellar code, is used to determine the local cooling rate.

\subsection{Post-processing Code}

Taking the exploding models, their explosive nucleosynthesis are calculated by a post-processing manner.
The time evolution of the density and the temperature are recorded for each Lagrangian mesh.
According to the record, the composition evolution of extended 300 isotopes,
which are the same as a network used in the stellar calculations, is calculated.
Decay process of the explosive yield is further considered by calculating
additional $10^{10}$ yr with a temperature of $10^3$ K and the density of $10^{-10}$ g cm$^{-3}$.
Comparisons with observations are made using solar-scaled values of
\begin{eqnarray}
	[i/j] = \log \Bigl( \frac{ n_i/n_j }{ n_{i, \odot}/n_{j, \odot} } \Bigl),
\end{eqnarray}
where $n_i$ is the number density of the $i$-th element.
Solar elemental abundance is taken from \citet{asplund+09}.

\section{Stellar evolution of PISN progenitors}

Evolution of a massive Pop III star having an initial mass of $M_{\rm ini} \in [100, 290]$ $M_\odot$
are calculated with three different rotation treatments.
The first model sequence is obtained for non-rotating models,
in which no rotational mixing, no rotational mass loss, and no centrifugal force are considered.
Besides, two sequences for rotating models are calculated;
the one with TS dynamo and the other without TS dynamo.
Hereafter they are referred to as magnetic- or non-magnetic (rotating) models, respectively.
The rigid rotation is applied for the initial rotation profile
with the rotation period of 30\% of the Kepler rotation at its surface.
Evolution of each model is calculated from the ZAMS stage
until the central temperature reaches $\sim$ $\log T_c \rm{[K]} = 9.2$.
In this section, evolutionary properties of PISN progenitors are discussed,
mainly focusing on how stellar rotation affects the results.

\subsection{160 $M_\odot$ models}

\begin{figure}[t]
	\includegraphics[width=\textwidth]{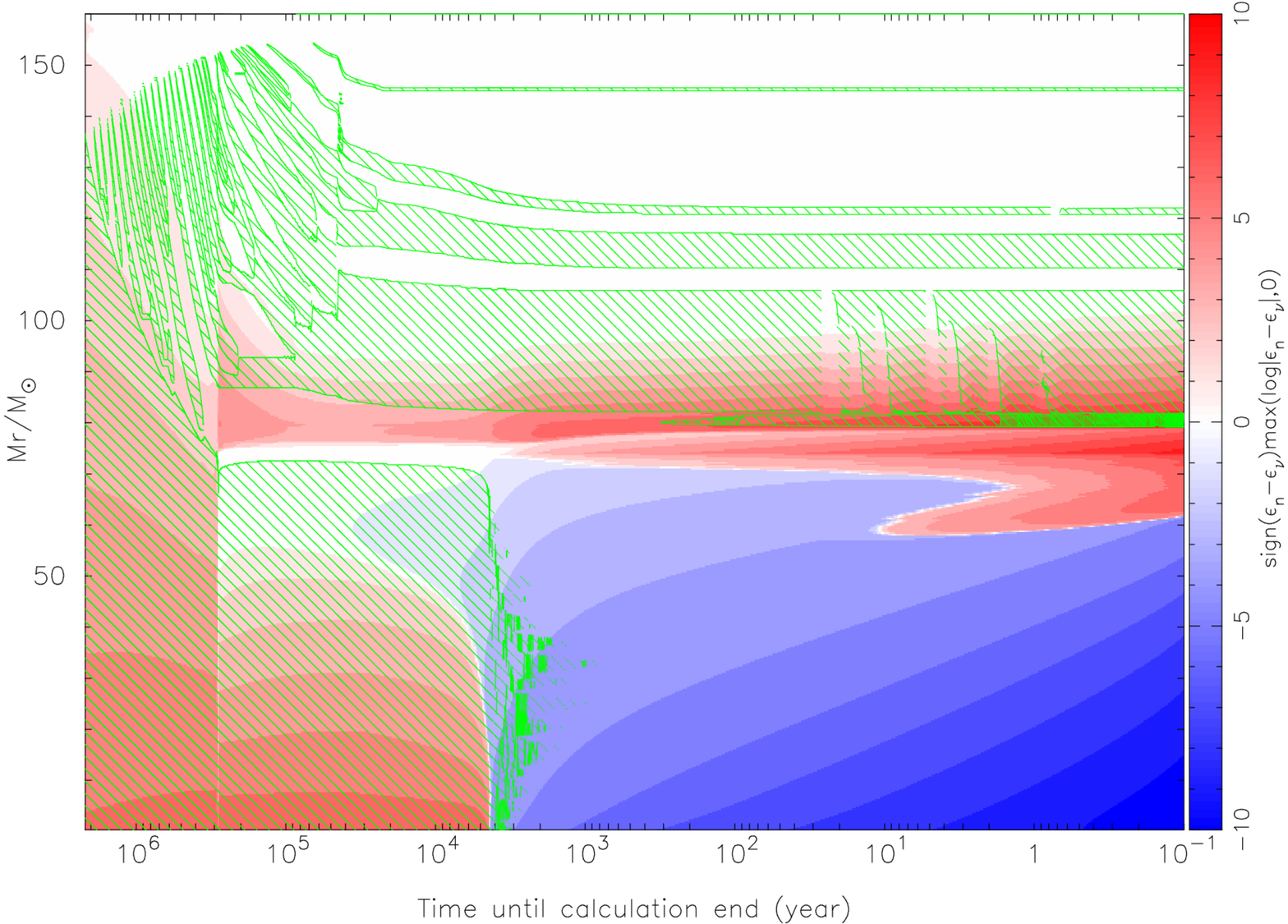}
	\caption{\footnotesize{Kippenhahn diagram of non-rotating 160 $M_\odot$ model.
	Green-hatched regions show the convective regions.
	Colors indicate the heating (red) or the cooling (blue) rates at the region.}}
	\label{fig-khd-m160o00}
\end{figure}
Evolution of a non-rotating PISN progenitor is simple.
As an example, evolution of convective regions are shown
for the non-rotating 160 $M_\odot$ model in Fig.\ref{fig-khd-m160o00}.
The star spends the core hydrogen burning stage for 1.88 Myr.
During this phase, the star develops the central convective region,
which firstly fills the inner 136 $M_\odot$ and gradually recedes in mass
forming a 79.1 $M_\odot$ He core.
The next core helium burning stage lasts $2.96 \times 10^5$ yr.
The star forms a roughly constant-mass convective helium burning core
of $\sim 71$ $M_\odot$ during this phase.
Finally the star forms a 74.0 $M_\odot$ CO core,
which is at the lower side of the CO core mass to explode as a PISN \citep{takahashi+16}.

\begin{figure}[t]
	\includegraphics[width=\textwidth]{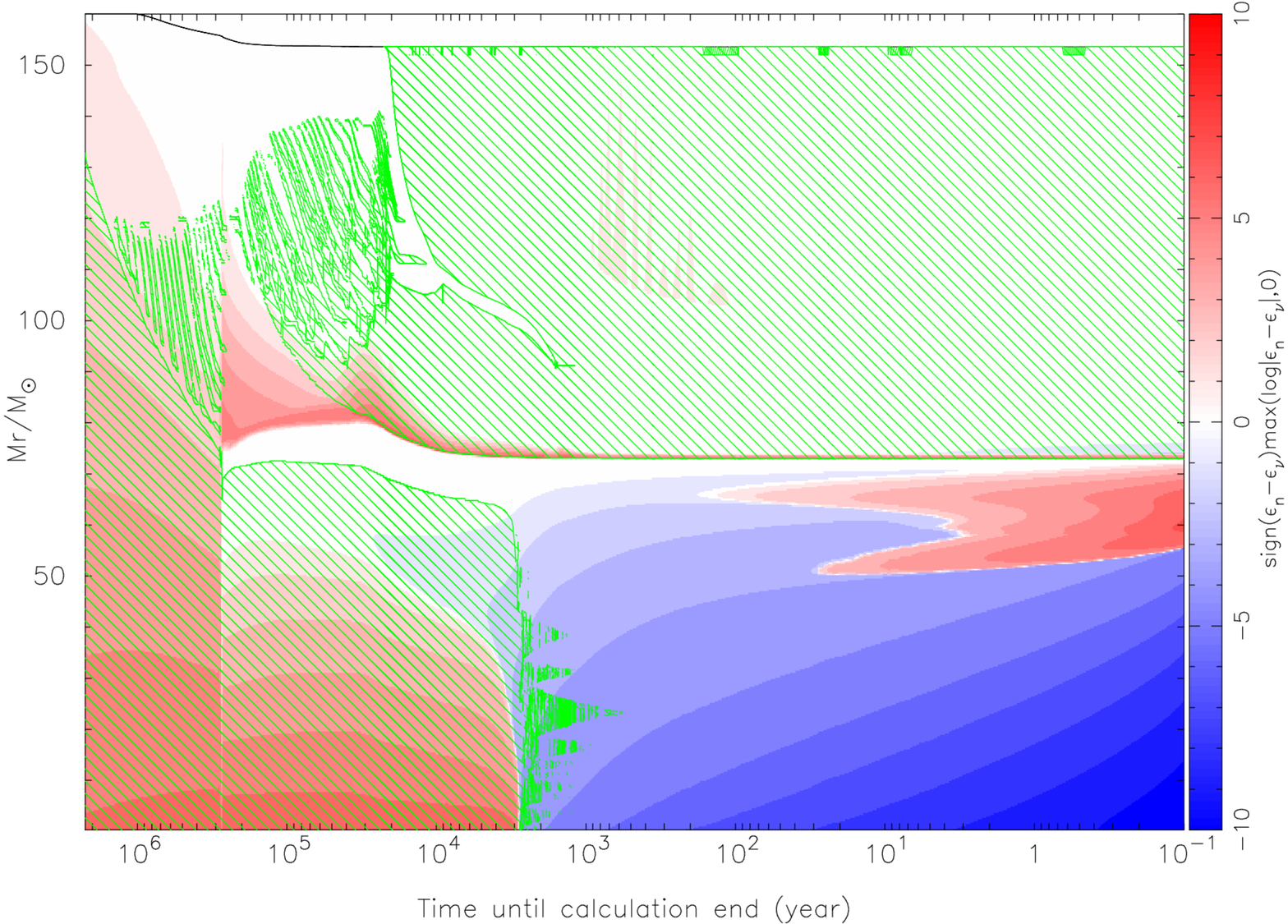}
	\caption{\footnotesize{Kippenhahn diagram of non-magnetic rotating 160 $M_\odot$ model.}}
	\label{fig-khd-m160o30n}
\end{figure}
Stellar rotation affects the evolution through mass loss
and rotational mixing for chemical species.
The centrifugal force can modify the hydrostatic structure in principle,
but the applied rotation speed in this work is too slow to have an influential effect.
Kippenhahn diagram of the non-magnetic rotating model of 160 $M_\odot$
is shown by Fig.\ref{fig-khd-m160o30n}.
Rotating 160 $M_\odot$ models lose parts of their envelope masses
by 6.42 $M_\odot$ for non-magnetic and by 10.69 $M_\odot$ for magnetic cases.
In these models, mass loss takes place at a latter half of the core hydrogen burning phase
and at the transition phase between the hydrogen burning and the helium burning phases.
Since the lost mass is small compared with the total envelope mass,
the mass loss has little effect on the evolution.
On the other hand, it may have an impact on the environment,
as the lost mass injects kinetic energy of $\sim 3 \times 10^{50}$ erg
into the ambient matter through this process.\footnote{In this estimate,
we assume the lost mass has a wind velocity of
$\sqrt{GM/R} = 1.74 \times 10^3$ km sec$^{-1}$.}
The lost material may sweep up the ambient matter,
triggering succeeding star formation.
Since the lost material has not been chemically processed,
the ambient matter mixed with the stellar wind retains its composition.
The stellar wind from a rotating Pop III star therefore may
enhance the local star formation rate of Pop III stars.

\begin{figure}[t]
	\includegraphics[width=\textwidth]{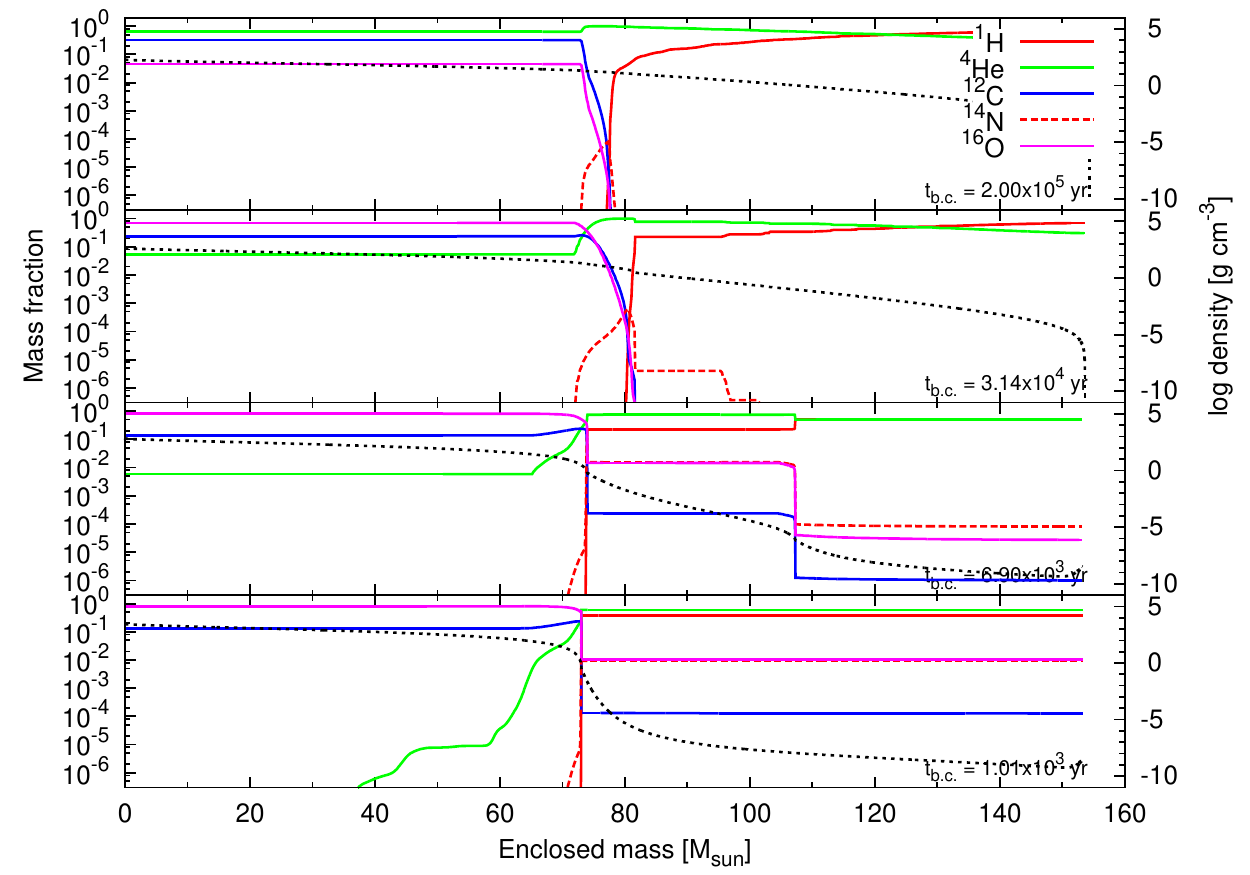}
	\caption{\footnotesize{Evolution of chemical distribution
	in the non-magnetic rotating 160 $M_\odot$ model
	are shown from top to bottom panels.
	The time before the collapse, $t_{\rm bc}$, is indicated for each panel.}}
	\label{fig-distevol-m160}
\end{figure}
The first consequence of rotational mixing is the extension of the He core,
which results in shifting to lower side of the mass range for PISN
\citep{chatzopoulos&wheeler12, yoon+12}.
During the core hydrogen burning phase, rotational mixing takes place
at the boundary region between convective core and the envelope.
This acts similar to convective overshooting,
transporting inner helium rich material to the outer region.
As a result, the magnetic 160 $M_\odot$ model
forms a larger He core of 83.0 $M_\odot$ than the non-rotating model.
Another direct consequence of the rotational mixing
is synthesis of primary nitrogen during the core helium burning phase.
Figure \ref{fig-distevol-m160} shows evolution of chemical distribution
during the core helium burning phase for the non-magnetic 160 $M_\odot$ model.
Over the core helium burning phase of $\sim 2 \times 10^5$ yr,
rotational mixing transports helium burning products of carbon and oxygen
from the inner convective region to the outer helium layer,
forming a CO-enriched tail in the layer.
As the edge of the tail permeates into the base of the hydrogen layer,
the helium burning products are transformed into $^{14}$N via CNO-cycle.
The non-magnetic rotating 160 $M_\odot$ model finally yields
$7.67 \times 10^{-1}$ $M_\odot$ of nitrogen in its outer regions than the CO core,
which is $\sim 4 \times 10^5$ times larger than the yield of the non-rotating model,
$1.97 \times 10^{-6}$ $M_\odot$.

However, reflecting the large uncertainty in the theory of rotational mixing,
results of the two rotating models with and without TS dynamo
differ both qualitatively and quantitatively.
First, the nitrogen yield in the magnetic model is $1.34 \times 10^{-5}$ $M_\odot$
and comparable to the non-rotating case.
This is because, the main mechanism that accounts for the rotational mixing
during the helium burning phase in the non-magnetic model is GSF instability.
Since GSF instability requires strong shear at the region to work effectively,
the efficient mixing does not take place in the helium layer of the model with TS dynamo,
in which magnetic torque effectively works to achieve the solid rotation.

The second difference is occurrence of the convective dredge-up in the non-magnetic model.
During the core helium burning phase, convection arises in the H/He boundary region.
The base of the shell convection moves inward for all models.
Furthermore, the convective base of the non-magnetic model
keeps its motion even after it reaches the edge of the hydrogen deficient region.
As a result, significant mass of the outer helium layer
is dredged into the shell convective region.
One consequence is that
the He core mass of the non-magnetic 160 $M_\odot$ model
finally becomes 72.9 $M_\odot$ and is even slightly smaller than the non-rotating model.
Another important consequence is boosting the nitrogen synthesis.
The reason why the convective dredge-up takes place is unclear.
However, it possibly relates to the rotational mixing in the helium layer,
by which the luminosity of the shell hydrogen burning is enhanced.
We note that a similar dredge-up episode also takes place in the work by \citet{yoon+12},
but for $>$ 200 $M_\odot$ non-rotating models.

Among the 160 $M_\odot$ models in this work, only the non-magnetic rotating model
forms an inflated red-giant envelope during the evolution.
This may be related to the fact that only the non-magnetic model experiences 
the dredge-up event and accompanying boosting of the shell hydrogen burning.
However, an important note here is that the theoretical estimate
for the envelope evolution of a massive Pop III star
significantly depends on numerical treatments of additional mixings,
such as overshooting, semi-convection, and rotational mixing.
The convective overshooting during the core hydrogen phase is one of the most influential:
with a somewhat larger overshoot parameter of 0.015, a non-rotating 160 $M_\odot$
develops an inflated envelope during its core helium burning phase.
Unless reliable calibration can be done for the massive Pop III stars,
it is needed to understand physical properties of these phenomena
before conducting a reasonable prediction on the radius.

\begin{figure}[t]
	\includegraphics[width=\textwidth]{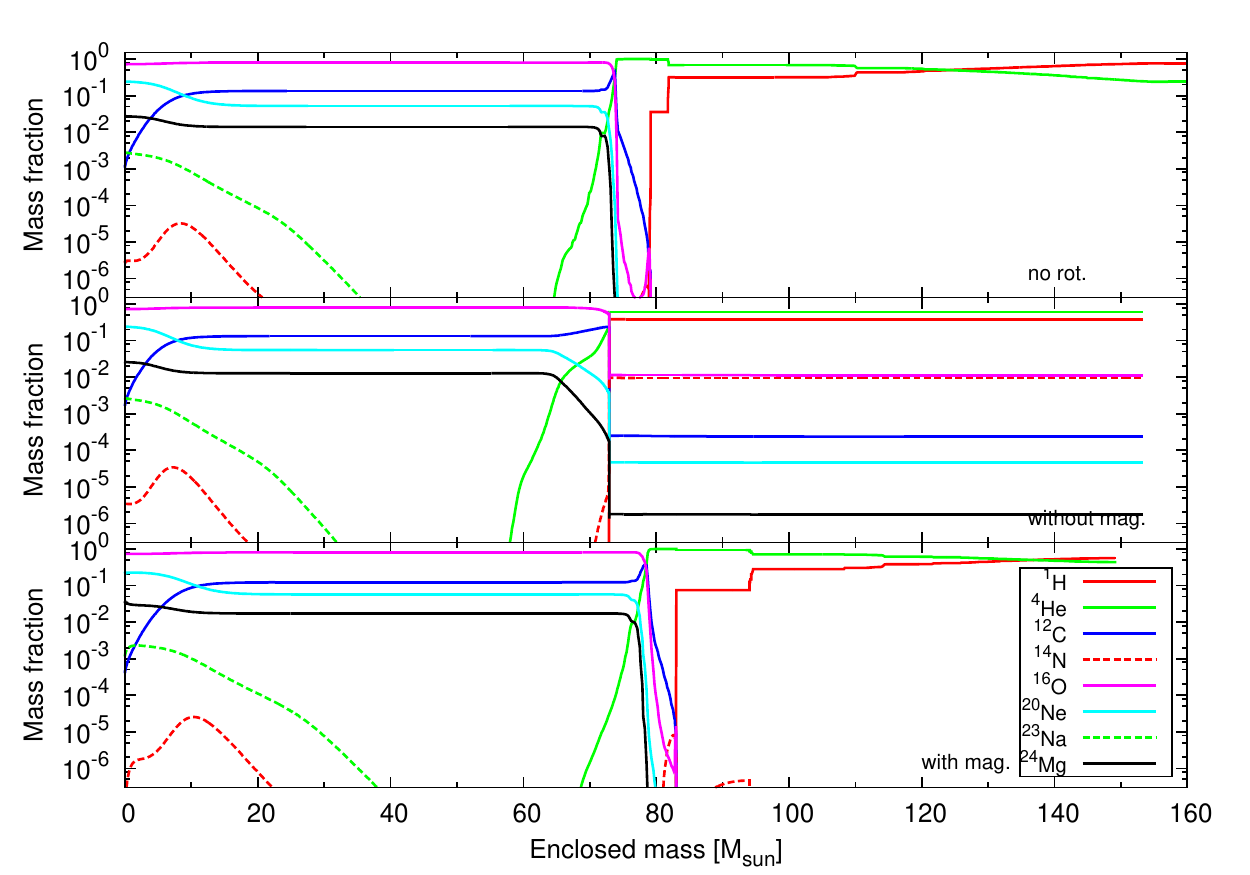}
	\caption{\footnotesize{Chemical distributions of 160 $M_\odot$ models
	at the end of evolution calculations.
	The top panel shows the result of non-rotating model,
	the middle is the non-magnetic model,
	and the bottom is the magnetic model.}}
	\label{fig-xdist-m160}
\end{figure}
Chemical distribution at the calculation end
is shown for 160 $M_\odot$ models in Fig.\ref{fig-xdist-m160}.
The overall characteristics of the non-rotating and the magnetic rotating model
are very similar except for the larger core mass and the slight lost of the envelope mass of the rotating model.
On the other hand, the non-magnetic rotating model has a nitrogen rich envelope,
and thus the model finally yields far more abundant nitrogen.
Chemical composition of the three CO cores are quite similar
except for the more diffused chemical distribution seen in the outer region of the core
for the non-magnetic rotating model.
The C/O ratios measured at homogeneous regions ($M_r = 40$ $M_\odot$)
are 0.168 for non-rotating, 0.164 for non-magnetic, and 0.152 for magnetic models.
Moreover, all of these cores have extremely small neutron excesses, 
$\eta \equiv \sum (N_i-Z_i) Y_i$, of $4.03 \times 10^{-7}$,
$5.22 \times 10^{-7}$, and $4.22 \times 10^{-7}$, respectively.
In conclusion, stellar rotation does not enhance the neutron excess of the CO core,
because nitrogen enhancement takes place only outside of the core.
Consequently, quite similar explosive yields result from the models
with different rotation treatments, which is reported later.

\subsection{Summary of the evolutionary calculations}

\begin{table}[t]
	\begin{center}
	\begin{tabular}{lcccccccccc}
\hline
\hline
	$M_{\rm ini}$	&	$M_{\rm fin}$			&
	$v_{\rm ini}$	&	$v_{\rm ini}/v_{\rm K}$	&
	$\tau_{\rm H}$	&	$\tau_{\rm He}$			&
	$M_{\rm c,He}$	&	$M_{\rm c,CO}$		&
	{\it Fate}		&	$E_{\rm tot}$	&	$M_{ ^{56} {\rm Ni} }$	\\
	$M_{\odot}$	&	$M_{\odot}$	&
	km sec$^{-1}$	&				&
	Myr			&	$10^5$ yr		&
	$M_{\odot}$	&	$M_{\odot}$	&
				&	10$^{51}$ erg	&	$M_{\odot}$	\\
\hline
	\multicolumn{8}{l}{Non-rotating models}	\\
	100	&	100.0	&	0	&	0	&	2.244	&	3.392	&	46.51	&	42.19	&	-		&	-	&	-	\\
	120	&	120.0	&	0	&	0	&	2.051	&	3.267	&	56.90	&	52.13	&	-		&	-	&	-	\\
	140	&	140.0	&	0	&	0	&	1.979	&	3.120	&	68.97	&	63.79	&	-		&	-	&	-	\\
	160	&	160.0	&	0	&	0	&	1.884	&	2.964	&	79.18	&	74.04	&	PPISN	&	(4.115)	&	(0.009)	\\
	170	&	170.0	&	0	&	0	&	1.842	&	3.075	&	84.40	&	77.59	&	PPISN	&	(8.722)	&	(0.127)	\\
	180	&	180.0	&	0	&	0	&	1.818	&	2.942	&	89.86	&	84.77	&	PISN		&	17.31	&	0.567	\\
	200	&	200.0	&	0	&	0	&	1.749	&	2.861	&	100.07	&	92.55	&	PISN		&	24.86	&	1.264	\\
	220	&	220.0	&	0	&	0	&	1.705	&	2.814	&	110.41	&	103.87	&	PISN		&	43.16	&	6.508	\\
	240	&	240.0	&	0	&	0	&	1.661	&	2.838	&	121.06	&	112.76	&	PISN		&	54.48	&	12.54	\\
	260	&	260.0	&	0	&	0	&	1.619	&	2.764	&	127.42	&	119.90	&	PISN		&	65.19	&	20.65	\\
	280	&	280.0	&	0	&	0	&	1.596	&	2.795	&	142.84	&	133.71	&	PISN		&	94.47	&	47.68	\\
	290	&	290.0	&	0	&	0	&	1.586	&	2.864	&	149.99	&	140.08	&	BH		&	-	&	-	\\
\hline
	\multicolumn{8}{l}{Non-magnetic rotating models}	\\
	100	&	97.05	&	638	&	0.30	&	2.355	&	3.216	&	45.96	&	44.48	&	-		&	-	&	-	\\
	120	&	114.78	&	624	&	0.30	&	2.160	&	3.071	&	59.29	&	54.96	&	-		&	-	&	-	\\
	140	&	134.62	&	687	&	0.30	&	2.043	&	3.063	&	68.47	&	64.45	&	PPISN	&	(4.374)	&	(0.009)	\\
	160	&	153.58	&	710	&	0.30	&	1.945	&	2.912	&	72.93	&	72.94	&	PISN		&	6.828	&	0.025	\\
	180	&	172.42	&	727	&	0.30	&	1.872	&	2.917	&	83.10	&	83.11	&	PISN		&	21.12	&	0.796	\\
	200	&	191.30	&	746	&	0.30	&	1.808	&	2.884	&	93.18	&	93.20	&	PISN		&	36.10	&	4.267	\\
	220	&	210.09	&	761	&	0.30	&	1.755	&	2.776	&	100.02	&	100.02	&	PISN		&	45.53	&	8.209	\\
	240	&	228.95	&	778	&	0.30	&	1.710	&	2.793	&	108.45	&	108.45	&	PISN		&	59.40	&	18.73	\\
	260	&	247.78	&	793	&	0.30	&	1.671	&	2.719	&	118.98	&	118.98	&	PISN		&	76.50	&	34.40	\\
	280	&	266.51	&	806	&	0.30	&	1.634	&	2.670	&	126.08	&	126.08	&	BH		&	-	&	-	\\
\hline
	\multicolumn{8}{l}{Magnetic rotating models}	\\
	100	&	94.73	&	637	&	0.30	&	2.453	&	3.299	&	51.91	&	47.99	&	-		&	-	&	-	\\
	120	&	111.31	&	624	&	0.30	&	2.253	&	3.123	&	63.74	&	59.70	&	-		&	-	&	-	\\
	140	&	131.24	&	688	&	0.30	&	2.110	&	3.028	&	72.66	&	68.39	&	PPISN	&	(1.168)	&	(0.264)	\\
	160	&	149.31	&	709	&	0.30	&	2.000	&	2.964	&	83.02	&	78.73	&	PISN		&	9.373	&	0.169	\\
	180	&	167.48	&	729	&	0.30	&	1.916	&	2.903	&	94.04	&	88.79	&	PISN		&	20.71	&	0.892	\\
	200	&	185.52	&	747	&	0.30	&	1.849	&	2.890	&	106.83	&	100.08	&	PISN		&	36.47	&	3.741	\\
	220	&	203.41	&	762	&	0.30	&	1.794	&	2.830	&	116.92	&	110.30	&	PISN		&	51.84	&	11.12	\\
	240	&	221.36	&	778	&	0.30	&	1.743	&	2.835	&	127.83	&	120.33	&	PISN		&	69.09	&	23.40	\\
	260	&	239.33	&	794	&	0.30	&	1.700	&	2.785	&	136.97	&	130.36	&	PISN		&	85.10	&	38.22	\\
	280	&	257.25	&	805	&	0.30	&	1.659	&	3.012	&	139.09	&	139.09	&	BH		&	-	&	-	\\
\hline
	\end{tabular}
	\end{center}
	\caption{\footnotesize{Model properties. $M_{\rm ini}$ and $M_{\rm fin}$ are the initial and final masses;
	$v_{\rm ini}$ and $v_{\rm K} \equiv \sqrt{GM/R}$ are the surface rotation velocity
	and the surface Kepler velocity at the zero age main sequence;
	$\tau_{\rm H}$ and $\tau_{\rm He}$ are lifetimes of core hydrogen and core helium burning stages;
	$M_{\rm c,He}$ and $M_{\rm c,CO}$ are masses of helium and carbon-oxygen cores.
	{\it Fate} is chosen from PPISN, PISN, or BH according to the explosion simulations.
	$E_{\rm tot}$ and $M_{ ^{56} {\rm Ni} }$, the explosion energy and the ejected $^{56} {\rm Ni}$ mass respectively,
	are shown for exploding models.} }
	\label{tab-summary-evol}
\end{table}

Properties obtained for 160 $M_\odot$ models are common for models with different masses.
In Table \ref{tab-summary-evol}, core masses, lifetimes, and nitrogen yields
at $\log T_c \rm{[K]} = 9.2$ are summarized.
Total masses of rotating models are slightly reduced by the rotation-induced mass loss,
while no mass loss is assumed for non-rotating models.
He core masses of magnetic rotating models are increased,
while non-magnetic rotating models reduce the core mass
through the convective dredge-up episode.
Despite the different core masses,
durations of the core hydrogen and the core helium burning phases
are almost independent from the rotation treatments.
This is because these durations only slightly depend
on the total and the core masses in this massive initial mass range.

Nitrogen yields of non-rotating models are small.
And only slight enhancements are seen in the magnetic rotating models,
except for the 280 M$_\odot$ model in which the dredge-up episode takes place.
The reason of the small enhancement is the same as in the 160 $M_\odot$ model:
the strong shear at the helium layer does not develop in the magnetic models.
On the other hand, non-magnetic rotating models yield significantly larger amount of nitrogen.
Because the dredge-up episode takes place at a later stage in the core helium burning phase,
the amount of nitrogen yields are smaller for a less massive models of 100--140 $M_\odot$,
but they are still far more abundant than non-rotating and magnetic rotating models.

\begin{figure}[t]
	\includegraphics[width=\textwidth]{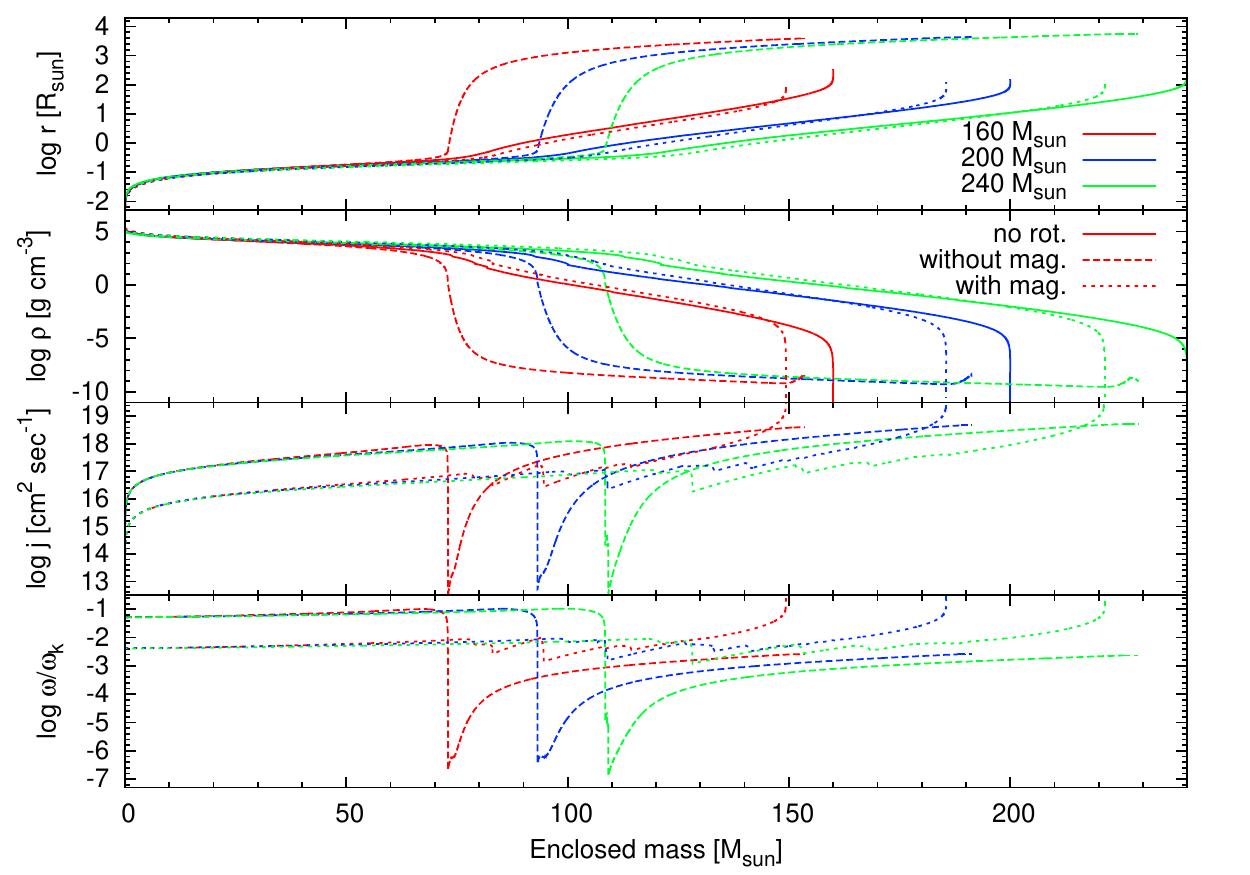}
	\caption{\footnotesize{Progenitor structures
	of 160 (red), 200 (blue), and 240 $M_\odot$ (green) models
	in terms of the radius (top), the density (second),
	the specific angular momentum (third), and the rotation period divided by
	the local Kepler rotation $\omega_{\rm K} \equiv \sqrt{ GM_r/r^3 }$ (bottom).
	Solid lines show results of non-rotating models.
	Dashed and dotted lines are rotating models without and with TS dynamo, respectively.
	}}
	\label{fig-prog}
\end{figure}
Finally, we show progenitor structures obtained in this work in Fig.\ref{fig-prog}.
Variations among the models with different rotation treatments can be seen
in the envelope structures.
That is, non-magnetic rotating models have dilute and inflated envelopes,
while envelopes in other models are more compact and dense.
Accordingly, the non-magnetic models show stronger decline in density at core edges.
Inner part of the cores, on the other hand, exhibit similar structures independent of the rotation treatments.

As for the internal rotation profile, existence of the strong shear at the core/envelope boundary
is the qualitative difference between the two sequences in rotating models.
Magnetic models more or less rotate uniformly.
This is due to the efficient angular momentum transfer by the TS dynamo,
which imposes rigid rotation on the star.
Since the central region of the star spins up as the star evolves,
outward transport of the angular momentum takes place.
Accordingly, magnetic models only retain specific angular momentum of
$\sim 10^{16}$ cm$^2$ sec$^{-1}$ in the core at the calculation end.
The corresponding core rotation rate becomes $\sim$ 1/100 of the Kepler rotation rate.
On the other hand, no instabilities considered in non-magnetic models
account for such an efficient transfer after the core helium burning phase.
As a result, the non-magnetic models keep specific angular momentum of
$\sim 10^{17}$--$10^{18}$ cm$^2$ sec$^{-1}$ even for the later stages of the evolution.
Accordingly, the angular velocity reaches $\sim$ 10\% of the Kepler rate at the calculation end.

The fast core rotation obtained in the non-magnetic models
may affect the further collapse and the explosion as a PISN,
although our explosion code does not handle the rotating flow.
Assuming that the local angular momentum is kept constant during core collapse,
the ratio between the local angular velocity and the Keplerian angular velocity
increases as $\omega / \omega_{\rm K} \propto r^{-1/2}$.
Assuming the highest density achieved during the collapse
is $\sim 10^{7}$ g cm$^{-3}$ for the most massive PISN model,
the core density increases by a factor of $\sim 10^{2}$
and correspondingly the core radius decreases by a factor of $\sim 10^{-2/3}$.
This means that the centrifugal force can be as high as $\sim 40\%$ of the local gravity
at its maximum during the explosion.
We let how rotating flow affects to the explosion open in this work, however,
further investigation with a multi-dimensional hydrodynamic code will be interesting.

\section{Initial mass range for PISNe}

Mechanism and dynamics of a PISN explosion have been
well investigated in previous works.
They can be summarized as follows.
A collapsing core is heated mainly by oxygen burning,
which initiates when the local temperature exceeds $\log T \rm{[K]} > 9.5$.
Due to the heating, a CO core with a mass of $\sim$ 65--130 $M_\odot$ 
reverses its motion and finally explodes.
On the other hand, a more massive CO core of $> \sim 130$ $M_\odot$ keeps collapsing,
because iron dissociation, which reduces the (non-relativistic) thermal energy and thus the pressure,
takes place at the high temperature central region of $\log T \rm{[K]} > 9.75$.

An accurate estimate of the initial mass range for PISN, however, is still difficult.
This is firstly because of the uncertain relation between the CO core mass and the initial mass.
The initial to core mass relation largely depends on
efficiencies of additional mixings during core hydrogen and core helium burning phases,
namely convective overshooting and rotation induced mixing,
both of which are constrained only poorly through observations.
In addition, here we firstly report that stars having the same core masses
can show different explodability when their envelope structures are different.

In this work, initial mass ranges for PISNe have become
180--280 $M_\odot$ for non-rotating models,
160--260 $M_\odot$ for magnetic models, and
160--260 $M_\odot$ for non-magnetic models.
The non-rotating models have larger minimum and maximum masses for PISNe,
while the two rotating sequences have the same mass range.
The corresponding CO core masses are
$\sim$ 84.7--133.7 $M_\odot$,
$\sim$ 78.7--130.3 $M_\odot$, and
$\sim$ 72.9--118.9 $M_\odot$, 
respectively for non-rotating, magnetic, and non-magnetic models.
We note that the CO core mass ranges of non-rotating and magnetic rotating models are consistent 
because the non-rotating 170 $M_\odot$ model, which is the largest mass to end up with
an incomplete explosion, has 77.5 $M_\odot$ CO core mass.
Therefore it can be said that the non-magnetic rotating models have
smaller minimum and maximum CO core masses for PISNe.

The shift of the initial mass range between the non-rotating and magnetic rotating models
is due to the core mass enhancement by rotation induced mixing.
This does not affect the explosion and thus the CO core mass ranges for PISN.
Meanwhile, the shift in the CO core mass range in non-magnetic rotating models
is resulted from the different envelope structure that these models have, that is,
the non-magnetic models develop inflated envelope during the evolutionary phases.

The reason of the lower shift of the minimum CO core will be because
the inflating envelope has a small binding energy that requires small explosion energy to be blown off.
This can be shown by comparing the non-magnetic 160 $M_\odot$ model,
which explodes with the explosion energy of $7.38 \times 10^{51}$ erg,
with the non-rotating 170 $M_\odot$ model,
which end up with an incomplete explosion but still has a positive total energy of $8.72 \times 10^{51}$ erg.
Here, it is noteworthy that the judgment of explosion can be inaccurate for low mass models,
although the discussion above will be qualitatively correct.
This is because there is no clear division between complete explosion
and incomplete explosion ejecting only the outer part of the star,
which is called pulsational-PISN (PPISN, \citealt{woosley17} and reference therein).
For example, the non-rotating 170 $M_\odot$ model firstly expands the whole part of the star
with a positive total energy of $8.72 \times 10^{51}$ erg, but later the central part turns back the motion
after $\sim$ 5,000 sec from the expansion.
The result thus can be affected from small changes in numerical treatments such as resolutions.

\begin{figure}[t]
	\includegraphics[width=\textwidth]{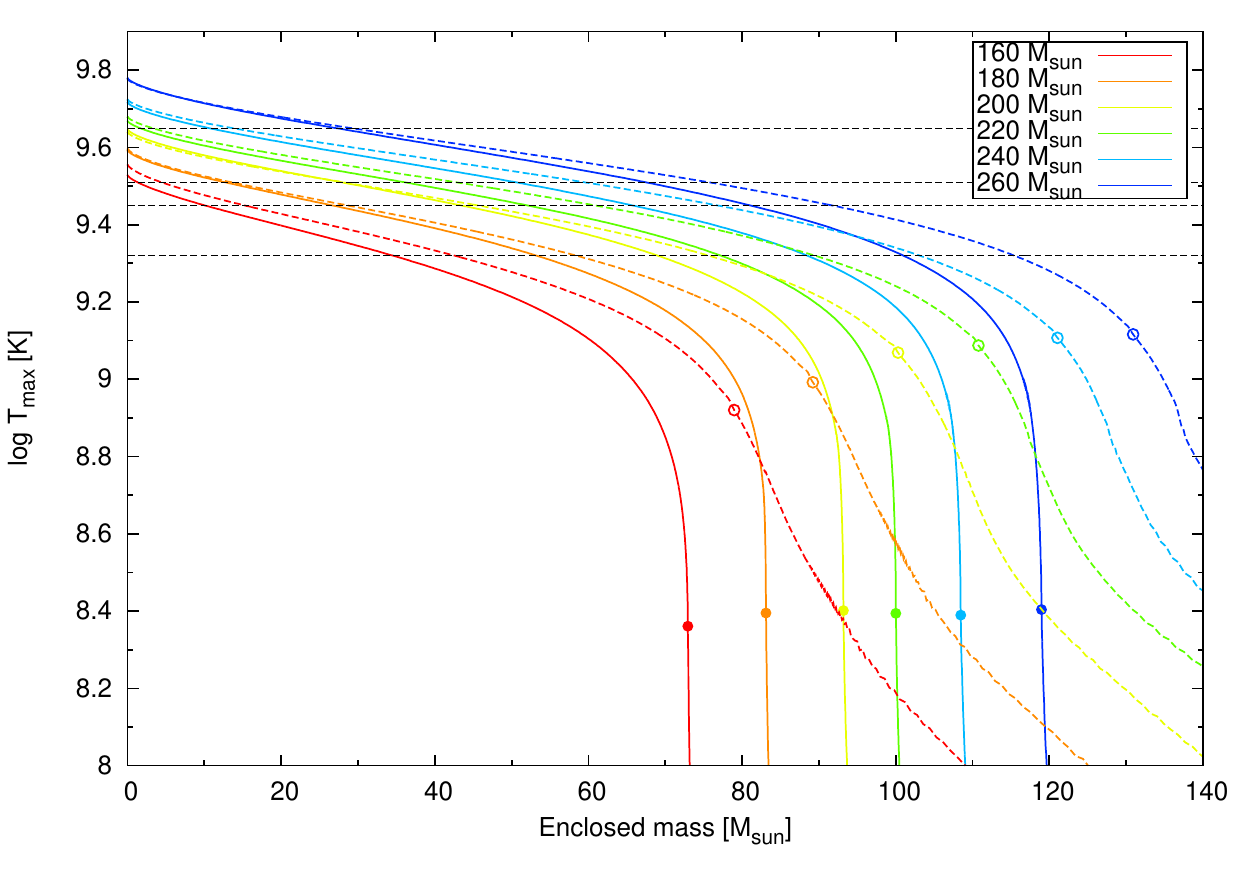}
	\caption{\footnotesize{Distributions of the local maximum temperature
	reaching during the explosion for the rotating models.
	Non-magnetic models are shown by solid lines
	and magnetic models are by dashed lines.
	Each point indicates the position of the CO core/He layer boundary.
	Black dashed lines show referential temperatures. See also Fig. \ref{fig-tmax}.}}
	\label{fig-tmax-o30}
\end{figure}
The maximum CO core mass for PISN is related to the maximum central temperature during the explosion.
This is because the mechanism to trigger the final collapse is the iron dissociation reactions,
which transforms the internal energy to the rest mass energy reducing the pressure,
and the reactions initiate with a high temperature of $\log T \rm{[K]} > 9.75$.
Therefore a massive model cannot stop the contracting motion
once the central temperature exceeds the critical value of $\log T \rm{[K]}s \sim 9.80$ \citep{takahashi+16}.
Figure \ref{fig-tmax-o30} shows distributions of maximum temperature during the explosion
for exploding rotating models.
As discussed above, non-magnetic models (shown by solid lines) develop
inflated envelopes during their evolution, whereas magnetic models (dashed lines) do not.
As a consequence, non-magnetic models develop steeper distributions of temperature and density
in their outer core regions, in order to connect with the low temperature base of the inflating envelope.
The figure shows that the steeper temperature distribution is kept during the explosion.
Consequently, a CO core with an inflated envelope has a smaller core mass than
a core surrounded by a deflated envelope for the same maximum central temperature.
This results in the smaller maximum core mass for models with inflated envelopes.

\section{Explosive nucleosynthesis in PISNe}

\subsection{Explosive nucleosynthesis in non-rotating models}

\begin{figure}[t]
	\includegraphics[width=\textwidth]{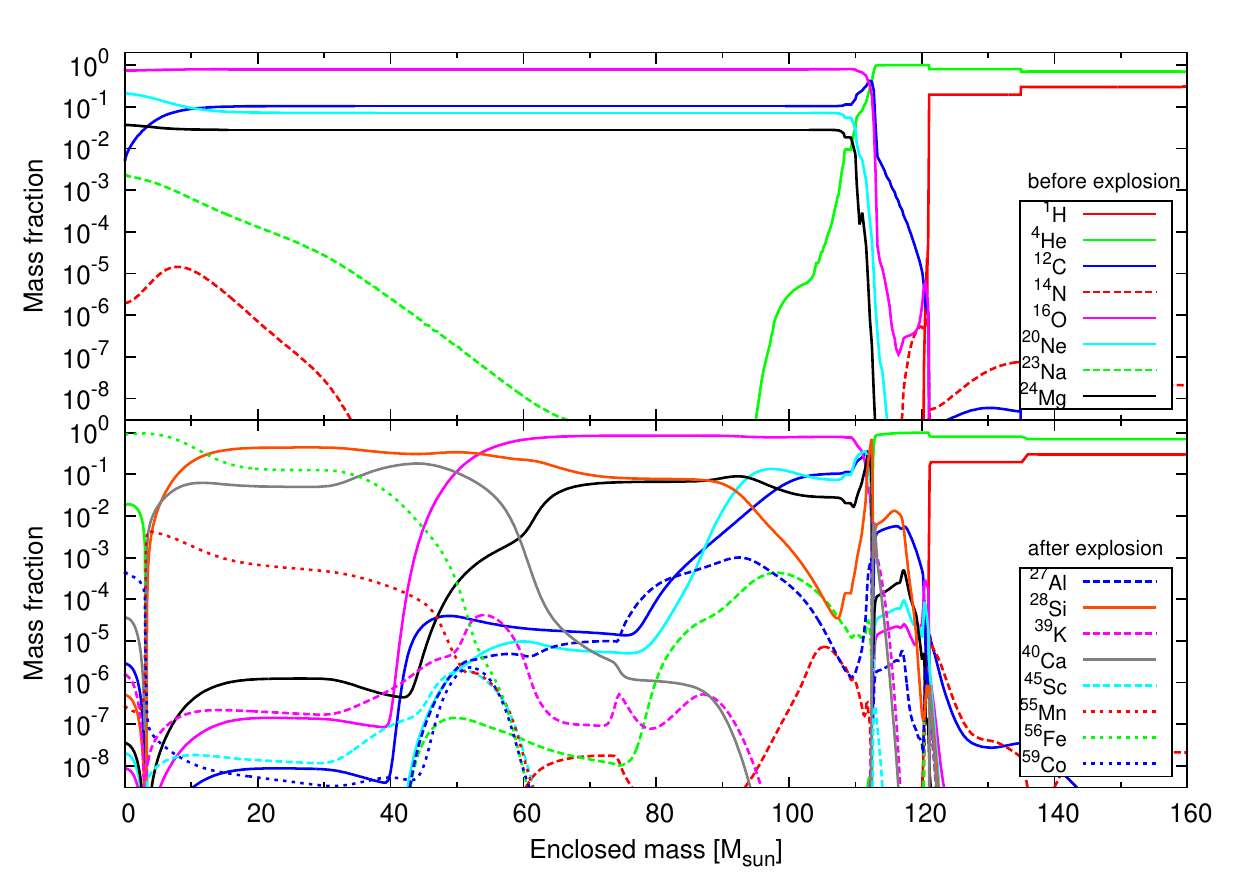}
	\caption{\footnotesize{
	Chemical distributions before and after the PISN explosion
	for the non-rotating 240 $M_\odot$ model.
	Decay calculation has been applied to the bottom panel.}}
	\label{fig-distexp-m240o00}
\end{figure}
\begin{table}[tbp]
	\begin{center}
	\caption{PISN Yields of the non-rotating 240 $M_\odot$ model \label{tab-yields-nonrot-m240}}
	\tabcolsep = 1mm

	\begin{tabular}{lccccccccc}
	\hline
	\hline
	Element	&	Yield			&	Element	&	Yield			&	Element	&	Yield			&	Element	&	Yield			&	Element	&	Yield			\\
	\hline                                   
  $^{}$p           &   5.712E+01 &  $^{}$d           &   5.318E-14 &  $^{3}$He         &   5.526E-04 &  $^{4}$He         &   7.220E+01 &  $^{6}$Li         &   1.632E-09 \\                        
  $^{7}$Li         &   6.881E-10 &  $^{9}$Be         &   2.633E-20 &  $^{10}$B         &   5.151E-15 &  $^{11}$B         &   2.658E-10 &  $^{12}$C         &   1.916E+00 \\                        
  $^{13}$C         &   3.366E-06 &  $^{14}$N         &   4.867E-05 &  $^{15}$N         &   2.444E-05 &  $^{16}$O         &   4.549E+01 &  $^{17}$O         &   1.377E-06 \\                        
  $^{18}$O         &   4.676E-08 &  $^{19}$F         &   1.152E-07 &  $^{20}$Ne        &   2.477E+00 &  $^{21}$Ne        &   2.379E-05 &  $^{22}$Ne        &   3.238E-05 \\                        
  $^{23}$Na        &   4.470E-03 &  $^{24}$Mg        &   2.918E+00 &  $^{25}$Mg        &   1.041E-03 &  $^{26}$Mg        &   2.993E-03 &  $^{27}$Al        &   1.337E-02 \\                        
  $^{28}$Si        &   2.203E+01 &  $^{29}$Si        &   2.057E-02 &  $^{30}$Si        &   2.395E-03 &  $^{31}$P         &   3.140E-03 &  $^{32}$S         &   1.678E+01 \\                        
  $^{33}$S         &   9.349E-03 &  $^{34}$S         &   8.008E-04 &  $^{36}$S         &   2.525E-09 &  $^{35}$Cl        &   3.526E-03 &  $^{37}$Cl        &   1.300E-03 \\                        
  $^{36}$Ar        &   3.704E+00 &  $^{38}$Ar        &   3.749E-04 &  $^{40}$Ar        &   2.627E-10 &  $^{39}$K         &   3.356E-03 &  $^{40}$K         &   2.430E-09 \\                        
  $^{41}$K         &   3.760E-04 &  $^{40}$Ca        &   4.259E+00 &  $^{42}$Ca        &   2.040E-05 &  $^{43}$Ca        &   2.022E-05 &  $^{44}$Ca        &   1.503E-03 \\                        
  $^{46}$Ca        &   9.004E-14 &  $^{48}$Ca        &   1.975E-13 &  $^{45}$Sc        &   1.664E-05 &  $^{46}$Ti        &   8.986E-06 &  $^{47}$Ti        &   1.216E-06 \\                        
  $^{48}$Ti        &   2.172E-02 &  $^{49}$Ti        &   6.228E-04 &  $^{50}$Ti        &   2.448E-11 &  $^{50}$V         &   8.710E-12 &  $^{51}$V         &   5.134E-04 \\                        
  $^{50}$Cr        &   6.368E-04 &  $^{52}$Cr        &   3.610E-01 &  $^{53}$Cr        &   1.402E-11 &  $^{54}$Cr        &   3.608E-10 &  $^{55}$Mn        &   4.202E-02 \\                        
  $^{54}$Fe        &   1.442E-01 &  $^{56}$Fe        &   1.254E+01 &  $^{57}$Fe        &   8.931E-02 &  $^{58}$Fe        &   3.389E-09 &  $^{59}$Co        &   9.443E-04 \\                        
  $^{58}$Ni        &   9.086E-02 &  $^{60}$Ni        &   2.360E-02 &  $^{61}$Ni        &   8.247E-04 &  $^{62}$Ni        &   4.654E-03 &  $^{64}$Ni        &   1.613E-15 \\                        
  $^{63}$Cu        &   1.847E-06 &  $^{65}$Cu        &   2.439E-06 &  $^{64}$Zn        &   1.861E-05 &  $^{66}$Zn        &   3.294E-05 &  $^{67}$Zn        &   2.638E-08 \\                        
  $^{68}$Zn        &   6.209E-09 &  $^{70}$Zn        &   1.093E-28 &  $^{69}$Ga        &   2.504E-09 &  $^{71}$Ga        &   1.940E-10 &  $^{70}$Ge        &   6.648E-08 \\                        
  $^{72}$Ge        &   9.794E-11 &  $^{73}$Ge        &   1.962E-13 &  $^{74}$Ge        &   6.135E-26 &  $^{75}$As        &   6.869E-16 &  $^{74}$Se        &   1.942E-14 \\                        
  $^{76}$Se        &   6.690E-19 &  $^{77}$Se        &   2.337E-22 &  $^{78}$Se        &   3.754E-27 &  $^{79}$Br        &   2.192E-28 \\                                                          
	\hline\\
	\end{tabular}
	\end{center}

	{\footnotesize {\bf Notes.} Yields of the non-rotating 240 $M_\odot$ model.
	Yields of other models are also tabulated in Appendix \ref{app-1}. }
\end{table}
As an example, change of the chemical distribution before and after the PISN explosion
is shown in Fig.\ref{fig-distexp-m240o00} for the non-rotating 240 $M_\odot$ model.
Besides the explosive nucleosynthesis,
decay calculation has been applied to show the distribution after the explosion.
The yields of the model are summarized in Table \ref{tab-yields-nonrot-m240}.

Alpha elements, $^{12}$C, $^{16}$O, $^{20}$Ne, and $^{24}$Mg,
which originally compose the CO core, remain at the edge region after the explosion.
All of these isotopes except for helium are synthesized during core helium burning.
At the inner core, odd-elements of $^{23}$Na and $^{27}$Al
have already synthesized before the explosion.
They are carbon burning products,
but soon are burnt by further burning processes.

The outer core region of $\sim$ 91--112 $M_\odot$ has
a low maximum temperature during the explosion of $\log T \rm{[K]} < 9.32$.
While original CO core materials remain to be the major yields of the region,
alpha particles are almost completely captured during the explosion
to synthesize an alpha-element of $^{28}$Si and moreover
odd-Z elements of $^{23}$Na and $^{27}$Al.

In the middle region of $\sim$ 62--91 $M_\odot$,
carbon and neon are burnt to synthesize abundant $^{28}$Si, $^{23}$Na, and $^{27}$Al
with the maximum temperature during the explosion of $\log T \rm{[K]} < 9.45$.
Furthermore, $^{24}$Mg burns as well in a inner region of $\sim$ 45--62 $M_\odot$,
producing intermediate-mass elements including $^{40}$Ca and
odd-Z elements of $^{39}$K and $^{45}$Sc.

The inner region of $\sim$ 3.1--45 $M_\odot$ has
a high maximum temperature during the explosion of $\log T \rm{[K]} > 9.51$.
The high temperature allows $^{16}$O to burn to synthesize
intermediate-mass elements from silicon to scandium
and iron-peak elements from titanium to iron.
The odd-Z iron-peak elements such as
$^{51}$V, which is firstly synthesized as $^{51}$Cr and $^{51}$Mn,
and $^{55}$Mn, which is as $^{55}$Fe and $^{55}$Co,
are produced in this region.

At the inner most region of $\sim 3.1$ $M_\odot$,
the maximum temperature during the explosion exceeds $\log T \rm{[K]} > 9.65$.
More than 90\% of the material are finally synthesized into $^{56}$Fe,
which is a decay product of $^{56}$ Ni.
Small amount of isotopes of cobalt (mostly in the form of $^{59}$Co),
nickel, and copper are synthesized in this region as well.

The CO core of a non-rotating 240 $M_\odot$ model is 
surrounded by the shell helium region of $\sim$ 112--121 $M_\odot$.
In spite of the low maximum temperature of $\log T \rm{[K]} <  9.13$,
various isotopes are synthesized in this region
as a result of alpha capture reactions.
Yields of most of these isotopes are actually far from abundant
compared with the yields from the inner CO core.
However, there are exceptions.
I.e., more than 90\% of yields of odd-Z elements of $^{35}$Cl and $^{39}$K
are produced at the base of the helium layer.

\begin{figure}[t]
	\includegraphics[width=\textwidth]{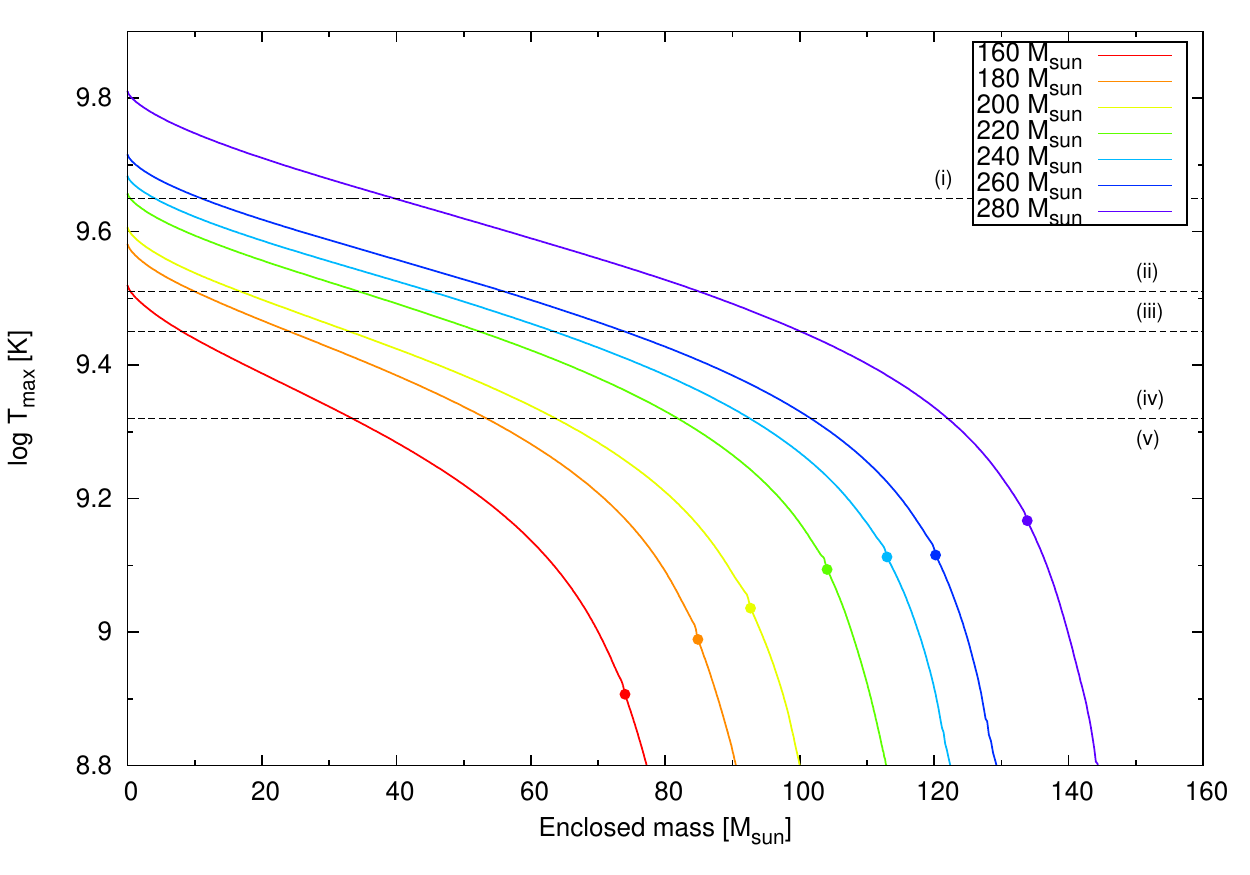}
	\caption{\footnotesize{Distributions of the local maximum temperature
	reaching during the explosion for the non-rotating models.
	Each point indicates the position of the CO core/He layer boundary.
	Dashed lines show referential temperatures. See text for definitions.}}
	\label{fig-tmax}
\end{figure}
In the end, explosive nucleosynthesis in a PISN yields results in 
mostly the same chemical abundance in a region with the same local maximum temperature.
The local maximum temperature during the explosion
is shown by Fig.\ref{fig-tmax} for exploding non-rotating models.
It is informative to divide a core into parts that have maximum temperatures of
(i) $\log T \rm{[K]} > 9.65$; a region yielding heavy iron-peak elements,
(ii) $\log T \rm{[K]} \in [9.51, 9.65]$; a region yielding oxygen burning products,
(iii) $\log T \rm{[K]} \in [9.45, 9.51]$; a region yielding intermediate-mass elements up to scandium,
(iv) $\log T \rm{[K]} \in [9.32, 9.45]$; a region yielding intermediate-mass elements up to silicon, and
(v) $\log T \rm{[K]} < 9.32$; a region yielding CO core materials.
Then it is the mass ratio among these regions in the CO core
that basically determines the abundance pattern of a PISN.
In addition to the yields from the core, yield in a helium layer can be important 
for some odd-Z elements.

\begin{figure}[t]
	\includegraphics[width=\textwidth]{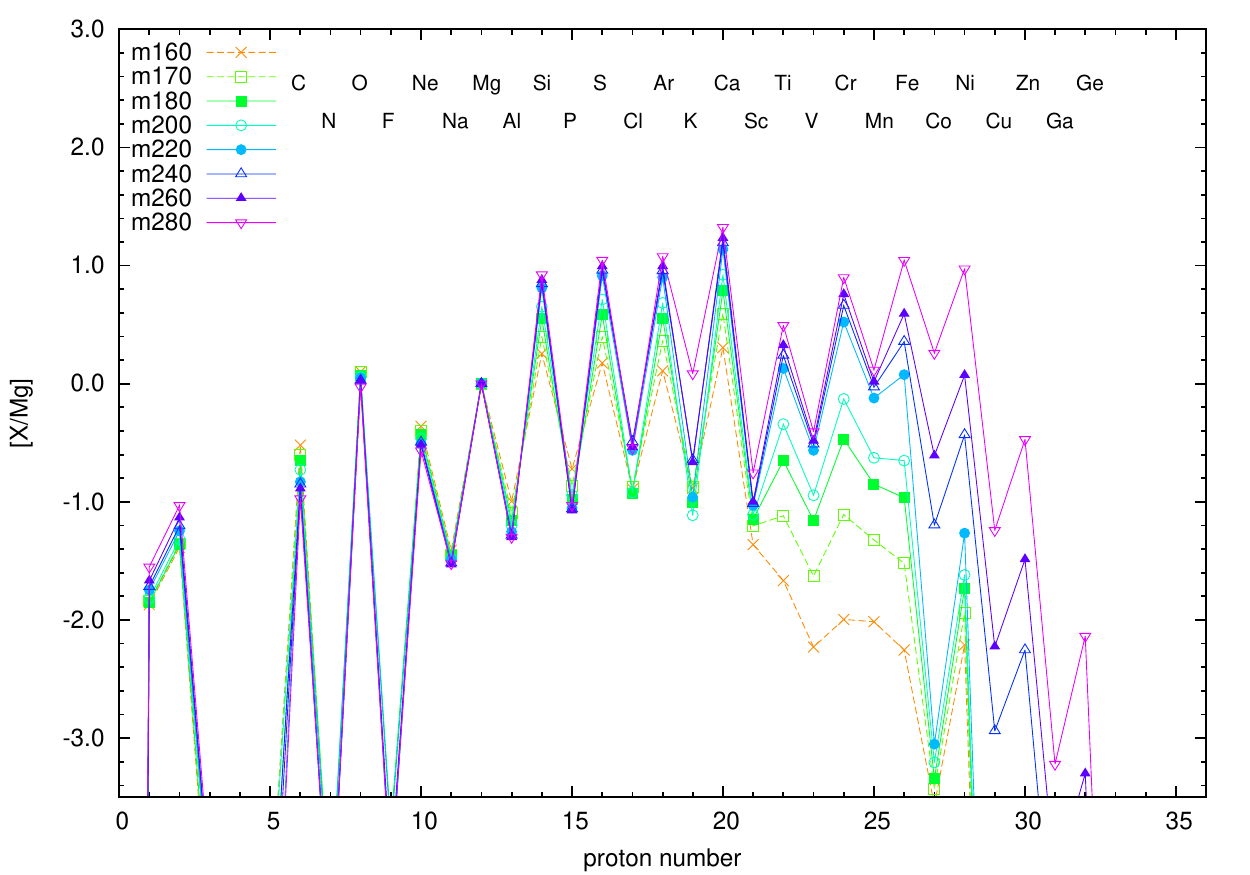}
	\caption{\footnotesize{Abundance patterns of PISN yields normalized by the magnesium yield.
	Results of non-rotating models are shown.
	Note that 160 and 170 $M_\odot$ models end up with incomplete explosions, so that
	the dotted lines are not correct abundance patterns.}}
	\label{fig-abundance}
\end{figure}
\begin{figure}[t]
	\includegraphics[width=\textwidth]{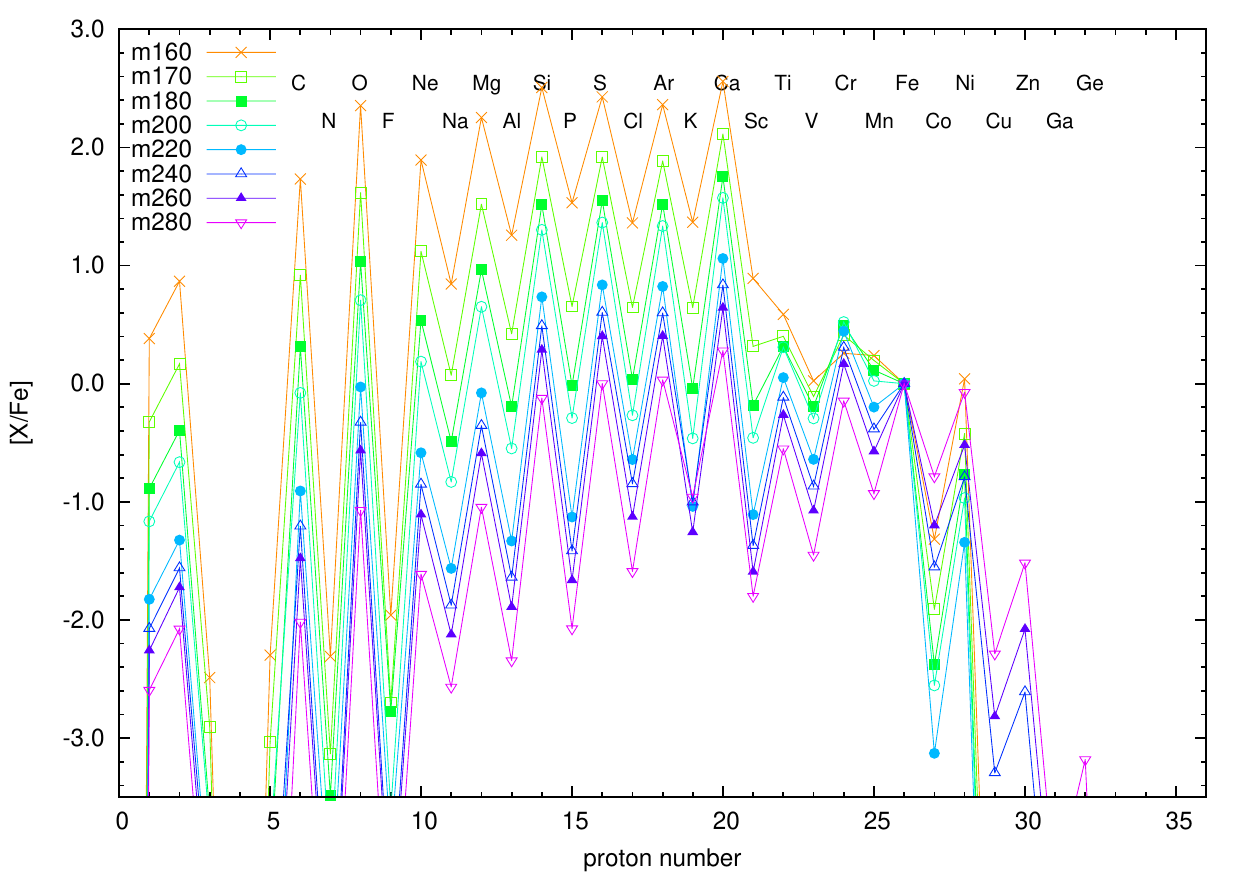}
	\caption{\footnotesize{Abundance patterns of non-rotating PISN yields normalized by the iron yield.}}
	\label{fig-abundance-fe}
\end{figure}
Fig.\ref{fig-abundance} shows abundance patterns of PISN yields of non-rotating models.
Note that results of 160 and 170 $M_\odot$ models are included in this figure,
although they do not show complete explosions.
We select the magnesium yield, rather than iron, as a denominator in the figure.
This is because all of PISN models eject abundant magnesium,
the magnesium yield has a small dependency on the CO core mass,
and moreover it is one of the most accessible element in observing MP stars.
The overall patterns are much scattered if iron yield is used as the reference instead (Fig.\ref{fig-abundance-fe}).

A well-known peculiarity of PISN yields can be seen
as the pronounced variance between odd-Z and even-Z elemental yields,
which discriminates PISN yields from the usual CCSN yields \citep{heger&woosley02, umeda&nomoto02}.
The odd-even variance is due to the low neutron excess, or the high $Y_e \sim 0.5$, of PISN ejecta.
Since the overturn from the collapse takes place within a short timescale,
electron capture reactions are too slow to change the core $Y_e$.
With the low neutron excess, the explosive nucleosynthesis favorably synthesizes even-Z elements.

The most important property for lighter elements from carbon to aluminum
is the similarity in the patterns among the models with different initial masses.
Scatters of the abundance ratios are especially small for four elements of
[O/Mg] = $-0.02$--0.06,
[Ne/Mg] = $-0.56$--$-0.42$,
[Na/Mg] = $-1.53$--$-1.45$, and
[Al/Mg] = $-1.30$--$-1.15$.
Especially, the small abundance ratios seen in odd-Z elements of sodium and aluminum
can be considered as representatives of the distinctive odd-even variation,
and to be used to discriminate PISN yields from others.

The opposite trends for the initial mass are found
between even- and odd-Z intermediate-mass elements.
That is, abundance ratios of even-Z elements
increase with increasing the mass of the progenitor, spanning, e.g., 
[Si/Mg] = 0.54--0.92 and
[Ca/Mg] = 0.78--1.32.
Contrastingly, rations of odd-Z elements basically decrease with increasing the mass, e.g.,
[P/Mg] = $-1.07$--$-0.98$ for the same mass range.
The reason why massive models of $>$ 220 $M_\odot$, or $>$ 240 $M_\odot$,
produce more abundant chlorine or potassium is due to the nucleosynthesis in helium layers.
These ratios range
[Cl/Mg] = $-0.92$--$-0.49$ and
[K/Mg] = $-1.11$--0.08.

Heavier elements from titanium to germanium are synthesized in the innermost region of the star.
Because less massive models explode without entering this high temperature regime,
the mass dependence of yields of these elements becomes large.
Thus, the lowest mass model yields smallest abundances of
[Fe/Mg] = $-0.96$,
[Co/Mg] = $-3.34$,
[Ni/Mg] = $-1.73$, and
[Zn/Mg] = $-8.82$.
In contrast, the highest mass model yields large amount of
[Fe/Mg] = 1.04,
[Co/Mg] = 0.25,
[Ni/Mg] = 0.97, and
[Zn/Mg] = $-0.47$.
Another characteristic pattern in this range is the steep decline above Z $>$ 28,
which can be indicated by the small abundance ratios of [Zn/Fe] or [Zn/Ni].
This is due to the low maximum temperature of the explosion \citep{umeda&nomoto02}.
For the same reason, even the most massive model produces heavy isotopes
only up to germanium (A $< \sim 70$) with [Ge/Mg] = $-2.13$,
and productions of heavier elements are negligible, [As, Se/Mg] $< -8$.

\subsection{Yields of rotating models}

Distributions of maximum temperature during the explosion
have already been shown in Fig.\ref{fig-tmax-o30} for rotating models.
Because of the similar envelope structure,
PISNe of magnetic rotating models yield
similar abundance patterns to non-rotating models.
On the other hand, non-magnetic rotating models have
low maximum temperature in the outer cores and the envelopes.
This affects the PISN nucleosynthesis.

\begin{figure}[t]
	\includegraphics[width=\textwidth]{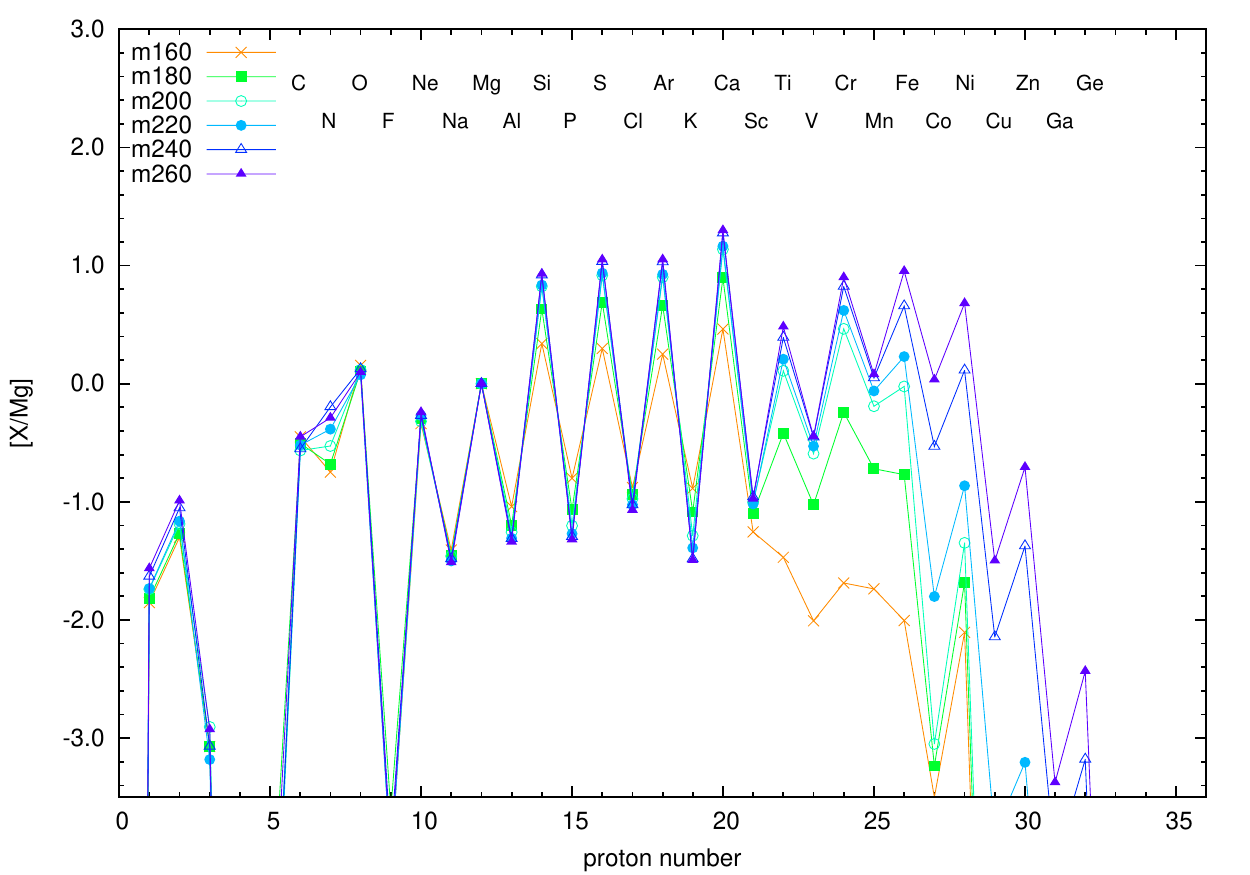}
	\caption{\footnotesize{Abundance patterns of non-magnetic rotating models.}}
	\label{fig-abundance-o30n}
\end{figure}
Figure \ref{fig-abundance-o30n} shows abundance patterns of PISN yields of
the non-magnetic rotating models.
Overall properties are still very similar to non-rotating results.
However, abundance ratios of odd-Z elements of [Cl/Mg] and [K/Mg] now
show clear decreasing trends towards increasing mass,
because elemental productions in a helium layer disappears in this case.
Besides, slightly higher abundances are obtained for lighter elements of carbon and neon.
The lower temperature at the core edge allows these elements to survive the explosion.
In addition to the high nitrogen abundance owing to the convective dredge-up episode during the evolutionary phase,
such slight modifications may characterize yields of PISNe from rotating progenitors.

In summary, the non-magnetic models;
[O/Mg] = 0.07--0.15,
[Ne/Mg] = $-0.33$--$-0.24$,
[Na/Mg] = $-1.51$--$-1.40$, and
[Al/Mg] = $-1.33$--$-1.04$ for lighter elements,
[Si/Mg] = 0.33--0.92,
[P/Mg] = $-1.31$--$-0.80$,
[Cl/Mg] = $-1.06$--$-0.87$,
[K/Mg] = $-1.49$--$-0.88$,
[Ca/Mg] = 0.46--1.29 for intermediate-mass elements, and
[Fe/Mg] = $-2.00$--0.95,
[Co/Mg] = $-3.23$--0.03,
[Ni/Mg] = $-1.68$--0.68,
[Zn/Mg] = $-8.72$--$-0.70$ for iron-peak elements.

\section{Comparison between PISN yields and surface abundances of MP stars}

\subsection{General trends of observed abundance ratios}

\begin{table}[tbp]
	\begin{center}
	\caption{Number of stars in {\it SAGA} database \label{tab-elements}}
	\tabcolsep = 1mm

	\begin{tabular}{lc}
	\hline
	\hline
	element	& number of stars \\
	\hline
	Na		&	2445\\
	Mg		&	3619	\\
	Al		&	2121	\\
	Si		&	2591	\\
	P		&	0	\\
	S		&	363	\\
	Cl		&	0	\\
	Ar		&	0	\\
	K		&	260	\\
	Ca		&	3617\\
	Sc		&	1389	\\
	Ti		&	3324	\\
	V		&	943	\\
	Cr		&	2413	\\
	Mn		&	1395\\
	Fe		&	4108	\\
	Co		&	1265	\\
	Ni		&	2808	\\
	Cu		&	600	\\
	Zn		&	1801\\
	\hline\\
	\end{tabular}
	\end{center}

	{\footnotesize {\bf Notes.} Numbers of stars compiled in {\it SAGA}
	database\footnote{Numbers are obtained by Nov. 8, 2017 updated version.},
	in which the abundance is observed for each element.}
\end{table}

The purpose of this work is to find abundance ratios
that is useful to discriminate a possible candidate of PISN children,
which we tentatively refer to as a PISN-MP
star\footnote{A PISN-MP star may be defined such that a high fraction,
say $>$ 90\%, of its metal is originated from PISNe \citep[e.g.,][]{karlsson+08}.}
from the other normal metal-poor stars.
Such ratios are desired to be selected from easily accessible elements by observations.
Table\ref{tab-elements} shows numbers of stars compiled in {\it SAGA} database,
for which the abundance of each element is determined.
This indicates that the most accessible elements in observations of MP stars
are magnesium, calcium, titanium, and iron, then
followed by sodium, aluminum, silicon, chromium, nickel, and zinc.
Among them, magnesium is selected as the base, or the denominator,
of the abundance ratios for the investigation in this work.
This is because the ratio facilitates comprehensive comparisons
of the theoretical yields of PISNe as shown above.

\begin{figure}[tbp]
	\includegraphics[width=\textwidth]{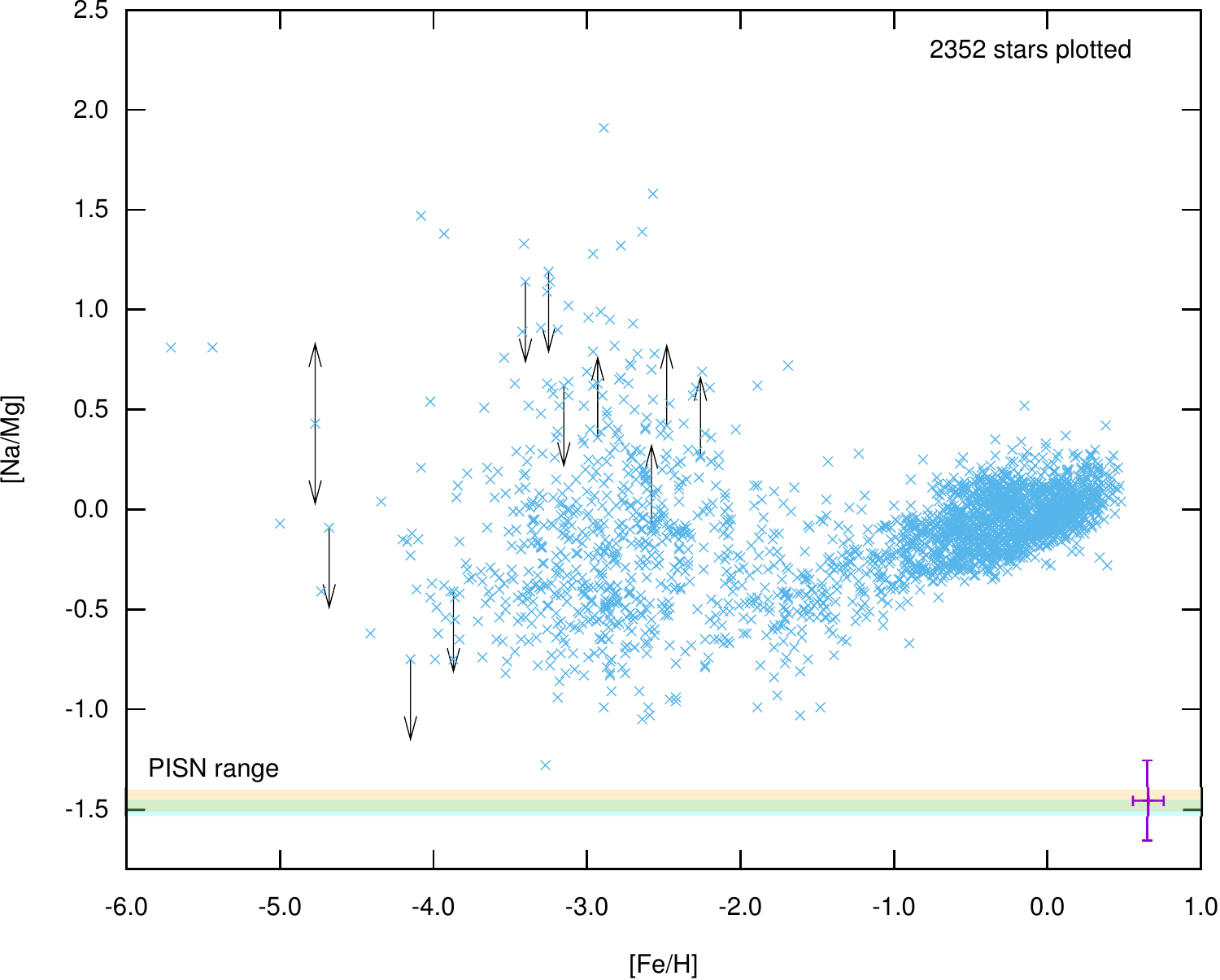}
	\caption{ \footnotesize{[Na/Mg] versus [Fe/H] collected from the {\it SAGA} database.
	Stellar data are plotted by points.
	The typical errors of $\pm$0.2 dex for [Na/Mg] and $\pm$0.1 dex for [Fe/H] are shown by the purple cross.
	The ranges of theoretical yields by changing the initial mass are shown by cyan and orange bands
	for non-rotating and non-magnetic results respectively.}
	\label{fig-Na-Fe} }
\end{figure}
Among the peculiar abundance patterns of the PISN yields,
the ratio of [Na/Mg] is found to be the most important
for comparison between the observations and the theoretical yields.
In Fig.\ref{fig-Na-Fe}, observed ratios are plotted as a function of [Fe/H]
and the theoretical range are overlaid as cyan (non-rotating) and orange (non-magnetic) bands.
Typical errors of $\pm$0.2 dex for [Na/Mg] and $\pm$0.1 dex for [Fe/H] are indicated by the purple cross. 
Although there are some EMP stars showing more than 2 dex enhancement in the [Na/Mg] ratio,
the main component including the decreasing trend toward low [Fe/H] has been well reproduced by
galactic chemical evolution models taking ejecta of core-collapse supernovae into account \citep[e.g.][]{kobayashi+06}.
Interestingly, despite the figure includes more than 2000 stellar data,
no stars are found within the theoretical band of
the narrow and nearly mass independent ratios of $-1.53$--$-1.40$.

Additional pollution of CCSN yields of $\sim$ 10\% in mass increases the [Na/Mg] ratio,
but this effect will be too weak to shift the ratio to explain the deviation.
PISN models in this work have [Na/Mg] $\sim$ -1.5 and $\sim$ 1.5\% in mass of the yield is magnesium.
Assuming a Pop III CCSN has [Na/Mg] $\sim$ -0.6 and $\sim$ 1\% in mass of the yield is magnesium
\citep{umeda&nomoto05, kobayashi+06}, the $\sim$ 10\% pollution merely increases the ratio up to [Na/Mg] $\sim$ -1.33.
Therefore we conclude that no PISN-MP stars are found from the currently observed MP stars,
which have been compiled in the {\it SAGA} database.
This results also indicates that a PISN-MP star can be discriminated by its small [Na/Mg] ratio.
Possibly the lack of stars in the low [Na/Mg] region is explained as an observational limit.
A PISN-MP star, if observation achieves to detect its low sodium abundance,
will be a rare exception passing this test.

Unfortunately, abundance ratios made by iron-peak elements of titanium, chromium, iron, and nickel
are found to be unprofitable for discrimination of a candidate of PISN children.
The fundamental reason is that the most commonly observed values of [X/Mg] = 0
is included in the wide band of the theoretical scattering.
For example, the theoretical range of [Ti/Mg] spans from $-1.46$ to $0.49$.
As a result, majority of observations is included in this theoretical band.
For other accessible elements of aluminum, silicon, calcium, and zinc,
moderate numbers (several to tens) of stars are found inside the theoretical bands.
Then, as a next step, comparisons using combinations of them are conducted.

\begin{figure}[tbp]
	\includegraphics[width=\textwidth]{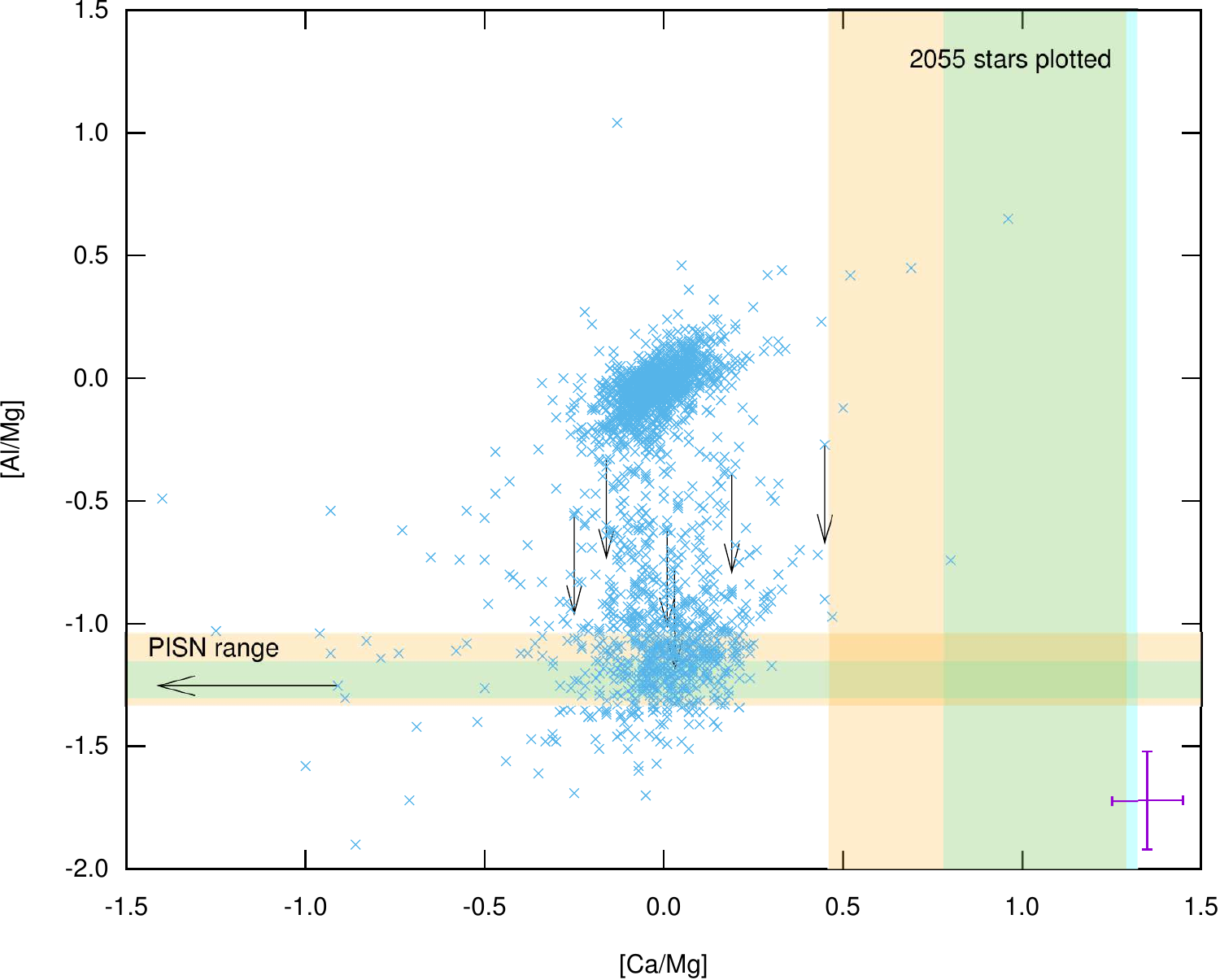}
	\caption{ \footnotesize{Same as Fig.~\ref{fig-Na-Fe} but for [Al/Mg] versus [Ca/Mg].
	The typical errors of $\pm$0.2 dex for [Al/Mg] and $\pm$0.1 dex for [Ca/Mg] are shown by the purple cross.}
	\label{fig-Al-Ca} }
\end{figure}
In Fig.\ref{fig-Al-Ca}, stellar data are plotted on a plane of [Ca/Mg] and [Al/Mg] .
As well as sodium, the main evolution sequences of calcium and aluminum
have been well explained by CCSN ejecta \citep{kobayashi+06}.
On the other hand, for the PISN yields,
no stars show comparable values of both ratios at the same time.
This is due to the large offset of the main stream of [Ca/Mg] from the theoretical band of PISN yields.
In other words, the requirement of [Ca/Mg] = 0.46 -- 1.32 is too large for the current stellar samples to agree with.
A PISN-MP star thus will show a much higher [Ca/Mg] ratio than the other normal MP stars.

\subsection{Detailed comparisons with metal-poor stars}

\begin{table}[t]
\caption{Stellar sample	 \label{tab-stars-pisn}}
\scriptsize
\centering
\tabcolsep = 1mm
\begin{tabular}{rl rrrrrrr lll}
\hline
\hline
\# & Object & [Fe/H] & [Na/Mg] & [Al/Mg] & [Si/Mg] & [Ca/Mg] & [Sc/Mg] & [Zn/Fe] & Reference &  \\
\hline
1 & BD-01\_2582 & -2.62 & -0.27 & -1.16 & 0.32 & 0.00 & -0.54 & 0.33 & 1. &  \\
2 & BS16085-0050 & -2.91 & - & -1.14 & 0.32 & -0.23 & -0.31 & - & \citet{honda+04} &  \\
3 & BS16928-053 & -3.07 & -0.15 & -1.15 & 0.93 & -0.04 & -0.41 & - & \citet{Lai+08} &  \\
4 & CS22180-014 & -2.86 & -0.54 & -1.26 & 0.22 & -0.02 & -0.51 & $<$ -2.12 & 1. &  \\
5 & CS22183-031 & -3.57 & - & -1.35 & 0.32 & -0.03 & -0.66 & 0.67 & 1. &  \\
6 & CS22186-002 & -2.72 & -0.52 & -1.14 & 0.29 & 0.11 & -0.39 & 0.23 & 1. &  \\
7 & CS22189-009 & -3.92 & -0.54 & -1.24 & 0.40 & -0.06 & -0.63 & 0.36 & 1. &  \\
8 & CS22877-011 & -3.23 & - & -1.26 & 0.40 & -0.06 & -0.34 & 0.28 & 1. &  \\
9 & CS22878-101 & -3.53 & - & -1.19 & 0.22 & -0.11 & -0.65 & 0.62 & 2. &  \\
10 & CS22879-103 & -2.16 & - & -1.12 & 0.23 & -0.06 & -0.48 & 0.38 & 2. &  \\
11 & CS22886-044 & -1.86 & - & -1.29 & 0.33 & 0.14 & -0.25 & 0.09 & 1. &  \\
12 & CS22888-002 & -2.93 & - & -1.26 & 0.40 & 0.05 & -0.51 & $<$ -2.13 & 1. &  \\
13 & CS22891-200 & -4.06 & - & -1.19 & 0.23 & -0.05 & -1.09 & $<$ -3.19 & 1. &  \\
14 & CS22893-005 & -2.99 & -0.56 & -1.05 & 0.27 & -0.01 & -0.47 & $<$ -2.48 & 1. &  \\
15 & CS22894-019 & -2.98 & - & -1.13 & 0.33 & 0.02 & -0.33 & $<$ -1.43 & 2. &  \\
16 & CS22894-049 & -2.84 & - & -1.14 & 0.31 & 0.02 & -0.36 & $<$ -2.13 & 1. &  \\
17 & CS22896-015 & -2.85 & -0.57 & -1.20 & 0.25 & -0.03 & -0.47 & 0.13 & 1. &  \\
18 & CS22896-110 & -2.85 & - & -1.13 & 0.27 & -0.03 & -0.57 & 0.22 & 1. &  \\
19 & CS22896-136 & -2.41 & - & -1.02 & 0.29 & 0.18 & -0.36 & 0.27 & 1. &  \\
20 & CS22898-047 & -3.51 & - & -1.25 & 0.28 & -0.02 & -0.71 & 0.66 & 1. &  \\
21 & CS22941-017 & -3.11 & -0.57 & -1.19 & 0.28 & -0.06 & -0.44 & 0.24 & 2. &  \\
22 & CS22942-002 & -3.61 & - & -1.31 & 0.26 & -0.16 & -0.62 & 0.53 & 1. &  \\
23 & CS22942-011 & -2.88 & -0.45 & -1.22 & 0.45 & -0.10 & -0.49 & 0.30 & 1. &  \\
24 & CS22943-095 & -2.52 & - & -1.12 & 0.28 & 0.11 & -0.31 & 0.42 & 2. &  \\
25 & CS22943-132 & -2.63 & - & -1.17 & 0.24 & -0.03 & 0.24 & 0.04 & 2. &  \\
26 & CS22944-032 & -3.22 & - & -1.36 & 0.30 & -0.09 & -0.59 & 0.30 & 1. &  \\
27 & CS22945-028 & -2.92 & -0.56 & -1.14 & 0.34 & -0.04 & -0.64 & 0.17 & 1. &  \\
28 & CS22947-187 & -2.58 & -0.40 & -1.14 & 0.24 & -0.05 & -0.45 & 0.27 & 1. &  \\
29 & CS22949-048 & -3.55 & - & -1.20 & 0.40 & -0.02 & -0.51 & 0.88 & 1. &  \\
30 & CS22950-046 & -4.12 & - & -1.21 & 0.50 & -0.09 & -0.69 & $<$ -3.23 & 1. &  \\
31 & CS22951-059 & -2.84 & - & -1.11 & 0.37 & -0.01 & -0.14 & 0.22 & 2. &  \\
32 & CS22953-003 & -3.13 & -0.42 & -1.19 & 0.55 & -0.06 & -0.62 & 0.37 & 1. & \\
33 & CS22956-050 & -3.67 & - & -1.24 & 0.49 & -0.03 & -0.85 & 0.59 & 1. &  \\
34 & CS22956-062 & -2.75 & -0.69 & -1.24 & 0.25 & -0.05 & -0.74 & $<$ -2.19 & 1. &  \\
35 & CS22956-114 & -3.19 & - & -1.13 & 0.20 & 0.02 & -0.51 & $<$ -2.48 & 1. &  \\
36 & CS22957-019 & -2.43 & - & -1.12 & 0.29 & 0.12 & -0.20 & 0.14 & 2. &  \\
37 & CS22957-022 & -3.28 & - & -1.09 & 0.34 & -0.06 & -0.56 & 0.44 & 1. &  \\
38 & CS22958-083 & -3.05 & -0.52 & -1.03 & 0.69 & -0.18 & -0.69 & 0.67 & 1. & \\
39 & CS22968-029 & -3.10 & - & -1.10 & 0.62 & 0.03 & -0.56 & $<$ -2.17 & 2. & \\
40 & CS29502-092 & -2.76 & -0.27 & -1.10 & 0.53 & -0.06 & -0.35 & -0.10 & 2. & \\
41 & CS29517-042 & -2.53 & - & -1.08 & 0.31 & 0.14 & -0.26 & 0.25 & 1. &  \\
42 & CS30312-059 & -3.41 & - & -1.28 & 0.45 & -0.05 & -0.67 & 0.49 & 1. &  \\
43 & CS30325-094 & -3.17 & - & -1.15 & 0.38 & -0.18 & -0.21 & 0.51 & \citet{Aoki+05} &  \\
44 & CS30339-073 & -3.93 & -0.38 & -1.26 & 0.37 & -0.08 & -0.66 & $<$ -2.74 & 1. &  \\
45 & G25-24 & -2.11 & - & -1.31 & 0.14 & 0.02 & -0.43 & -0.11 & 1. &  \\
46 & HD110184 & -2.52 & - & -1.03 & 0.30 & 0.10 & -0.25 & 0.06 & \citet{honda+04}, \citet{Roederer+10} \\
47 & HD126587 & -3.29 & -0.55 & -1.15 & 0.23 & -0.08 & -0.59 & 0.43 & 1. &  \\
48 & HD175606 & -2.39 & -0.40 & -1.04 & 0.20 & 0.16 & -0.26 & 0.37 & 1. &  \\
49 & HE0048-6408 & -3.75 & -0.39 & -1.45 & 0.17 & -0.16 & -0.65 & - & \citet{Placco+14} &  \\
50 & HE0056-3022 & -3.77 & -0.35 & -1.26 & 0.41 & -0.06 & -0.75 & 0.49 & 1. &  \\
\hline
&&&&&&&&&&
\end{tabular}\\{\footnotesize {\bf Notes.} References are:
1. \citet{roederer+14a},
2. \citet{roederer+14b}, and
3. \citet{jacobson+15}.}
\end{table}
\begin{table}[t]

\caption{Stellar sample}
\scriptsize
\centering
\tabcolsep = 1mm
\begin{tabular}{rl rrrrrrr lll}
\hline
\hline
\# & Object & [Fe/H] & [Na/Mg] & [Al/Mg] & [Si/Mg] & [Ca/Mg] & [Sc/Mg] & [Zn/Fe] & Reference &  \\
\hline
51 & HE0057-4541 & -2.36 & - & -1.08 & 0.31 & -0.10 & -0.37 & - & \citet{SiqueiraMello+14} &  \\
52 & HE0105-6141 & -2.58 & - & -1.07 & 0.31 & 0.01 & -0.20 & - & \citet{SiqueiraMello+14} &  \\
53 & HE0109-4510 & -2.96 & -0.39 & -1.20 & 0.45 & -0.03 & -0.22 & $<$ -2.16 & \citet{Hansen+15} &  \\
54 & HE0302-3417A & -3.70 & - & -1.37 & 0.50 & -0.09 & -0.51 & 0.42 & \citet{hollek+11} &  \\
55 & HE1320-2952 & -3.69 & -0.31 & -1.00 & 0.28 & -0.09 & -0.42 & - & \citet{Yong+13a} &  \\
56 & HE2302-2154A & -3.88 & - & -1.30 & 0.26 & 0.01 & -0.50 & 0.70 & \citet{hollek+11} &  \\
57 & SDSSJ082511+163459 & -3.22 & - & -1.00 & 0.23 & 0.19 & - & - & \citet{Caffau+11b} &  \\
58 & SMSSJ0106-5244 & -3.79 & -0.27 & -1.16 & 0.37 & -0.10 & -0.75 & - & 3. &  \\
59 & SMSSJ0224-5346 & -3.40 & 0.18 & -1.01 & 0.38 & -0.32 & -0.59 & - & 3. &  \\
60 & SMSSJ0342-2842 & -2.33 & -0.12 & -1.25 & 0.31 & 0.01 & -0.42 & -0.33 & 3. &  \\
61 & SMSSJ0617-6007 & -2.72 & 0.02 & -1.04 & 0.36 & -0.05 & -0.59 & 0.12 & 3. &  \\
62 & SMSSJ0702-6004 & -2.62 & 0.15 & -1.04 & 0.27 & -0.10 & -0.72 & 0.20 & 3. &  \\
63 & SMSSJ1358-1509 & -2.58 & 0.08 & -1.06 & 0.62 & 0.03 & -0.48 & - & 3. &  \\
64 & SMSSJ1511-1821 & -2.71 & - & -1.24 & 0.36 & -0.05 & -0.47 & - & 3. &  \\
65 & SMSSJ1750-4146 & -2.76 & 0.03 & -1.23 & 0.34 & 0.03 & -0.58 & - & 3. &  \\
66 & SMSSJ1757-4548 & -2.46 & -0.11 & -1.09 & 0.25 & -0.06 & -0.57 & - & 3. &  \\
67 & SMSSJ1832-3434 & -3.00 & -0.03 & -1.07 & 0.28 & -0.08 & -0.72 & - & 3. &  \\
68 & SMSSJ1905-2149 & -3.11 & 0.00 & -1.19 & 0.43 & 0.01 & -0.48 & - & 3. &  \\
69 & SMSSJ1944-7205 & -2.43 & -0.07 & -1.24 & 0.42 & -0.04 & -0.76 & 0.11 & 3. &  \\
70 & SMSSJ2002-5331 & -3.22 & 0.06 & -1.48 & 0.24 & -0.30 & -0.37 & 0.71 & 3. &  \\
71 & SMSSJ2158-6513 & -3.41 & 1.33 & -1.41 & 0.49 & 0.07 & -0.60 & - & 3. &  \\
\hline
 & Aoki star &  &  &  &  &  &  &  &  &  \\
\hline
72 & SDSSJ0018-0939 & -2.46 & -0.39 & 0.00 & 0.29 & 0.43 &  $<$ -0.20 & $<$ 1.39 & \citet{aoki+14} &  \\
\hline
&&&&&&&&&&
\end{tabular}\\{\footnotesize {\bf Notes.} References are:
1. \citet{roederer+14a},
2. \citet{roederer+14b}, and
3. \citet{jacobson+15}.}
\end{table}

\begin{figure}[tbp]
	\includegraphics[width=\textwidth]{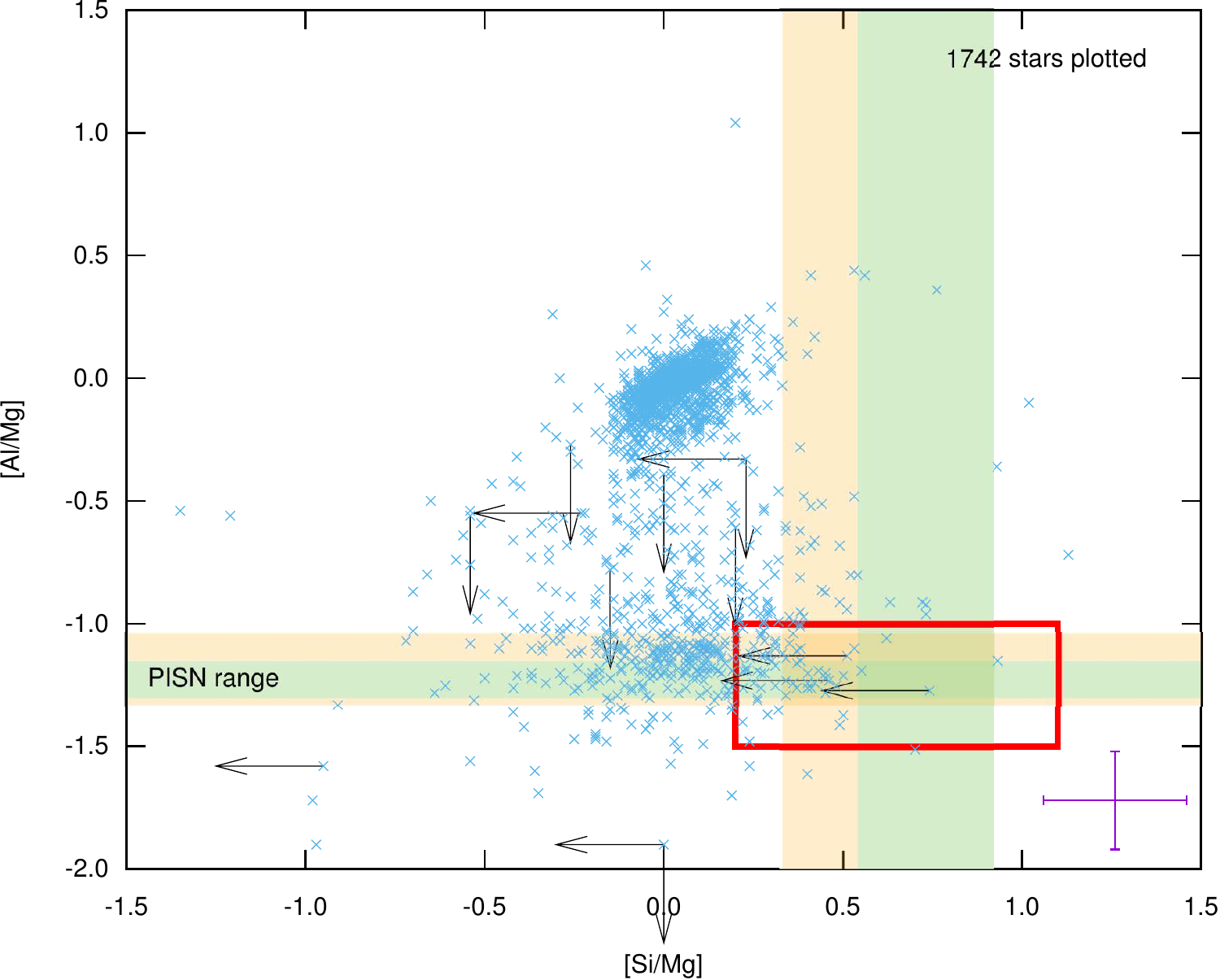}
	\caption{ \footnotesize{Same as Fig.~\ref{fig-Al-Ca} but for [Al/Mg] versus [Si/Mg].
	The typical errors of $\pm$0.2 dex for [Al/Mg] and [Si/Mg] are shown by the purple cross.
	Stars in the orange box are selected for the comparison with theoretical yields.}
	\label{fig-Al-Si} }
\end{figure}
In order to specify other characteristic abundance patterns of PISN yields,
more detailed comparisons with non-rotating PISN model yields have been conducted for 72 MP stars.
They are summarized in Table\ref{tab-stars-pisn} and in Appendix.
Most (71 out of 72) of those MP stars are selected according to the low [Al/Mg] and the high [Si/Mg] ratios,
which place inside the red box of ([Al/Mg], [Si/Mg]) = ($0.2$, $-1.0$) -- ($1.1$, $-1.5$) shown in Fig.~\ref{fig-Al-Si}.
In addition, the abundance pattern of SDSS J0018-0939 \citep{aoki+14} is analyzed,
which is characterized by the low [$\alpha$/Fe] ratios of [C, Mg, Si/Fe]
and by the exceptionally small [Co/Ni] ratio.
Since the metallicity of the stars, [Fe/H], is not utilized during the selection,
the sample includes metal-poor stars with metallicity as large as $-1.86$.
As most of the previous works except for \citet{aoki+14} have considered
only EMP stars of [Fe/H] $\lesssim$ $-3$ to be compared with PISN yield patterns,
this high maximum metallicity characterizes the sample of this work.
Most of these relatively-metal-rich stars will show the results of
not one-time but multiple metal pollutions in their abundances.
However, the wide range in metallicity can be rather adequate for the comparison with PISN yields,
as some theoretical works suggest that high metallicity of $\sim$ $10^{-3}$ Z$_\odot$ is reachable
by a one-shot PISN in the early universe due to the large metal production \citep{karlsson+08, greif+10}.

\begin{figure}[tbp]
	\includegraphics[width=\textwidth]{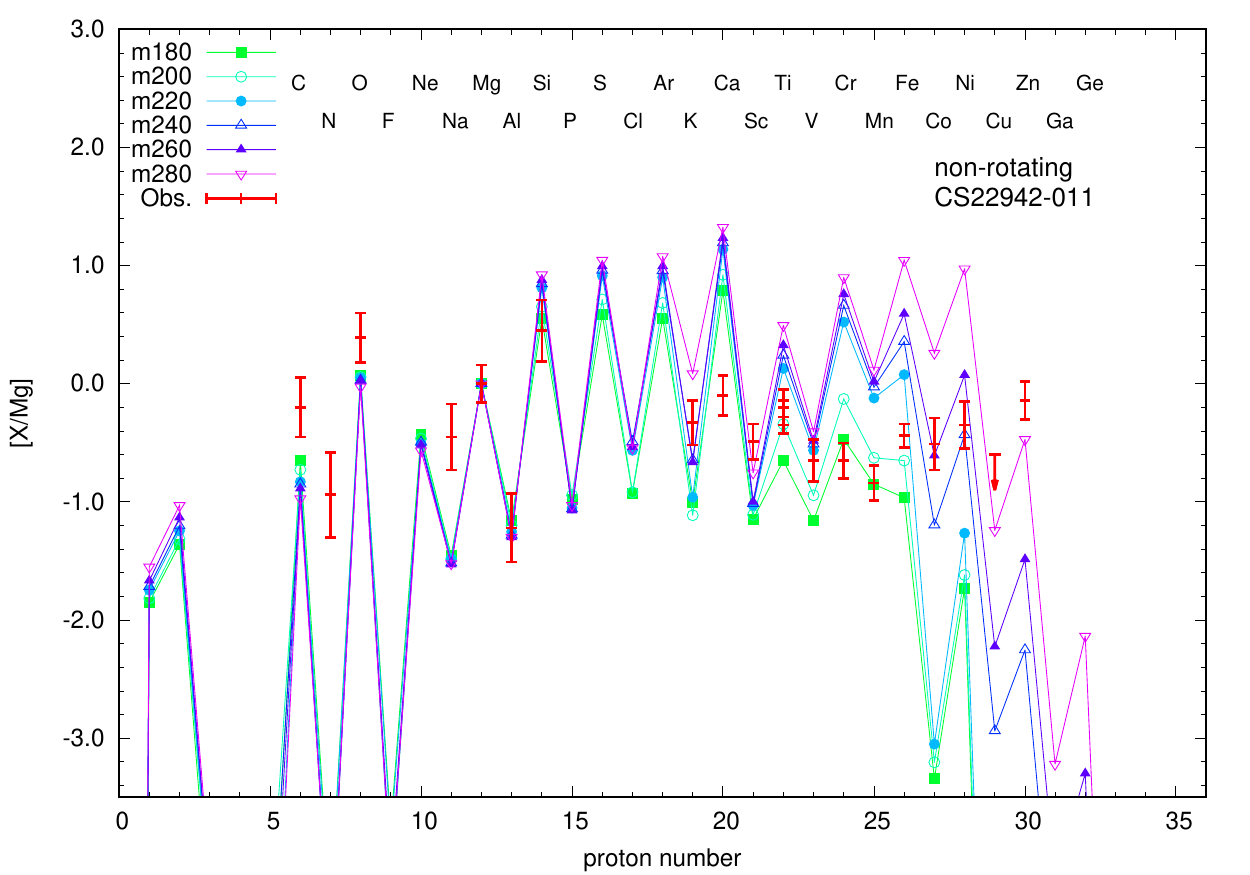}
	\caption{ \footnotesize{The abundance pattern of \#23 CS22942-011.
	Red thick crosses show observed values,
	while PISN abundances obtained from non-rotating models are shown by thin lines.}
	\label{fig-stars-PISN1} }
\end{figure}
As an example, the surface abundance pattern of \#23 CS22942-011 is shown in Fig.\ref{fig-stars-PISN1}.
The observation is compared with theoretical models of non-rotating PISN abundances.
First of all, as a confirmation of the earlier findings,
the figure shows that this MP star neither has the low [Na/Mg] nor the high [Ca/Mg] to match with the theory.
In addition, abundance ratios of [Sc/Mg] and [Zn/Fe] are found to be informative.
The observed ratios of [Sc/Mg] $= -0.49$ and [Zn/Fe] $= 0.53$
are higher than theoretical predictions, indicating that the MP star is not a PISN-MP star.

A merit of using these abundance ratios is their good accessibilities:
[Sc/Mg] is obtained for 71 stars including one star with an upper-limit, and
[Zn/Fe] is obtained for 66 stars including 12 stars with an upper-limit as well.
Moreover, theoretical predictions give low upper-limits of [Sc/Mg] $<$ $-0.75$ and [Zn/Fe] $<$ $-1.51$,
while only 9 stars in the sample (\#13, 20, 33, 34, 50, 58, 62, 67, and 69) have [Sc/Mg] $<$ $-0.7$
and all zinc-detected stars have higher [Zn/Fe] than the theoretical limit.
Therefore these ratios can be used as the constraints similar to [Na/Mg],
though the high accessibilities may be owing to the high occupancy in the sample
of the observation using high resolution spectroscopy at the Magellan Telescopes \citep{roederer+14b, roederer+14a}.

A note for the scandium abundance is that
a Pop III CCSN yield also fails to reproduce the observed value of [Sc/Fe] $\sim$ 0 \citep[e.g.,][]{kobayashi+06}.
In other word, a normal CCSN model produces too little amount of scandium to match with the observation.
Hence, origin of scandium in MP stars is somewhat uncertain,
while the too small production of scandium may be solved
by considering ejection of high entropy material in a jet-induced explosion \citep{tominaga09, tominaga+14}.
Nevertheless a result that PISN models yield too small scandium to explain observations is still valid,
because the theoretical prediction of PISN yields is robust
thanks to the clear understanding of the explosion mechanism.

\subsection{SDSS J0018-0939}
SDSS J0018-0939 is a metal-poor main-sequence star
with a metallicity of [Fe/H] = $-2.46$ discovered by \citet{aoki+13}.
\citet{aoki+14} further observe the distinctive abundance pattern,
which is characterized by the low [$\alpha$/Fe] ratios of [C, Mg, Si/Fe]
and by the exceptionally small [Co/Ni] $= -0.68$ ratio.
Despite the star has the relatively large metallicity,
they assume that the star possesses primitive chemical abundances
based on the low abundances of neutron-rich elements of [Sr/Fe] $<$ $-1.8$ and [Ba/Fe] $<$ $-1.3$.
One explanation given in their work is a single nucleosynthesis by
a very massive star occurring in the early universe.
They compare two theoretical yields in this line with the observation;
a Pop III 1000 M$_\odot$ CCSN model exploded with $6.67 \times 10^{53}$ erg \citep{ohkubo+06}
and a Pop III PISN model with a 130 M$_\odot$ He core \citep{umeda&nomoto02},
and discuss that the low [C, Mg/Fe] and the low [Co/Ni] can be explained by these very massive models.

\begin{figure}[tbp]
	\includegraphics[width=\textwidth]{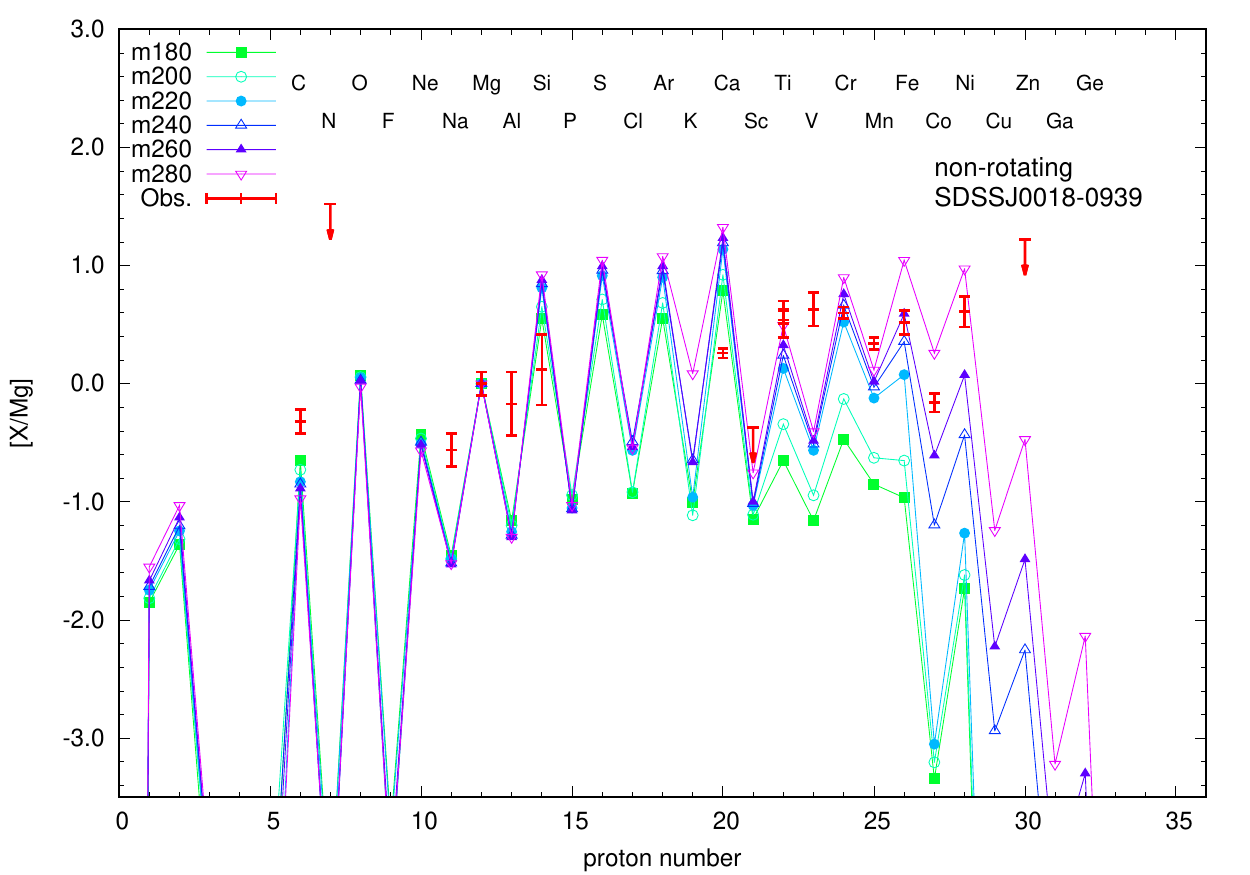}
	\caption{ \footnotesize{The abundance pattern of \#72 SDSS J0018-0939.
	Red thick crosses show observed values,
	while PISN abundances obtained from non-rotating models are shown by thin lines.}
	\label{fig-stars-PISN2} }
\end{figure}
Comparison between the stellar abundances of SDSS J0018-0939 \citep{aoki+14}
and non-rotating PISN abundances has been made in Fig. \ref{fig-stars-PISN2}.
The most important result in this comparison is the smaller [Ca/Mg] $= 0.43$
than the theoretical lower-limit of [Ca/Mg] $> 0.46$.
This result will already exclude the possibility that this MP star is a PISN-MP star.
Considering the large uncertainties in both theoretical modeling and observation,
PISN from the least massive progenitor may be adequate to explain the small [Ca/Mg] ratio.
However, the large stellar abundances of
the iron-peak elements of [(Cr, Co, Ni)/Mg] are then totally inconsistent with the theory,
because the most massive progenitor is required for the large abundances in contrast.
Another inconsistency is the higher abundance ratios of [(Na, Al, V)/Mg] than the theoretical models.
From these results, we conclude that the abundance pattern of SDSS J0018-0939 is not compatible with PISN models.

\section{Discussion and Conclusion}

Existence of PISNe in the early universe, if it is confirmed, can be a direct proof of
not only the hydrodynamical instability due to the electron-positron pair production,
which is a fundamental consequence in the theory of very massive star evolution,
but also the prediction of the initial mass function in the early universe,
which is estimated to have a high-mass peak of $\sim$ 100 $M_\odot$.
The confirmation can be done by detecting a PISN-MP star
from extremely- and very-metal-poor stars existing in our galaxy.
We have shown that characteristic abundance ratios of low [Na/Mg] and high [Ca/Mg]
will be fundamentally useful for the detection of the PISN-MP star.
Similarly, a PISN-MP star will show high [Si/Mg] and low [(Al, Sc, Zn)/Mg] ratios.
It is noteworthy that ratios of [X/Fe] are not as informative as [X/Mg]
since iron production in PISNe is highly dependent on the progenitor mass.
Stellar rotation may trigger effective internal mixing during the stellar life.
We have also shown that the internal mixing does not affect the explosive nucleosynthesis in PISN,
but can induce efficient nitrogen production.
Therefore the ratio [N/Mg] can be regarded as the first indicator of the stellar rotation of the PISN progenitor.

Through comparing our theoretical abundances with big sample of surface abundances of MP stars
compiled in the {\it SAGA} database, we have demonstrated that no PISN-MP star is included
in the currently observed MP stars.
Moreover, we have concluded that a VMP star SDSS J0018-0939 is not a PISN-MP star
because of the inconsistencies in the abundance pattern, especially the low calcium abundance
and the high sodium, aluminum, and vanadium abundances.
This result might be already problematic for the theoretical prediction that
$\sim$ 25\% of first stars in number will explode as a PISN
and thus $\sim$ 1/400 of MP stars with [Ca/H] $<$ $-2$ are PISN-MP stars.

One possible explanation for the none detection discussed in the literature
is that the observational bias towards the low calcium abundance in the MP stellar sample.
If a PISN-MP star has a metallicity of $Z = 10^{-3} Z_\odot$,
which may be a reasonable estimate of the metal contents of a second generation star \citep[e.g.,][]{wise&abel08, greif+10},
and if $\sim$ 1--3\% of the total metal in mass is calcium, which is derived from our calculation,
the corresponding calcium abundance becomes [Ca/H] = $-2.34$--$-1.86$.
Therefore the PISN-MP star is actually an order-of-magnitude calcium-richer than the majority of EMP stars,
which have [Fe/H] $<$ $-3$ in definition and thus [Ca/H] $\lesssim -3$. 
In our sample, 541 stars are included in the range of [Ca/H] = $-2.4$--$-1.8$,
and the number reduces to 237, if we require the [Na/Mg] data in addition.
Probably the number is too small to detect the PISN-MP star.

Another possible reason is the overestimate in number fraction of PISN progenitors in the first stars.
For this aspect, fragmentation of the star-forming gas cloud
will be the most important physics missing in the work by \citet{hirano+14, hirano+15}.
Gas fragmentation induces the formation of a binary and multiple-star system, and
there are some reasons to expect smaller mass stars are formed in such a system \citep[see discussions in][]{hirano+14}.
Because there are more sinks of gas than in a single system, a total accreted gas onto one star will be reduced.
Besides, with a smaller mass accretion rate in a multiple-star system,
the Kelvin-Helmholtz contraction begins at an earlier phase in which the proto-stellar mass is still small.
Since gas accretion onto the proto-star is prohibited after the star starts to radiate UV photons,
the early Kelvin-Helmholtz contraction limits the stellar mass to be small.
Therefore, by considering the fragmentation in the first star formation,
the peak mass in the initial mass distribution will be shifted to the lower side,
as well as the total number of the low mass first stars increases.
These effects can reduce the number fraction of PISN progenitors in the first stars.
It is noteworthy that the high fraction of being a binary or multiple-star system at its birth
has been observationally proven for massive OB stars in the local universe \citep[e.g.][]{sana+12}.
It may be reasonable to assume the similar frequency to the first stars,
since forming the binary system is the simplest way to get rid of
the high specific angular momentum of the star-forming gas cloud.

Finally, it is important to examine the reliability of the theory of evolution and explosion of the PISN progenitor.
Let us consider the most simple case first:
is it robust to estimate that a non-rotating massive first star
forms a CO core of about a half of its initial mass?

As far as the static evolution of a star is considered,
the one-dimensional stellar structure will be determined
as a solution of 4th order ordinary differential equations,
in which time evolution is expressed through time-evolving chemical and entropy profiles.
In the radiative (non-convective) layers, matter mixing will be negligible
and the radiative transfer to determine the entropy profile will be well approximated by a diffusion equation.
Then the first uncertainty will be incorporated in the treatment of stellar convection.
Although the current understanding of stellar convection is poor,
it seems that there is no significant uncertainty that affects the formation of a massive CO core.
Because the lifetime of the core burning phase in both cases is much longer than the timescale of convective mixing, 
and because the efficiency of heat transfer by the convective mixing will be sufficiently high,
it is reasonable to assume the constant chemical and entropy profiles inside the convective core.
Hence the only remaining uncertainty is the convective criterion.
In fact, there is a long-standing question on which convective criterion
of the Schwartzschild or the Ledoux criterions describe stellar convection.
However, this uncertainty will not be so much affective to change the current prediction qualitatively.
Convective overshoot can increase the CO core mass,
and conversely strong magnetic field, which may not exist in a first generation star though,
may suppress core growth \citep{petermann+15}.
However they only shift the core mass to the initial mass relation and
existence of a massive CO core does not change.
After all, chemical and entropy structures,
and thus the overall stellar structures and evolution until core carbon burning,
will be well described by the present evolutionary calculations.

Another important assumption here is that the wind mass loss of a massive first star is ineffective.
This will be a reasonable assumption since the line driven wind,
which potentially explains the heavy wind mass loss of OB and WR stars in the local universe,
will not act on the metal-free surface of the massive first star.
Surface metal pollution by rotational mixing has been proposed
to trigger the line driven wind during main sequence phase.
However, some negative results have been obtained by different authors:
the dredge-up during main sequence phase is not so efficient \citep{ekstroem+08b} and moreover
the efficiency of wind acceleration by the CNO elements are too weak to drive an efficient wind \citep{krticka&kubat09}.
Although some mechanisms, such as the vibrational instability \citep{baraffe+01,sonoi&umeda12,moriya&langer15},
may trigger a mass loss from a Pop III red-giant envelope,
this only strips the envelope of the star and the CO core will keep its mass.
Therefore it will be reasonable to assume that at least a massive CO core can be formed in a massive first star
and furthermore a total mass of a massive first star will be nearly conserved during its evolution.
The remaining possibility may be an interplay between
a large luminosity near the Eddington limit and a fast rotation near the Keplerian value.
A 150 M$_\odot$ exploratory model in \citet{ekstroem+08a} indeed loses significant mass of $\sim 50$ M$_\odot$
due to the $\Omega-\Gamma$ limit during the post-main-sequence phase.
However, such a highly effective mass loss has not taken place in fast rotating models in \citet{yoon+12}.
Further investigation on the enhancement of the $\Omega-\Gamma$ limit
and mass loss rate of WR surface with an extremely low metallicity will be needed to draw a conclusion.

After the formation of a massive CO core,
the core collapses due to the hydrodynamical instability of the electron-positron pair creation.
The only assumption of this instability is
the fast enough reactions of creation and annihilation of electron-position pairs,
which will be robust for the high density stellar environment.
In the collapsing core, carbon, neon, and oxygen are burnt to heat the surroundings.
Probably the reaction rates have big uncertainties.
However, the total rest mass energy emitted through the reactions should be determinate,
and therefore the final fate of the explosion will be robust.

Although multi-dimensionality during the explosion may affect the explosion,
strong convection does not arise in the collapsing massive CO core unlike in the case of a core-collapse supernova.
This is because the core collapse and the following expansion in a PISN take place with a dynamical timescale,
and this is too short to develop a strong convection, which requires a time at least several times of the dynamical time.
Multi-dimensional simulations resolve a growing instability at the core edge region \citep{Chatzopoulos+13,chen+14b,gilmer+17},
however this instability sets in as a result of the explosion and does not affect the explosion itself.
Another possibility is a fast rotation of a CO core at its collapse.
As we have demonstrated in this work,
if there is no effective angular momentum transport during core helium burning phase,
the remaining angular momentum in the CO core can be sufficiently large
to support the core by the centrifugal force during the core collapse.
Moreover, the rotating flow may create a strong shear during the collapse,
which may develop an instability to mix a fresh nuclear fuel into the core center.
Both effects will reduce the energy of the explosion.
Therefore, by considering the fast rotation of the massive CO core,
it is possible to expect that the initial mass range for PISN with incomplete mass ejection,
i.e., so-called pulsational-PISN, to be extended to more massive side.
This can reduce the possibility to have a PISN in the early universe.

In conclusion, PISN is an inevitable fate in first stars under a current understanding of stellar evolution.
Therefore, a PISN-MP stars will be found from a future survey,
in which sufficient number of mildly calcium rich and extremely sodium poor MP stars are observed.
Not only PISN but also pulsational-PISN is a highly possible fate for massive first stars in the early universe.
We have shown that a massive first star with a dense envelope tends to eject only outer part of the star during the explosion.
A fast rotating CO core may also end up in the incomplete explosion. 
So far, explosive yields of pulsational-PISNe have not been investigated in detail.
This will be an important next step for the future work.\\

K. T. was supported by Japan Society for the Promotion of Science (JSPS) Overseas Research Fellowships.
This work was supported in part by JSPS KAKENHI grant Nos. 26400271, 26104007, 17H01130, and 17K05380.

\bibliography{biblio}

\begin{thebibliography}{}
\expandafter\ifx\csname natexlab\endcsname\relax\def\natexlab#1{#1}\fi

\bibitem[{{Aoki} {et~al.}(2014){Aoki}, {Tominaga}, {Beers}, {Honda}, \&
  {Lee}}]{aoki+14}
{Aoki}, W., {Tominaga}, N., {Beers}, T.~C., {Honda}, S., \& {Lee}, Y.~S. 2014,
  Science, 345, 912

\bibitem[{{Aoki} {et~al.}(2005){Aoki}, {Honda}, {Beers}, {Kajino}, {Ando},
  {Norris}, {Ryan}, {Izumiura}, {Sadakane}, \& {Takada-Hidai}}]{Aoki+05}
{Aoki}, W., {Honda}, S., {Beers}, T.~C., {et~al.} 2005, \apj, 632, 611

\bibitem[{{Aoki} {et~al.}(2013){Aoki}, {Beers}, {Lee}, {Honda}, {Ito},
  {Takada-Hidai}, {Frebel}, {Suda}, {Fujimoto}, {Carollo}, \&
  {Sivarani}}]{aoki+13}
{Aoki}, W., {Beers}, T.~C., {Lee}, Y.~S., {et~al.} 2013, \aj, 145, 13

\bibitem[{{Asplund} {et~al.}(2009){Asplund}, {Grevesse}, {Sauval}, \&
  {Scott}}]{asplund+09}
{Asplund}, M., {Grevesse}, N., {Sauval}, A.~J., \& {Scott}, P. 2009, \araa, 47,
  481

\bibitem[{{Baraffe} {et~al.}(2001){Baraffe}, {Heger}, \&
  {Woosley}}]{baraffe+01}
{Baraffe}, I., {Heger}, A., \& {Woosley}, S.~E. 2001, \apj, 550, 890

\bibitem[{{Barkat} {et~al.}(1967){Barkat}, {Rakavy}, \& {Sack}}]{barkat+67}
{Barkat}, Z., {Rakavy}, G., \& {Sack}, N. 1967, Physical Review Letters, 18,
  379

\bibitem[{{Bonifacio} {et~al.}(2012){Bonifacio}, {Caffau}, {Venn}, \&
  {Lambert}}]{bonifacio+12}
{Bonifacio}, P., {Caffau}, E., {Venn}, K.~A., \& {Lambert}, D.~L. 2012, \aap,
  544, A102

\bibitem[{{Bromm} \& {Larson}(2004)}]{bromm&larson04}
{Bromm}, V., \& {Larson}, R.~B. 2004, \araa, 42, 79

\bibitem[{{Caffau} {et~al.}(2011){Caffau}, {Bonifacio}, {Fran{\c c}ois},
  {Spite}, {Spite}, {Zaggia}, {Ludwig}, {Monaco}, {Sbordone}, {Cayrel},
  {Hammer}, {Randich}, {Hill}, \& {Molaro}}]{Caffau+11b}
{Caffau}, E., {Bonifacio}, P., {Fran{\c c}ois}, P., {et~al.} 2011, \aap, 534,
  A4

\bibitem[{{Caughlan} \& {Fowler}(1988)}]{caughlan&fowler88}
{Caughlan}, G.~R., \& {Fowler}, W.~A. 1988, Atomic Data and Nuclear Data
  Tables, 40, 283

\bibitem[{{Chatzopoulos} {et~al.}(2015){Chatzopoulos}, {van Rossum}, {Craig},
  {Whalen}, {Smidt}, \& {Wiggins}}]{chatzopoulos+15}
{Chatzopoulos}, E., {van Rossum}, D.~R., {Craig}, W.~J., {et~al.} 2015, \apj,
  799, 18

\bibitem[{{Chatzopoulos} \& {Wheeler}(2012)}]{chatzopoulos&wheeler12}
{Chatzopoulos}, E., \& {Wheeler}, J.~C. 2012, \apj, 748, 42

\bibitem[{{Chatzopoulos} {et~al.}(2013{\natexlab{a}}){Chatzopoulos}, {Wheeler},
  \& {Couch}}]{chatzopoulos+13b}
{Chatzopoulos}, E., {Wheeler}, J.~C., \& {Couch}, S.~M. 2013{\natexlab{a}},
  \apj, 776, 129

\bibitem[{{Chatzopoulos} {et~al.}(2013{\natexlab{b}}){Chatzopoulos}, {Wheeler},
  \& {Couch}}]{Chatzopoulos+13}
---. 2013{\natexlab{b}}, \apj, 776, 129

\bibitem[{{Chen} {et~al.}(2014){Chen}, {Heger}, {Woosley}, {Almgren}, \&
  {Whalen}}]{chen+14b}
{Chen}, K.-J., {Heger}, A., {Woosley}, S., {Almgren}, A., \& {Whalen}, D.~J.
  2014, \apj, 792, 44

\bibitem[{{Chieffi} \& {Limongi}(2013)}]{Chieffi&Limongi13}
{Chieffi}, A., \& {Limongi}, M. 2013, \apj, 764, 21

\bibitem[{{Cohen} {et~al.}(2013){Cohen}, {Christlieb}, {Thompson}, {McWilliam},
  {Shectman}, {Reimers}, {Wisotzki}, \& {Kirby}}]{cohen+13}
{Cohen}, J.~G., {Christlieb}, N., {Thompson}, I., {et~al.} 2013, \apj, 778, 56

\bibitem[{{Cyburt} {et~al.}(2010){Cyburt}, {Amthor}, {Ferguson}, {Meisel},
  {Smith}, {Warren}, {Heger}, {Hoffman}, {Rauscher}, {Sakharuk}, {Schatz},
  {Thielemann}, \& {Wiescher}}]{Cyburt+10}
{Cyburt}, R.~H., {Amthor}, A.~M., {Ferguson}, R., {et~al.} 2010, \apjs, 189,
  240

\bibitem[{{Denissenkov} \& {Pinsonneault}(2007)}]{Denissenkov&Pinsonneault07}
{Denissenkov}, P.~A., \& {Pinsonneault}, M. 2007, \apj, 655, 1157

\bibitem[{{Ekstr{\"o}m} {et~al.}(2008{\natexlab{a}}){Ekstr{\"o}m}, {Meynet},
  {Chiappini}, {Hirschi}, \& {Maeder}}]{ekstroem+08b}
{Ekstr{\"o}m}, S., {Meynet}, G., {Chiappini}, C., {Hirschi}, R., \& {Maeder},
  A. 2008{\natexlab{a}}, \aap, 489, 685

\bibitem[{{Ekstr{\"o}m} {et~al.}(2008{\natexlab{b}}){Ekstr{\"o}m}, {Meynet}, \&
  {Maeder}}]{ekstroem+08a}
{Ekstr{\"o}m}, S., {Meynet}, G., \& {Maeder}, A. 2008{\natexlab{b}}, in
  American Institute of Physics Conference Series, Vol. 990, First Stars III,
  ed. B.~W. {O'Shea} \& A.~{Heger}, 220--224

\bibitem[{{Ekstr{\"o}m} {et~al.}(2012){Ekstr{\"o}m}, {Georgy}, {Eggenberger},
  {Meynet}, {Mowlavi}, {Wyttenbach}, {Granada}, {Decressin}, {Hirschi},
  {Frischknecht}, {Charbonnel}, \& {Maeder}}]{ekstroem+12}
{Ekstr{\"o}m}, S., {Georgy}, C., {Eggenberger}, P., {et~al.} 2012, \aap, 537,
  A146

\bibitem[{{Endal} \& {Sofia}(1978)}]{endal&sofia78}
{Endal}, A.~S., \& {Sofia}, S. 1978, \apj, 220, 279

\bibitem[{{Frebel} \& {Norris}(2015)}]{frebel&norris15}
{Frebel}, A., \& {Norris}, J.~E. 2015, \araa, 53, 631

\bibitem[{{Gal-Yam} {et~al.}(2009){Gal-Yam}, {Mazzali}, {Ofek}, {Nugent},
  {Kulkarni}, {Kasliwal}, {Quimby}, {Filippenko}, {Cenko}, {Chornock},
  {Waldman}, {Kasen}, {Sullivan}, {Beshore}, {Drake}, {Thomas}, {Bloom},
  {Poznanski}, {Miller}, {Foley}, {Silverman}, {Arcavi}, {Ellis}, \&
  {Deng}}]{gal-yam+09}
{Gal-Yam}, A., {Mazzali}, P., {Ofek}, E.~O., {et~al.} 2009, \nat, 462, 624

\bibitem[{{Georgy} {et~al.}(2012){Georgy}, {Ekstr{\"o}m}, {Meynet}, {Massey},
  {Levesque}, {Hirschi}, {Eggenberger}, \& {Maeder}}]{georgy+12}
{Georgy}, C., {Ekstr{\"o}m}, S., {Meynet}, G., {et~al.} 2012, \aap, 542, A29

\bibitem[{{Georgy} {et~al.}(2013){Georgy}, {Ekstr{\"o}m}, {Eggenberger},
  {Meynet}, {Haemmerl{\'e}}, {Maeder}, {Granada}, {Groh}, {Hirschi}, {Mowlavi},
  {Yusof}, {Charbonnel}, {Decressin}, \& {Barblan}}]{georgy+13}
{Georgy}, C., {Ekstr{\"o}m}, S., {Eggenberger}, P., {et~al.} 2013, \aap, 558,
  A103

\bibitem[{{Gilmer} {et~al.}(2017){Gilmer}, {Kozyreva}, {Hirschi},
  {Fr{\"o}hlich}, \& {Yusof}}]{gilmer+17}
{Gilmer}, M.~S., {Kozyreva}, A., {Hirschi}, R., {Fr{\"o}hlich}, C., \& {Yusof},
  N. 2017, \apj, 846, 100

\bibitem[{{Greif} {et~al.}(2010){Greif}, {Glover}, {Bromm}, \&
  {Klessen}}]{greif+10}
{Greif}, T.~H., {Glover}, S.~C.~O., {Bromm}, V., \& {Klessen}, R.~S. 2010,
  \apj, 716, 510

\bibitem[{{Hansen} {et~al.}(2015){Hansen}, {Hansen}, {Christlieb}, {Beers},
  {Yong}, {Bessell}, {Frebel}, {Garc{\'{\i}}a P{\'e}rez}, {Placco}, {Norris},
  \& {Asplund}}]{Hansen+15}
{Hansen}, T., {Hansen}, C.~J., {Christlieb}, N., {et~al.} 2015, \apj, 807, 173

\bibitem[{{Heger} {et~al.}(2000){Heger}, {Langer}, \& {Woosley}}]{heger+00}
{Heger}, A., {Langer}, N., \& {Woosley}, S.~E. 2000, \apj, 528, 368

\bibitem[{{Heger} \& {Woosley}(2002)}]{heger&woosley02}
{Heger}, A., \& {Woosley}, S.~E. 2002, \apj, 567, 532

\bibitem[{{Hirano} {et~al.}(2014){Hirano}, {Hosokawa}, {Yoshida}, {Umeda},
  {Omukai}, {Chiaki}, \& {Yorke}}]{hirano+14}
{Hirano}, S., {Hosokawa}, T., {Yoshida}, N., {et~al.} 2014, \apj, 781, 60

\bibitem[{{Hirano} {et~al.}(2015){Hirano}, {Zhu}, {Yoshida}, {Spergel}, \&
  {Yorke}}]{hirano+15}
{Hirano}, S., {Zhu}, N., {Yoshida}, N., {Spergel}, D., \& {Yorke}, H.~W. 2015,
  \apj, 814, 18

\bibitem[{{Hollek} {et~al.}(2011){Hollek}, {Frebel}, {Roederer}, {Sneden},
  {Shetrone}, {Beers}, {Kang}, \& {Thom}}]{hollek+11}
{Hollek}, J.~K., {Frebel}, A., {Roederer}, I.~U., {et~al.} 2011, \apj, 742, 54

\bibitem[{{Honda} {et~al.}(2004){Honda}, {Aoki}, {Kajino}, {Ando}, {Beers},
  {Izumiura}, {Sadakane}, \& {Takada-Hidai}}]{honda+04}
{Honda}, S., {Aoki}, W., {Kajino}, T., {et~al.} 2004, \apj, 607, 474

\bibitem[{{Itoh} {et~al.}(1989){Itoh}, {Adachi}, {Nakagawa}, {Kohyama}, \&
  {Munakata}}]{itoh+89}
{Itoh}, N., {Adachi}, T., {Nakagawa}, M., {Kohyama}, Y., \& {Munakata}, H.
  1989, \apj, 339, 354

\bibitem[{{Itoh} {et~al.}(1996){Itoh}, {Hayashi}, {Nishikawa}, \&
  {Kohyama}}]{itoh+96}
{Itoh}, N., {Hayashi}, H., {Nishikawa}, A., \& {Kohyama}, Y. 1996, \apjs, 102,
  411

\bibitem[{{Jacobson} {et~al.}(2015){Jacobson}, {Keller}, {Frebel}, {Casey},
  {Asplund}, {Bessell}, {Da Costa}, {Lind}, {Marino}, {Norris}, {Pe{\~n}a},
  {Schmidt}, {Tisserand}, {Walsh}, {Yong}, \& {Yu}}]{jacobson+15}
{Jacobson}, H.~R., {Keller}, S., {Frebel}, A., {et~al.} 2015, \apj, 807, 171

\bibitem[{{Karlsson} {et~al.}(2008){Karlsson}, {Johnson}, \&
  {Bromm}}]{karlsson+08}
{Karlsson}, T., {Johnson}, J.~L., \& {Bromm}, V. 2008, \apj, 679, 6

\bibitem[{{Kasen} {et~al.}(2011){Kasen}, {Woosley}, \& {Heger}}]{kasen+11}
{Kasen}, D., {Woosley}, S.~E., \& {Heger}, A. 2011, \apj, 734, 102

\bibitem[{{Keller} {et~al.}(2014){Keller}, {Bessell}, {Frebel}, {Casey},
  {Asplund}, {Jacobson}, {Lind}, {Norris}, {Yong}, {Heger}, {Magic}, {da
  Costa}, {Schmidt}, \& {Tisserand}}]{keller+14}
{Keller}, S.~C., {Bessell}, M.~S., {Frebel}, A., {et~al.} 2014, \nat, 506, 463

\bibitem[{{Kobayashi} {et~al.}(2006){Kobayashi}, {Umeda}, {Nomoto}, {Tominaga},
  \& {Ohkubo}}]{kobayashi+06}
{Kobayashi}, C., {Umeda}, H., {Nomoto}, K., {Tominaga}, N., \& {Ohkubo}, T.
  2006, \apj, 653, 1145

\bibitem[{{Kozyreva} {et~al.}(2014{\natexlab{a}}){Kozyreva}, {Blinnikov},
  {Langer}, \& {Yoon}}]{kozyreva+14a}
{Kozyreva}, A., {Blinnikov}, S., {Langer}, N., \& {Yoon}, S.-C.
  2014{\natexlab{a}}, \aap, 565, A70

\bibitem[{{Kozyreva} {et~al.}(2014{\natexlab{b}}){Kozyreva}, {Yoon}, \&
  {Langer}}]{kozyreva+14b}
{Kozyreva}, A., {Yoon}, S.-C., \& {Langer}, N. 2014{\natexlab{b}}, \aap, 566,
  A146

\bibitem[{{Krti{\v c}ka} \& {Kub{\'a}t}(2009)}]{krticka&kubat09}
{Krti{\v c}ka}, J., \& {Kub{\'a}t}, J. 2009, \aap, 493, 585

\bibitem[{{Lai} {et~al.}(2008){Lai}, {Bolte}, {Johnson}, {Lucatello}, {Heger},
  \& {Woosley}}]{Lai+08}
{Lai}, D.~K., {Bolte}, M., {Johnson}, J.~A., {et~al.} 2008, \apj, 681, 1524

\bibitem[{{Langer}(1998)}]{langer98}
{Langer}, N. 1998, \aap, 329, 551

\bibitem[{{Langer} {et~al.}(2007){Langer}, {Norman}, {de Koter}, {Vink},
  {Cantiello}, \& {Yoon}}]{langer+07}
{Langer}, N., {Norman}, C.~A., {de Koter}, A., {et~al.} 2007, \aap, 475, L19

\bibitem[{{Maeder} \& {Meynet}(2000)}]{maeder&meynet00}
{Maeder}, A., \& {Meynet}, G. 2000, \aap, 361, 159

\bibitem[{{Maeder} \& {Meynet}(2004)}]{Maeder&Meynet04}
---. 2004, \aap, 422, 225

\bibitem[{{Maeder} \& {Zahn}(1998)}]{Maeder&Zahn98}
{Maeder}, A., \& {Zahn}, J.-P. 1998, \aap, 334, 1000

\bibitem[{{Meynet} \& {Maeder}(2000)}]{meynet&maeder00}
{Meynet}, G., \& {Maeder}, A. 2000, \aap, 361, 101

\bibitem[{{Moriya} \& {Langer}(2015)}]{moriya&langer15}
{Moriya}, T.~J., \& {Langer}, N. 2015, \aap, 573, A18

\bibitem[{{Nomoto} {et~al.}(2013){Nomoto}, {Kobayashi}, \&
  {Tominaga}}]{nomoto+13}
{Nomoto}, K., {Kobayashi}, C., \& {Tominaga}, N. 2013, \araa, 51, 457

\bibitem[{{Ohkubo} {et~al.}(2006){Ohkubo}, {Umeda}, {Maeda}, {Nomoto},
  {Suzuki}, {Tsuruta}, \& {Rees}}]{ohkubo+06}
{Ohkubo}, T., {Umeda}, H., {Maeda}, K., {et~al.} 2006, \apj, 645, 1352

\bibitem[{{Petermann} {et~al.}(2015){Petermann}, {Langer}, {Castro}, \&
  {Fossati}}]{petermann+15}
{Petermann}, I., {Langer}, N., {Castro}, N., \& {Fossati}, L. 2015, \aap, 584,
  A54

\bibitem[{{Pinsonneault} {et~al.}(1989){Pinsonneault}, {Kawaler}, {Sofia}, \&
  {Demarque}}]{pinsonneault+89}
{Pinsonneault}, M.~H., {Kawaler}, S.~D., {Sofia}, S., \& {Demarque}, P. 1989,
  \apj, 338, 424

\bibitem[{{Placco} {et~al.}(2014){Placco}, {Frebel}, {Beers}, {Christlieb},
  {Lee}, {Kennedy}, {Rossi}, \& {Santucci}}]{Placco+14}
{Placco}, V.~M., {Frebel}, A., {Beers}, T.~C., {et~al.} 2014, \apj, 781, 40

\bibitem[{{Rakavy} {et~al.}(1967){Rakavy}, {Shaviv}, \& {Zinamon}}]{rakavy+67}
{Rakavy}, G., {Shaviv}, G., \& {Zinamon}, Z. 1967, \apj, 150, 131

\bibitem[{{Ren} {et~al.}(2012){Ren}, {Christlieb}, \& {Zhao}}]{ren+12}
{Ren}, J., {Christlieb}, N., \& {Zhao}, G. 2012, Research in Astronomy and
  Astrophysics, 12, 1637

\bibitem[{{Roederer} {et~al.}(2010){Roederer}, {Cowan}, {Karakas}, {Kratz},
  {Lugaro}, {Simmerer}, {Farouqi}, \& {Sneden}}]{Roederer+10}
{Roederer}, I.~U., {Cowan}, J.~J., {Karakas}, A.~I., {et~al.} 2010, \apj, 724,
  975

\bibitem[{{Roederer} {et~al.}(2014{\natexlab{a}}){Roederer}, {Preston},
  {Thompson}, {Shectman}, \& {Sneden}}]{roederer+14b}
{Roederer}, I.~U., {Preston}, G.~W., {Thompson}, I.~B., {Shectman}, S.~A., \&
  {Sneden}, C. 2014{\natexlab{a}}, \apj, 784, 158

\bibitem[{{Roederer} {et~al.}(2014{\natexlab{b}}){Roederer}, {Preston},
  {Thompson}, {Shectman}, {Sneden}, {Burley}, \& {Kelson}}]{roederer+14a}
{Roederer}, I.~U., {Preston}, G.~W., {Thompson}, I.~B., {et~al.}
  2014{\natexlab{b}}, \aj, 147, 136

\bibitem[{{Sana} {et~al.}(2012){Sana}, {de Mink}, {de Koter}, {Langer},
  {Evans}, {Gieles}, {Gosset}, {Izzard}, {Le Bouquin}, \&
  {Schneider}}]{sana+12}
{Sana}, H., {de Mink}, S.~E., {de Koter}, A., {et~al.} 2012, Science, 337, 444

\bibitem[{{Scannapieco} {et~al.}(2005){Scannapieco}, {Madau}, {Woosley},
  {Heger}, \& {Ferrara}}]{scannapieco+05}
{Scannapieco}, E., {Madau}, P., {Woosley}, S., {Heger}, A., \& {Ferrara}, A.
  2005, \apj, 633, 1031

\bibitem[{{Siqueira Mello} {et~al.}(2014){Siqueira Mello}, {Hill}, {Barbuy},
  {Spite}, {Spite}, {Beers}, {Caffau}, {Bonifacio}, {Cayrel}, {Fran{\c c}ois},
  {Schatz}, \& {Wanajo}}]{SiqueiraMello+14}
{Siqueira Mello}, C., {Hill}, V., {Barbuy}, B., {et~al.} 2014, \aap, 565, A93

\bibitem[{{Smidt} {et~al.}(2015){Smidt}, {Whalen}, {Chatzopoulos}, {Wiggins},
  {Chen}, {Kozyreva}, \& {Even}}]{smidt+15}
{Smidt}, J., {Whalen}, D.~J., {Chatzopoulos}, E., {et~al.} 2015, \apj, 805, 44

\bibitem[{{Sonoi} \& {Umeda}(2012)}]{sonoi&umeda12}
{Sonoi}, T., \& {Umeda}, H. 2012, \mnras, 421, L34

\bibitem[{{Spruit}(2002)}]{spruit02}
{Spruit}, H.~C. 2002, \aap, 381, 923

\bibitem[{{Stacy} {et~al.}(2011){Stacy}, {Bromm}, \& {Loeb}}]{stacy+11b}
{Stacy}, A., {Bromm}, V., \& {Loeb}, A. 2011, \mnras, 413, 543

\bibitem[{{Stacy} {et~al.}(2013){Stacy}, {Greif}, {Klessen}, {Bromm}, \&
  {Loeb}}]{stacy+13}
{Stacy}, A., {Greif}, T.~H., {Klessen}, R.~S., {Bromm}, V., \& {Loeb}, A. 2013,
  \mnras, 431, 1470

\bibitem[{{Steigman}(2007)}]{steigman07}
{Steigman}, G. 2007, Annual Review of Nuclear and Particle Science, 57, 463

\bibitem[{{Suda} {et~al.}(2011){Suda}, {Yamada}, {Katsuta}, {Komiya},
  {Ishizuka}, {Aoki}, \& {Fujimoto}}]{suda+11}
{Suda}, T., {Yamada}, S., {Katsuta}, Y., {et~al.} 2011, \mnras, 412, 843

\bibitem[{{Suda} {et~al.}(2008){Suda}, {Katsuta}, {Yamada}, {Suwa}, {Ishizuka},
  {Komiya}, {Sorai}, {Aikawa}, \& {Fujimoto}}]{suda+08}
{Suda}, T., {Katsuta}, Y., {Yamada}, S., {et~al.} 2008, \pasj, 60, 1159

\bibitem[{Suda {et~al.}(2017)Suda, Hidaka, Aoki, Katsuta, Yamada, Fujimoto,
  Ohtani, Masuyama, Noda, \& Wada}]{suda+17}
Suda, T., Hidaka, J., Aoki, W., {et~al.} 2017, Publications of the Astronomical
  Society of Japan, 69, 76

\bibitem[{{Sumiyoshi} {et~al.}(2000){Sumiyoshi}, {Suzuki}, {Otsuki},
  {Terasawa}, \& {Yamada}}]{sumiyoshi+00}
{Sumiyoshi}, K., {Suzuki}, H., {Otsuki}, K., {Terasawa}, M., \& {Yamada}, S.
  2000, \pasj, 52, 601

\bibitem[{{Takahashi} {et~al.}(2014){Takahashi}, {Umeda}, \&
  {Yoshida}}]{takahashi+14}
{Takahashi}, K., {Umeda}, H., \& {Yoshida}, T. 2014, \apj, 794, 40

\bibitem[{{Takahashi} {et~al.}(2016){Takahashi}, {Yoshida}, {Umeda},
  {Sumiyoshi}, \& {Yamada}}]{takahashi+16}
{Takahashi}, K., {Yoshida}, T., {Umeda}, H., {Sumiyoshi}, K., \& {Yamada}, S.
  2016, \mnras, 456, 1320

\bibitem[{{Tominaga}(2009)}]{tominaga09}
{Tominaga}, N. 2009, \apj, 690, 526

\bibitem[{{Tominaga} {et~al.}(2014){Tominaga}, {Iwamoto}, \&
  {Nomoto}}]{tominaga+14}
{Tominaga}, N., {Iwamoto}, N., \& {Nomoto}, K. 2014, \apj, 785, 98

\bibitem[{{Umeda} \& {Nomoto}(2002)}]{umeda&nomoto02}
{Umeda}, H., \& {Nomoto}, K. 2002, \apj, 565, 385

\bibitem[{{Umeda} \& {Nomoto}(2005)}]{umeda&nomoto05}
---. 2005, \apj, 619, 427

\bibitem[{{Whalen} {et~al.}(2013){Whalen}, {Even}, {Frey}, {Smidt}, {Johnson},
  {Lovekin}, {Fryer}, {Stiavelli}, {Holz}, {Heger}, {Woosley}, \&
  {Hungerford}}]{whalen+13}
{Whalen}, D.~J., {Even}, W., {Frey}, L.~H., {et~al.} 2013, \apj, 777, 110

\bibitem[{{Wise} \& {Abel}(2008)}]{wise&abel08}
{Wise}, J.~H., \& {Abel}, T. 2008, \apj, 685, 40

\bibitem[{{Woosley}(2017)}]{woosley17}
{Woosley}, S.~E. 2017, \apj, 836, 244

\bibitem[{{Yamada}(1997)}]{yamada97}
{Yamada}, S. 1997, \apj, 475, 720

\bibitem[{{Yamada} {et~al.}(1999){Yamada}, {Janka}, \& {Suzuki}}]{yamada+99}
{Yamada}, S., {Janka}, H.-T., \& {Suzuki}, H. 1999, \aap, 344, 533

\bibitem[{{Yamada} {et~al.}(2013){Yamada}, {Suda}, {Komiya}, {Aoki}, \&
  {Fujimoto}}]{yamada+13}
{Yamada}, S., {Suda}, T., {Komiya}, Y., {Aoki}, W., \& {Fujimoto}, M.~Y. 2013,
  \mnras, 436, 1362

\bibitem[{{Yong} {et~al.}(2013{\natexlab{a}}){Yong}, {Norris}, {Bessell},
  {Christlieb}, {Asplund}, {Beers}, {Barklem}, {Frebel}, \& {Ryan}}]{Yong+13a}
{Yong}, D., {Norris}, J.~E., {Bessell}, M.~S., {et~al.} 2013{\natexlab{a}},
  \apj, 762, 26

\bibitem[{{Yong} {et~al.}(2013{\natexlab{b}}){Yong}, {Norris}, {Bessell},
  {Christlieb}, {Asplund}, {Beers}, {Barklem}, {Frebel}, \& {Ryan}}]{yong+13b}
---. 2013{\natexlab{b}}, \apj, 762, 27

\bibitem[{{Yoon} {et~al.}(2012){Yoon}, {Dierks}, \& {Langer}}]{yoon+12}
{Yoon}, S.-C., {Dierks}, A., \& {Langer}, N. 2012, \aap, 542, A113

\bibitem[{{Yoon} {et~al.}(2010){Yoon}, {Woosley}, \& {Langer}}]{yoon+10}
{Yoon}, S.-C., {Woosley}, S.~E., \& {Langer}, N. 2010, \apj, 725, 940

\bibitem[{{Yoshida} {et~al.}(2014){Yoshida}, {Okita}, \& {Umeda}}]{Yoshida+14}
{Yoshida}, T., {Okita}, S., \& {Umeda}, H. 2014, \mnras, 438, 3119

\bibitem[{{Yoshida} \& {Umeda}(2011)}]{Yoshida&Umeda11}
{Yoshida}, T., \& {Umeda}, H. 2011, \mnras, 412, L78

\bibitem[{{Yusof} {et~al.}(2013){Yusof}, {Hirschi}, {Meynet}, {Crowther},
  {Ekstr{\"o}m}, {Frischknecht}, {Georgy}, {Abu Kassim}, \&
  {Schnurr}}]{Yusof+13}
{Yusof}, N., {Hirschi}, R., {Meynet}, G., {et~al.} 2013, \mnras, 433, 1114

\bibitem[{{Zahn} {et~al.}(2007){Zahn}, {Brun}, \& {Mathis}}]{Zahn+07}
{Zahn}, J.-P., {Brun}, A.~S., \& {Mathis}, S. 2007, \aap, 474, 145

\end{thebibliography}

\appendix

\section{PISN yields} \label{app-1}

\begin{table}[tbp]
	\begin{center}
	\caption{PISN Yields of the non-rotating models \label{tab-yields-nonrot1}}
	\tabcolsep = 1mm

\begin{tabular}{lccccccccc}
	\hline
	\hline
	Element	&	Yield			&	Element	&	Yield			&	Element	&	Yield			&	Element	&	Yield			&	Element	&	Yield			\\
	\hline
	\multicolumn{5}{l}{$M_{\rm ini}$ = 180 $M_\odot$} \\
	\hline                   
  $^{}$p           &   4.408E+01 &  $^{}$d           &   4.326E-14 &  $^{3}$He         &   4.128E-04 &  $^{4}$He         &   5.298E+01 &  $^{6}$Li         &   1.065E-09 \\                        
  $^{7}$Li         &   5.751E-11 &  $^{9}$Be         &   5.553E-20 &  $^{10}$B         &   5.179E-15 &  $^{11}$B         &   2.037E-10 &  $^{12}$C         &   3.187E+00 \\                        
  $^{13}$C         &   8.841E-06 &  $^{14}$N         &   1.179E-04 &  $^{15}$N         &   6.491E-05 &  $^{16}$O         &   5.213E+01 &  $^{17}$O         &   4.313E-05 \\                        
  $^{18}$O         &   1.323E-07 &  $^{19}$F         &   3.463E-07 &  $^{20}$Ne        &   3.036E+00 &  $^{21}$Ne        &   6.076E-05 &  $^{22}$Ne        &   4.131E-05 \\                        
  $^{23}$Na        &   5.458E-03 &  $^{24}$Mg        &   3.066E+00 &  $^{25}$Mg        &   1.710E-03 &  $^{26}$Mg        &   3.837E-03 &  $^{27}$Al        &   1.870E-02 \\                        
  $^{28}$Si        &   1.167E+01 &  $^{29}$Si        &   2.540E-02 &  $^{30}$Si        &   3.448E-03 &  $^{31}$P         &   3.976E-03 &  $^{32}$S         &   7.452E+00 \\
\end{tabular}
	\end{center}
\end{table}
\begin{table}[tbp]
	\begin{center}
	\caption{PISN Yields of the non-rotating models \label{tab-yields-nonrot2}}
	\tabcolsep = 1mm

	\begin{tabular}{lccccccccc}
	\hline
	\hline
	Element	&	Yield			&	Element	&	Yield			&	Element	&	Yield			&	Element	&	Yield			&	Element	&	Yield			\\
	\hline
	\multicolumn{5}{l}{$M_{\rm ini}$ = 220 $M_\odot$} \\
	\hline                   
  $^{}$p           &   5.297E+01 &  $^{}$d           &   9.782E-16 &  $^{3}$He         &   5.044E-04 &  $^{4}$He         &   6.509E+01 &  $^{6}$Li         &   1.487E-09 \\                        
  $^{7}$Li         &   3.479E-10 &  $^{9}$Be         &   2.964E-23 &  $^{10}$B         &   9.798E-19 &  $^{11}$B         &   2.227E-11 &  $^{12}$C         &   1.995E+00 \\                        
  $^{13}$C         &   6.065E-06 &  $^{14}$N         &   9.708E-05 &  $^{15}$N         &   5.945E-05 &  $^{16}$O         &   4.736E+01 &  $^{17}$O         &   1.117E-04 \\                        
  $^{18}$O         &   9.165E-08 &  $^{19}$F         &   2.658E-07 &  $^{20}$Ne        &   2.392E+00 &  $^{21}$Ne        &   2.879E-05 &  $^{22}$Ne        &   3.392E-05 \\                        
  $^{23}$Na        &   4.791E-03 &  $^{24}$Mg        &   2.892E+00 &  $^{25}$Mg        &   1.155E-03 &  $^{26}$Mg        &   3.202E-03 &  $^{27}$Al        &   1.422E-02 \\                        
  $^{28}$Si        &   2.022E+01 &  $^{29}$Si        &   2.220E-02 &  $^{30}$Si        &   2.379E-03 &  $^{31}$P         &   3.182E-03 &  $^{32}$S         &   1.506E+01 \\
\end{tabular}
	\end{center}
\end{table}
\begin{table}[tbp]
	\begin{center}
	\caption{PISN Yields of the non-rotating models \label{tab-yields-nonrot3}}
	\tabcolsep = 1mm

	\begin{tabular}{lccccccccc}
	\hline
	\hline
	Element	&	Yield			&	Element	&	Yield			&	Element	&	Yield			&	Element	&	Yield			&	Element	&	Yield			\\
	\hline
	\multicolumn{5}{l}{$M_{\rm ini}$ = 260 $M_\odot$} \\
	\hline                   
  $^{}$p           &   6.123E+01 &  $^{}$d           &   9.960E-14 &  $^{3}$He         &   5.995E-04 &  $^{4}$He         &   8.125E+01 &  $^{6}$Li         &   1.771E-09 \\                        
  $^{7}$Li         &   6.023E-10 &  $^{9}$Be         &   3.690E-20 &  $^{10}$B         &   1.752E-14 &  $^{11}$B         &   1.013E-10 &  $^{12}$C         &   1.678E+00 \\                        
  $^{13}$C         &   4.880E-06 &  $^{14}$N         &   5.207E-05 &  $^{15}$N         &   3.317E-05 &  $^{16}$O         &   4.299E+01 &  $^{17}$O         &   4.533E-06 \\                        
  $^{18}$O         &   3.935E-08 &  $^{19}$F         &   1.376E-07 &  $^{20}$Ne        &   2.251E+00 &  $^{21}$Ne        &   2.099E-05 &  $^{22}$Ne        &   3.096E-05 \\                        
  $^{23}$Na        &   4.149E-03 &  $^{24}$Mg        &   2.790E+00 &  $^{25}$Mg        &   9.489E-04 &  $^{26}$Mg        &   2.766E-03 &  $^{27}$Al        &   1.231E-02 \\                        
  $^{28}$Si        &   2.260E+01 &  $^{29}$Si        &   1.871E-02 &  $^{30}$Si        &   2.156E-03 &  $^{31}$P         &   2.911E-03 &  $^{32}$S         &   1.740E+01 \\
\end{tabular}
	\end{center}
\end{table}

\begin{table}[tbp]
	\begin{center}
	\caption{PISN Yields of the non-magnetic rotating models \label{tab-yields-nonmag1}}
	\tabcolsep = 1mm

	\begin{tabular}{lccccccccc}
	\hline
	\hline
	Element	&	Yield			&	Element	&	Yield			&	Element	&	Yield			&	Element	&	Yield			&	Element	&	Yield			\\
	\hline
	\multicolumn{5}{l}{$M_{\rm ini}$ = 160 $M_\odot$} \\
	\hline                   
  $^{}$p           &   3.648E+01 &  $^{}$d           &   4.307E-15 &  $^{3}$He         &   7.438E-05 &  $^{4}$He         &   5.064E+01 &  $^{6}$Li         &   7.866E-18 \\                        
  $^{7}$Li         &   3.278E-08 &  $^{9}$Be         &   2.182E-20 &  $^{10}$B         &   3.307E-18 &  $^{11}$B         &   2.168E-11 &  $^{12}$C         &   4.208E+00 \\                        
  $^{13}$C         &   7.259E-03 &  $^{14}$N         &   7.656E-01 &  $^{15}$N         &   8.942E-04 &  $^{16}$O         &   5.303E+01 &  $^{17}$O         &   1.034E-03 \\                        
  $^{18}$O         &   1.453E-06 &  $^{19}$F         &   1.420E-07 &  $^{20}$Ne        &   3.115E+00 &  $^{21}$Ne        &   7.586E-05 &  $^{22}$Ne        &   5.043E-05 \\                        
  $^{23}$Na        &   5.050E-03 &  $^{24}$Mg        &   2.544E+00 &  $^{25}$Mg        &   1.594E-03 &  $^{26}$Mg        &   3.696E-03 &  $^{27}$Al        &   2.040E-02 \\                        
  $^{28}$Si        &   5.981E+00 &  $^{29}$Si        &   2.281E-02 &  $^{30}$Si        &   5.138E-03 &  $^{31}$P         &   4.981E-03 &  $^{32}$S         &   3.202E+00 \\
\end{tabular}
	\end{center}
\end{table}
\begin{table}[tbp]
	\begin{center}
	\caption{PISN Yields of the non-magnetic rotating models \label{tab-yields-nonmag2}}
	\tabcolsep = 1mm

	\begin{tabular}{lccccccccc}
	\hline
	\hline
	Element	&	Yield			&	Element	&	Yield			&	Element	&	Yield			&	Element	&	Yield			&	Element	&	Yield			\\
	\hline
	\multicolumn{5}{l}{$M_{\rm ini}$ = 200 $M_\odot$} \\
	\hline                   
  $^{}$p           &   4.453E+01 &  $^{}$d           &   7.548E-15 &  $^{3}$He         &   1.071E-04 &  $^{4}$He         &   6.166E+01 &  $^{6}$Li         &   5.952E-17 \\                        
  $^{7}$Li         &   4.522E-08 &  $^{9}$Be         &   3.701E-19 &  $^{10}$B         &   9.210E-16 &  $^{11}$B         &   1.699E-11 &  $^{12}$C         &   3.015E+00 \\                        
  $^{13}$C         &   1.122E-02 &  $^{14}$N         &   1.194E+00 &  $^{15}$N         &   1.324E-03 &  $^{16}$O         &   4.584E+01 &  $^{17}$O         &   1.575E-03 \\                        
  $^{18}$O         &   2.046E-06 &  $^{19}$F         &   1.313E-07 &  $^{20}$Ne        &   3.058E+00 &  $^{21}$Ne        &   4.225E-05 &  $^{22}$Ne        &   3.947E-05 \\                        
  $^{23}$Na        &   4.205E-03 &  $^{24}$Mg        &   2.391E+00 &  $^{25}$Mg        &   9.374E-04 &  $^{26}$Mg        &   2.667E-03 &  $^{27}$Al        &   1.167E-02 \\                        
  $^{28}$Si        &   1.714E+01 &  $^{29}$Si        &   1.842E-02 &  $^{30}$Si        &   1.586E-03 &  $^{31}$P         &   1.863E-03 &  $^{32}$S         &   1.251E+01 \\
\end{tabular}
	\end{center}
\end{table}
\begin{table}[tbp]
	\begin{center}
	\caption{PISN Yields of the non-magnetic rotating models \label{tab-yields-nonmag3}}
	\tabcolsep = 1mm

	\begin{tabular}{lccccccccc}
	\hline
	\hline
	Element	&	Yield			&	Element	&	Yield			&	Element	&	Yield			&	Element	&	Yield			&	Element	&	Yield			\\
	\hline
	\multicolumn{5}{l}{$M_{\rm ini}$ = 240 $M_\odot$} \\
	\hline                   
  $^{}$p           &   5.189E+01 &  $^{}$d           &   9.257E-15 &  $^{3}$He         &   1.383E-04 &  $^{4}$He         &   7.631E+01 &  $^{6}$Li         &   6.256E-18 \\                        
  $^{7}$Li         &   2.808E-08 &  $^{9}$Be         &   9.316E-20 &  $^{10}$B         &   3.439E-17 &  $^{11}$B         &   2.337E-11 &  $^{12}$C         &   2.807E+00 \\                        
  $^{13}$C         &   2.443E-02 &  $^{14}$N         &   2.325E+00 &  $^{15}$N         &   9.473E-04 &  $^{16}$O         &   4.224E+01 &  $^{17}$O         &   3.694E-03 \\                        
  $^{18}$O         &   3.779E-06 &  $^{19}$F         &   1.685E-07 &  $^{20}$Ne        &   3.099E+00 &  $^{21}$Ne        &   6.437E-05 &  $^{22}$Ne        &   4.536E-05 \\                        
  $^{23}$Na        &   3.626E-03 &  $^{24}$Mg        &   2.171E+00 &  $^{25}$Mg        &   7.880E-04 &  $^{26}$Mg        &   2.263E-03 &  $^{27}$Al        &   9.318E-03 \\                        
  $^{28}$Si        &   1.943E+01 &  $^{29}$Si        &   1.448E-02 &  $^{30}$Si        &   1.214E-03 &  $^{31}$P         &   1.364E-03 &  $^{32}$S         &   1.487E+01 \\
\end{tabular}
	\end{center}
\end{table}

\begin{table}[tbp]
	\begin{center}
	\caption{PISN Yields of the magnetic rotating models \label{tab-yields-mag1}}
	\tabcolsep = 1mm

	\begin{tabular}{lccccccccc}
	\hline
	\hline
	Element	&	Yield			&	Element	&	Yield			&	Element	&	Yield			&	Element	&	Yield			&	Element	&	Yield			\\
	\hline
	\multicolumn{5}{l}{$M_{\rm ini}$ = 160 $M_\odot$} \\
	\hline                   
  $^{}$p           &   3.057E+01 &  $^{}$d           &   1.059E-15 &  $^{3}$He         &   9.180E-06 &  $^{4}$He         &   5.225E+01 &  $^{6}$Li         &   5.405E-18 \\                        
  $^{7}$Li         &   5.770E-11 &  $^{9}$Be         &   9.439E-21 &  $^{10}$B         &   8.130E-17 &  $^{11}$B         &   1.973E-10 &  $^{12}$C         &   3.492E+00 \\                        
  $^{13}$C         &   2.746E-06 &  $^{14}$N         &   8.994E-05 &  $^{15}$N         &   3.078E-05 &  $^{16}$O         &   5.347E+01 &  $^{17}$O         &   2.448E-06 \\                        
  $^{18}$O         &   1.676E-06 &  $^{19}$F         &   4.624E-07 &  $^{20}$Ne        &   3.194E+00 &  $^{21}$Ne        &   5.502E-05 &  $^{22}$Ne        &   4.736E-05 \\                        
  $^{23}$Na        &   5.352E-03 &  $^{24}$Mg        &   2.984E+00 &  $^{25}$Mg        &   1.733E-03 &  $^{26}$Mg        &   3.968E-03 &  $^{27}$Al        &   2.150E-02 \\                        
  $^{28}$Si        &   8.162E+00 &  $^{29}$Si        &   2.469E-02 &  $^{30}$Si        &   5.936E-03 &  $^{31}$P         &   5.966E-03 &  $^{32}$S         &   4.832E+00 \\
\end{tabular}
	\end{center}
\end{table}
\begin{table}[tbp]
	\begin{center}
	\caption{PISN Yields of the magnetic rotating models \label{tab-yields-mag2}}
	\tabcolsep = 1mm

	\begin{tabular}{lccccccccc}
	\hline
	\hline
	Element	&	Yield			&	Element	&	Yield			&	Element	&	Yield			&	Element	&	Yield			&	Element	&	Yield			\\
	\hline
	\multicolumn{5}{l}{$M_{\rm ini}$ = 200 $M_\odot$} \\
	\hline                   
  $^{}$p           &   3.873E+01 &  $^{}$d           &   1.180E-14 &  $^{3}$He         &   1.535E-05 &  $^{4}$He         &   6.287E+01 &  $^{6}$Li         &   1.052E-16 \\                        
  $^{7}$Li         &   4.119E-10 &  $^{9}$Be         &   2.017E-20 &  $^{10}$B         &   1.277E-15 &  $^{11}$B         &   5.876E-11 &  $^{12}$C         &   2.224E+00 \\                        
  $^{13}$C         &   3.656E-06 &  $^{14}$N         &   7.803E-05 &  $^{15}$N         &   3.813E-05 &  $^{16}$O         &   4.947E+01 &  $^{17}$O         &   4.218E-05 \\                        
  $^{18}$O         &   7.732E-08 &  $^{19}$F         &   1.520E-07 &  $^{20}$Ne        &   2.555E+00 &  $^{21}$Ne        &   3.465E-05 &  $^{22}$Ne        &   3.390E-05 \\                        
  $^{23}$Na        &   4.850E-03 &  $^{24}$Mg        &   2.953E+00 &  $^{25}$Mg        &   1.232E-03 &  $^{26}$Mg        &   3.308E-03 &  $^{27}$Al        &   1.501E-02 \\                        
  $^{28}$Si        &   1.879E+01 &  $^{29}$Si        &   2.408E-02 &  $^{30}$Si        &   3.195E-03 &  $^{31}$P         &   4.345E-03 &  $^{32}$S         &   1.374E+01 \\
\end{tabular}
	\end{center}
\end{table}
\begin{table}[tbp]
	\begin{center}
	\caption{PISN Yields of the magnetic rotating models \label{tab-yields-mag3}}
	\tabcolsep = 1mm

	\begin{tabular}{lccccccccc}
	\hline
	\hline
	Element	&	Yield			&	Element	&	Yield			&	Element	&	Yield			&	Element	&	Yield			&	Element	&	Yield			\\
	\hline
	\multicolumn{5}{l}{$M_{\rm ini}$ = 240 $M_\odot$} \\
	\hline                   
  $^{}$p           &   4.703E+01 &  $^{}$d           &   2.492E-15 &  $^{3}$He         &   2.558E-05 &  $^{4}$He         &   7.455E+01 &  $^{6}$Li         &   5.427E-18 \\                        
  $^{7}$Li         &   4.150E-10 &  $^{9}$Be         &   3.591E-22 &  $^{10}$B         &   9.517E-19 &  $^{11}$B         &   7.650E-11 &  $^{12}$C         &   1.632E+00 \\                        
  $^{13}$C         &   2.418E-06 &  $^{14}$N         &   4.909E-05 &  $^{15}$N         &   2.580E-05 &  $^{16}$O         &   4.117E+01 &  $^{17}$O         &   1.512E-05 \\                        
  $^{18}$O         &   4.155E-08 &  $^{19}$F         &   9.189E-08 &  $^{20}$Ne        &   2.134E+00 &  $^{21}$Ne        &   2.100E-05 &  $^{22}$Ne        &   3.209E-05 \\                        
  $^{23}$Na        &   3.932E-03 &  $^{24}$Mg        &   2.626E+00 &  $^{25}$Mg        &   9.028E-04 &  $^{26}$Mg        &   2.632E-03 &  $^{27}$Al        &   1.150E-02 \\                        
  $^{28}$Si        &   2.236E+01 &  $^{29}$Si        &   1.732E-02 &  $^{30}$Si        &   2.195E-03 &  $^{31}$P         &   2.833E-03 &  $^{32}$S         &   1.738E+01 \\
\end{tabular}
	\end{center}
\end{table}

\section{Comparison of metal-poor stellar abundances with PISN yields} \label{app-2}

\begin{figure}[tbp]
	\begin{minipage}{0.5\textwidth}
		\includegraphics[width=\textwidth]{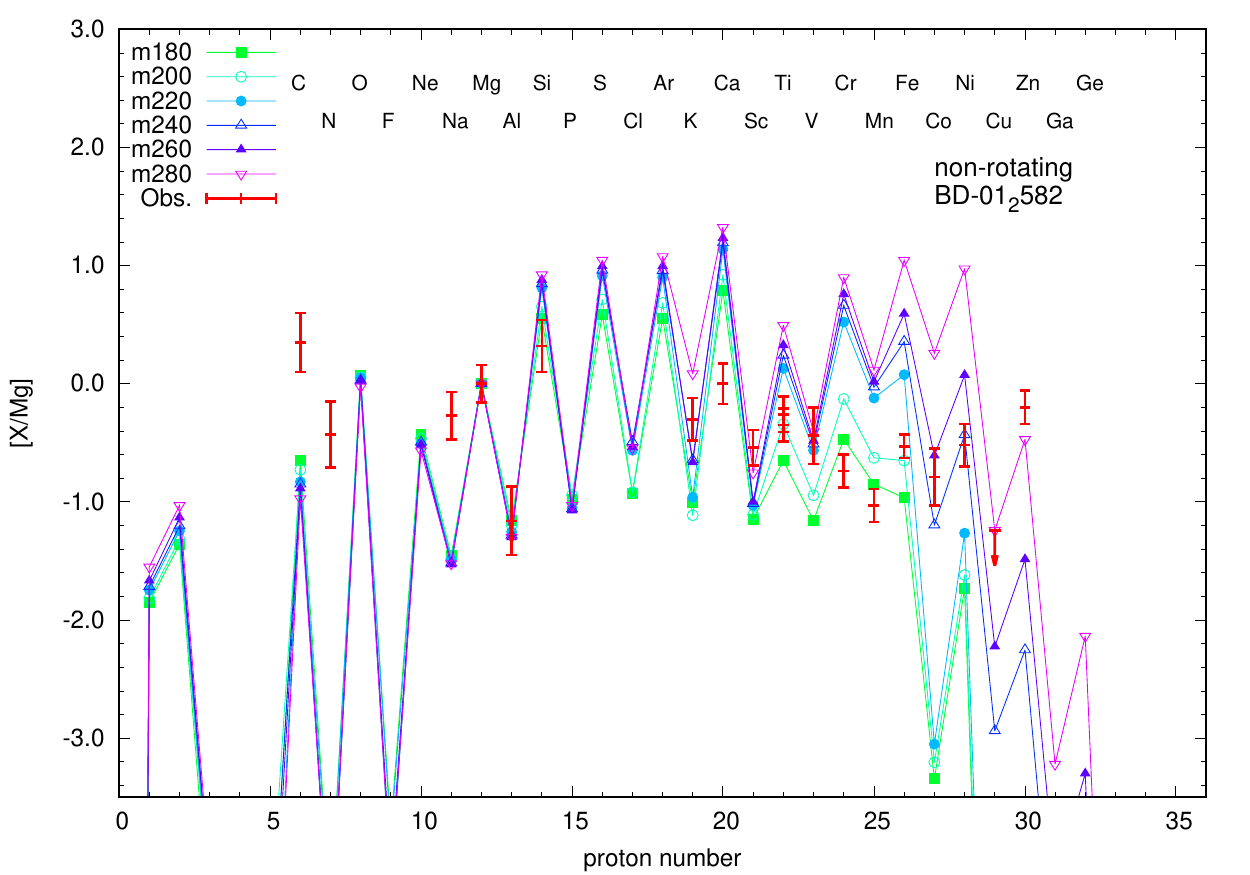}
	\end{minipage}
	\begin{minipage}{0.5\textwidth}
		\includegraphics[width=\textwidth]{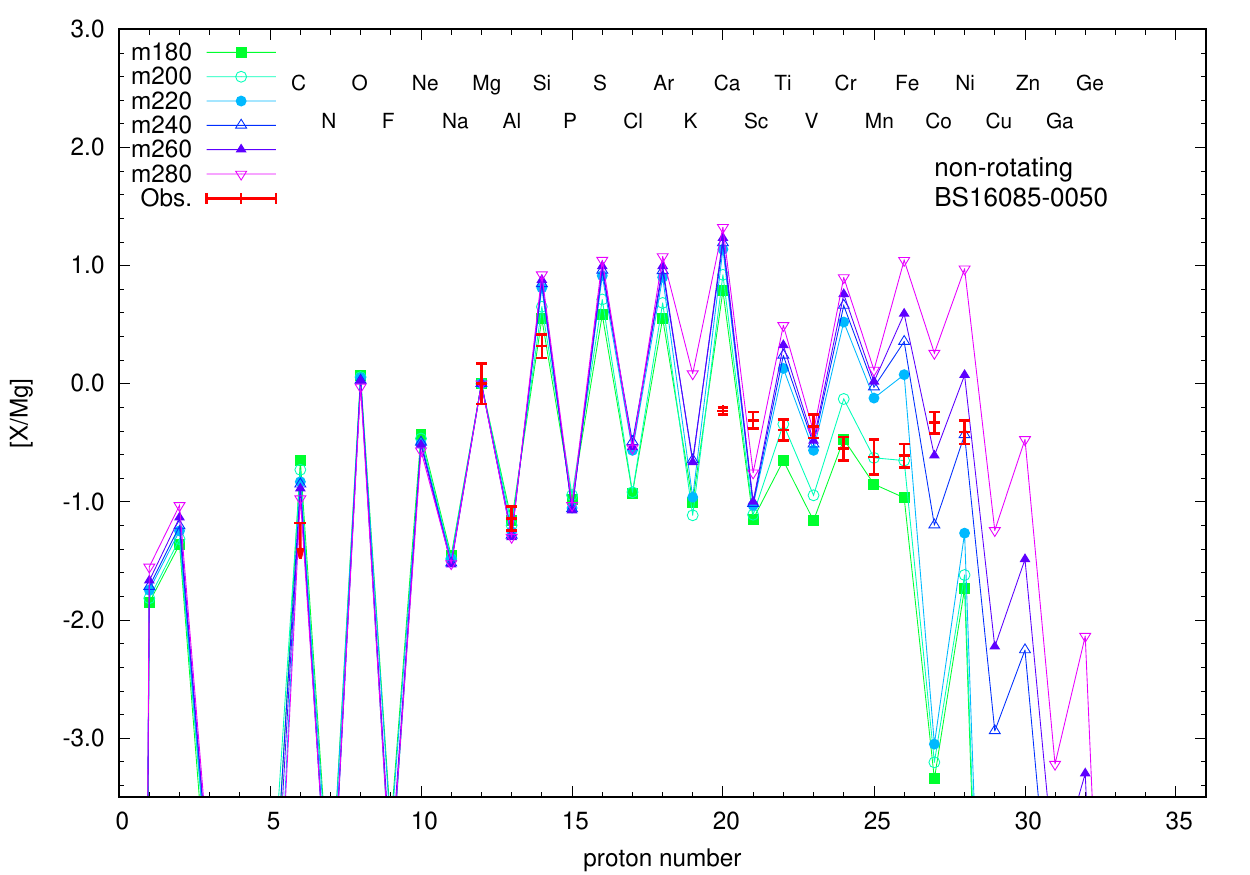}
	\end{minipage}

	\begin{minipage}{0.5\textwidth}
		\includegraphics[width=\textwidth]{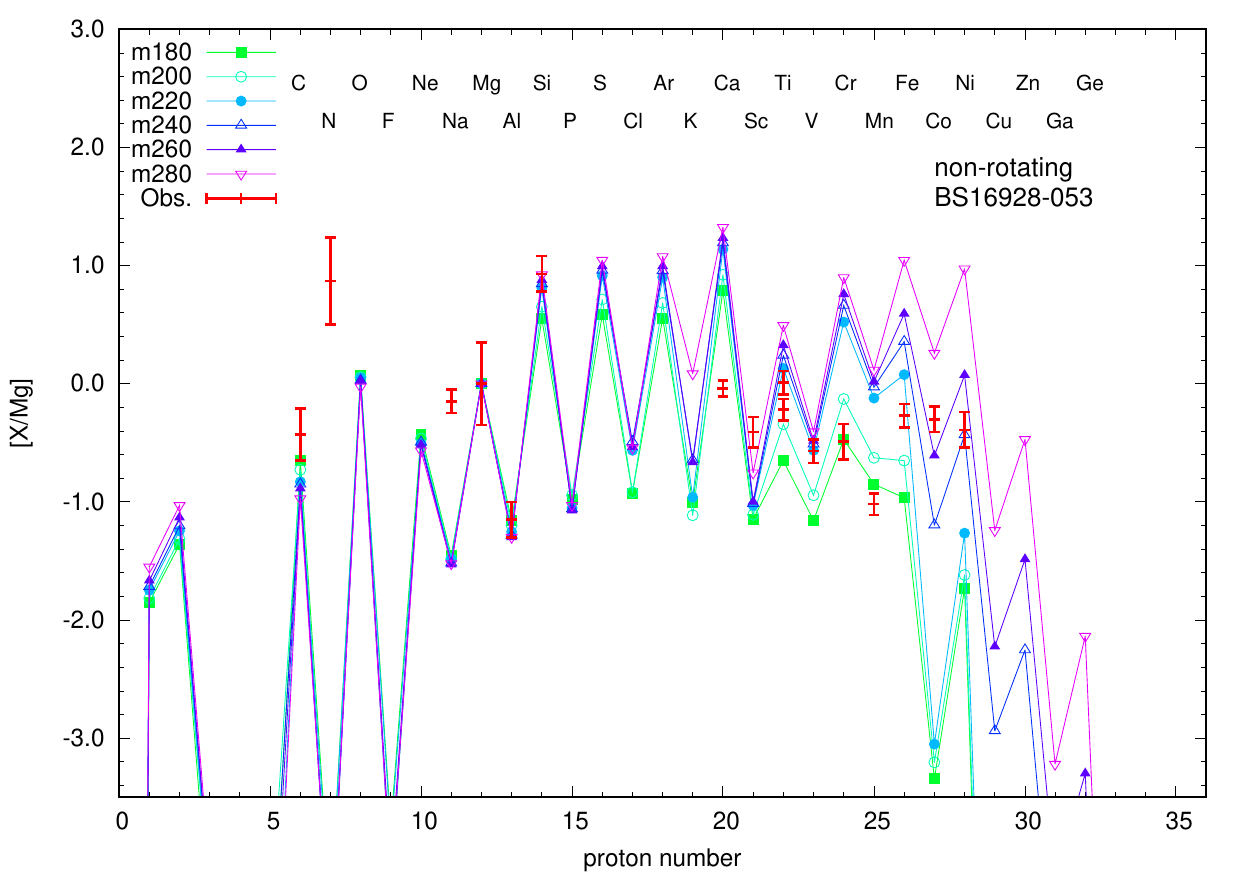}
	\end{minipage}
	\begin{minipage}{0.5\textwidth}
		\includegraphics[width=\textwidth]{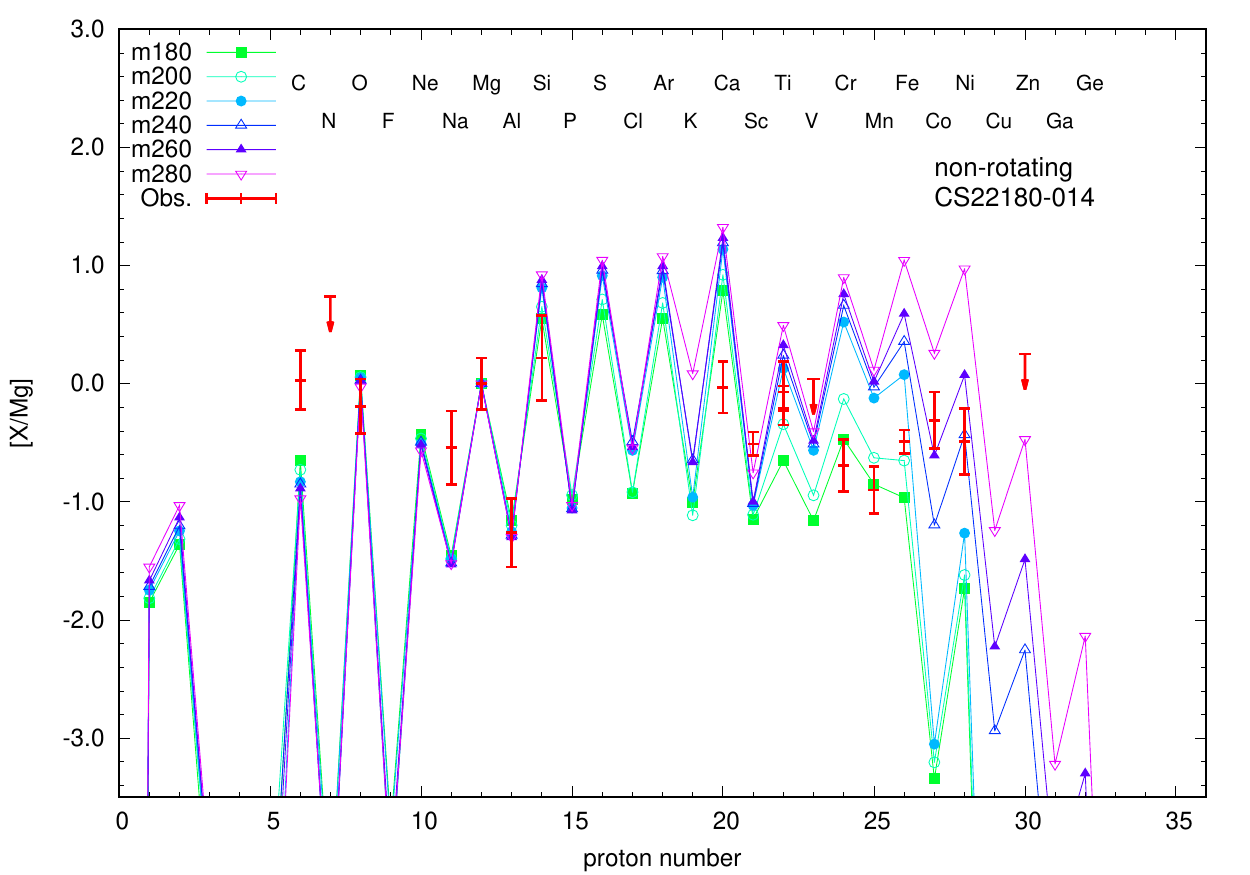}
	\end{minipage}

	\begin{minipage}{0.5\textwidth}
		\includegraphics[width=\textwidth]{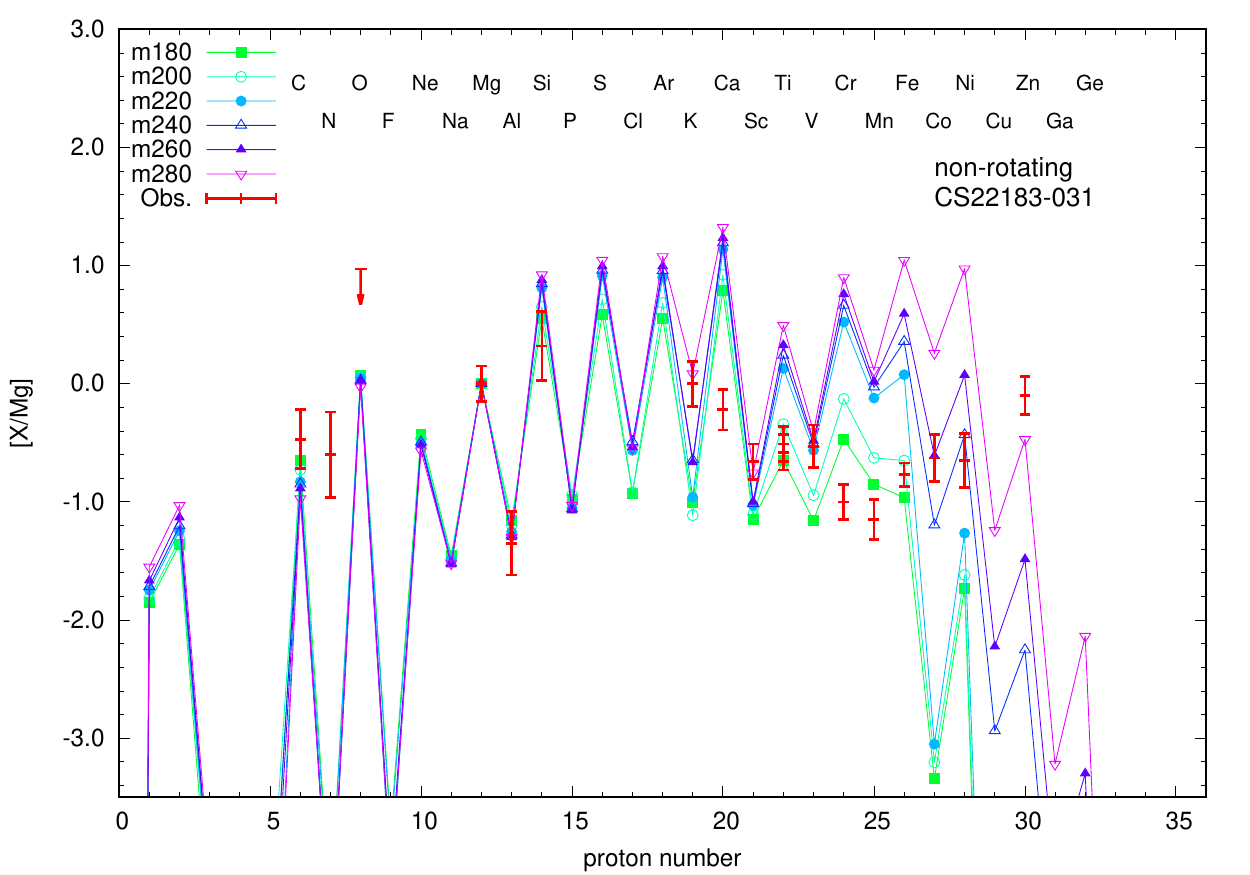}
	\end{minipage}
	\begin{minipage}{0.5\textwidth}
		\includegraphics[width=\textwidth]{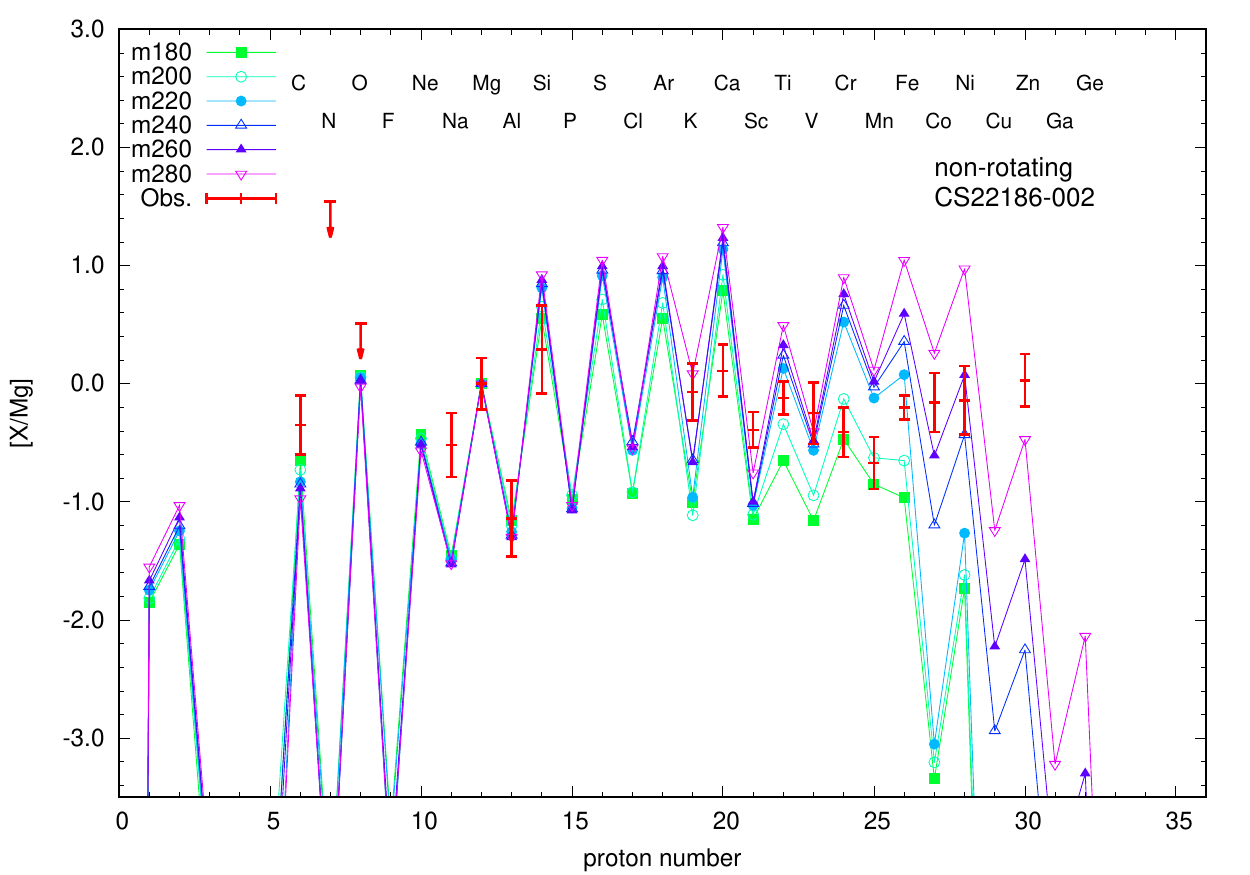}
	\end{minipage}
	
	\caption{ \footnotesize{The same as Fig.\ref{fig-stars-PISN1} but for MP stars of \#1--6.}}
\end{figure}

\begin{figure}[tbp]
	\begin{minipage}{0.5\textwidth}
		\includegraphics[width=\textwidth]{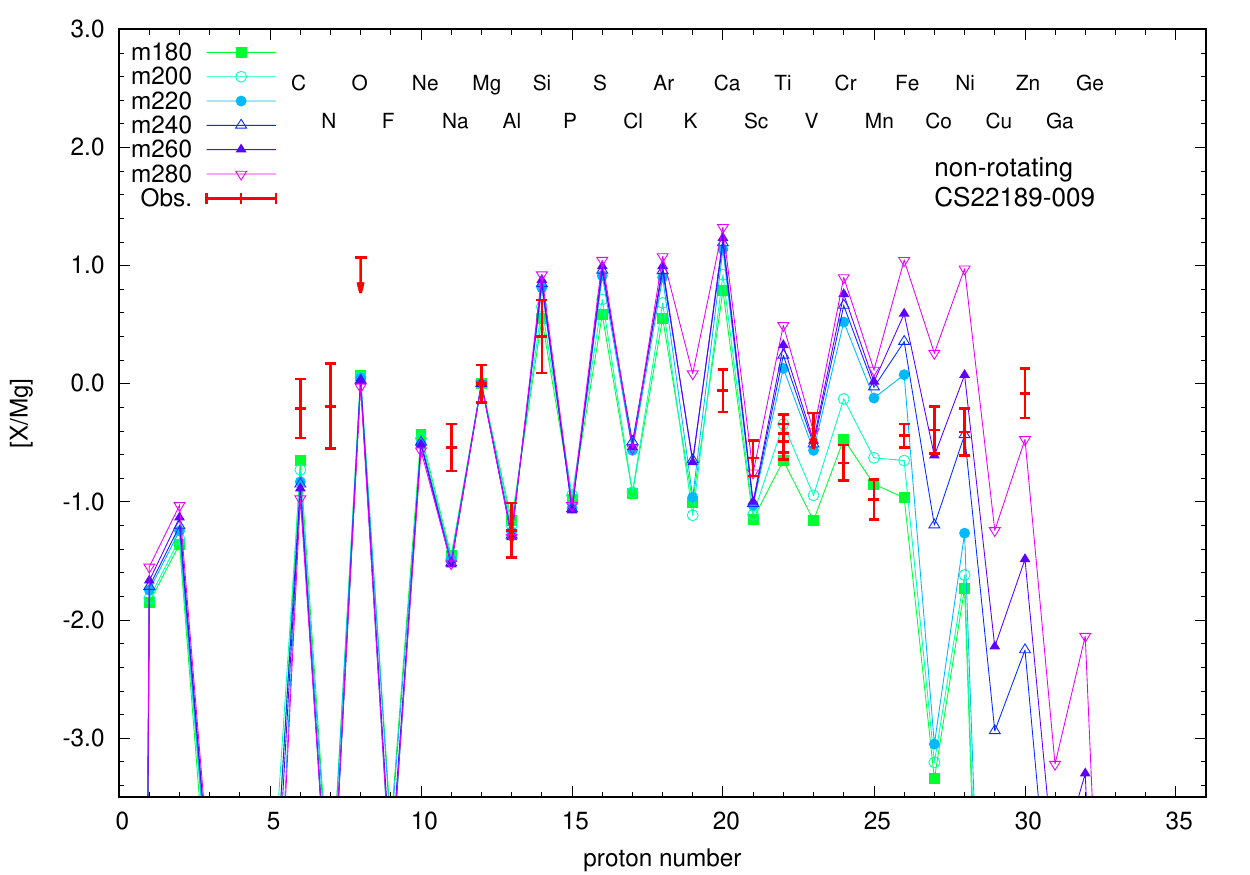}
	\end{minipage}
	\begin{minipage}{0.5\textwidth}
		\includegraphics[width=\textwidth]{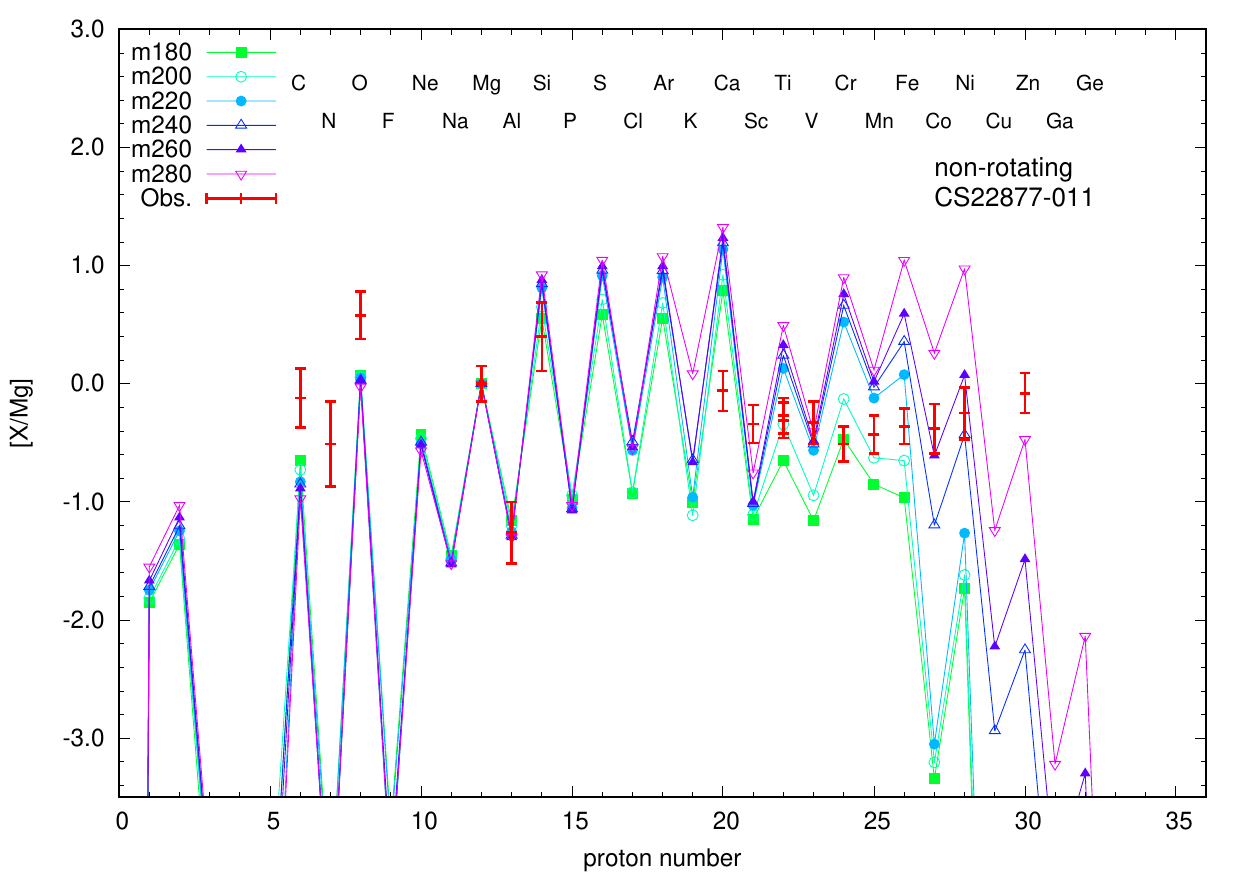}
	\end{minipage}

	\begin{minipage}{0.5\textwidth}
		\includegraphics[width=\textwidth]{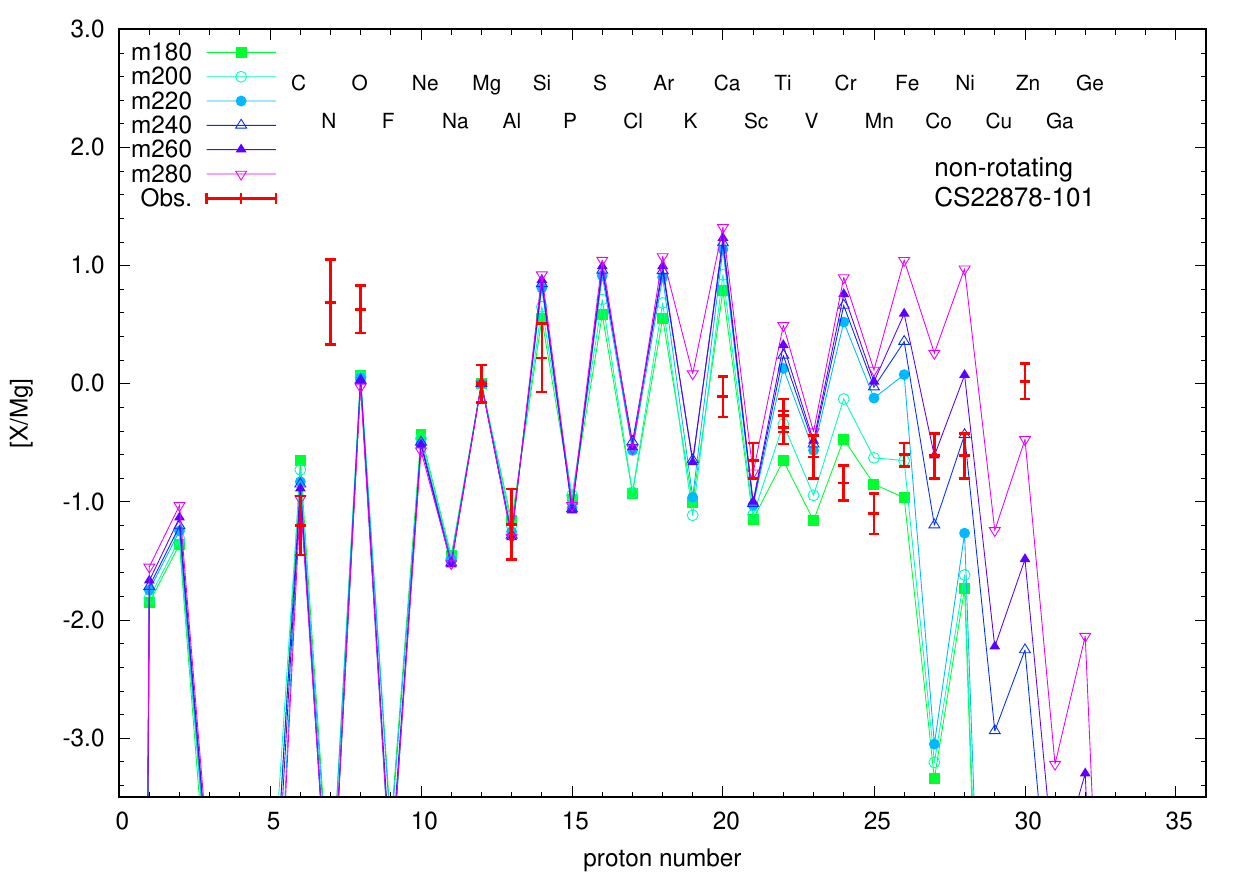}
	\end{minipage}
	\begin{minipage}{0.5\textwidth}
		\includegraphics[width=\textwidth]{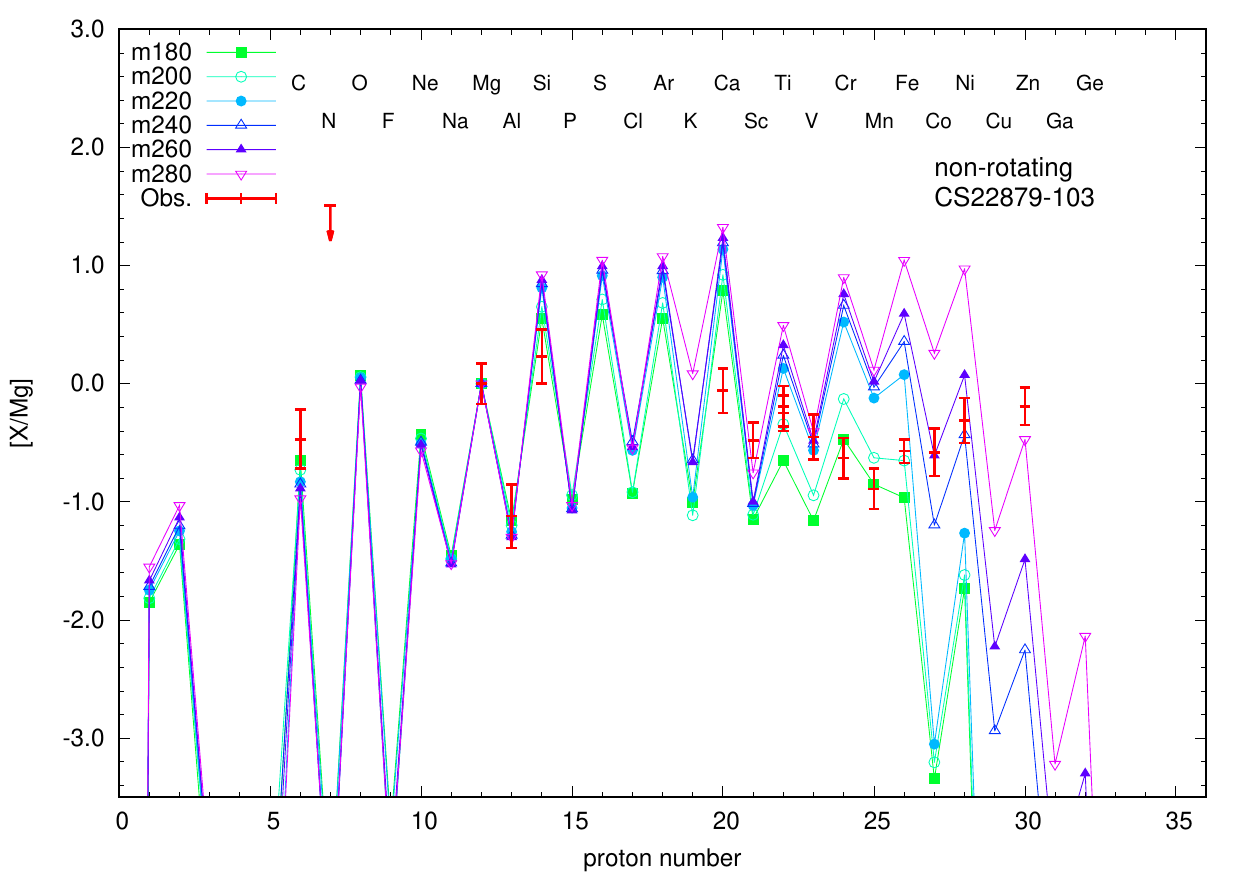}
	\end{minipage}

	\begin{minipage}{0.5\textwidth}
		\includegraphics[width=\textwidth]{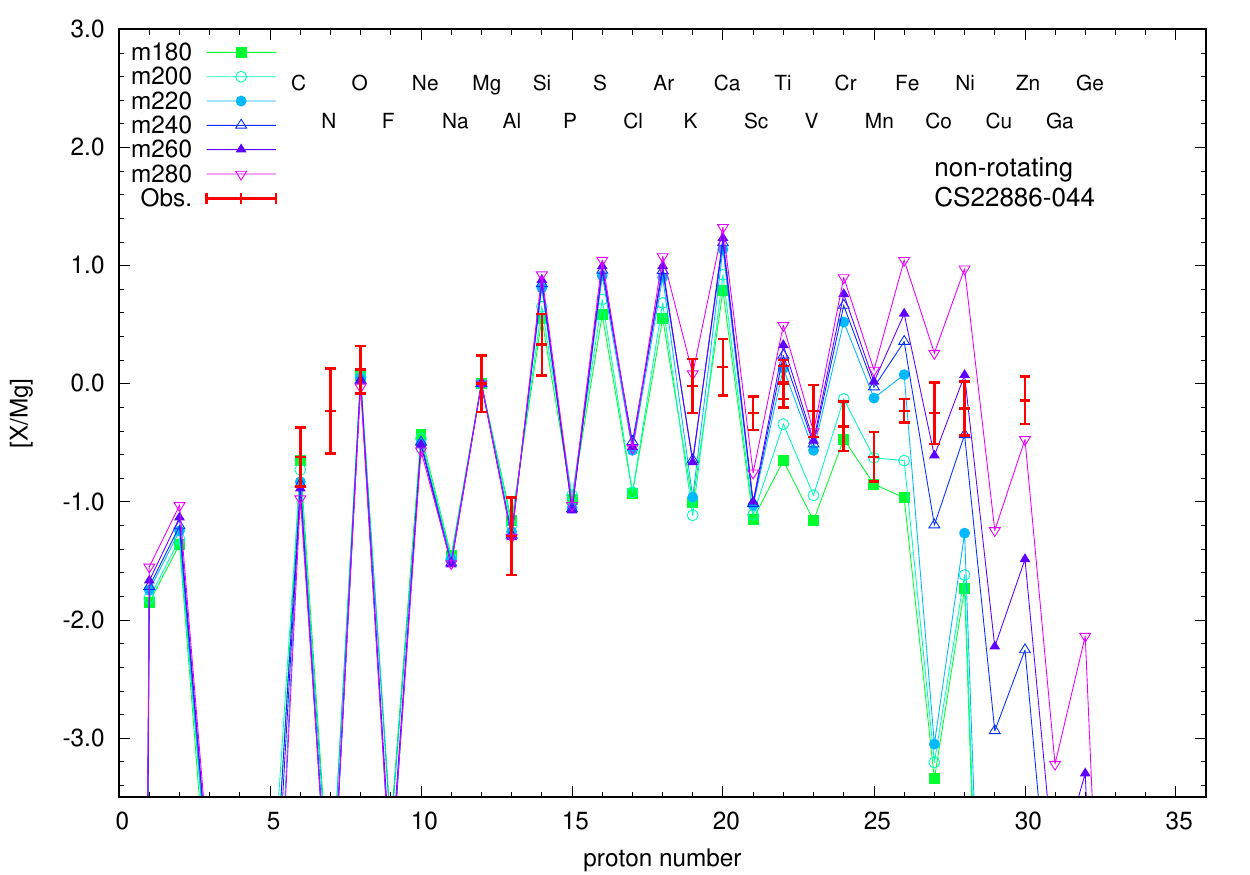}
	\end{minipage}
	\begin{minipage}{0.5\textwidth}
		\includegraphics[width=\textwidth]{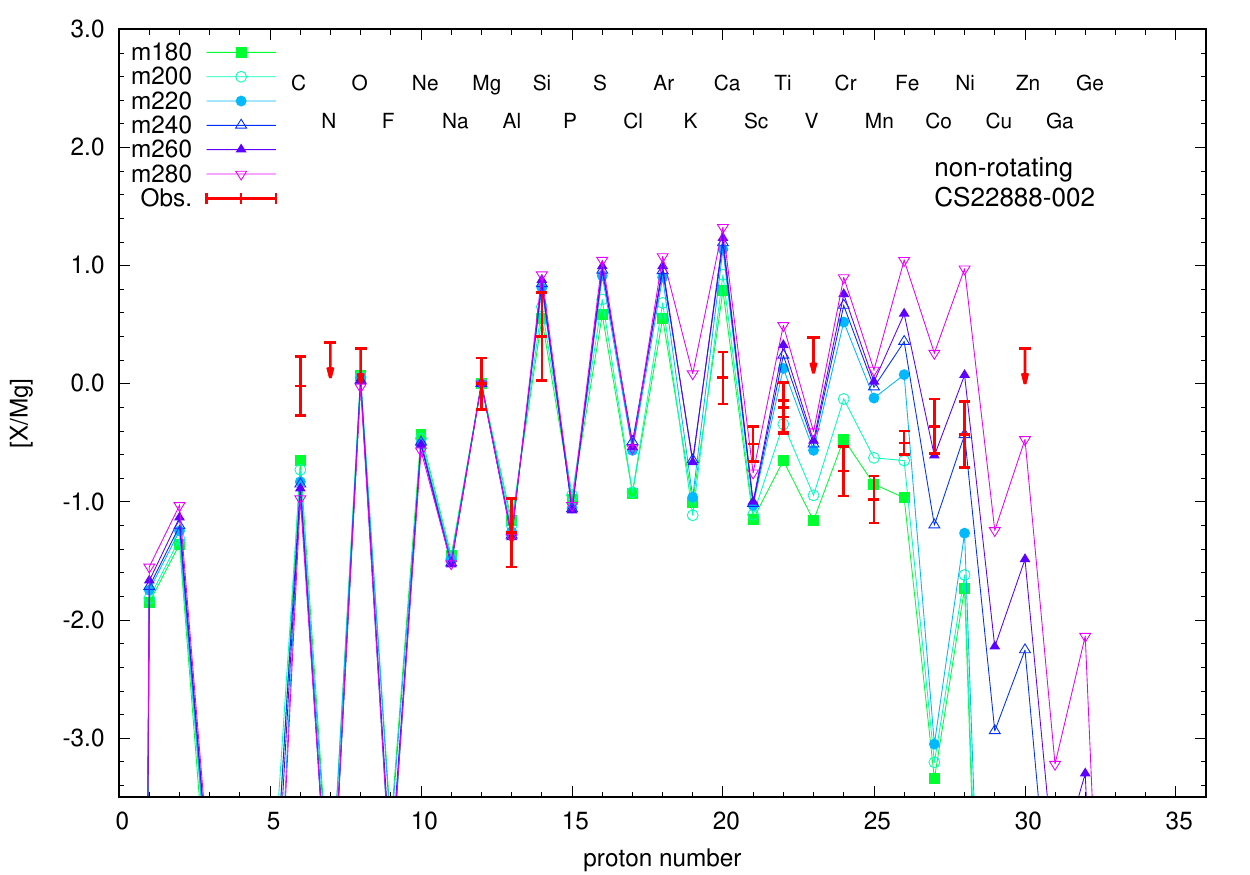}
	\end{minipage}
	
	\caption{ \footnotesize{The same as Fig.\ref{fig-stars-PISN1} but for MP stars of \#7--12.}}
\end{figure}

\begin{figure}[tbp]
	\begin{minipage}{0.5\textwidth}
		\includegraphics[width=\textwidth]{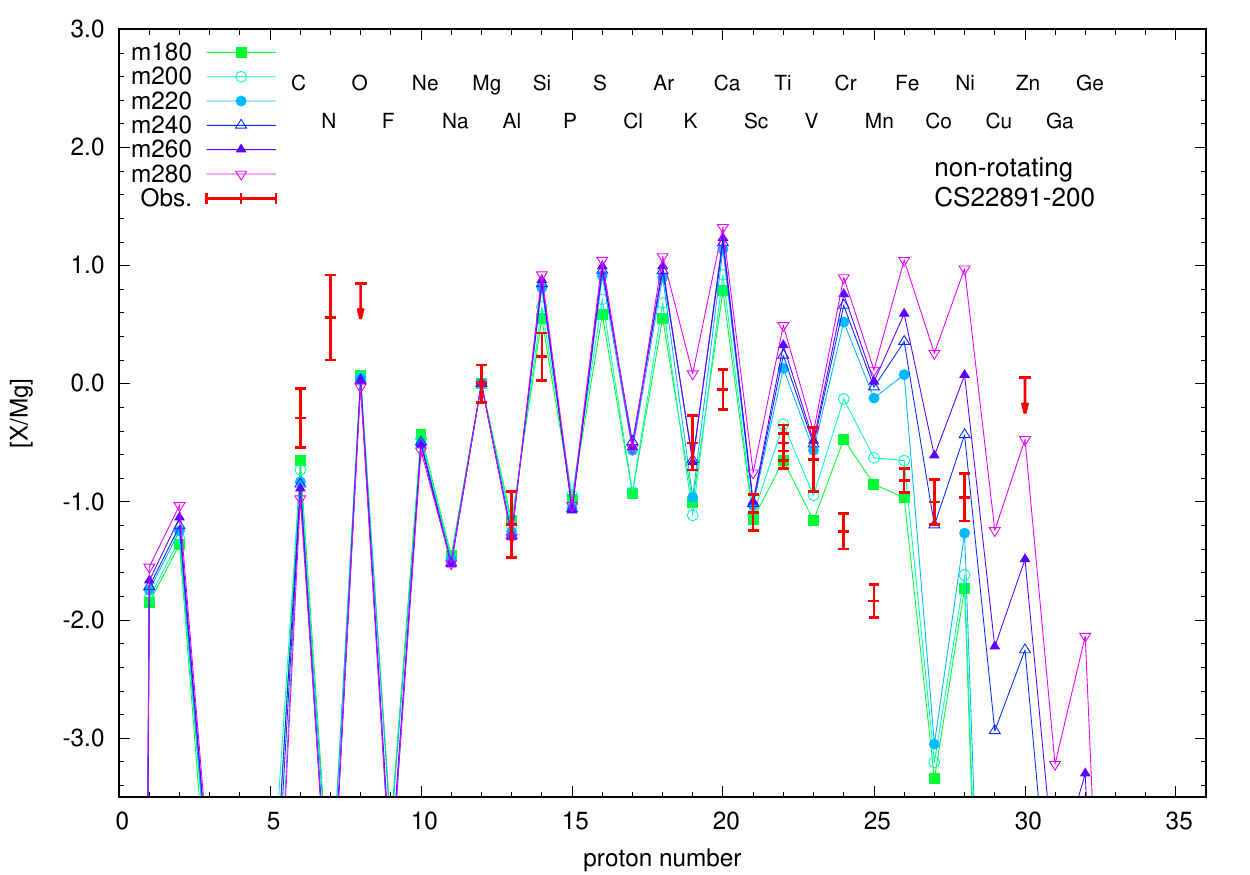}
	\end{minipage}
	\begin{minipage}{0.5\textwidth}
		\includegraphics[width=\textwidth]{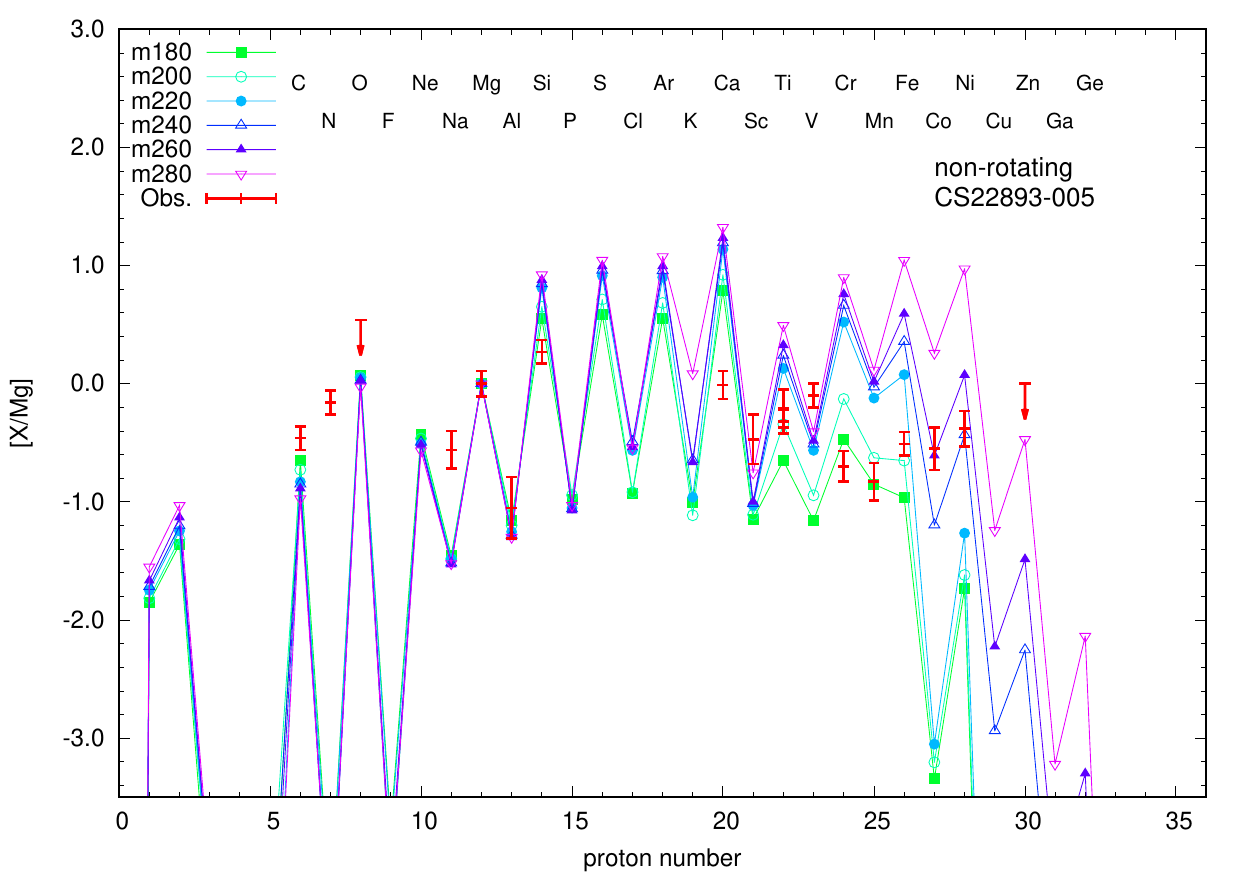}
	\end{minipage}

	\begin{minipage}{0.5\textwidth}
		\includegraphics[width=\textwidth]{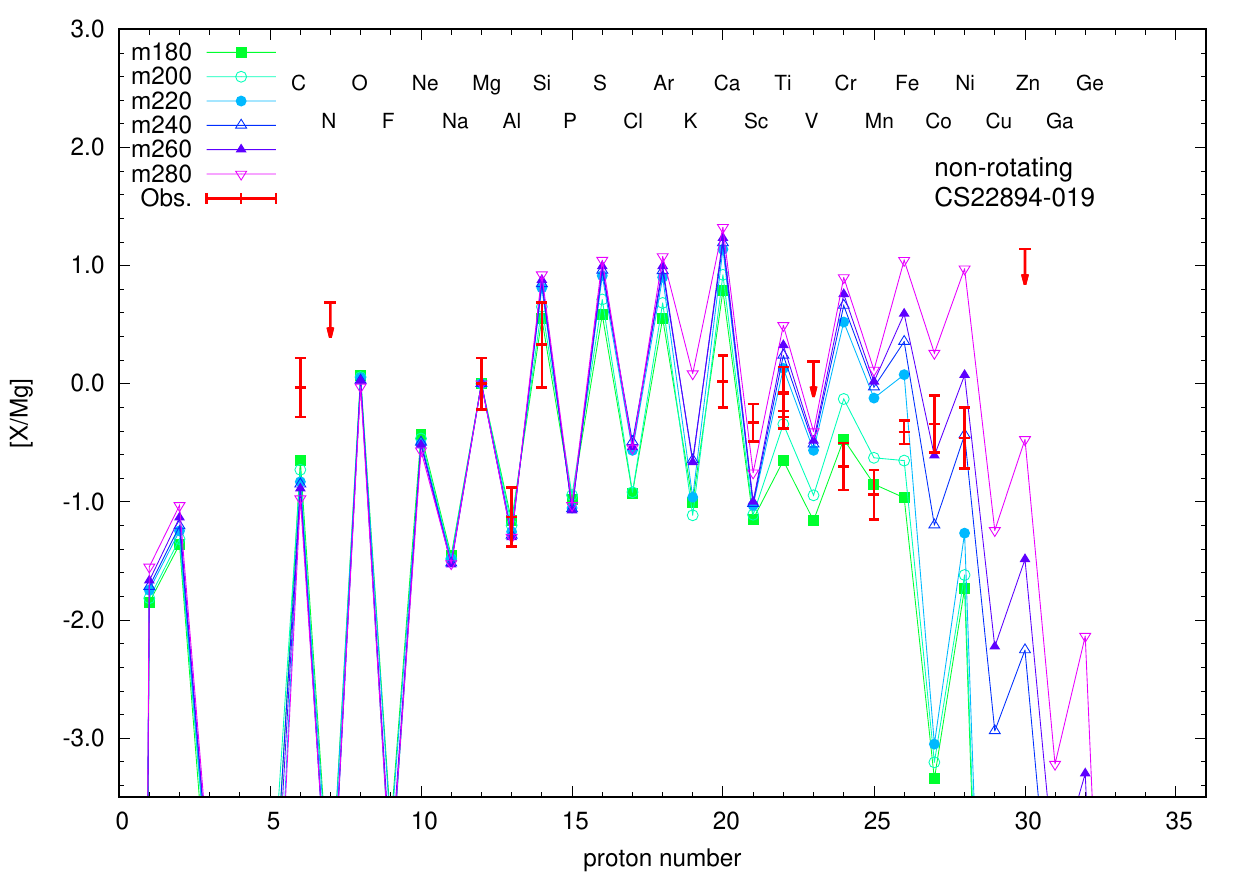}
	\end{minipage}
	\begin{minipage}{0.5\textwidth}
		\includegraphics[width=\textwidth]{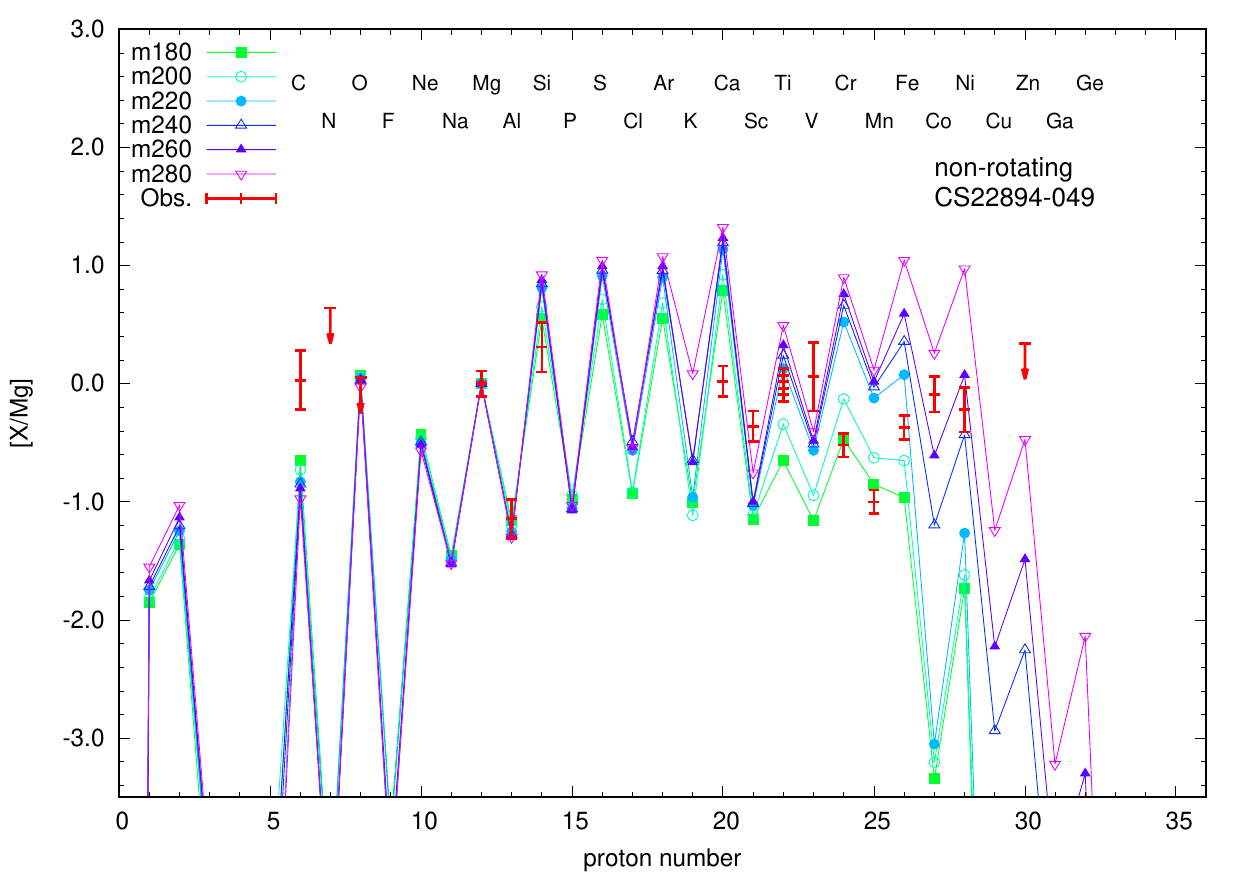}
	\end{minipage}

	\begin{minipage}{0.5\textwidth}
		\includegraphics[width=\textwidth]{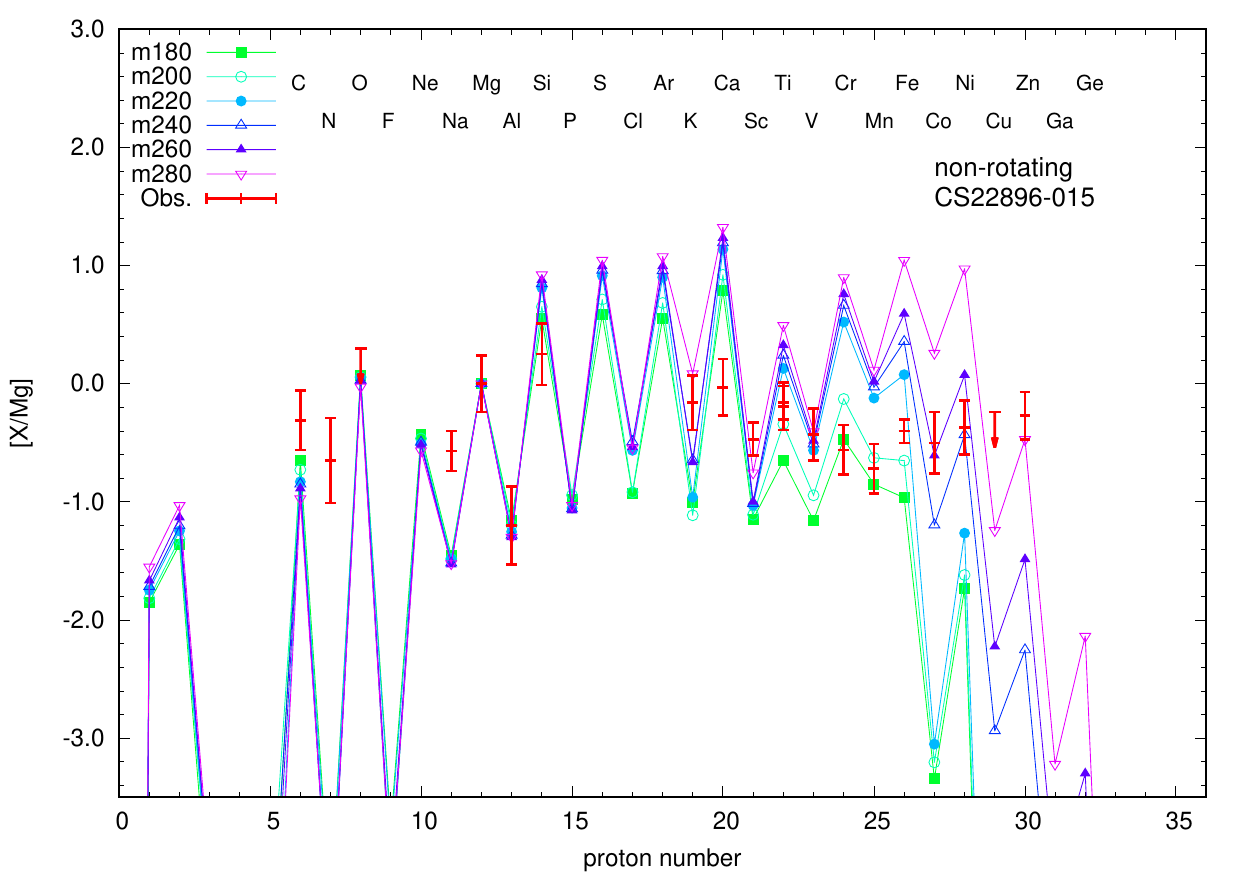}
	\end{minipage}
	\begin{minipage}{0.5\textwidth}
		\includegraphics[width=\textwidth]{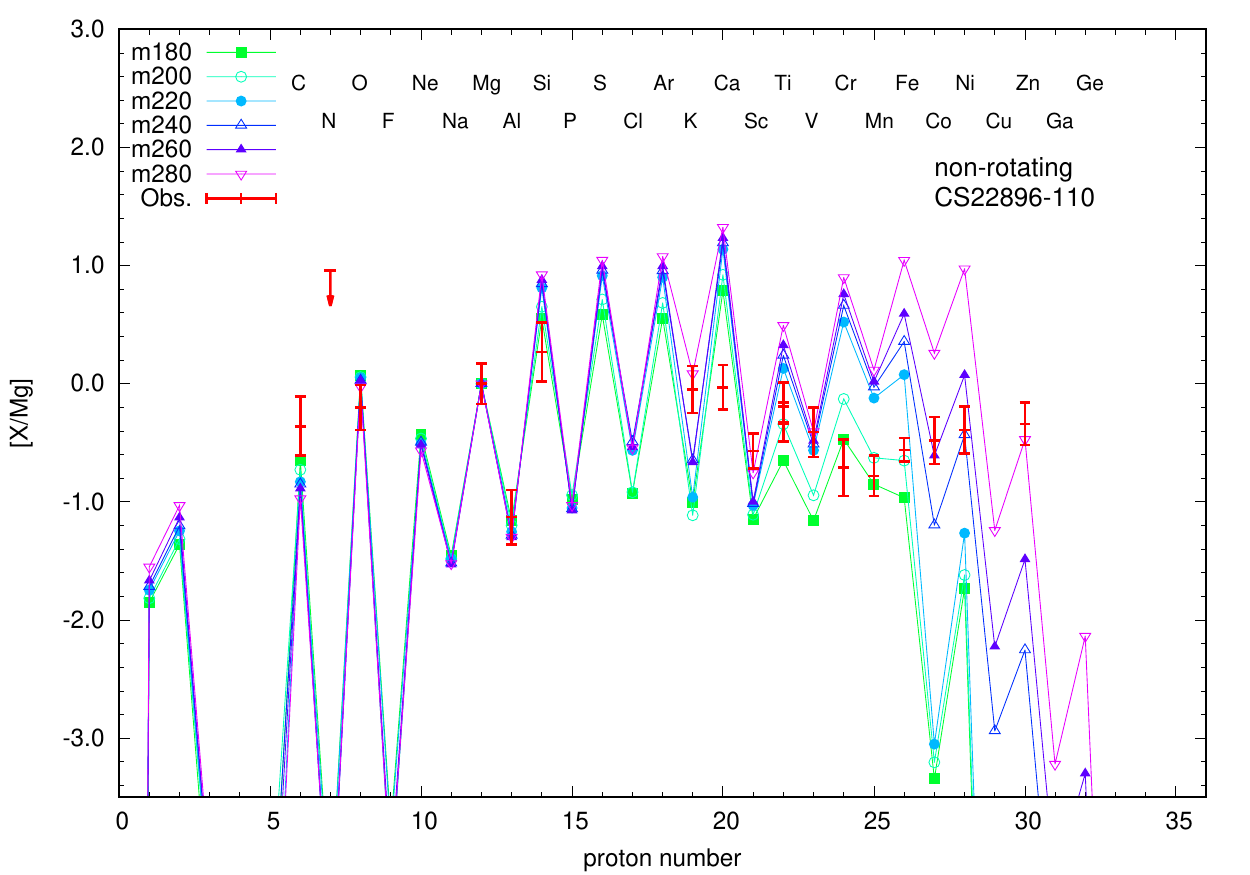}
	\end{minipage}
	
	\caption{ \footnotesize{The same as Fig.\ref{fig-stars-PISN1} but for MP stars of \#13--18.}}
\end{figure}

\begin{figure}[tbp]
	\begin{minipage}{0.5\textwidth}
		\includegraphics[width=\textwidth]{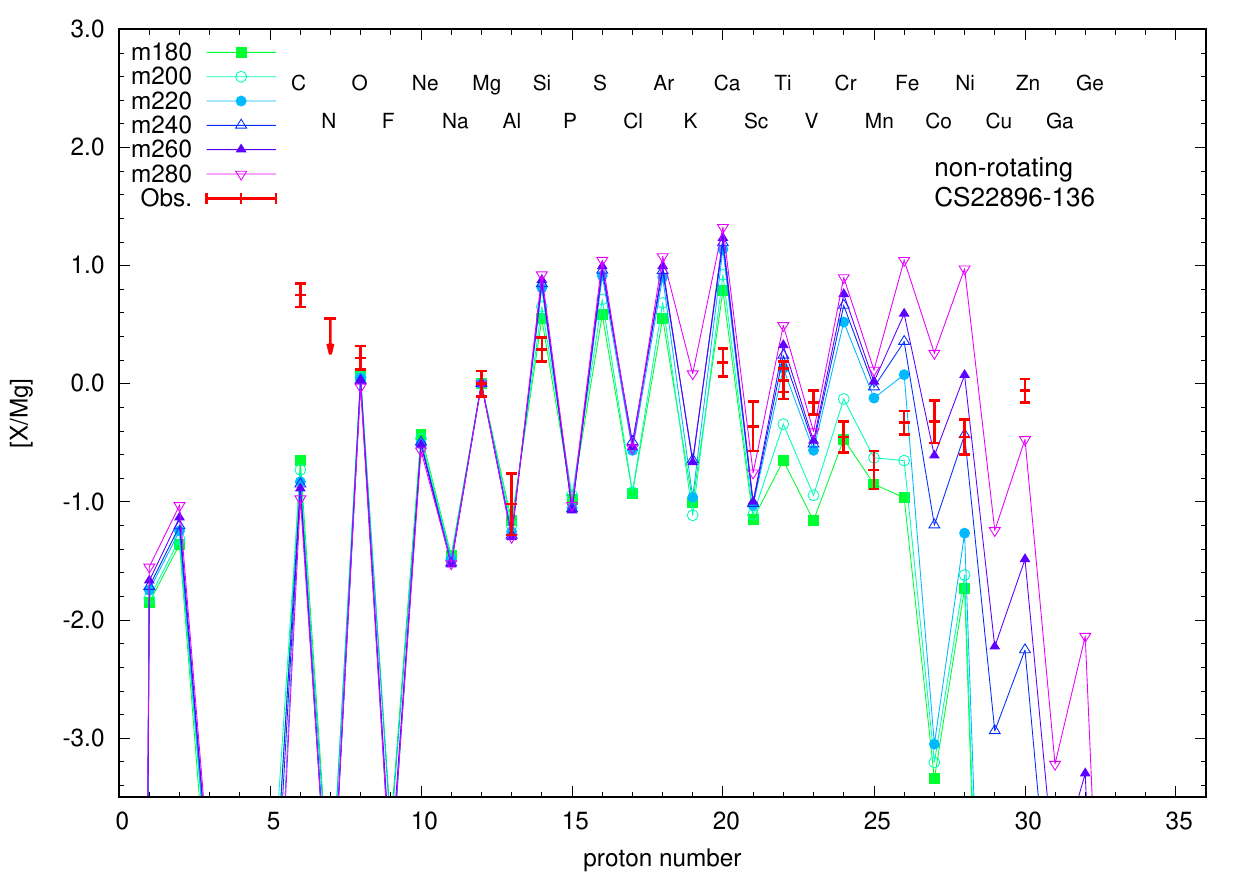}
	\end{minipage}
	\begin{minipage}{0.5\textwidth}
		\includegraphics[width=\textwidth]{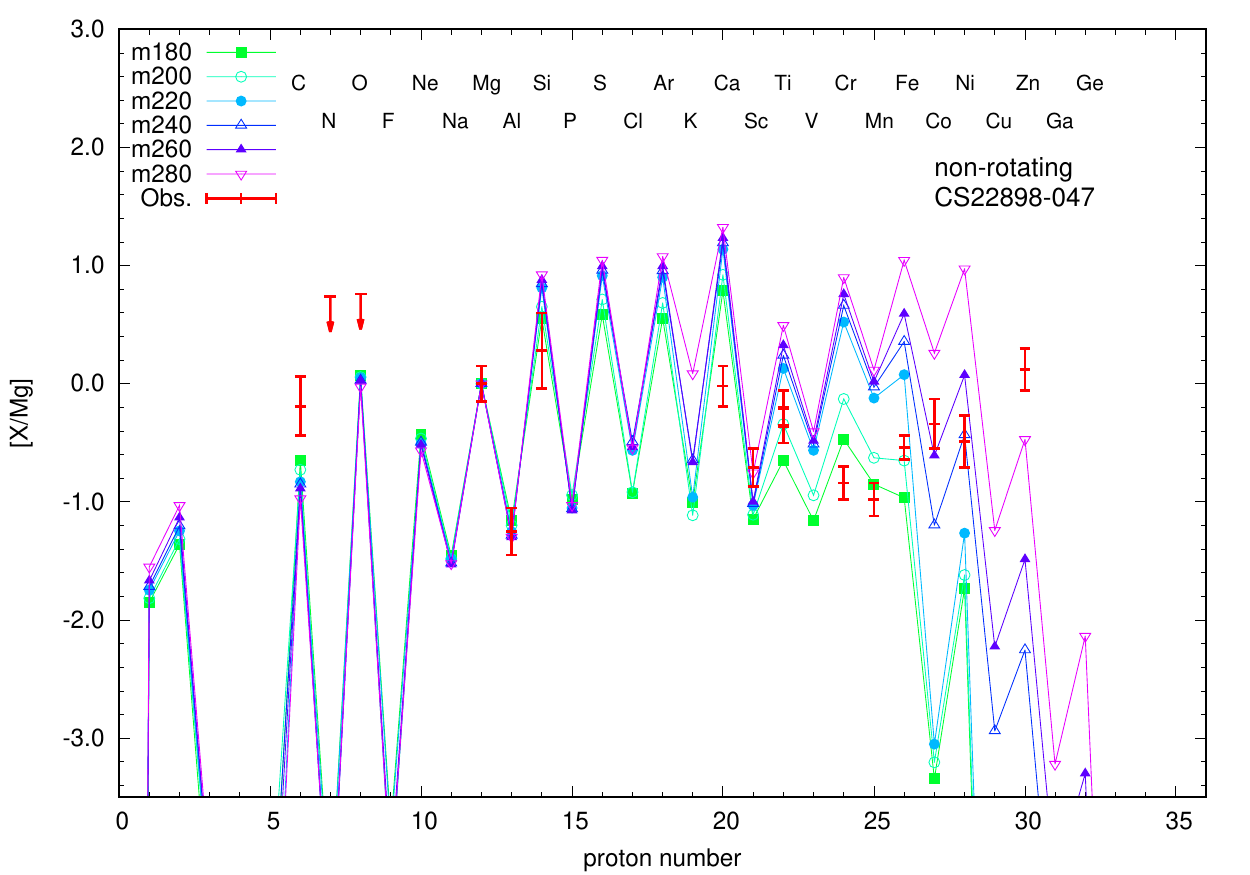}
	\end{minipage}

	\begin{minipage}{0.5\textwidth}
		\includegraphics[width=\textwidth]{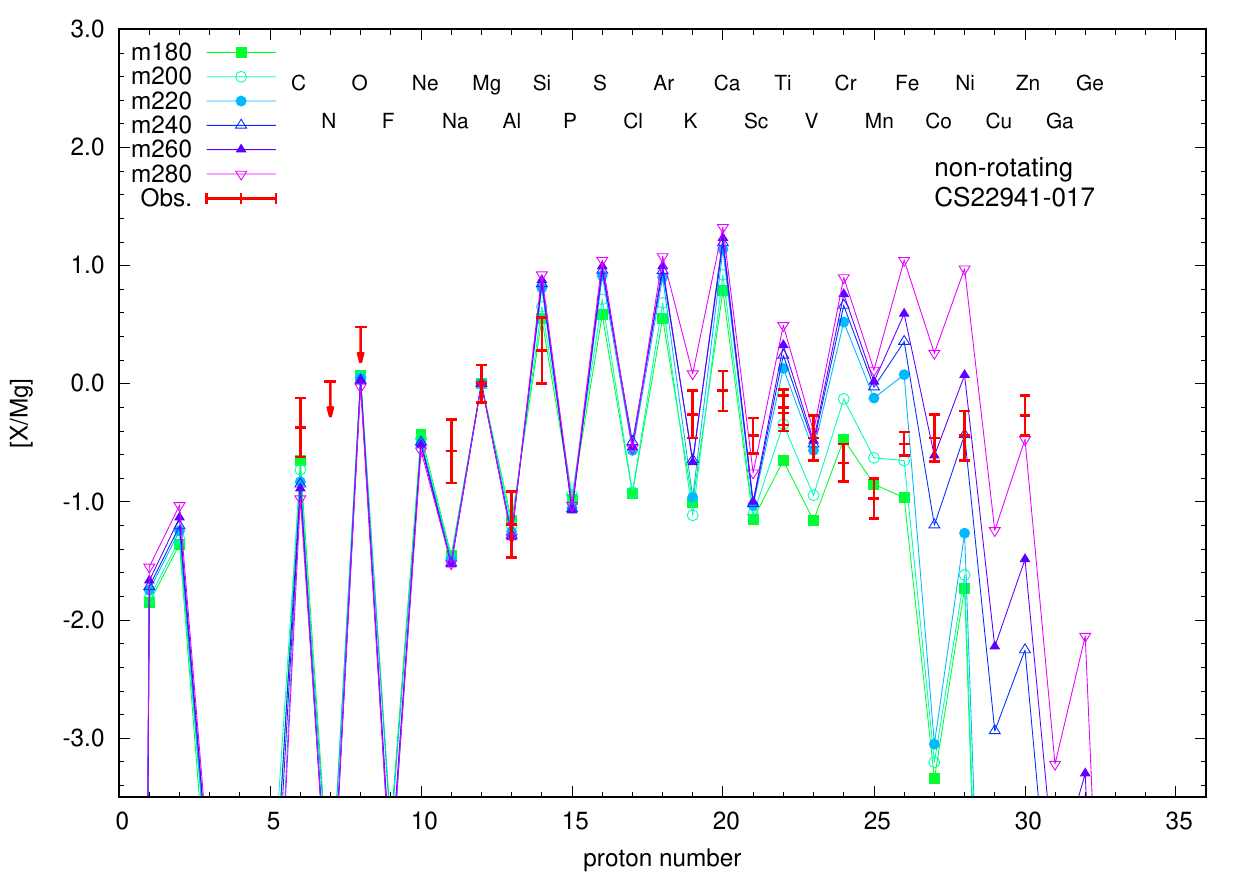}
	\end{minipage}
	\begin{minipage}{0.5\textwidth}
		\includegraphics[width=\textwidth]{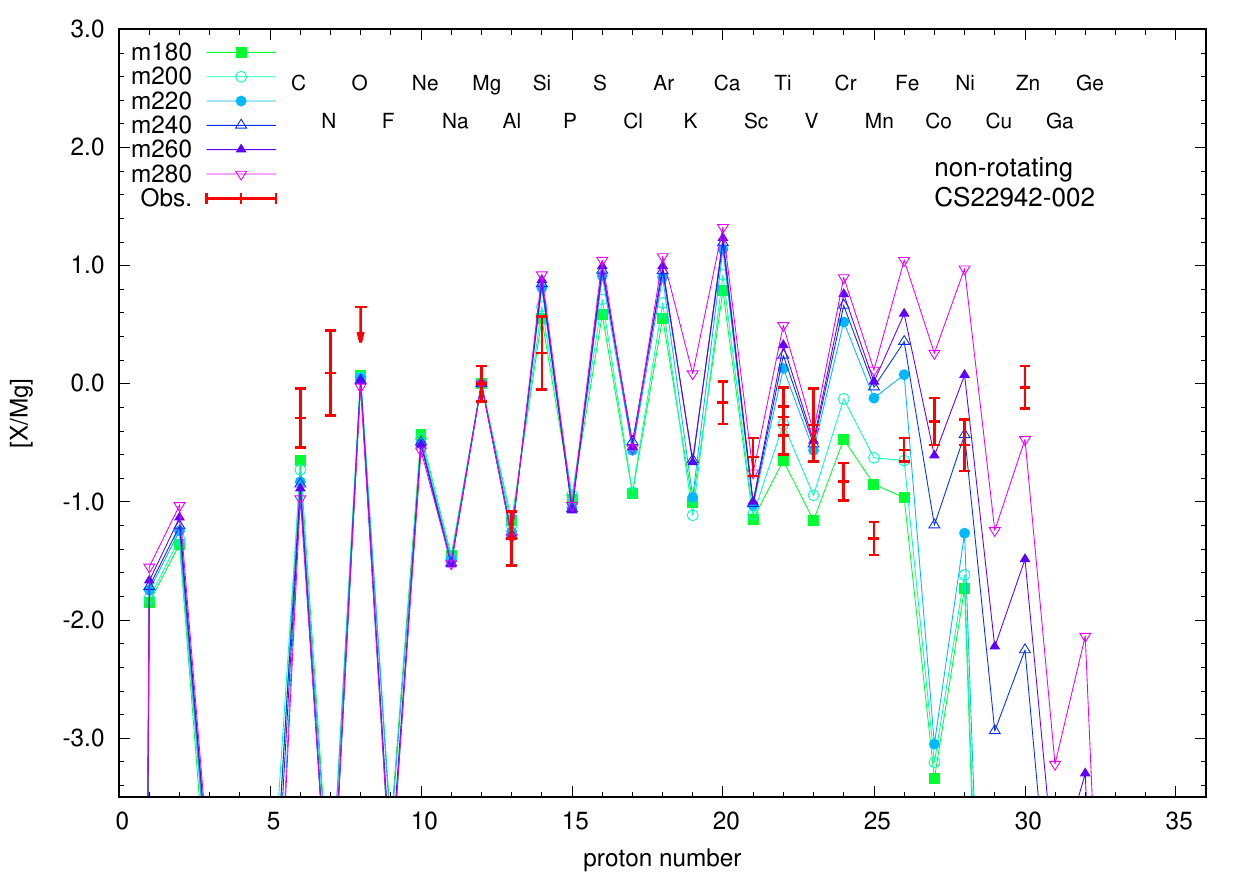}
	\end{minipage}

	\begin{minipage}{0.5\textwidth}
		\includegraphics[width=\textwidth]{stars_CS22942-011_o00.pdf}
	\end{minipage}
	\begin{minipage}{0.5\textwidth}
		\includegraphics[width=\textwidth]{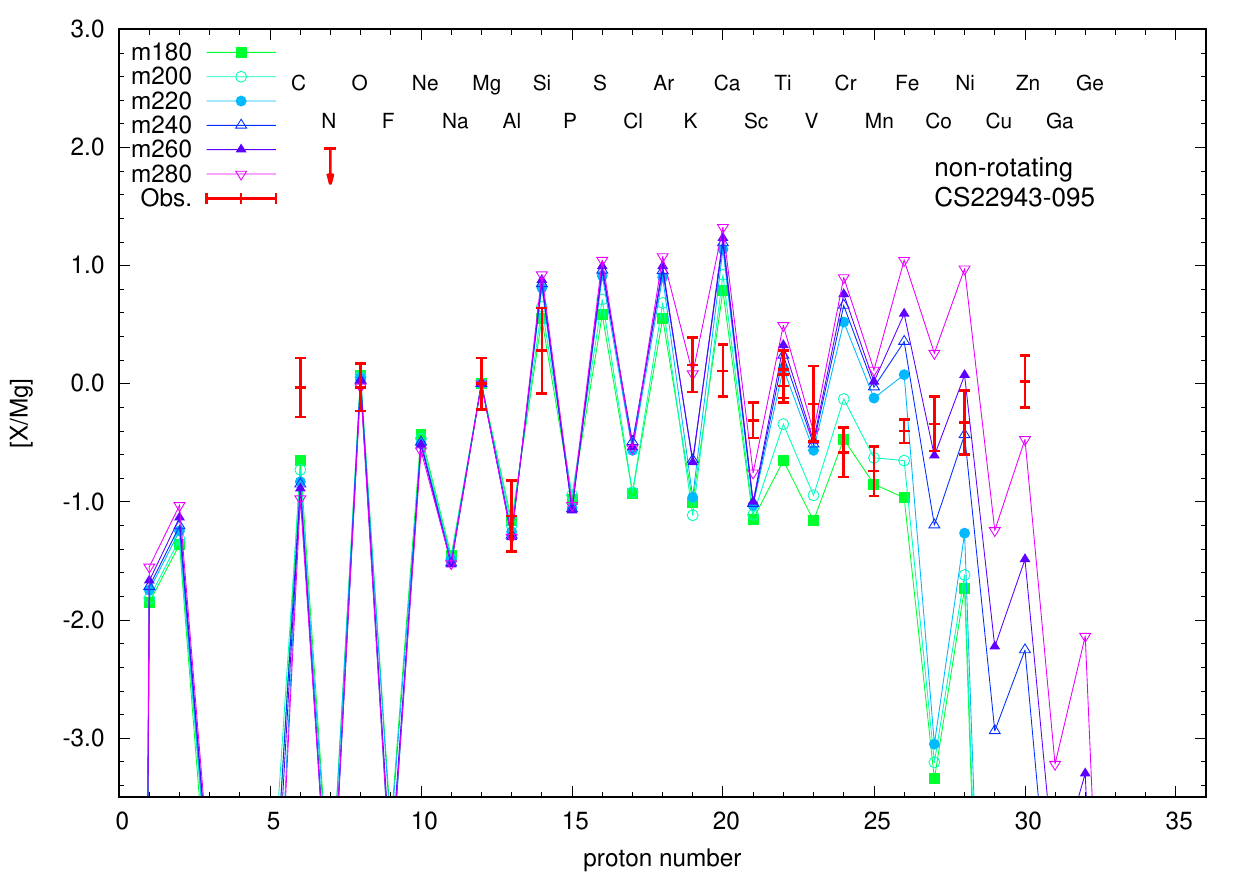}
	\end{minipage}
	
	\caption{ \footnotesize{The same as Fig.\ref{fig-stars-PISN1} but for MP stars of \#19--24.}}
\end{figure}

\begin{figure}[tbp]
	\begin{minipage}{0.5\textwidth}
		\includegraphics[width=\textwidth]{stars_CS22943-095_o00.pdf}
	\end{minipage}
	\begin{minipage}{0.5\textwidth}
		\includegraphics[width=\textwidth]{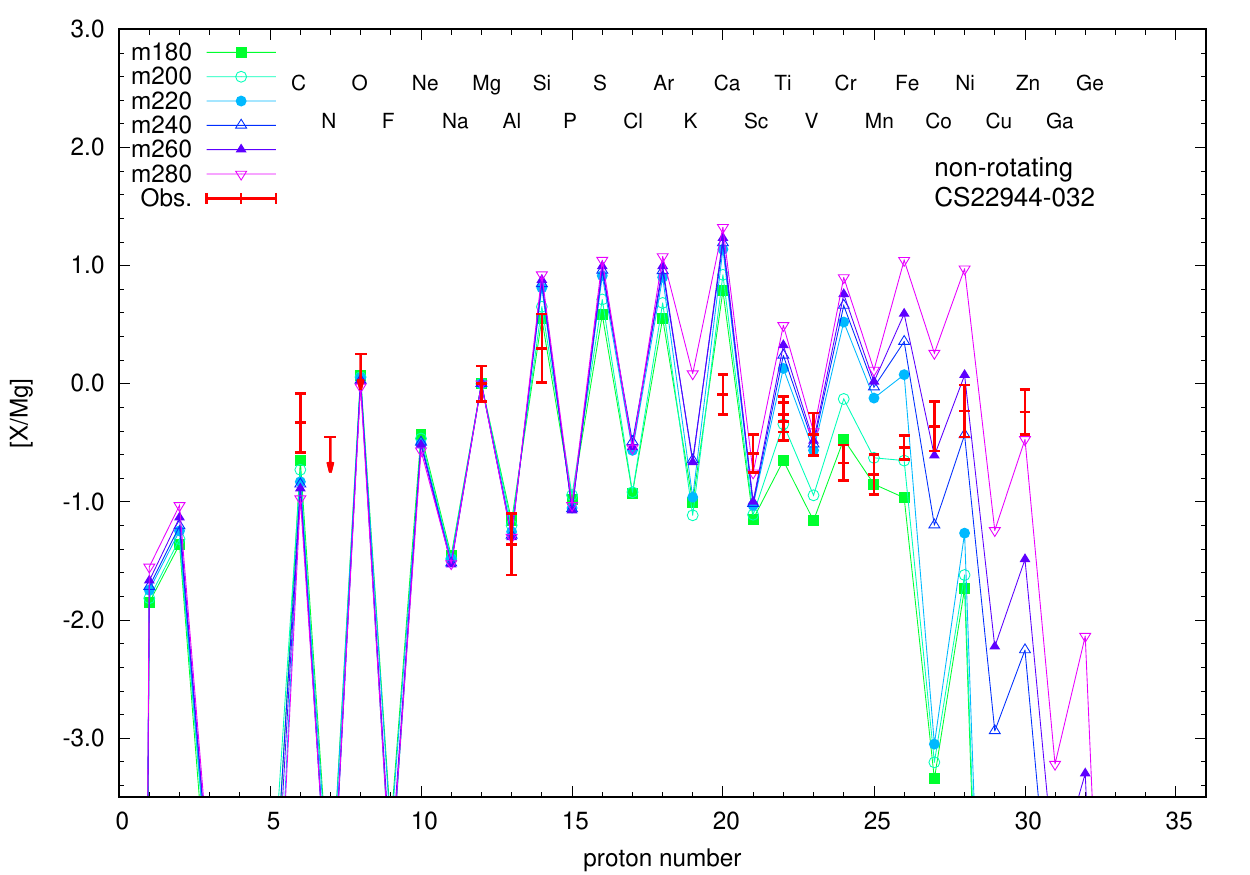}
	\end{minipage}

	\begin{minipage}{0.5\textwidth}
		\includegraphics[width=\textwidth]{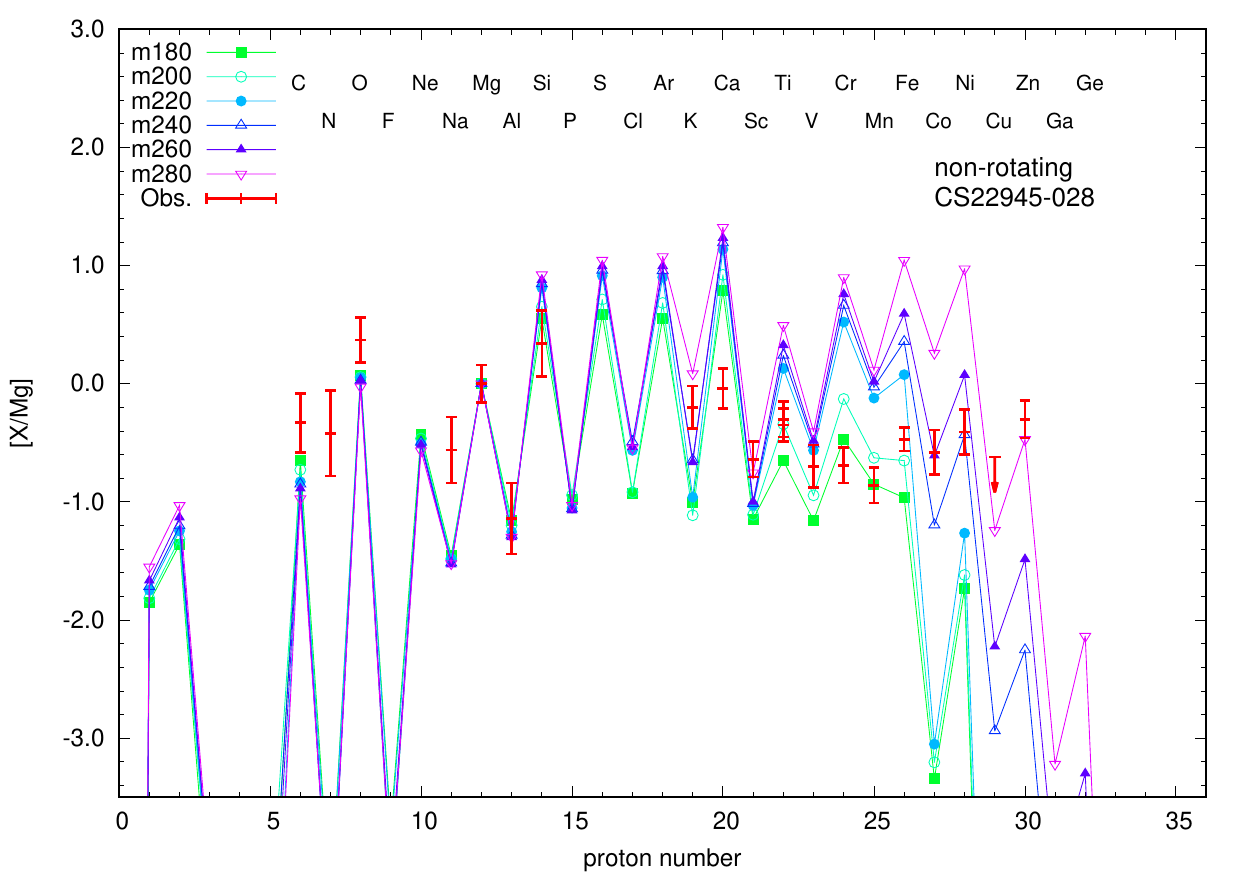}
	\end{minipage}
	\begin{minipage}{0.5\textwidth}
		\includegraphics[width=\textwidth]{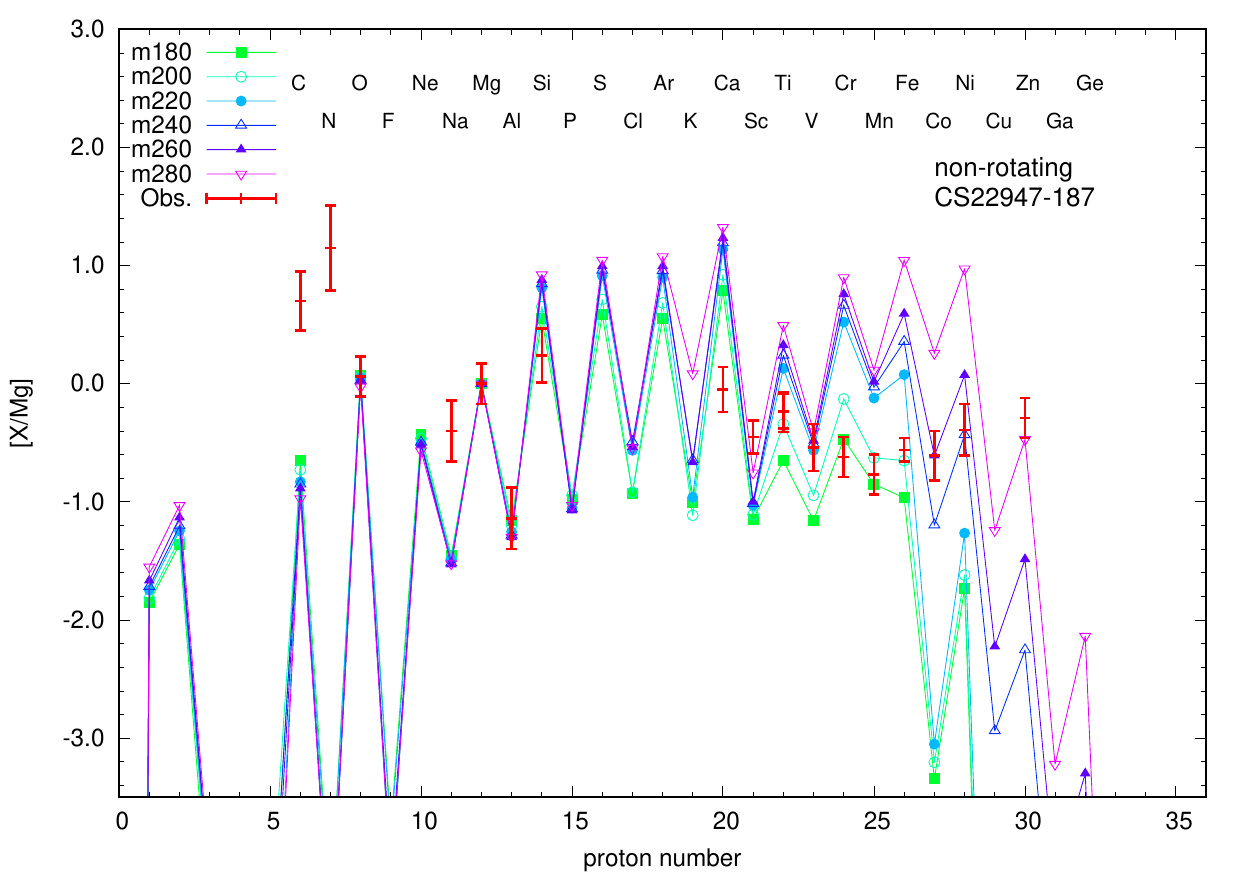}
	\end{minipage}

	\begin{minipage}{0.5\textwidth}
		\includegraphics[width=\textwidth]{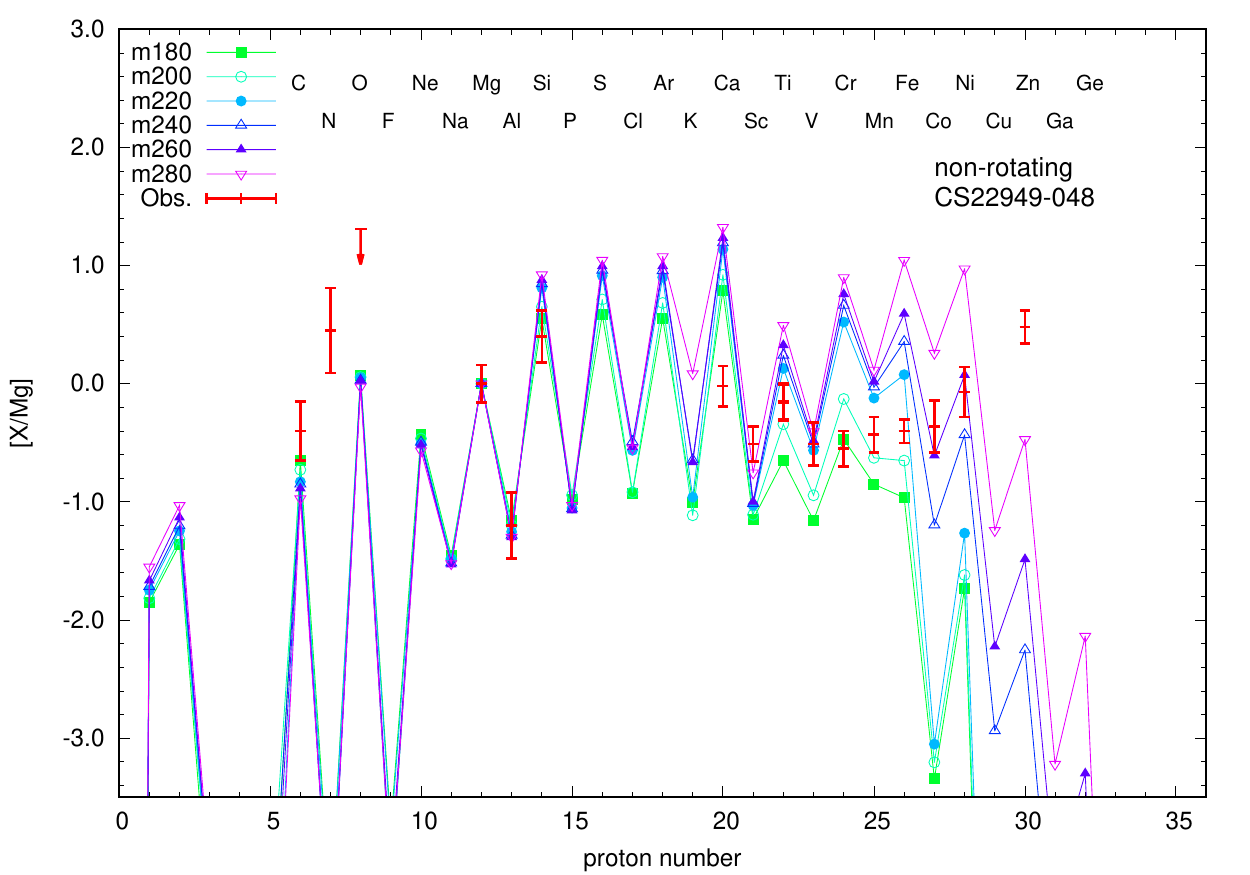}
	\end{minipage}
	\begin{minipage}{0.5\textwidth}
		\includegraphics[width=\textwidth]{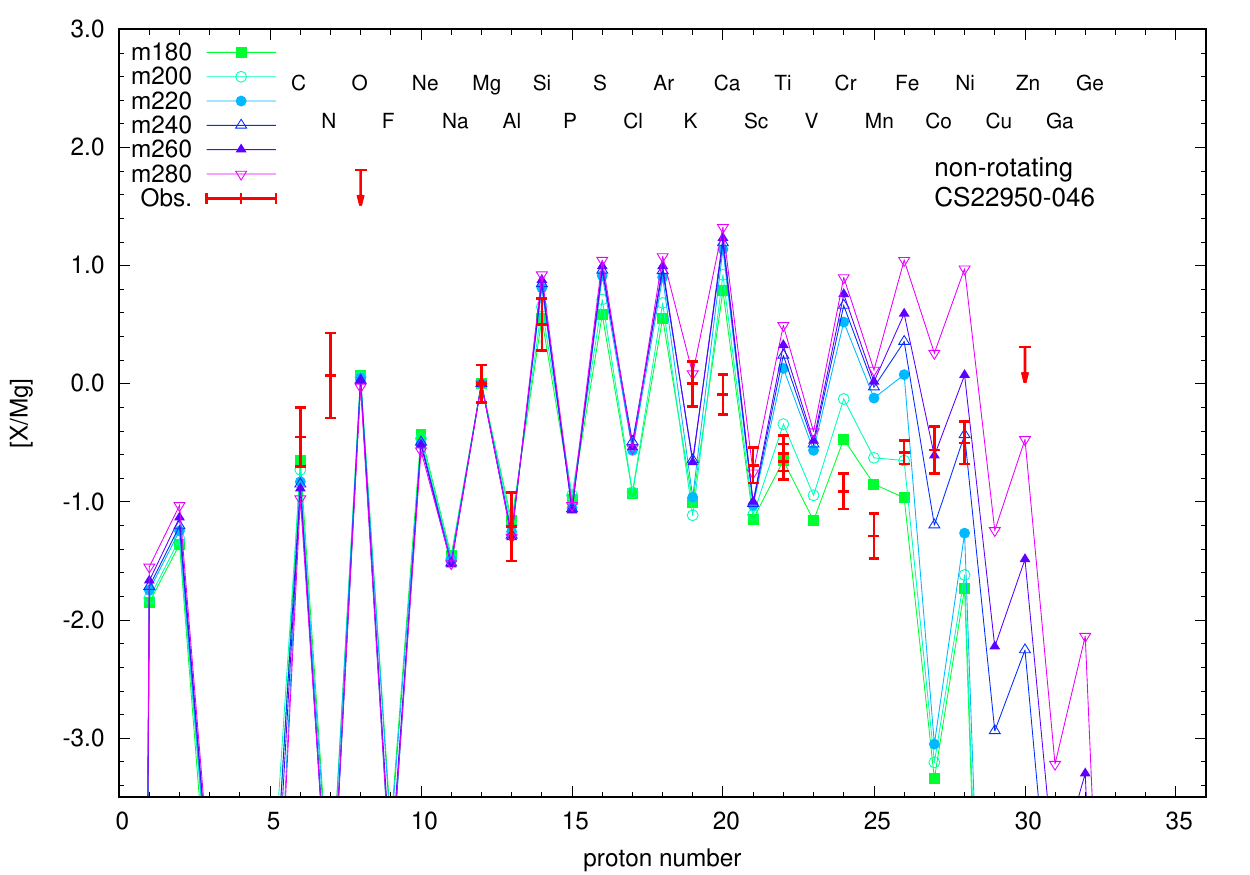}
	\end{minipage}
	
	\caption{ \footnotesize{The same as Fig.\ref{fig-stars-PISN1} but for MP stars of \#25--30.}}
\end{figure}

\begin{figure}[tbp]
	\begin{minipage}{0.5\textwidth}
		\includegraphics[width=\textwidth]{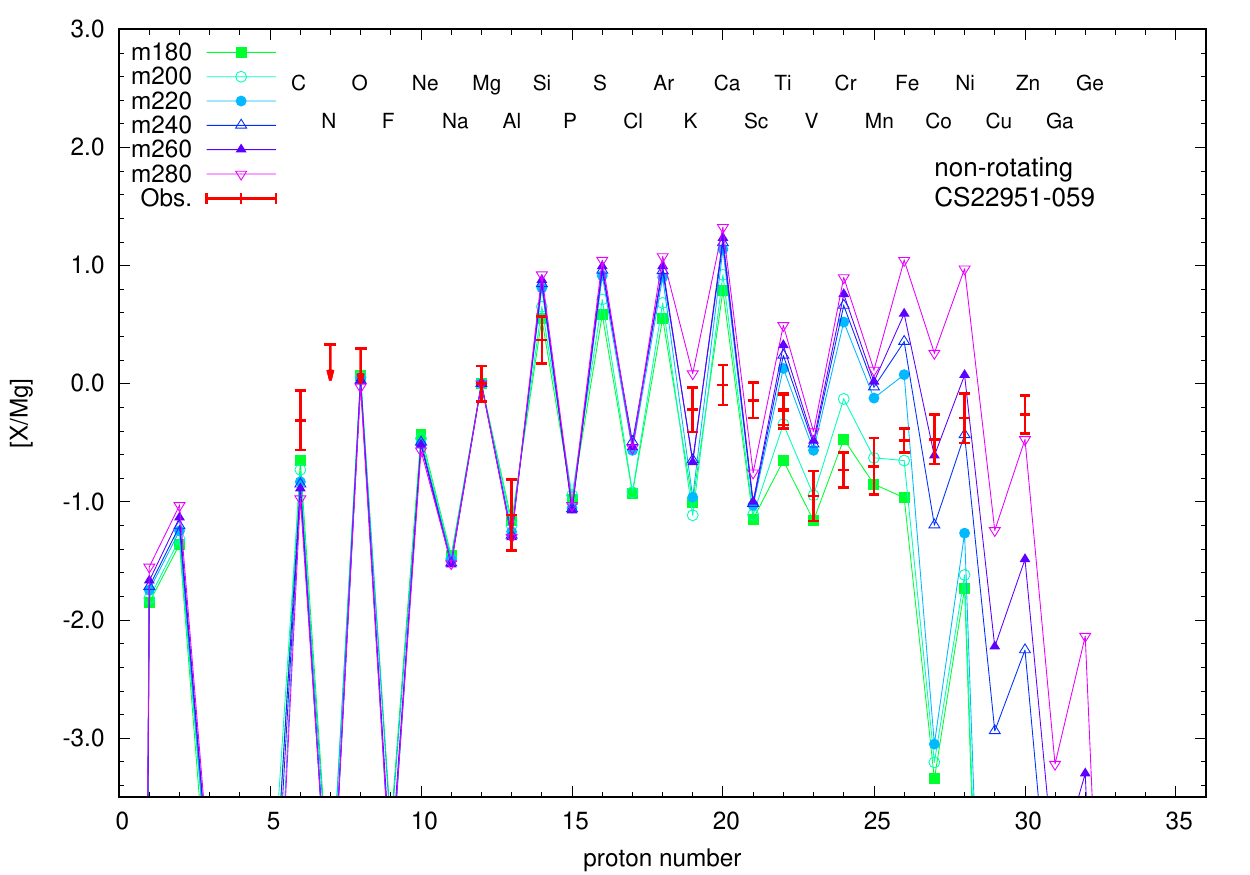}
	\end{minipage}
	\begin{minipage}{0.5\textwidth}
		\includegraphics[width=\textwidth]{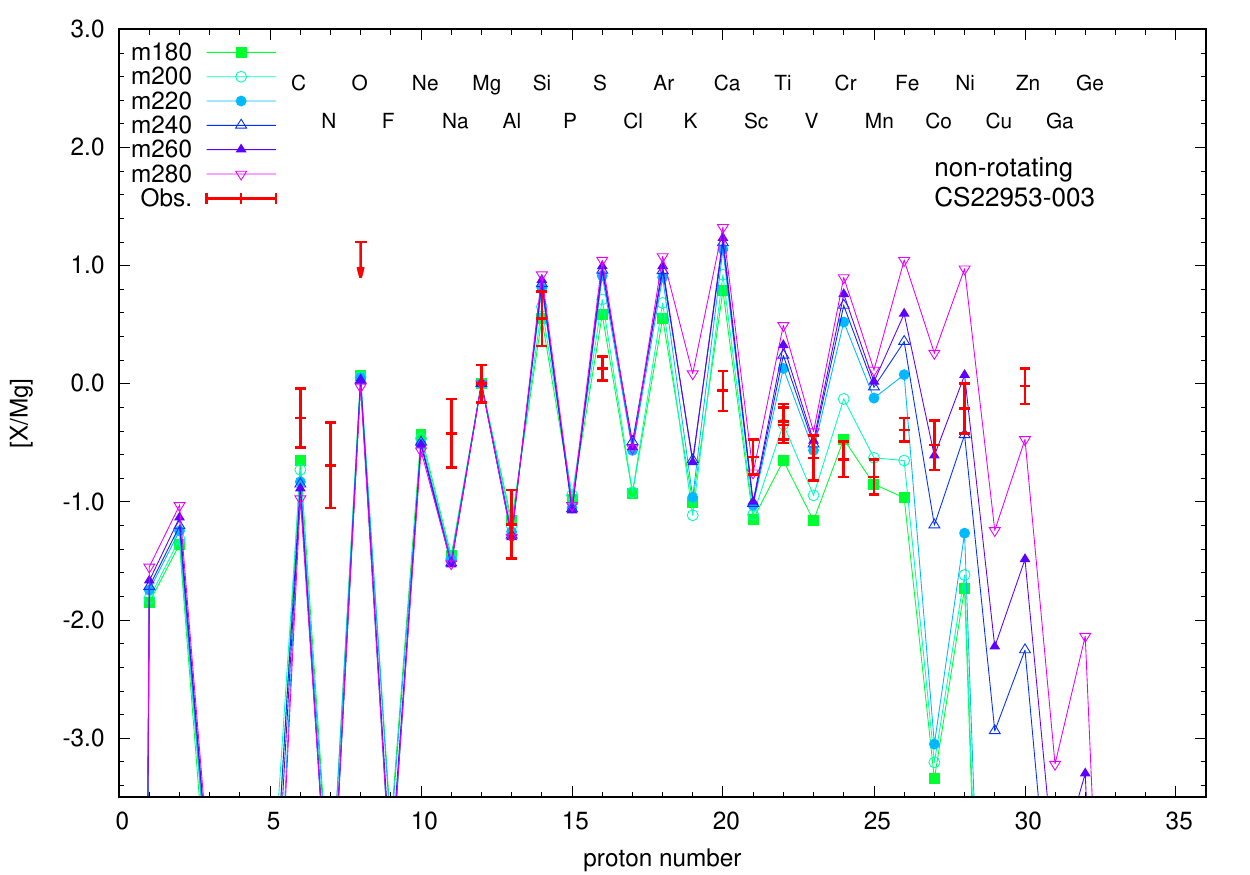}
	\end{minipage}

	\begin{minipage}{0.5\textwidth}
		\includegraphics[width=\textwidth]{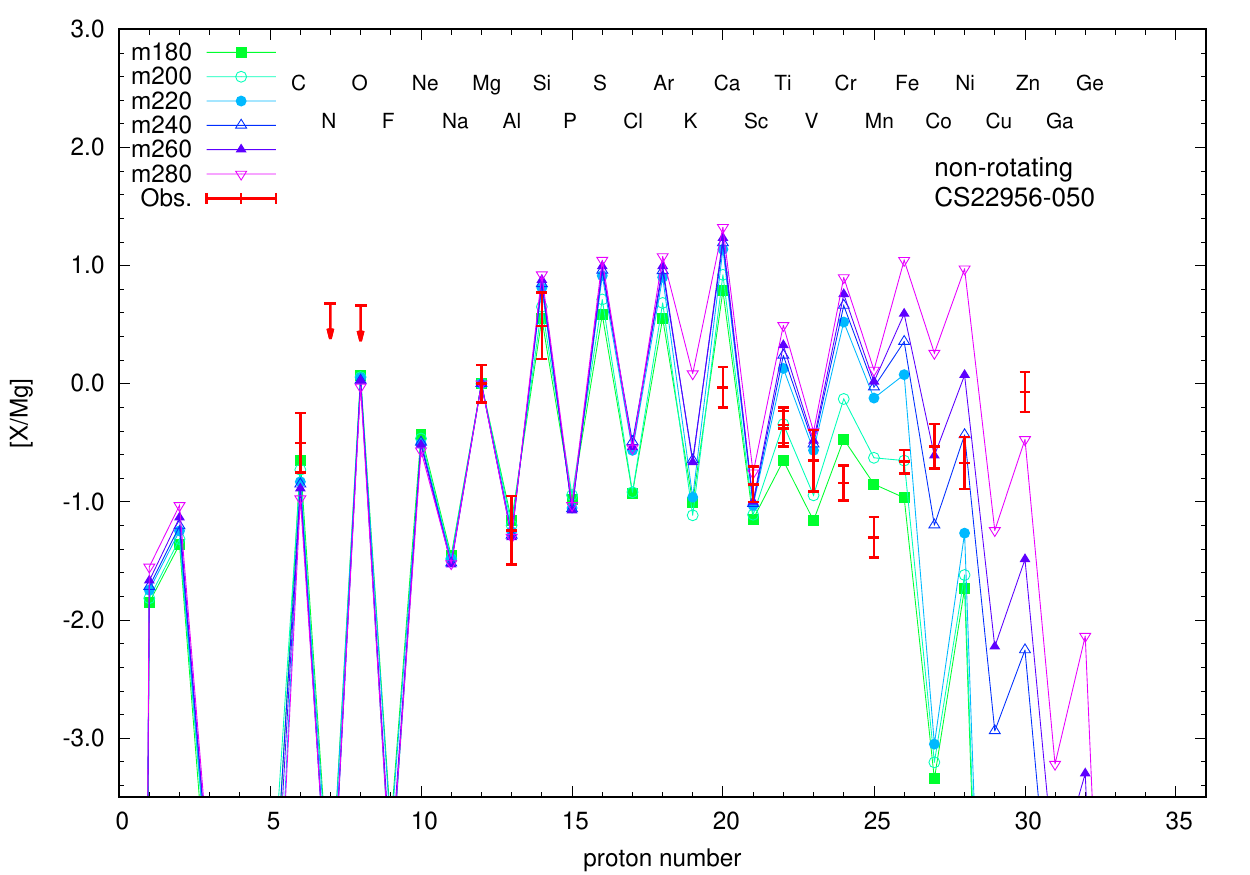}
	\end{minipage}
	\begin{minipage}{0.5\textwidth}
		\includegraphics[width=\textwidth]{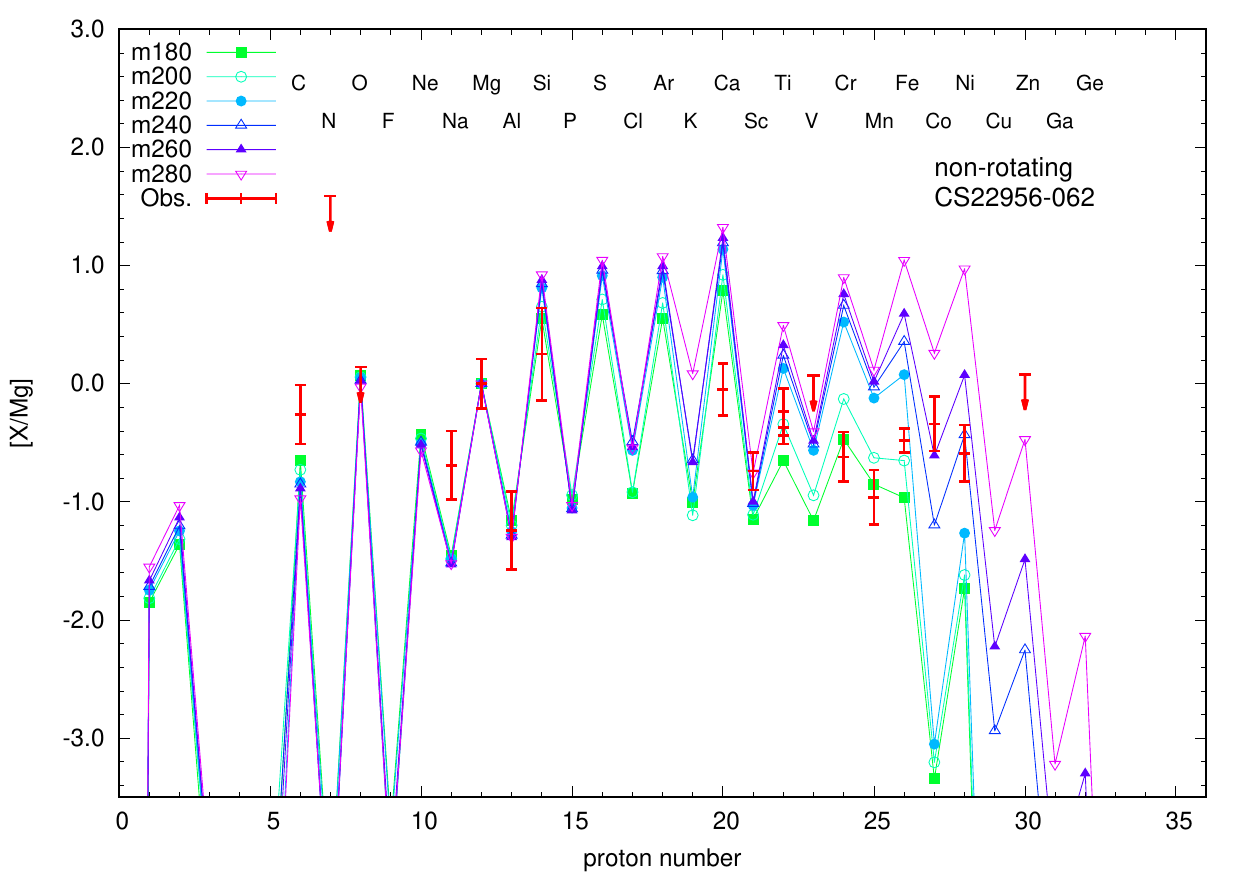}
	\end{minipage}

	\begin{minipage}{0.5\textwidth}
		\includegraphics[width=\textwidth]{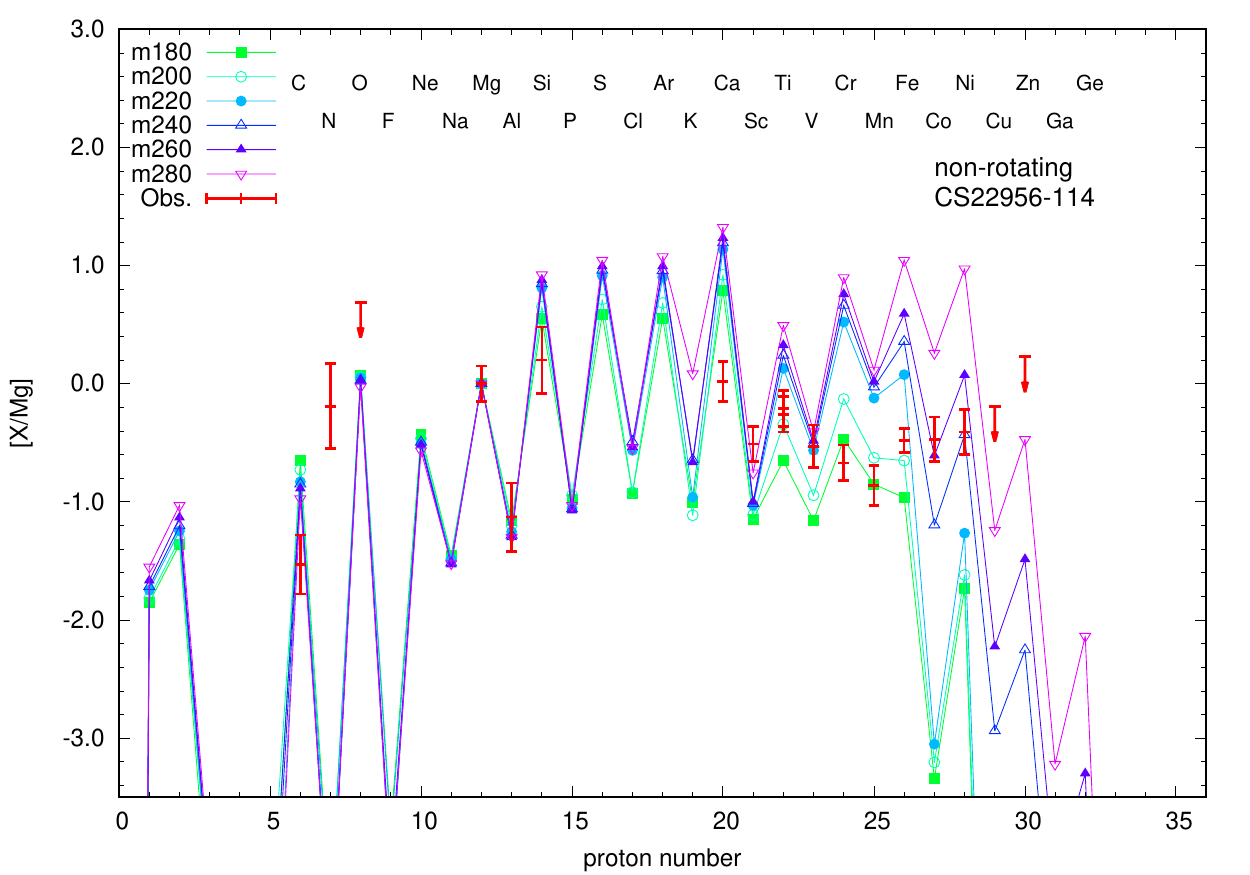}
	\end{minipage}
	\begin{minipage}{0.5\textwidth}
		\includegraphics[width=\textwidth]{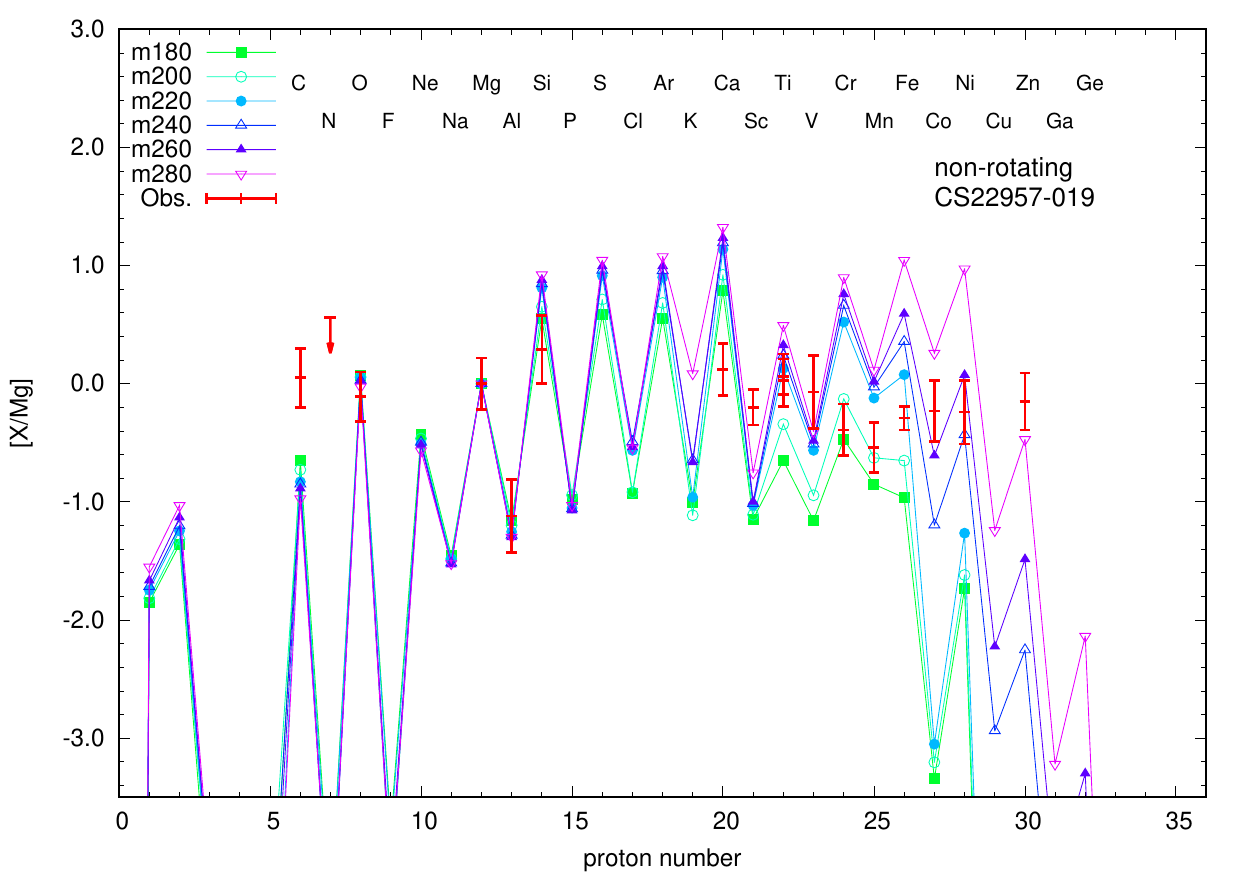}
	\end{minipage}
	
	\caption{ \footnotesize{The same as Fig.\ref{fig-stars-PISN1} but for MP stars of \#31--36.}}
\end{figure}

\begin{figure}[tbp]
	\begin{minipage}{0.5\textwidth}
		\includegraphics[width=\textwidth]{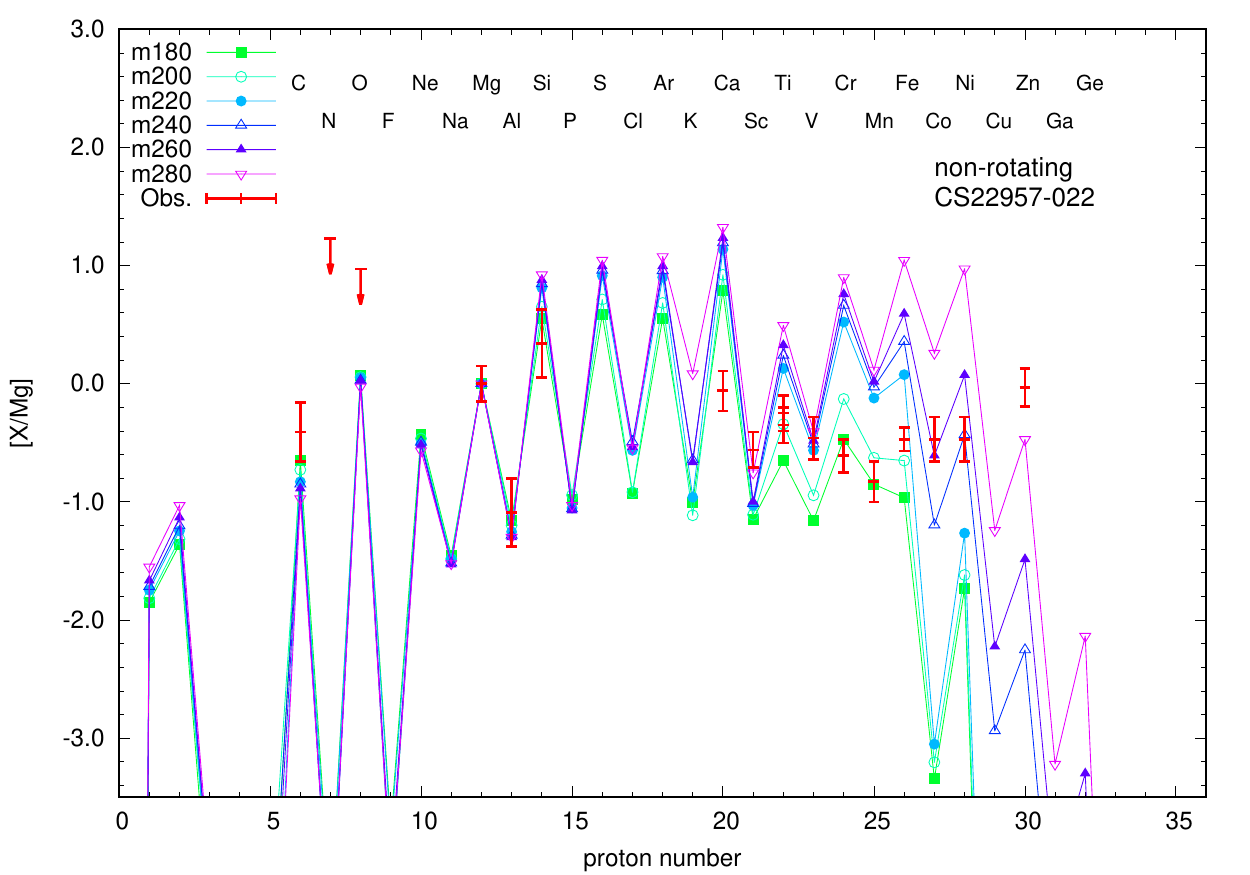}
	\end{minipage}
	\begin{minipage}{0.5\textwidth}
		\includegraphics[width=\textwidth]{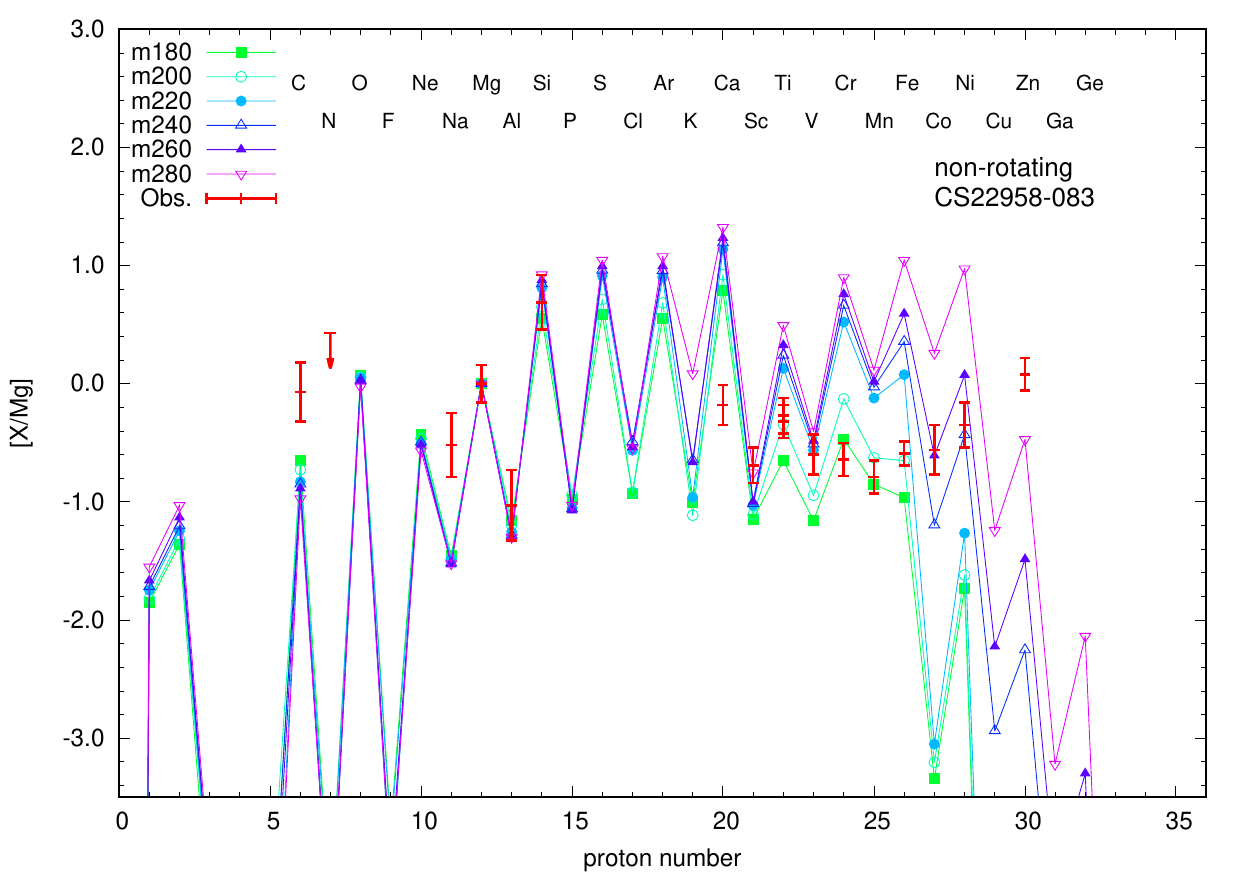}
	\end{minipage}

	\begin{minipage}{0.5\textwidth}
		\includegraphics[width=\textwidth]{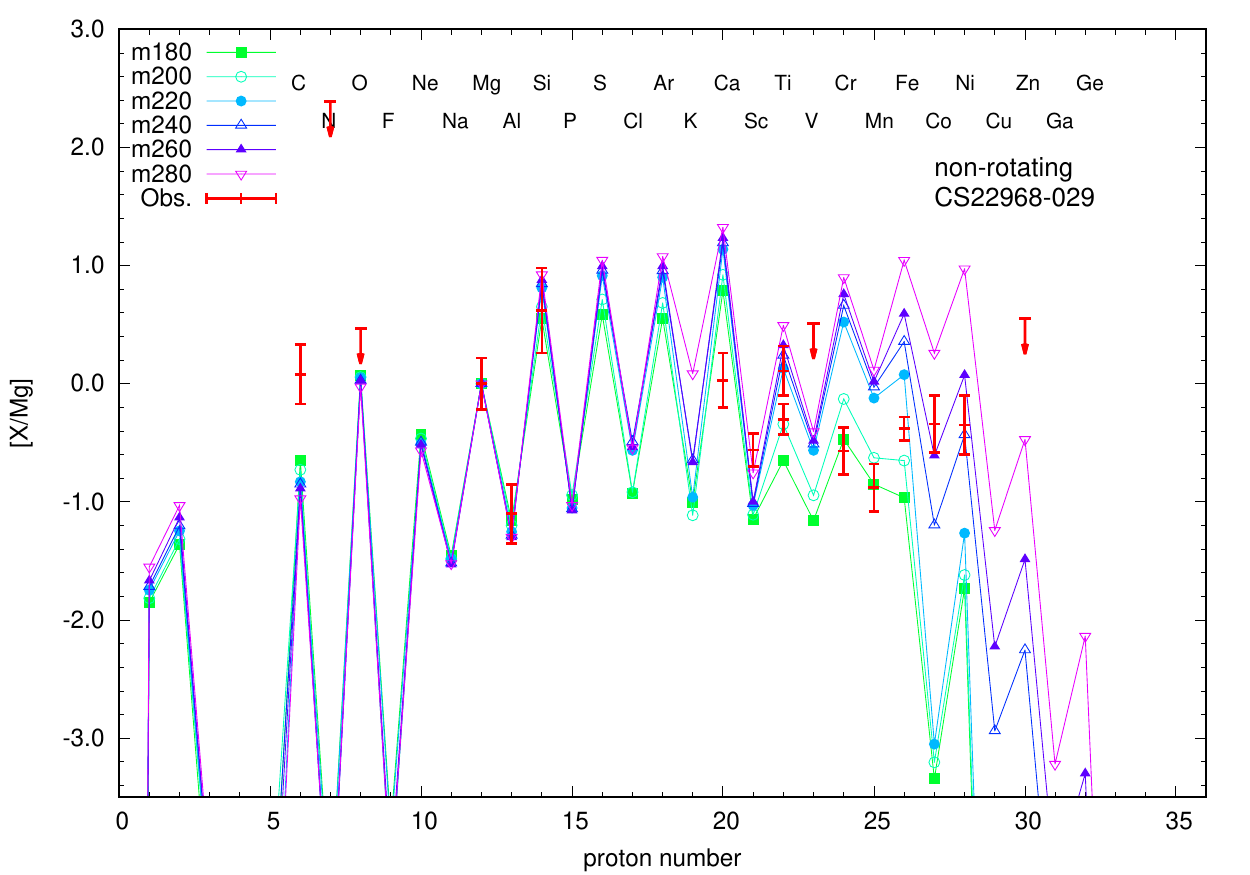}
	\end{minipage}
	\begin{minipage}{0.5\textwidth}
		\includegraphics[width=\textwidth]{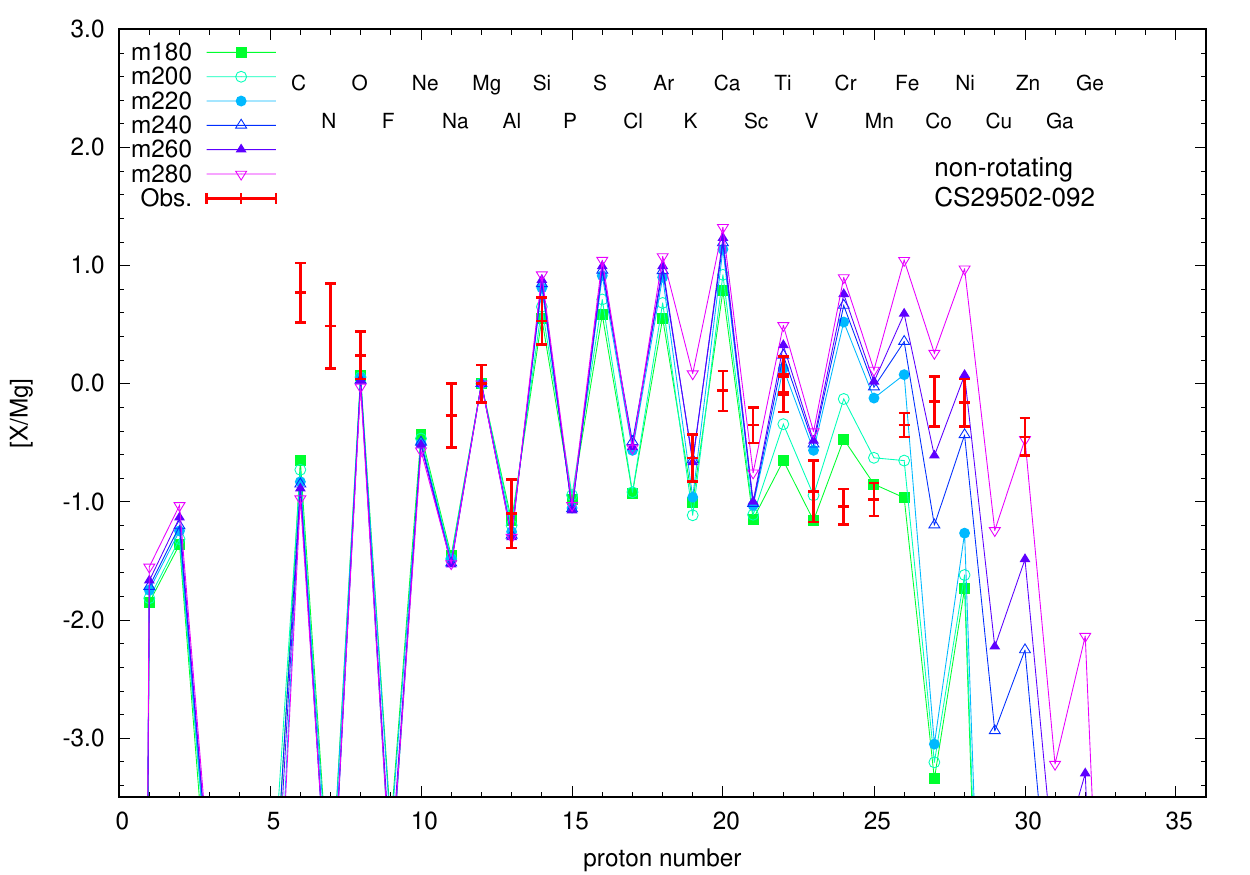}
	\end{minipage}

	\begin{minipage}{0.5\textwidth}
		\includegraphics[width=\textwidth]{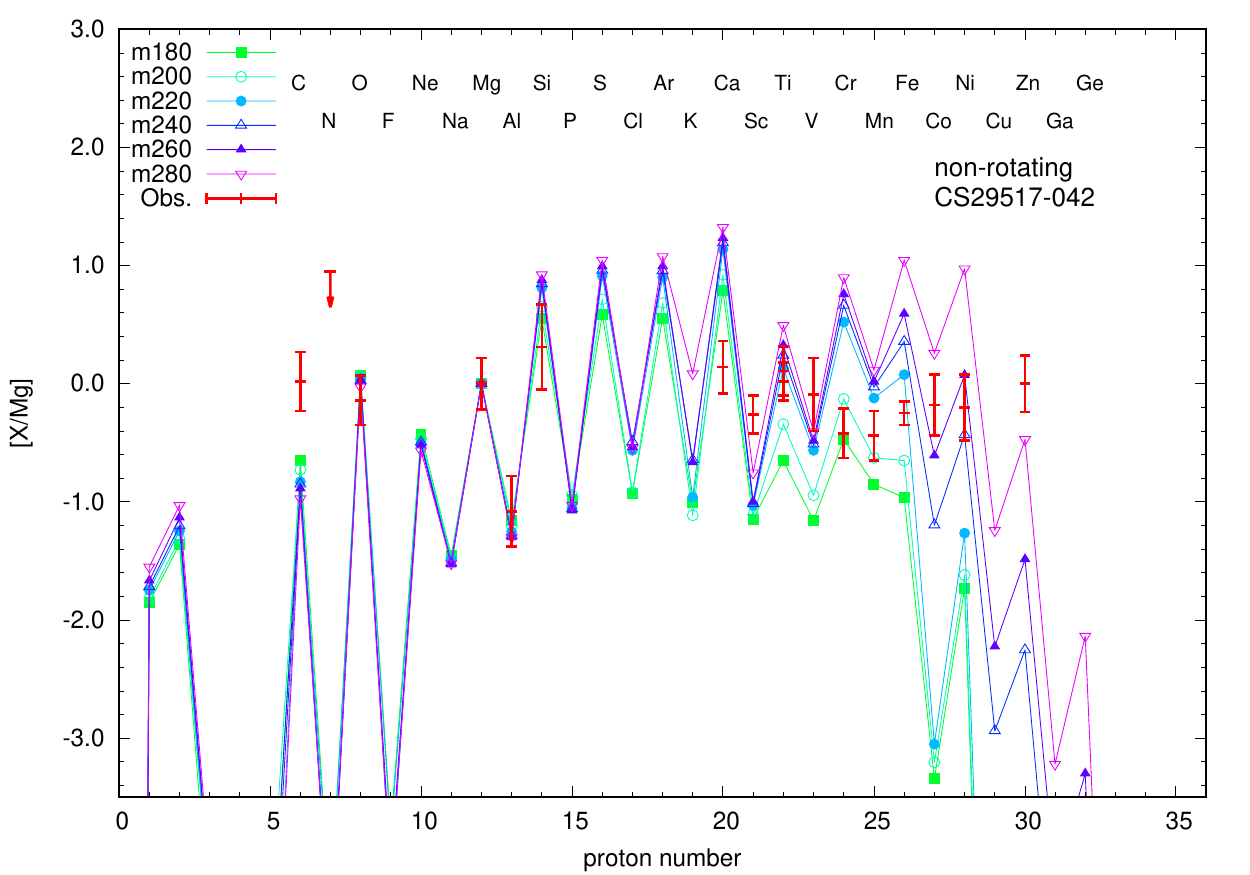}
	\end{minipage}
	\begin{minipage}{0.5\textwidth}
		\includegraphics[width=\textwidth]{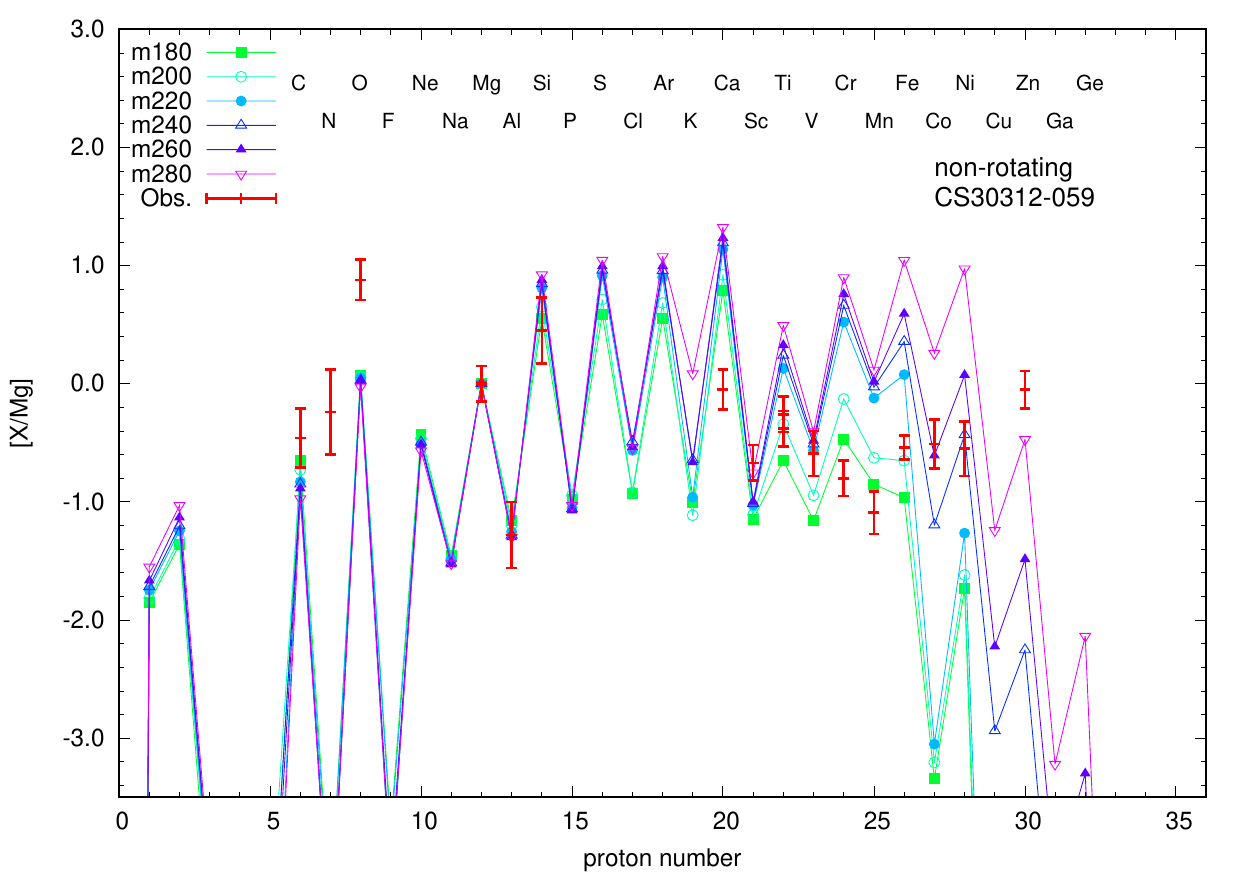}
	\end{minipage}
	
	\caption{ \footnotesize{The same as Fig.\ref{fig-stars-PISN1} but for MP stars of \#37--42.}}
\end{figure}

\begin{figure}[tbp]
	\begin{minipage}{0.5\textwidth}
		\includegraphics[width=\textwidth]{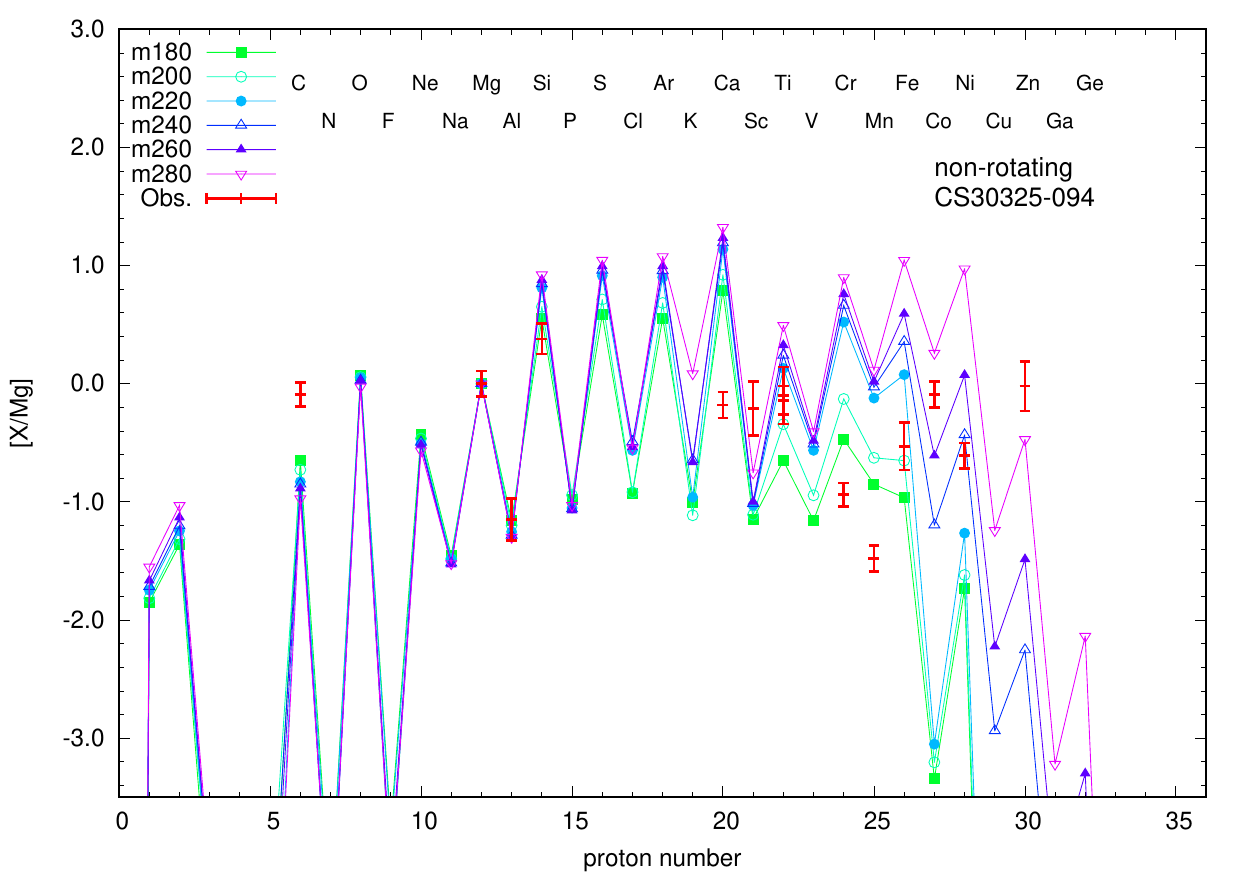}
	\end{minipage}
	\begin{minipage}{0.5\textwidth}
		\includegraphics[width=\textwidth]{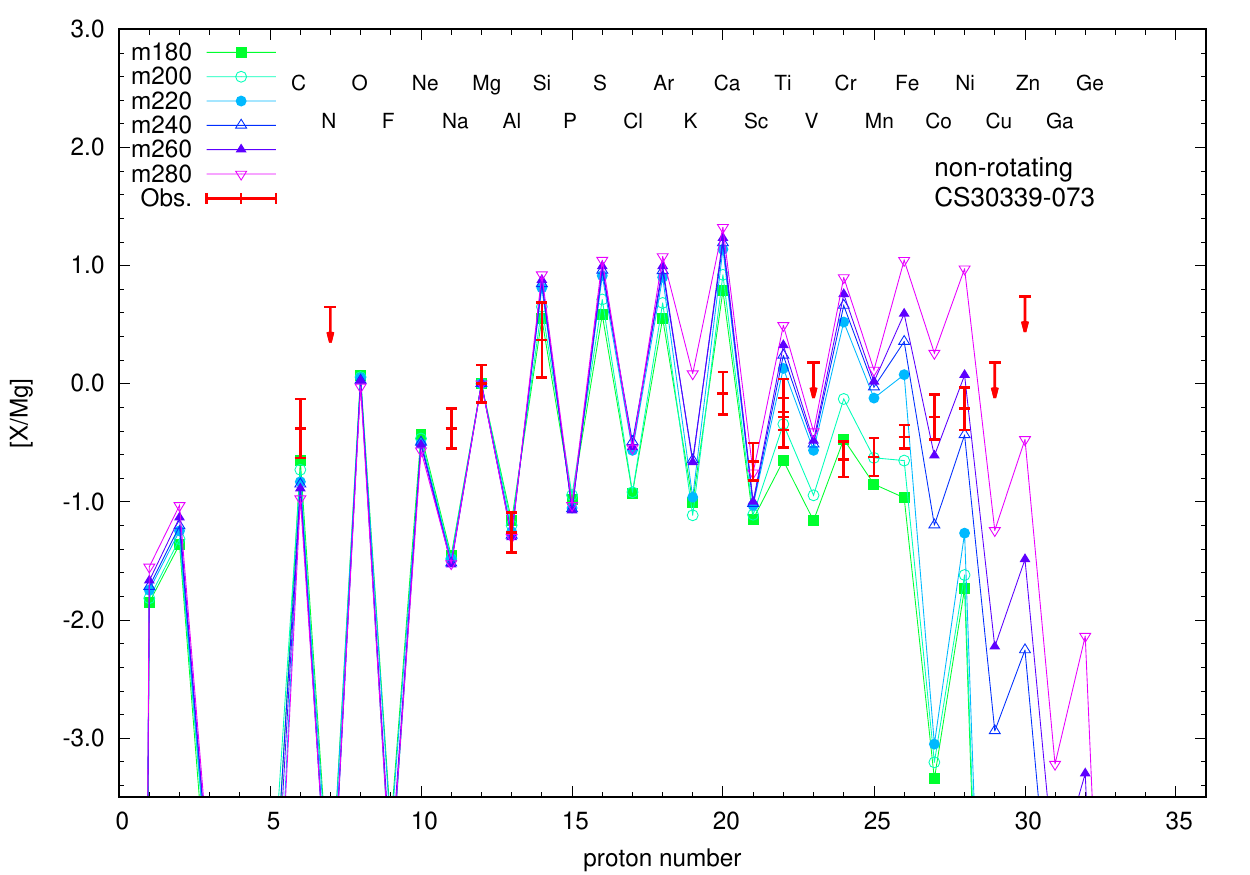}
	\end{minipage}

	\begin{minipage}{0.5\textwidth}
		\includegraphics[width=\textwidth]{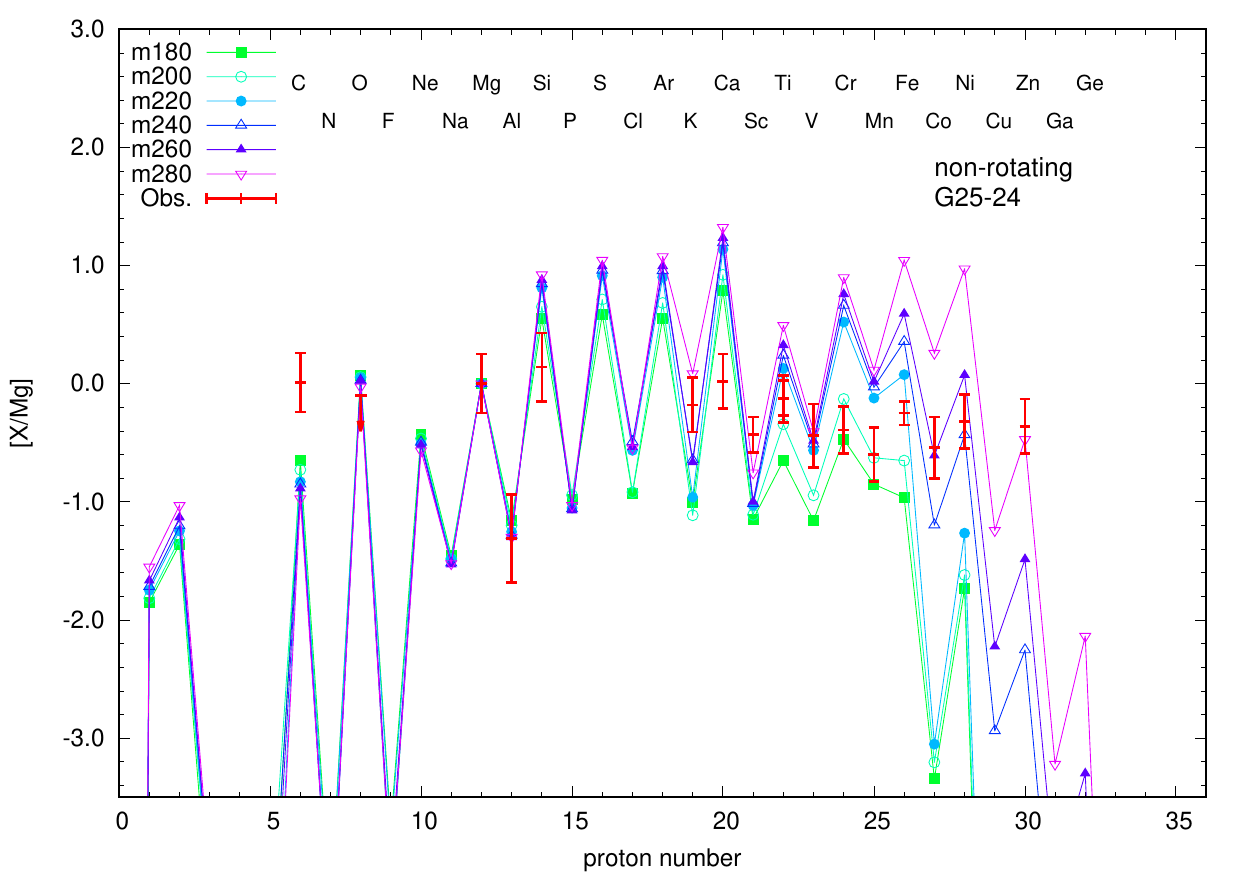}
	\end{minipage}
	\begin{minipage}{0.5\textwidth}
		\includegraphics[width=\textwidth]{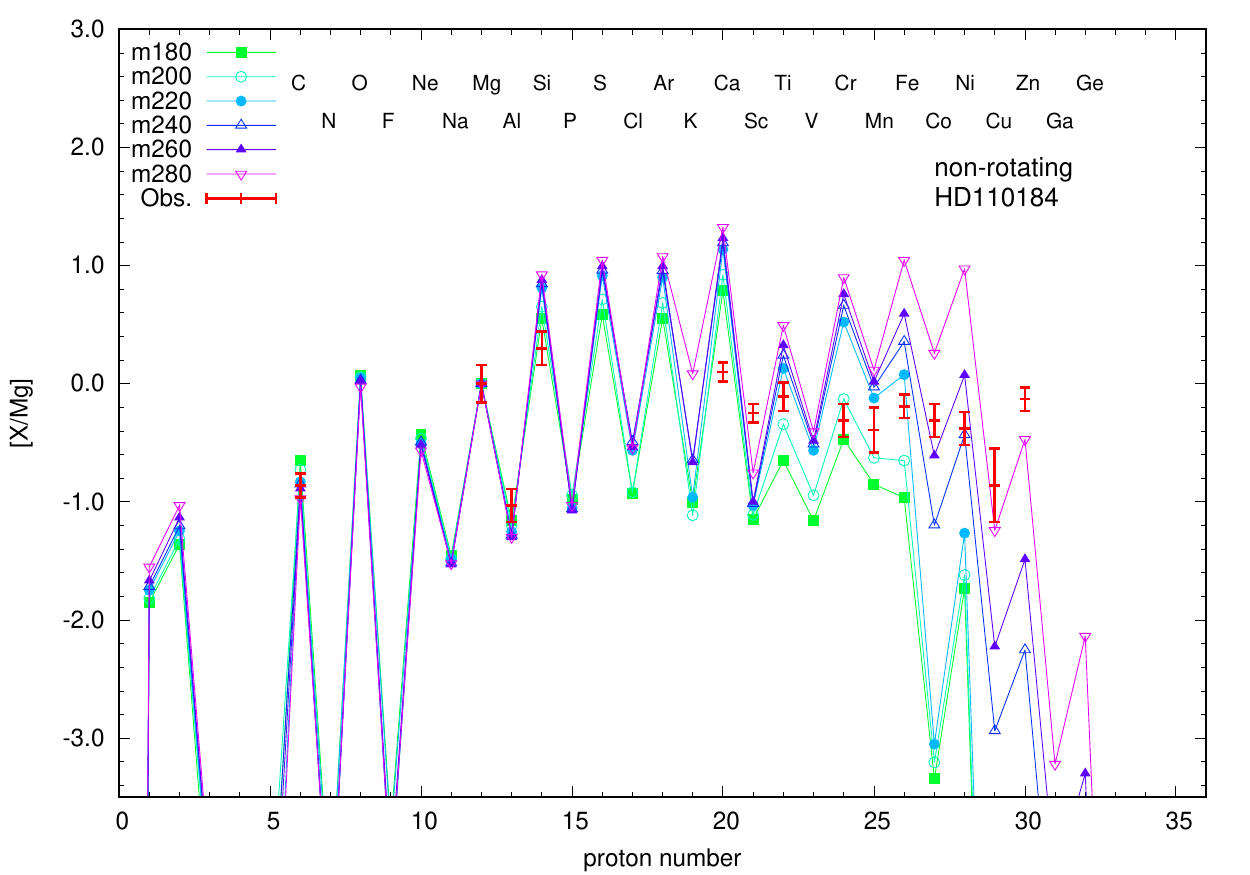}
	\end{minipage}

	\begin{minipage}{0.5\textwidth}
		\includegraphics[width=\textwidth]{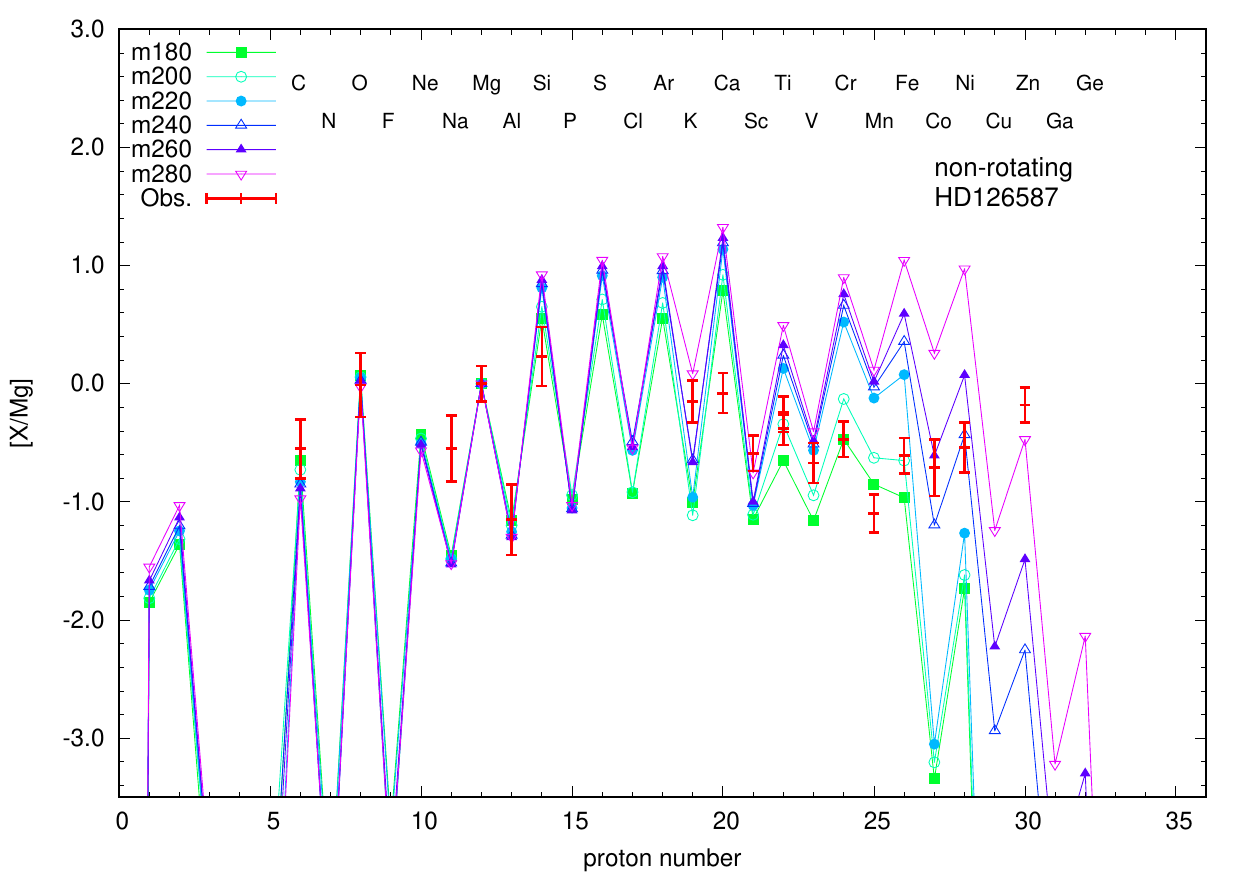}
	\end{minipage}
	\begin{minipage}{0.5\textwidth}
		\includegraphics[width=\textwidth]{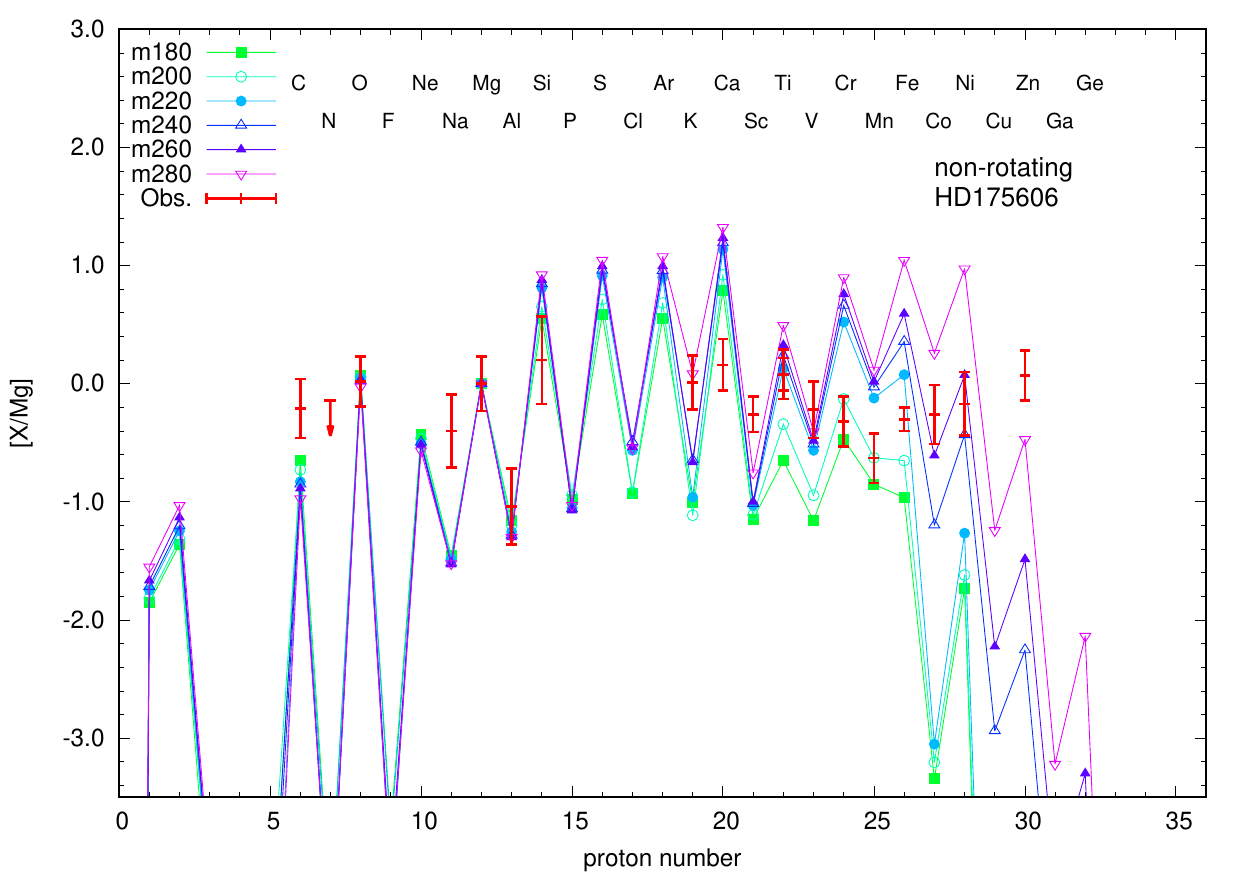}
	\end{minipage}
	
	\caption{ \footnotesize{The same as Fig.\ref{fig-stars-PISN1} but for MP stars of \#43--48.}}
\end{figure}

\begin{figure}[tbp]
	\begin{minipage}{0.5\textwidth}
		\includegraphics[width=\textwidth]{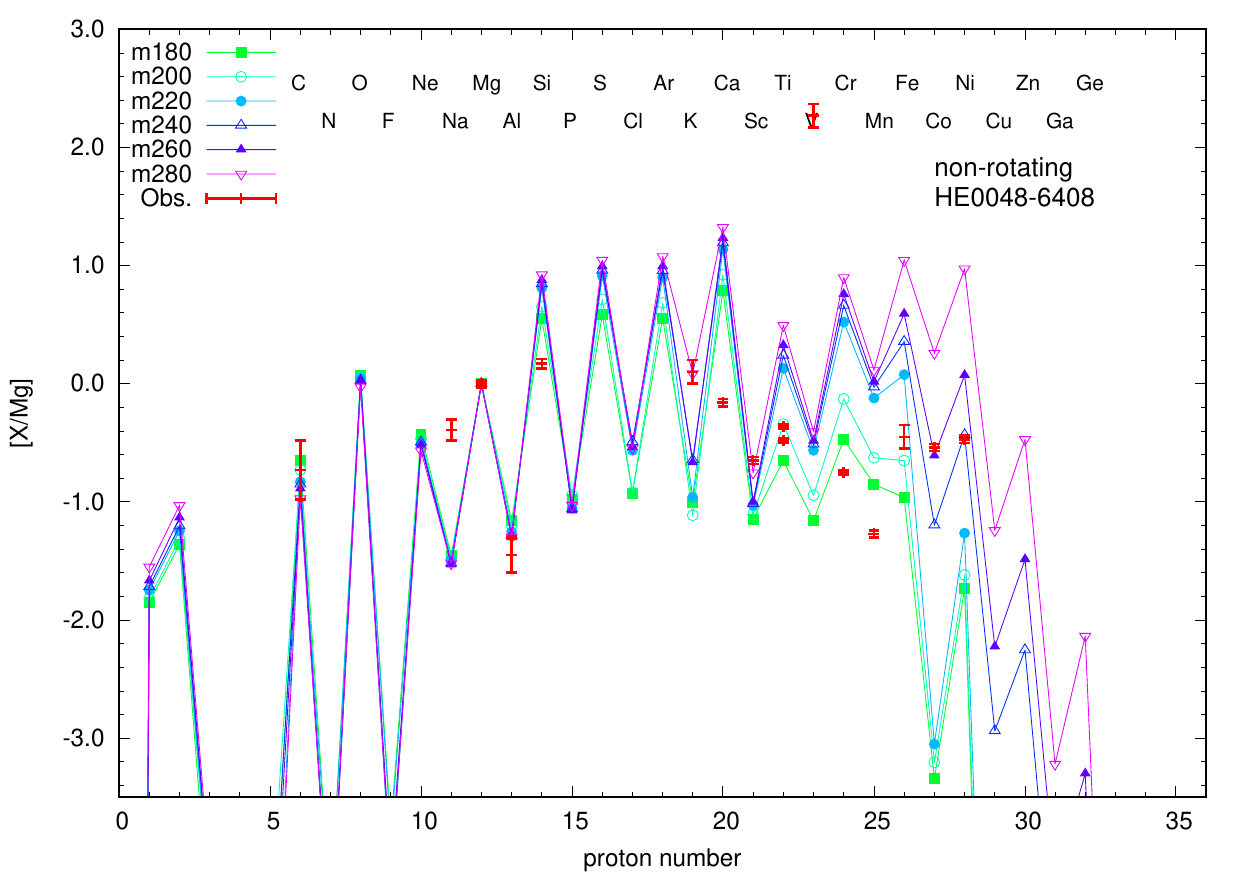}
	\end{minipage}
	\begin{minipage}{0.5\textwidth}
		\includegraphics[width=\textwidth]{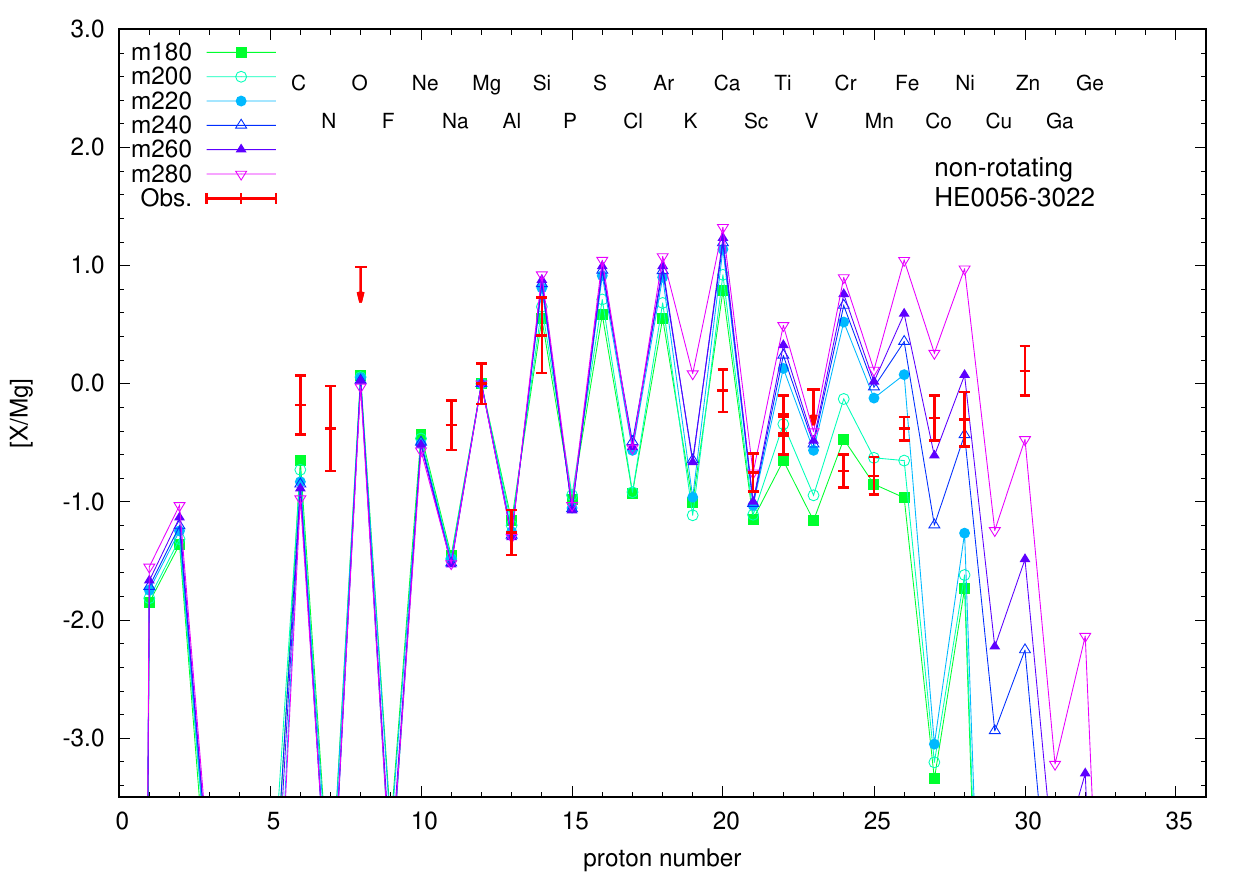}
	\end{minipage}

	\begin{minipage}{0.5\textwidth}
		\includegraphics[width=\textwidth]{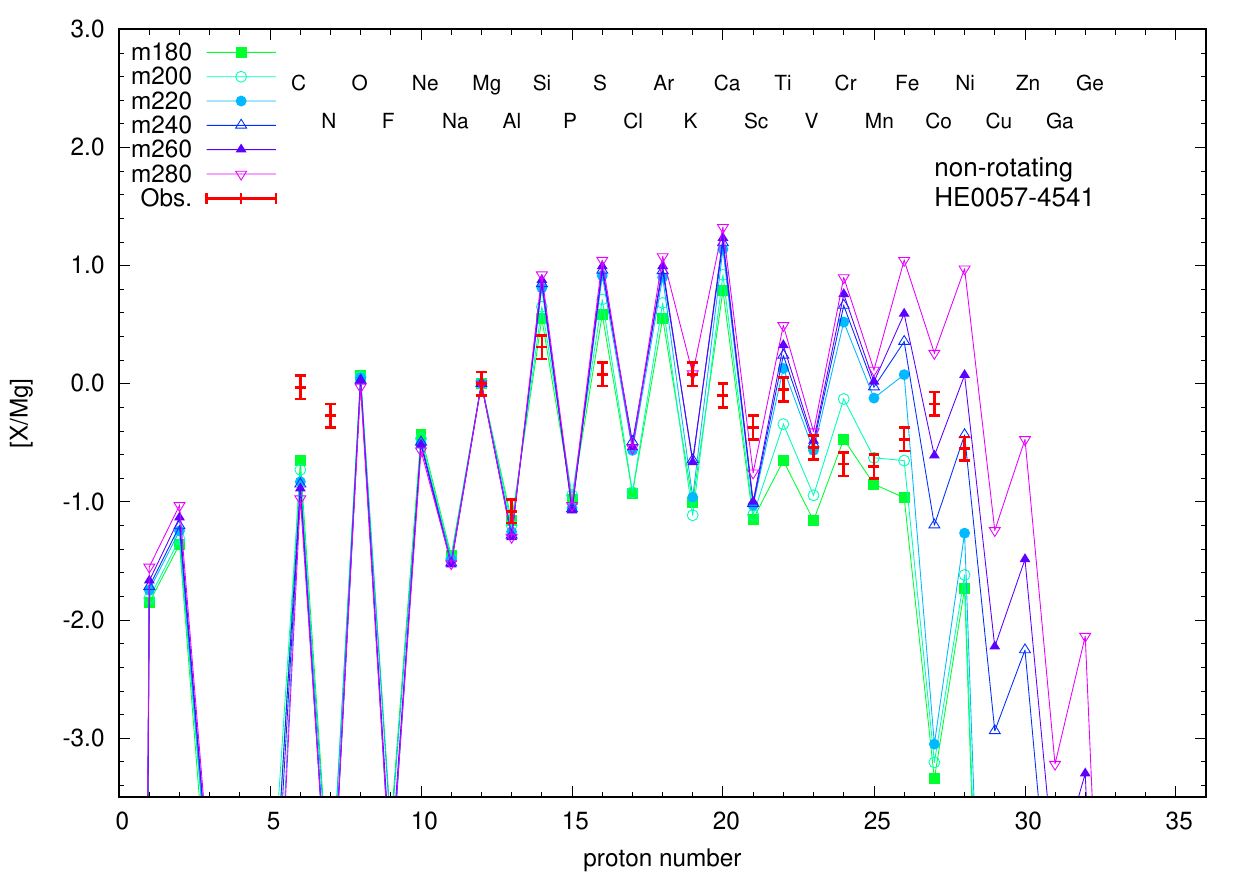}
	\end{minipage}
	\begin{minipage}{0.5\textwidth}
		\includegraphics[width=\textwidth]{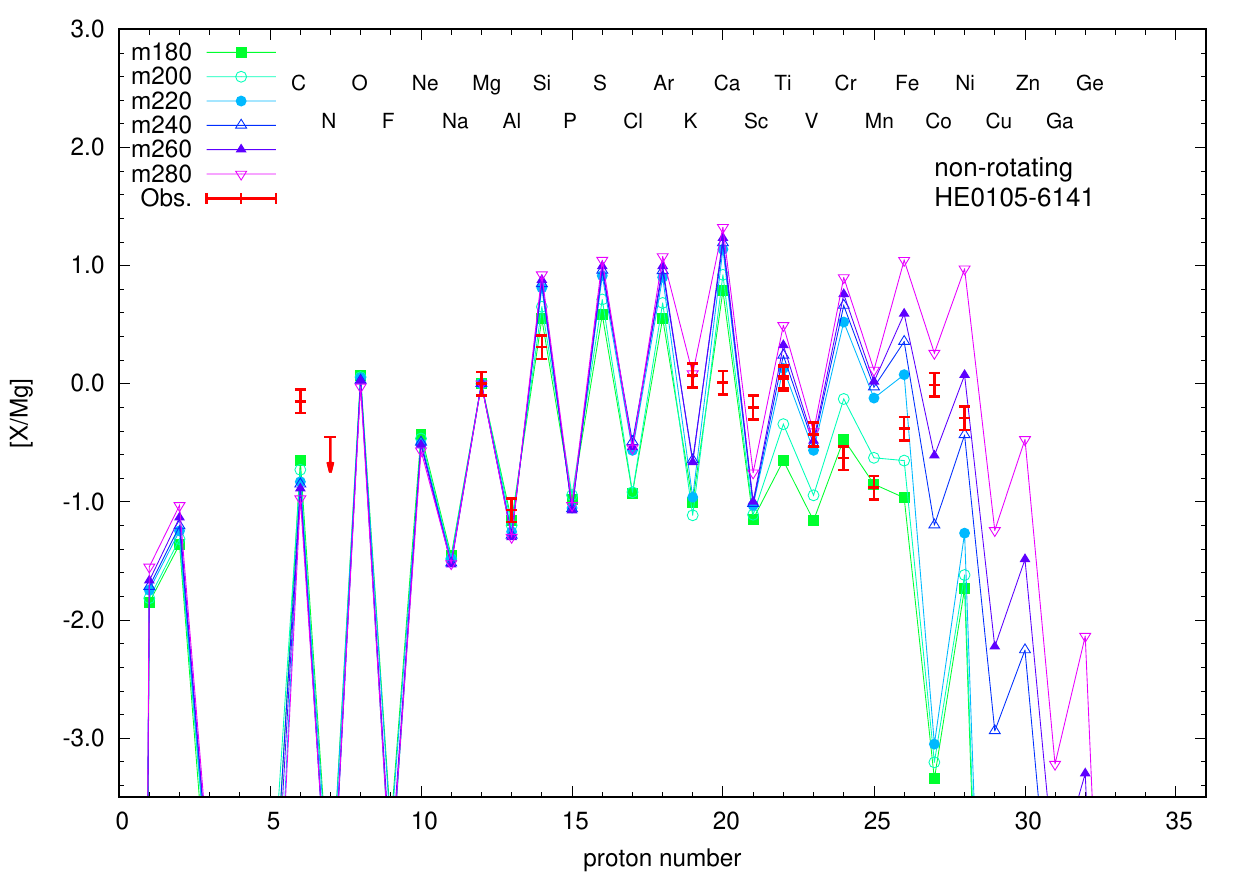}
	\end{minipage}

	\begin{minipage}{0.5\textwidth}
		\includegraphics[width=\textwidth]{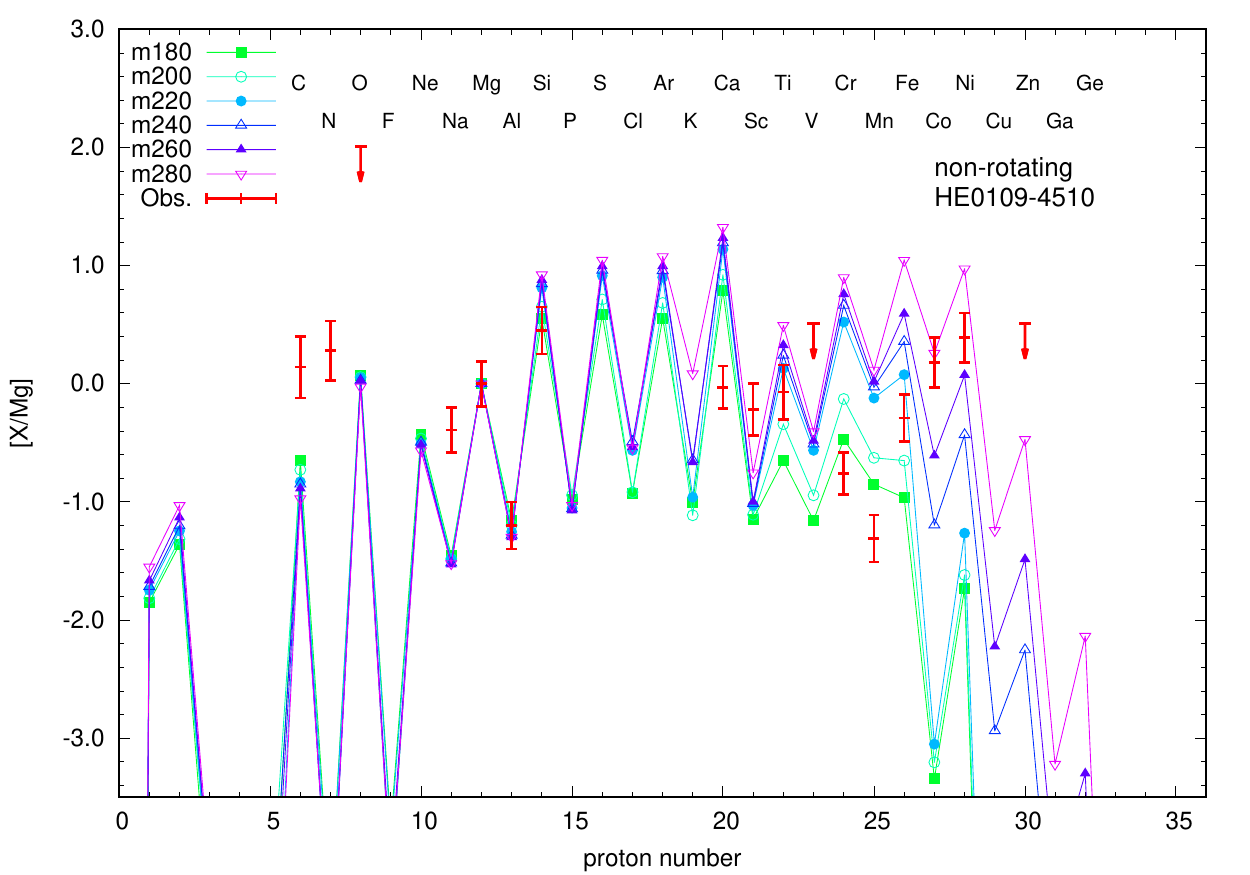}
	\end{minipage}
	\begin{minipage}{0.5\textwidth}
		\includegraphics[width=\textwidth]{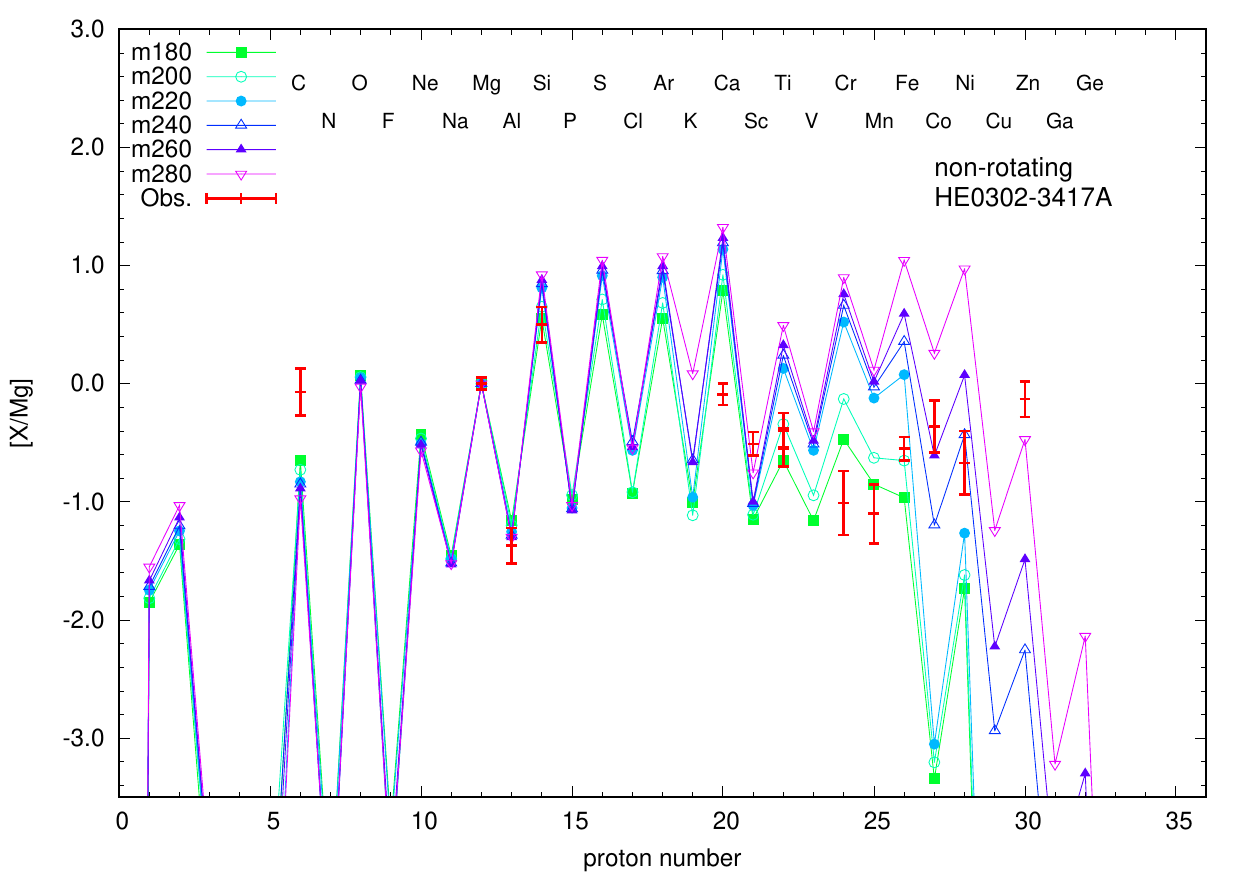}
	\end{minipage}
	
	\caption{ \footnotesize{The same as Fig.\ref{fig-stars-PISN1} but for MP stars of \#49--54.}}
\end{figure}

\begin{figure}[tbp]
	\begin{minipage}{0.5\textwidth}
		\includegraphics[width=\textwidth]{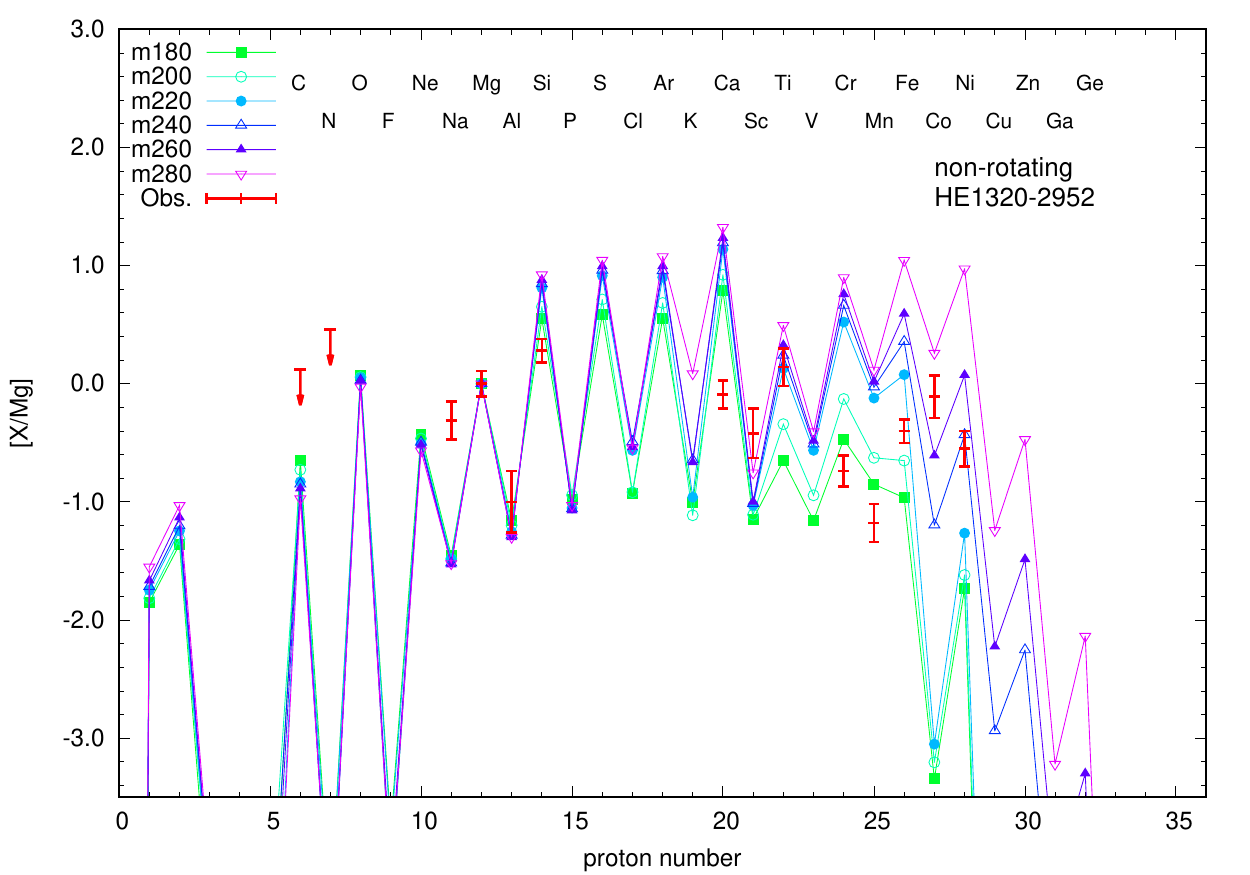}
	\end{minipage}
	\begin{minipage}{0.5\textwidth}
		\includegraphics[width=\textwidth]{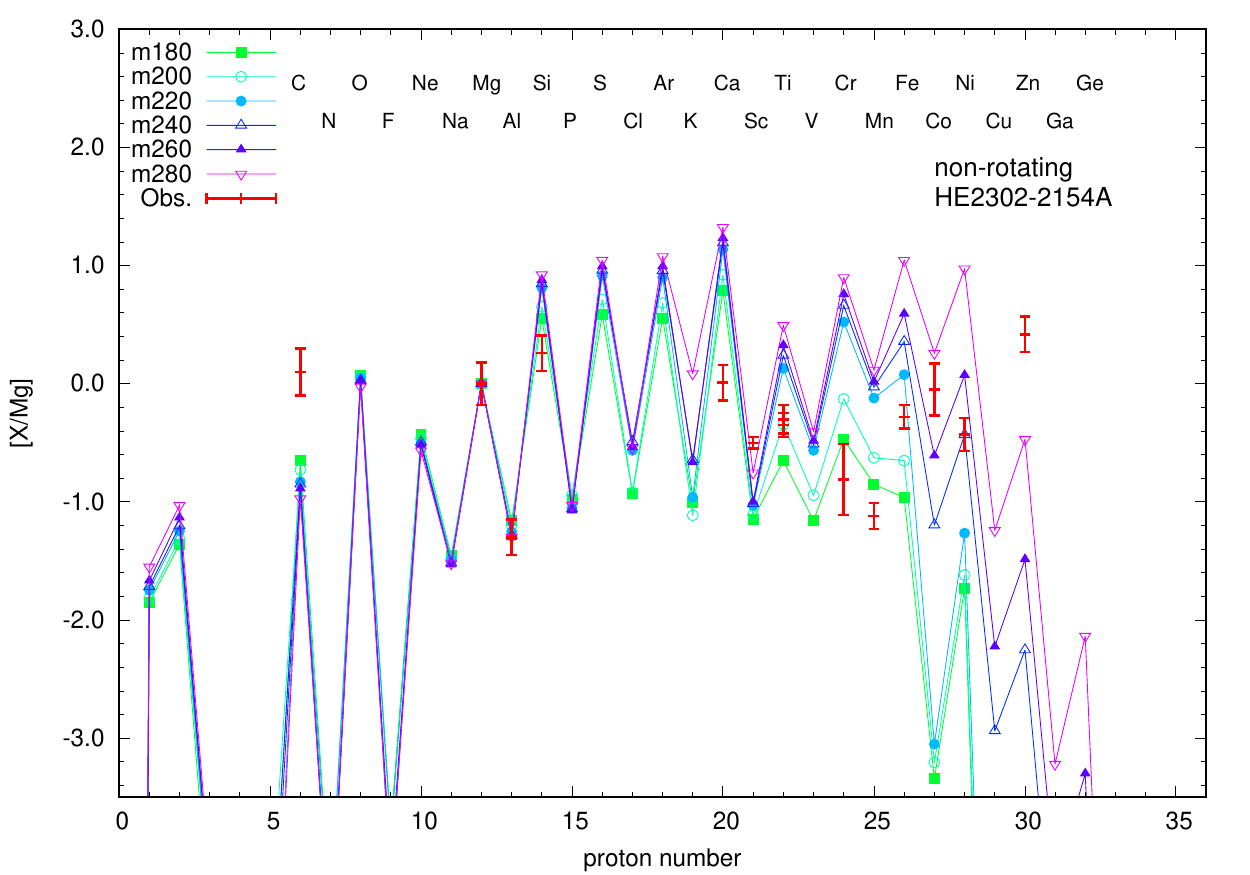}
	\end{minipage}

	\begin{minipage}{0.5\textwidth}
		\includegraphics[width=\textwidth]{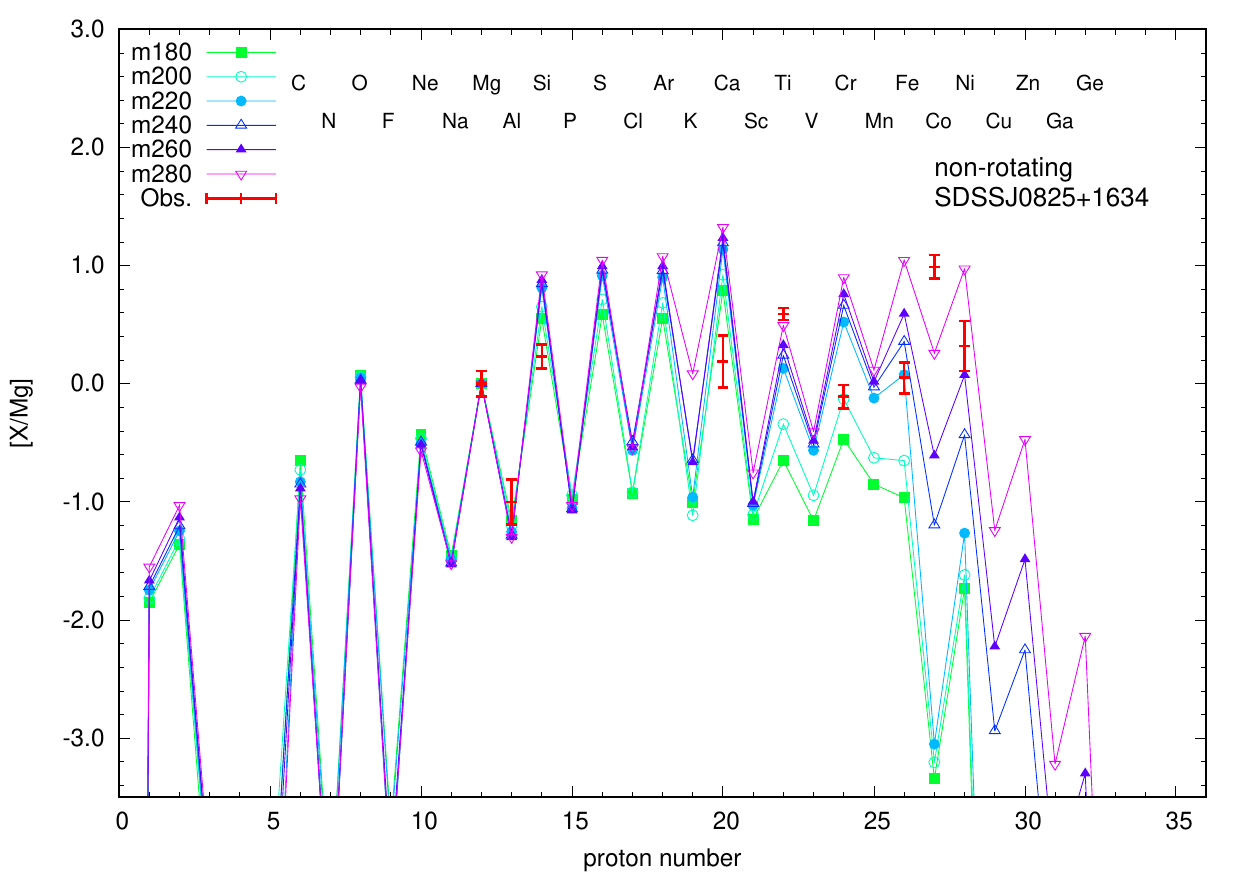}
	\end{minipage}
	\begin{minipage}{0.5\textwidth}
		\includegraphics[width=\textwidth]{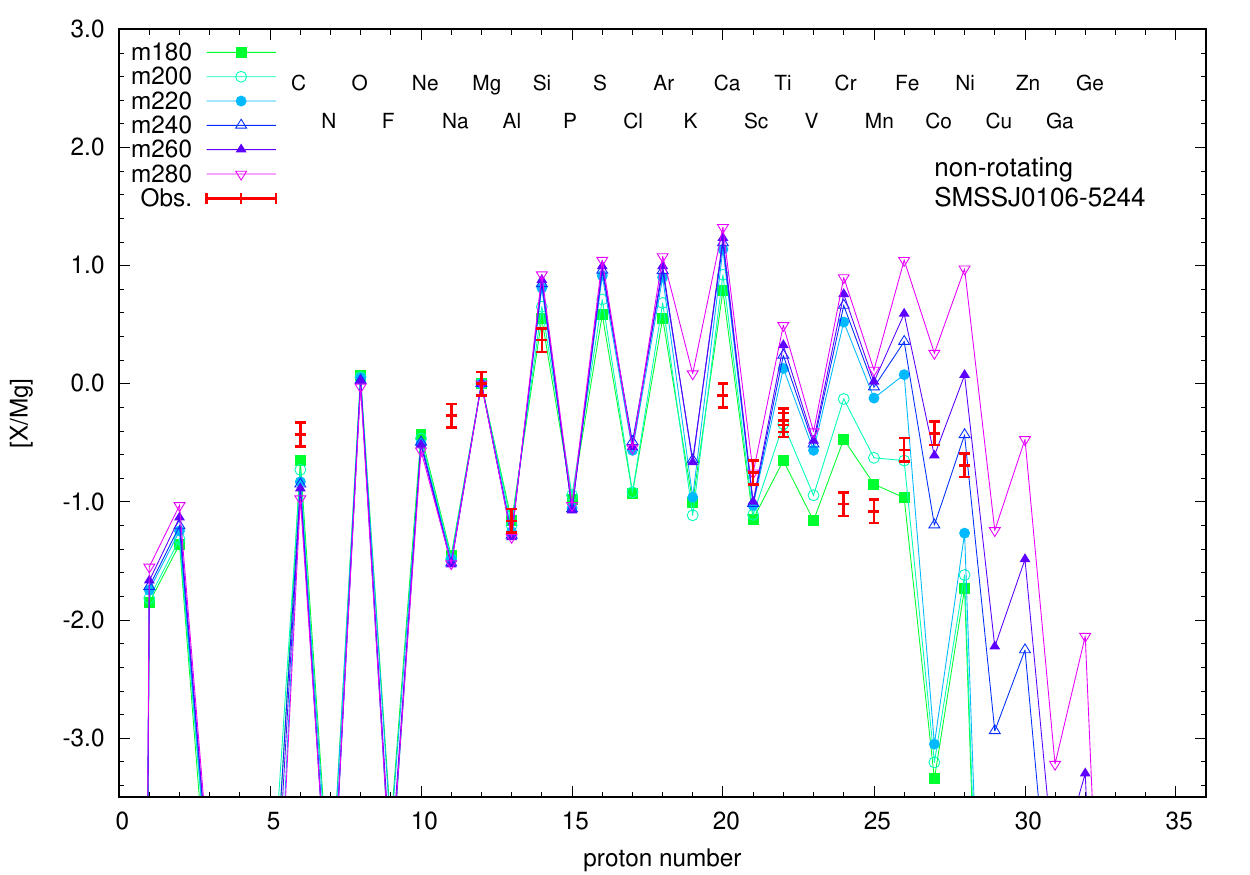}
	\end{minipage}

	\begin{minipage}{0.5\textwidth}
		\includegraphics[width=\textwidth]{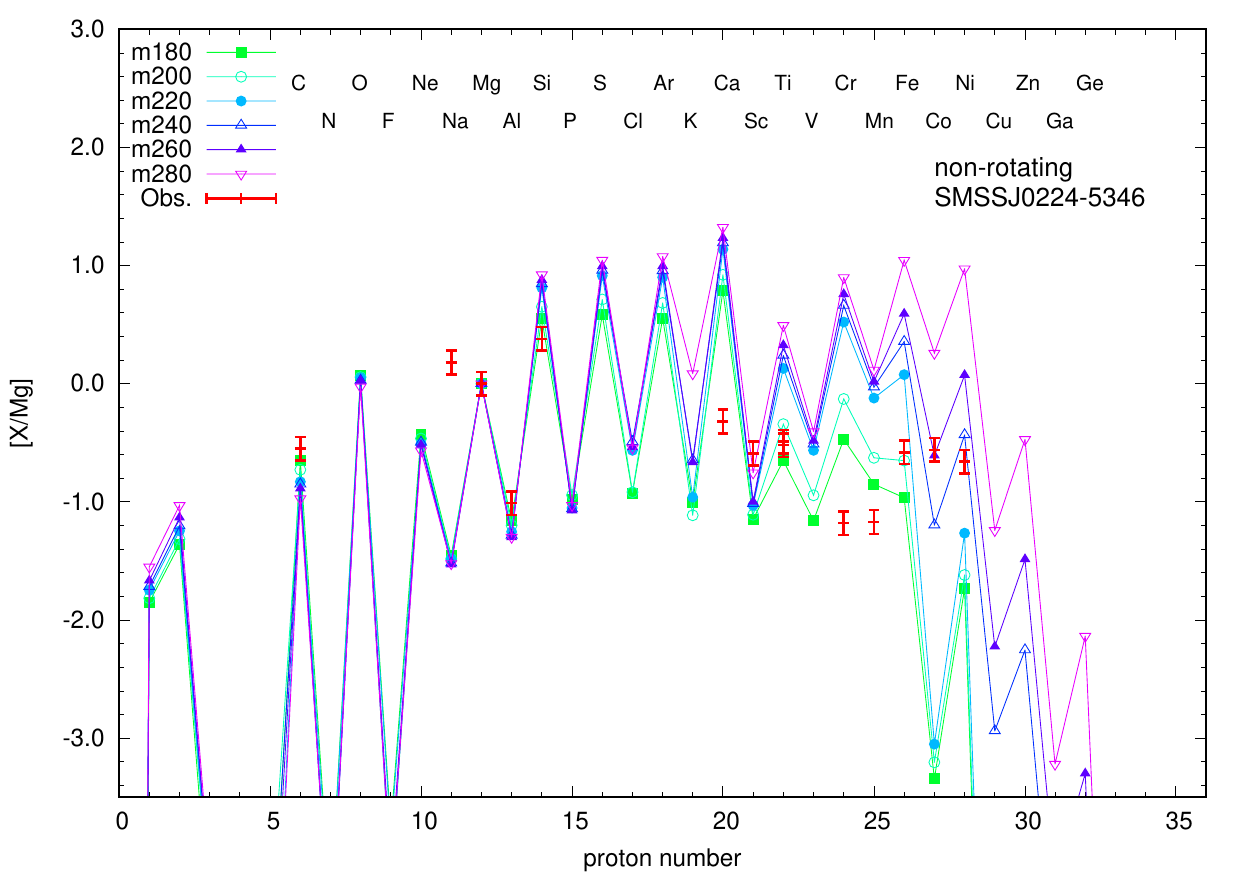}
	\end{minipage}
	\begin{minipage}{0.5\textwidth}
		\includegraphics[width=\textwidth]{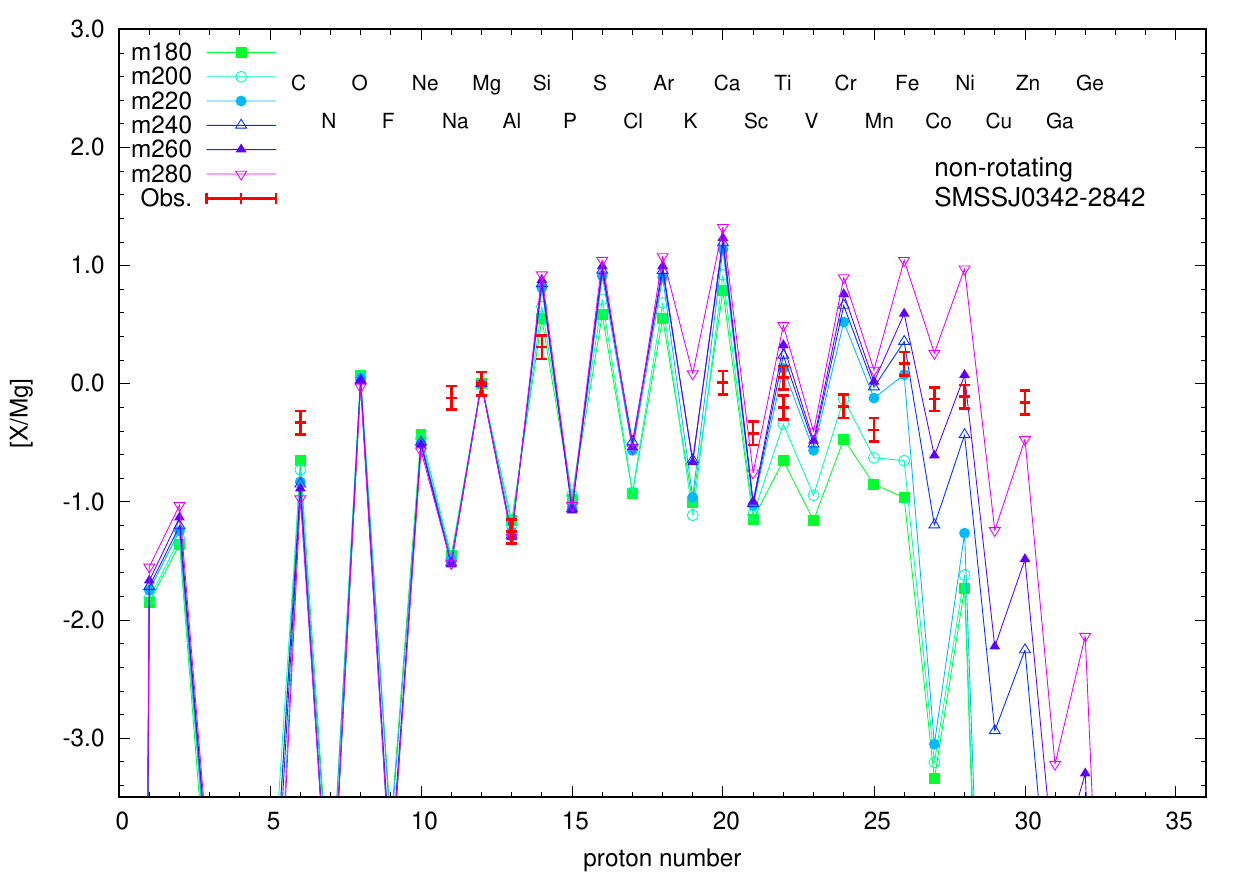}
	\end{minipage}
	
	\caption{ \footnotesize{The same as Fig.\ref{fig-stars-PISN1} but for MP stars of \#55--60.}}
\end{figure}

\begin{figure}[tbp]
	\begin{minipage}{0.5\textwidth}
		\includegraphics[width=\textwidth]{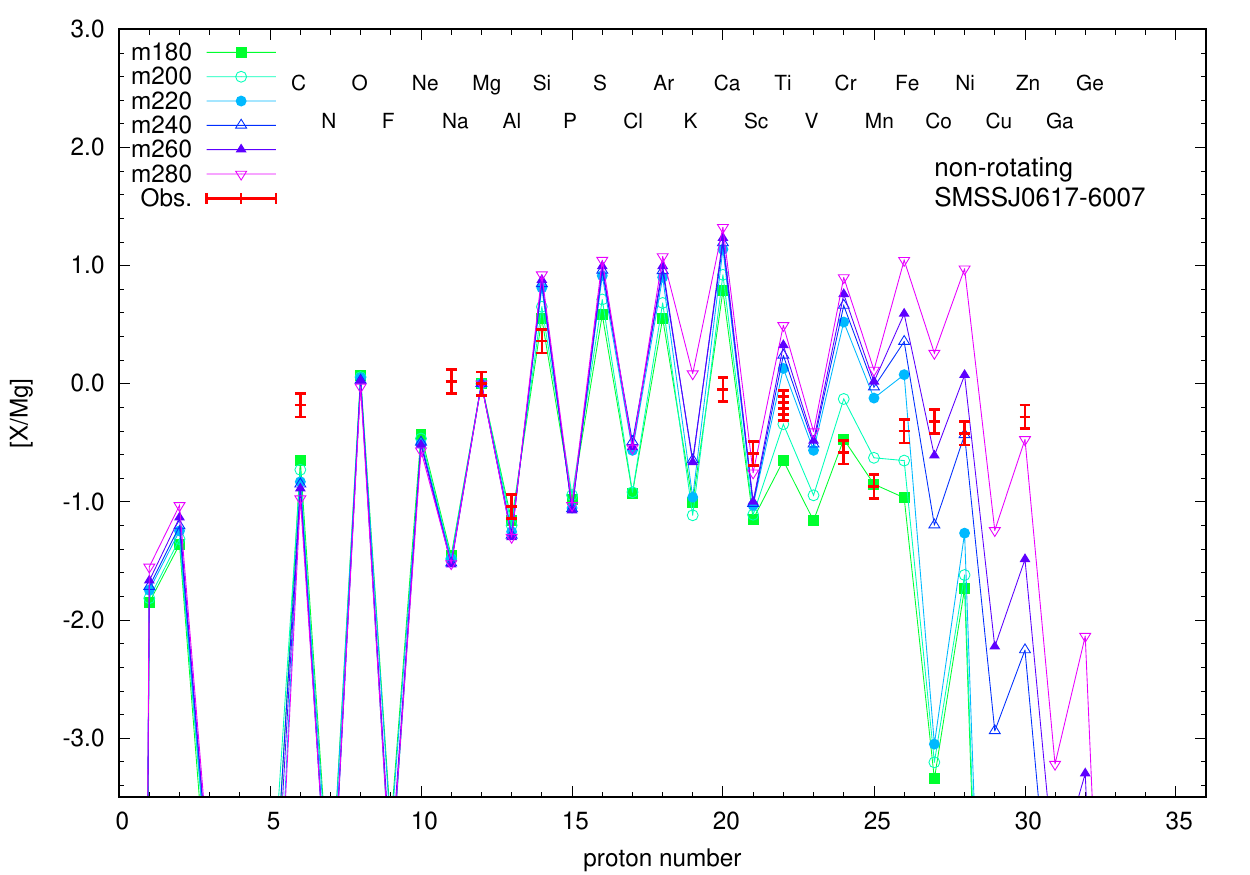}
	\end{minipage}
	\begin{minipage}{0.5\textwidth}
		\includegraphics[width=\textwidth]{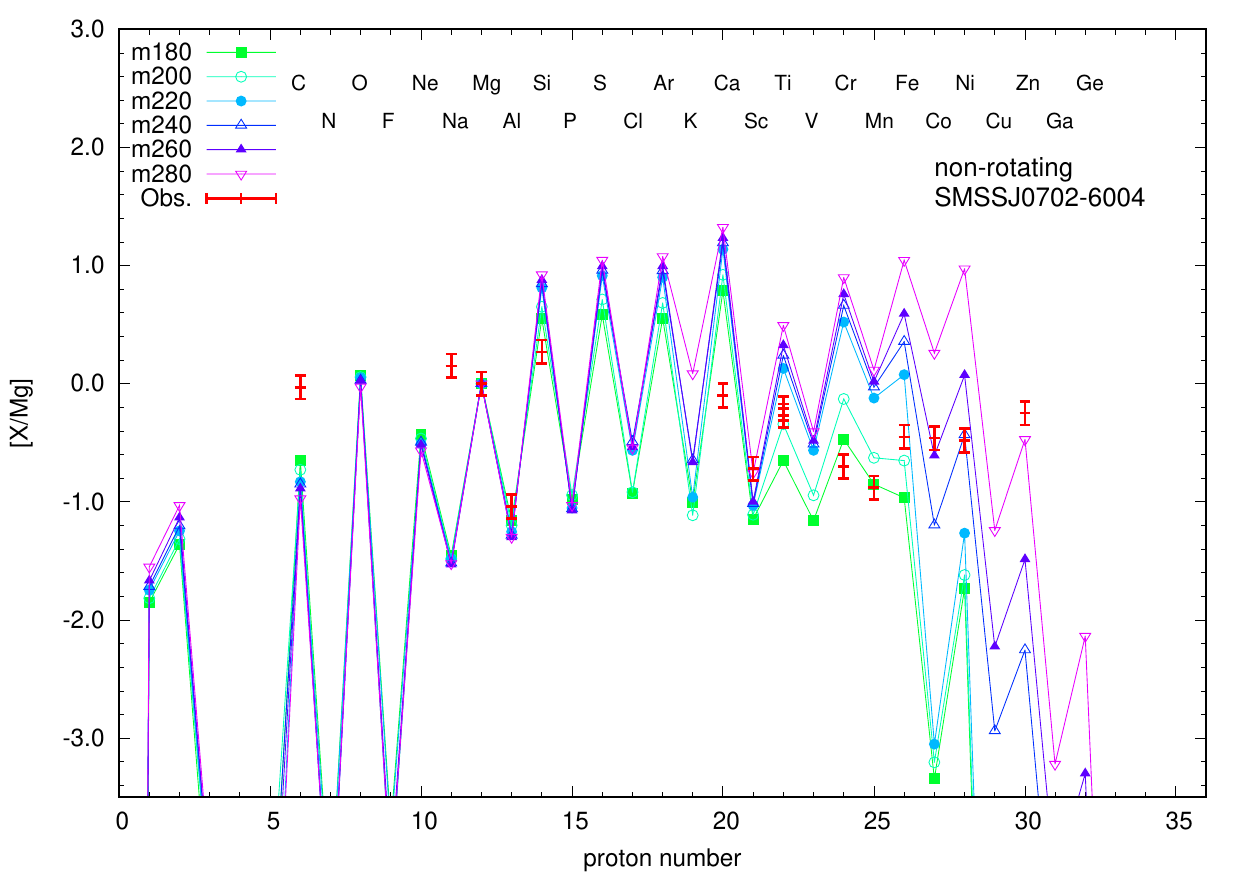}
	\end{minipage}

	\begin{minipage}{0.5\textwidth}
		\includegraphics[width=\textwidth]{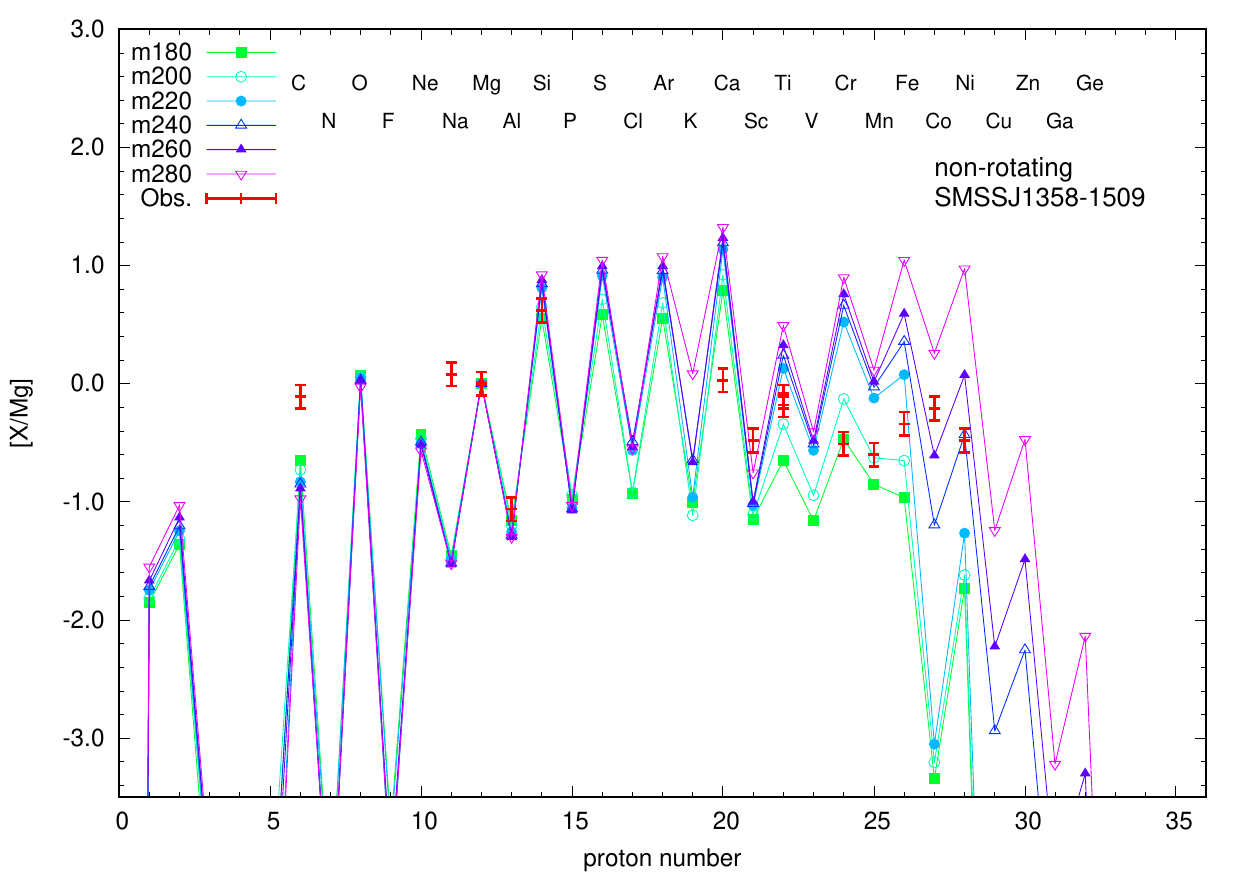}
	\end{minipage}
	\begin{minipage}{0.5\textwidth}
		\includegraphics[width=\textwidth]{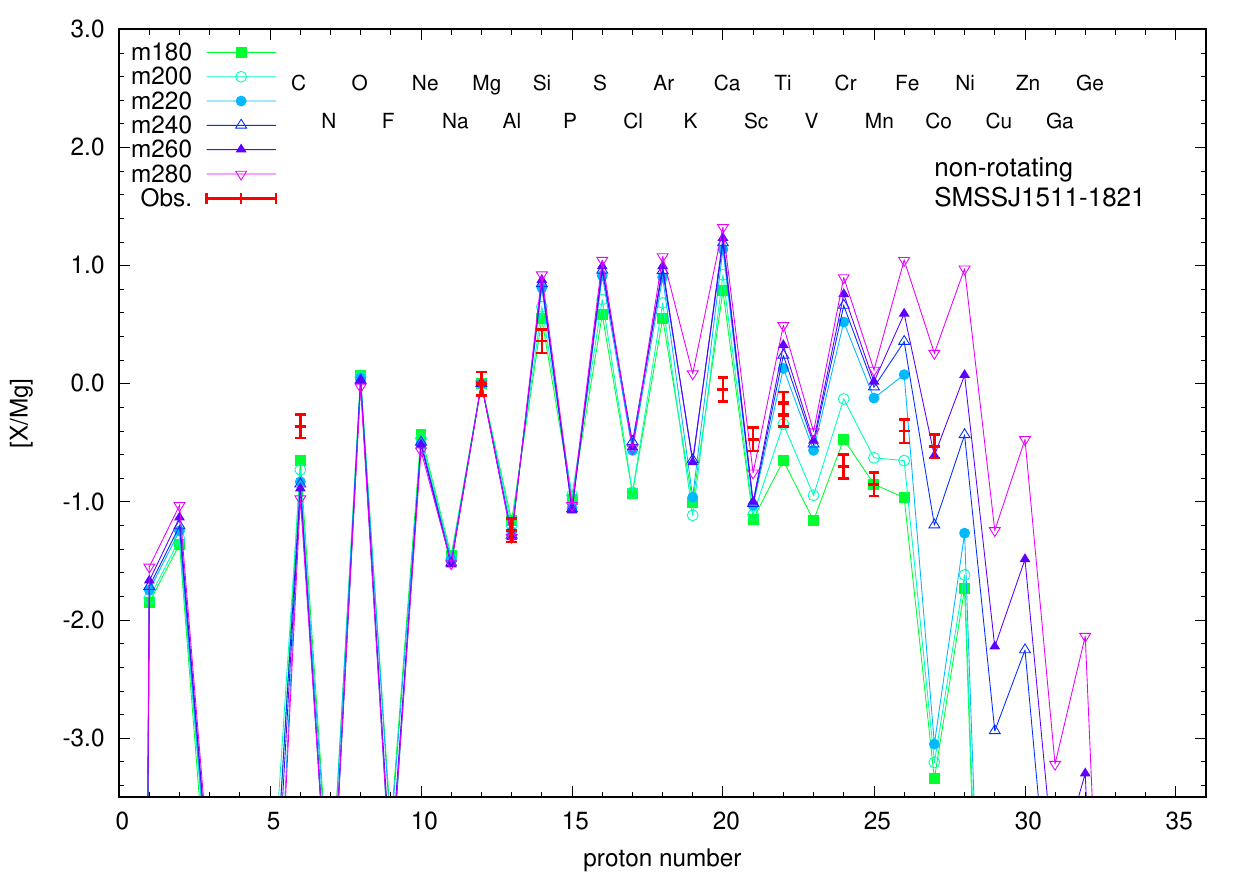}
	\end{minipage}

	\begin{minipage}{0.5\textwidth}
		\includegraphics[width=\textwidth]{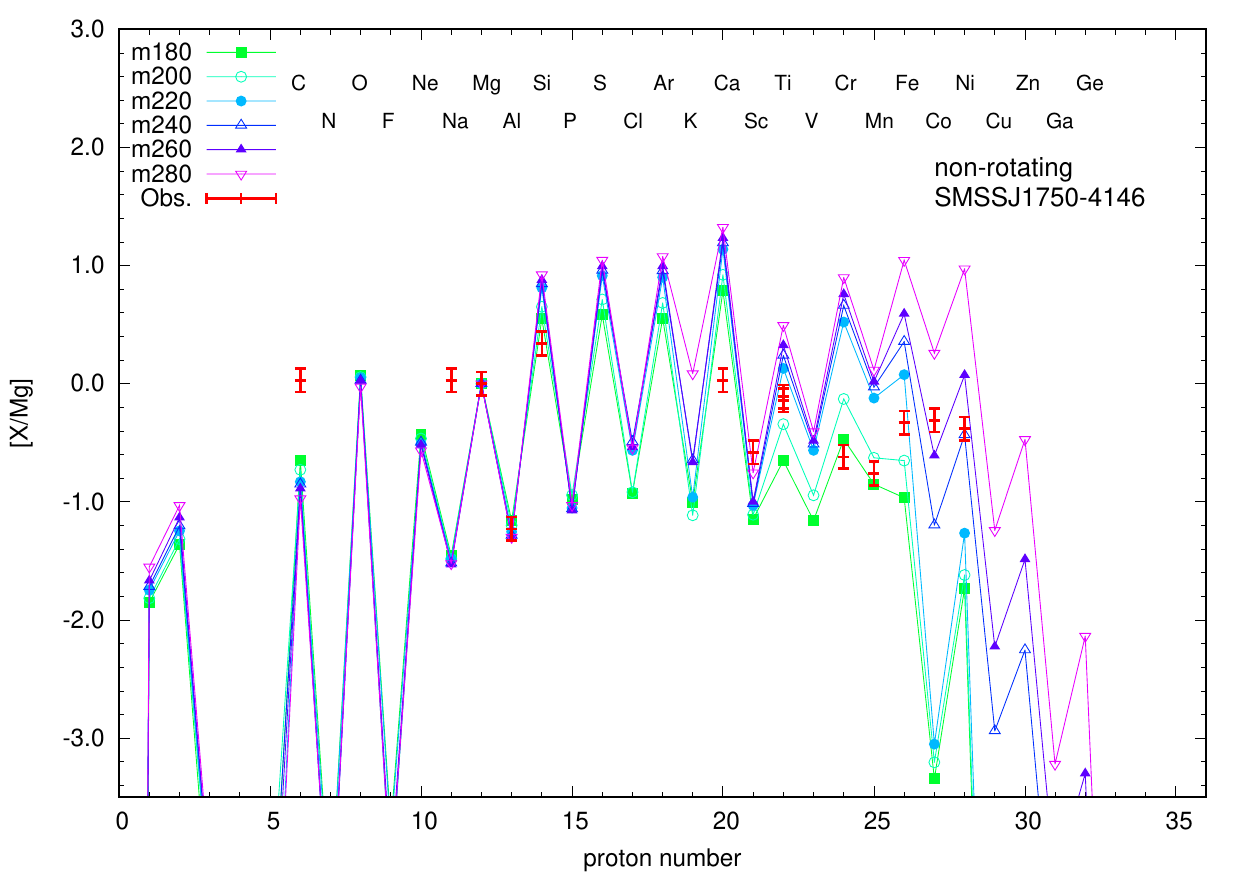}
	\end{minipage}
	\begin{minipage}{0.5\textwidth}
		\includegraphics[width=\textwidth]{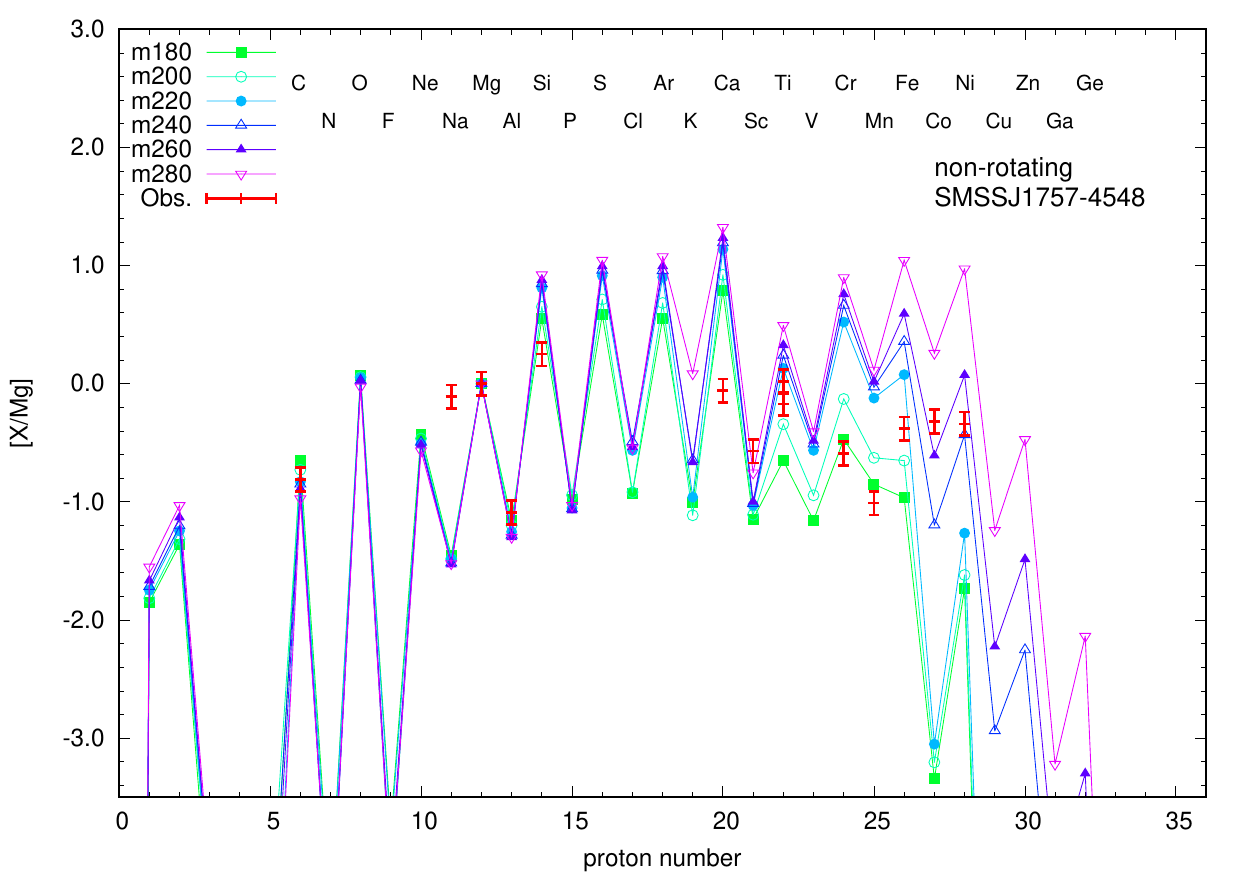}
	\end{minipage}
	
	\caption{ \footnotesize{The same as Fig.\ref{fig-stars-PISN1} but for MP stars of \#61--66.}}
\end{figure}

\begin{figure}[tbp]
	\begin{minipage}{0.5\textwidth}
		\includegraphics[width=\textwidth]{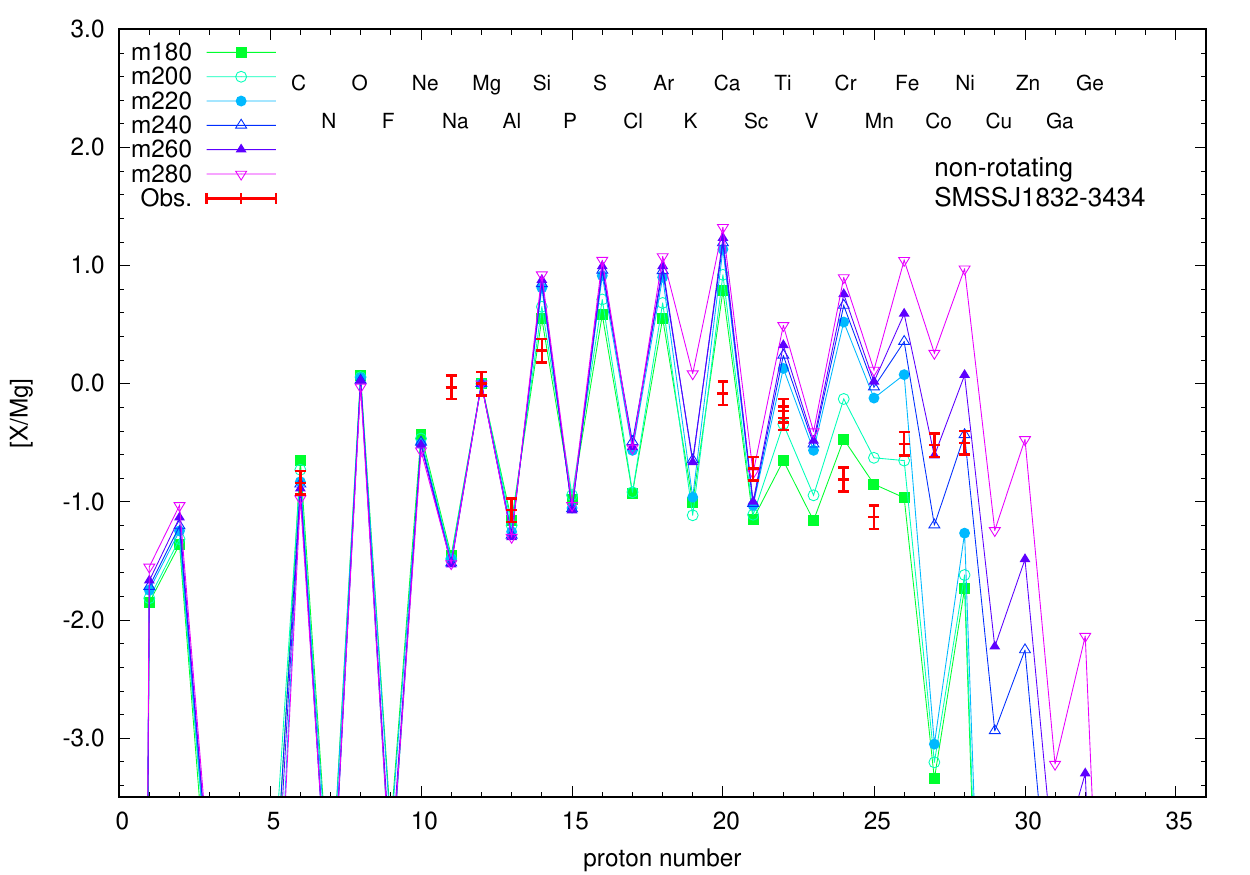}
	\end{minipage}
	\begin{minipage}{0.5\textwidth}
		\includegraphics[width=\textwidth]{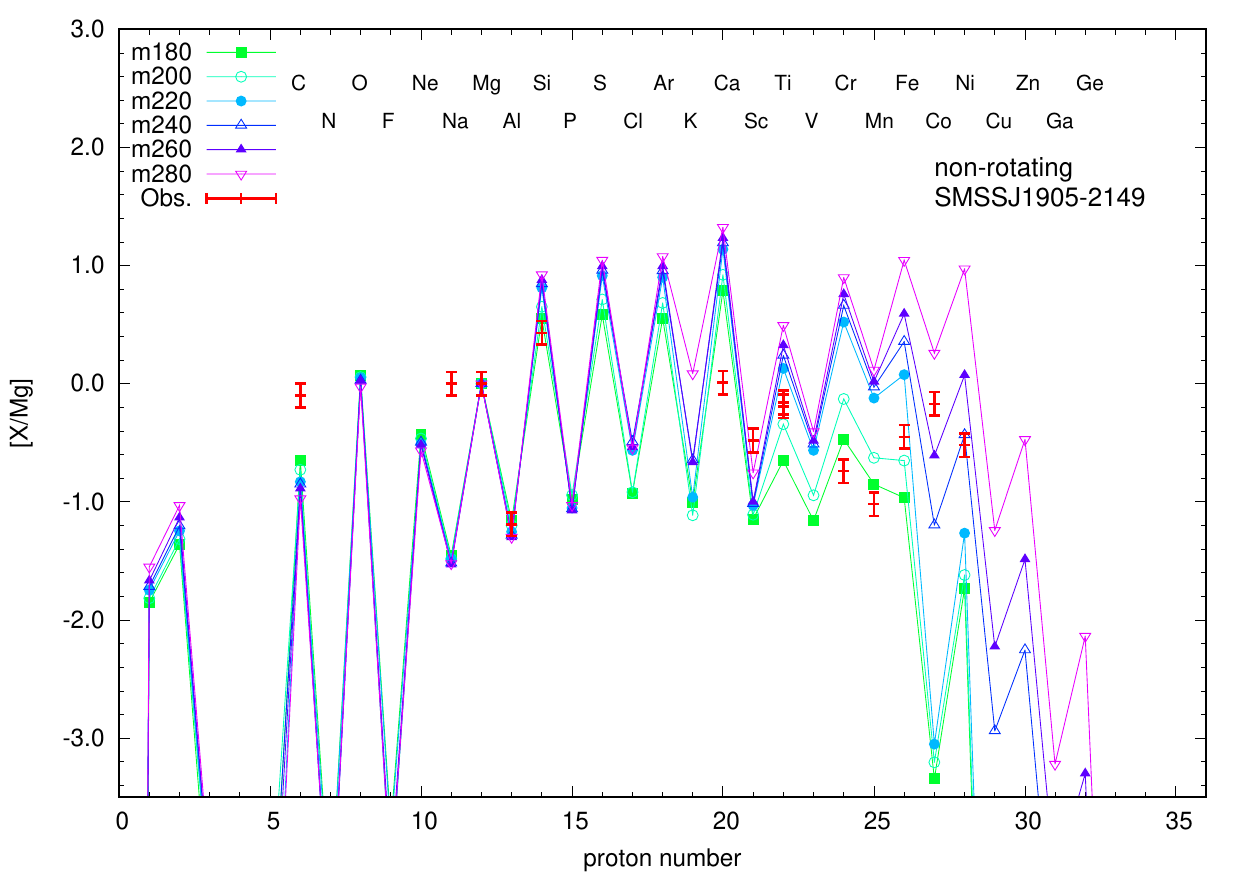}
	\end{minipage}

	\begin{minipage}{0.5\textwidth}
		\includegraphics[width=\textwidth]{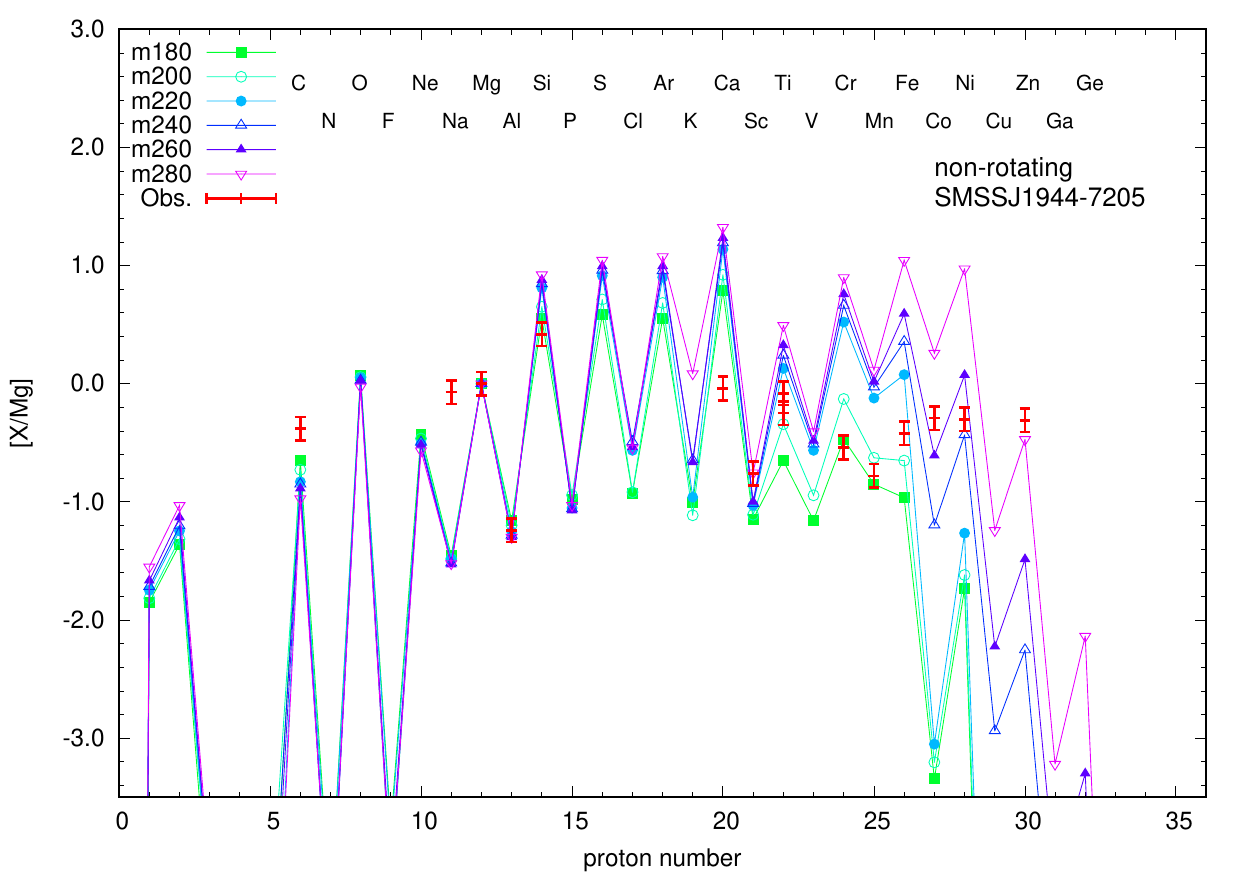}
	\end{minipage}
	\begin{minipage}{0.5\textwidth}
		\includegraphics[width=\textwidth]{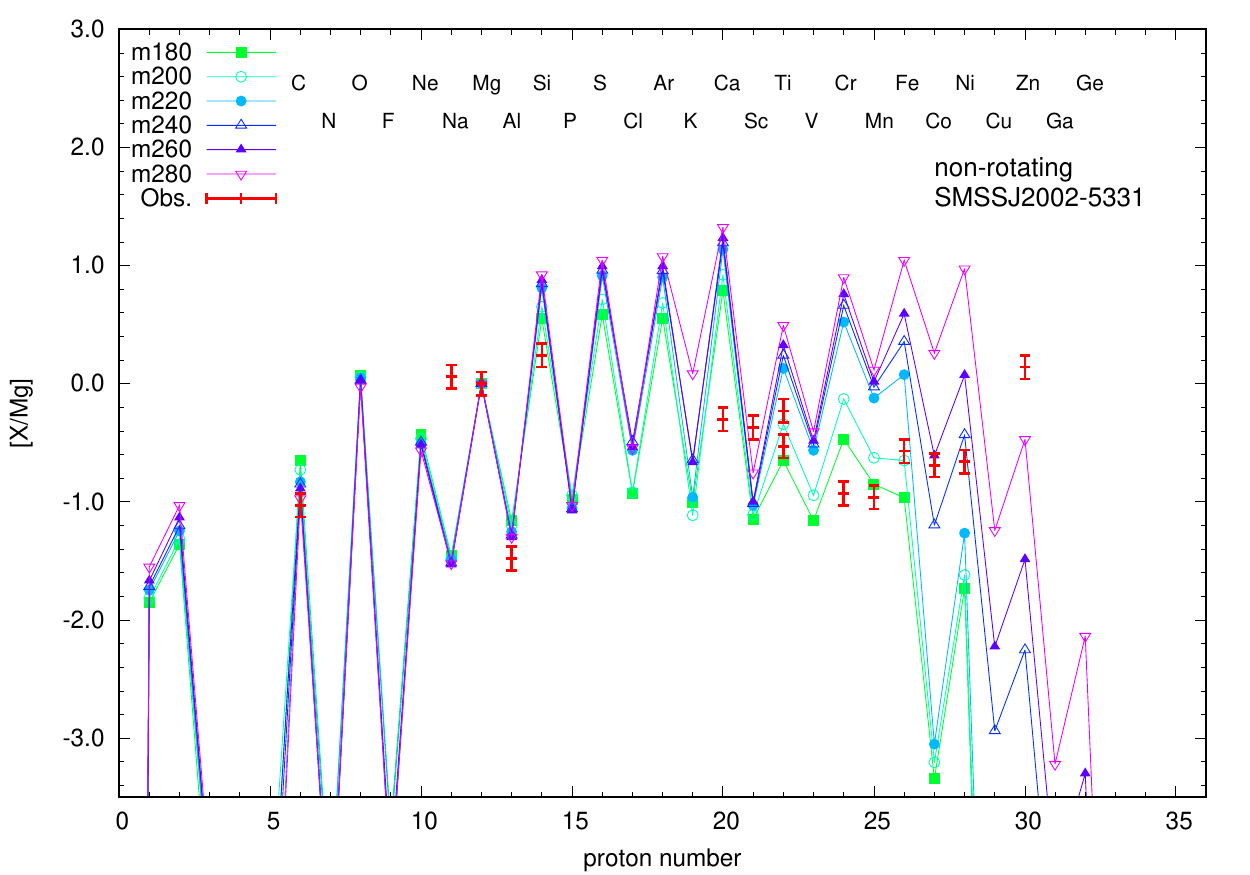}
	\end{minipage}

	\begin{minipage}{0.5\textwidth}
		\includegraphics[width=\textwidth]{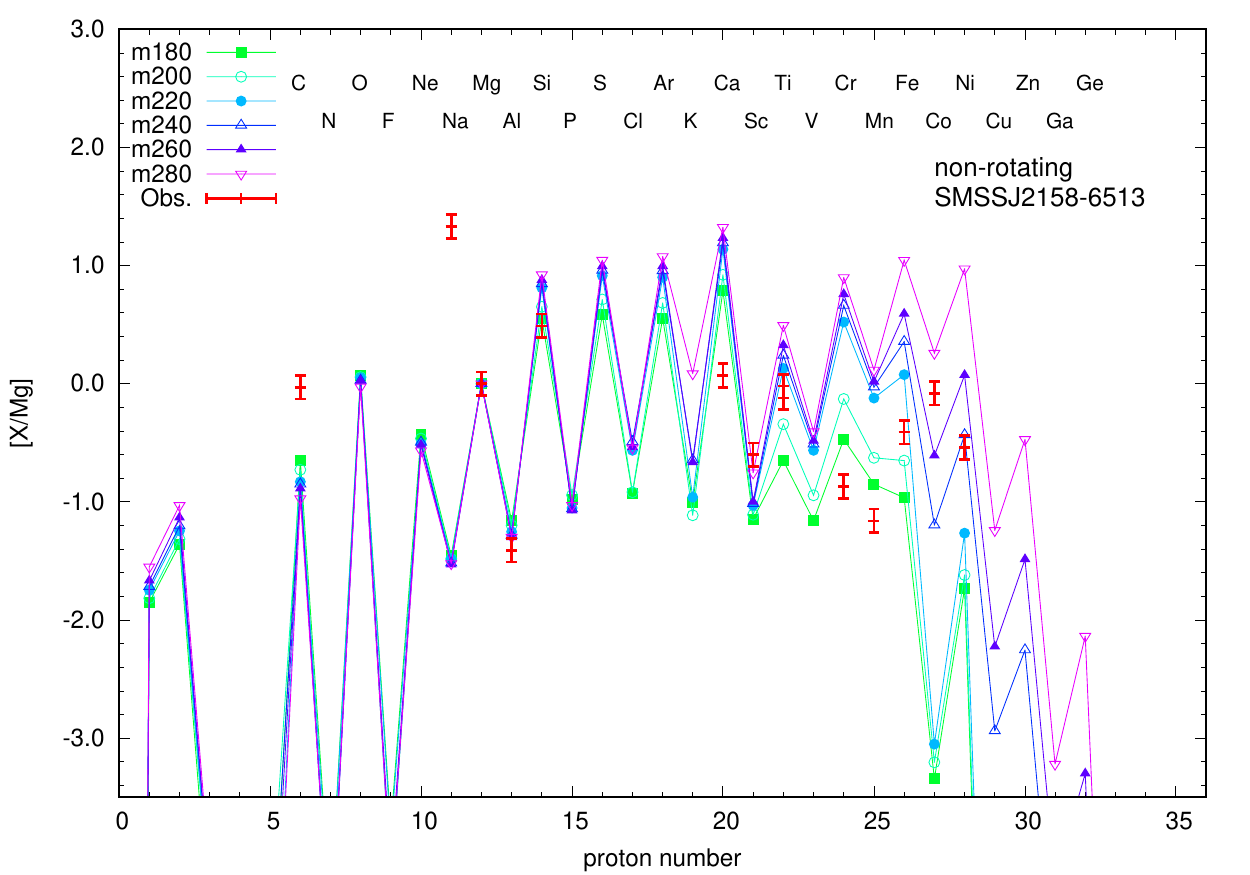}
	\end{minipage}
	\begin{minipage}{0.5\textwidth}
		\includegraphics[width=\textwidth]{stars_SDSSJ0018-0939_o00.pdf}
	\end{minipage}
	
	\caption{ \footnotesize{The same as Fig.\ref{fig-stars-PISN1} but for MP stars of \#67--72.}}
\end{figure}

\end{document}